\documentclass[onecolumn,nofootinbib]{revtex4-2}
\usepackage{tikz}
\usetikzlibrary{decorations,decorations.pathmorphing,decorations.markings,calc,snakes,positioning,fixedpointarithmetic,shapes,shapes.geometric,patterns}

\usepackage{etoolbox}
\AtBeginEnvironment{tikzpicture}{\catcode`$=3 } % $ % was important for auctex preview??

\tikzset{alignmid/.style={baseline={([yshift=-.5ex]current bounding box.center)}}} % adjust pictures vertically
\tikzset{every picture/.append style=alignmid}

\tikzset{
bottomzigzag/.style={postaction={draw,decorate, decoration={zigzag,amplitude=1pt,segment length=3pt,raise=1pt}}},
zigzag/.style={draw,decorate, decoration={zigzag,amplitude=1pt,segment length=3pt}},
rc/.style=rounded corners,
}

\tikzset{
    -|/.style={to path={-| (\tikztotarget)}},
    |-/.style={to path={|- (\tikztotarget)}},
}

\usepackage{ifthen}
\usepackage{xparse,pgffor}

\tikzset{
mark/.code={
\tikzset{postaction={/network/mark/.cd,#1,/tikz/.cd,decorate,decoration={name=markings,mark=at position \netmarkpos with{%+\netmarkposoff} with{
\begin{scope}[netmarktrafo]
\netmarkcode
\end{scope}
}}}}
\def\netmarkpos{0.5}%\pgfdecoratedpathlength}
},
}

\def\netmarkpos{0.5}%\pgfdecoratedpathlength}
\def\netmarkposoff{0cm}
\def\netmarkcode{}

\tikzset{
netmarktrafo/.style={},
netmarkstyle/.style={solid,semithick,sharp corners},
}

\pgfkeys{
/network/mark/.cd,
sty/.code={
\tikzset{netmarkstyle/.style={#1}}
},
asty/.code={
\tikzset{netmarkstyle/.append style={#1}}
},
f/.style={asty=fill},
p/.code={ % relative positioning on decorated line
\def\netmarkpos{#1}%\pgfdecoratedpathlength}
},
poff/.code={ % absolute shift on decorated line in cm
\def\netmarkposoff{#1cm}
},
e/.code={ % mark at the End of line
\def\netmarkpos{\pgfdecoratedpathlength-0.005cm-\netmarkposoff}

\tikzset{netmarktrafo/.append style={shift={(-\netmarkwidth,0)}}}
},
s/.code={ % mark at the Start of line
\def\netmarkpos{0.005cm+\netmarkposoff}

\tikzset{netmarktrafo/.append style={shift={(\netmarkwidth,0)},xscale=-1,yscale=-1}}
},
a/.code={ % mark After end of line
\def\netmarkpos{\pgfdecoratedpathlength-0.005cm}

\tikzset{netmarktrafo/.append style={xscale=-1,shift={(-\netmarkwidth,0)}}}
},
b/.code={ % mark Before start of line
\def\netmarkpos{0.005cm}

\tikzset{netmarktrafo/.append style={xscale=-1,shift={(\netmarkwidth,0),yscale=-1}}}
},
-/.code={ % flip horizontally (mark backwards)
\tikzset{netmarktrafo/.append style={xscale=-1}}
},
r/.code={ % flip vertically (mark on the Right instead of left)
\tikzset{netmarktrafo/.append style={yscale=-1}}
},
sideoff/.code={ % shift sideways
\tikzset{netmarktrafo/.append style={shift={(0,#1)}}}
},
slab/.code={ % label next to the line
\def\netmarkwidth{0}
\def\netmarkcode{
\node[inner sep=0.04cm,netmarkstyle,draw=none] (mylabelwidthtest) at (0,0){\phantom{#1}};
\path let \p1=(mylabelwidthtest.north east), \p2=(mylabelwidthtest.south east), \n1 = {max(abs(\y1),abs(\y2))} in node[inner sep=0.04cm,netmarkstyle] at (0,\n1) {#1};
}
},
lab/.code={ % label on the line
\def\netmarkwidth{0}
\def\netmarkcode{
\node[inner sep=0.04cm,anchor=\netmarkanchor] (mylabelwidthtest) at (0,0) {\phantom{#1}};
\draw[white] (mylabelwidthtest.\pgfdecoratedangle)--(mylabelwidthtest.\pgfdecoratedangle+180);
\node[inner sep=0.04cm,anchor=\netmarkanchor,netmarkstyle] at (0,0) {#1};
}
},
rotlab/.code={ % label on the line rotated with the line
\def\netmarkwidth{0}
\def\netmarkcode{
\node[inner sep=0.04cm,fill=white,transform shape,rotate=90,anchor=\netmarkrotanchor,netmarkstyle] (mydecorationnodename) at (0,0) {#1};
}
},
mrotlab/.style={ % small math label on the line rotated with the line
rotlab={$\scriptstyle{#1}$}
},
arr/.code={ % arrow
\def\netmarkwidth{0.04}
\def\netmarkcode{\draw[netmarkstyle] (-0.04,0.08)--(0.04,0)--(-0.04,-0.08);}
},
arrbar/.code={ % arrow
\def\netmarkwidth{0.08}
\def\netmarkcode{\draw[netmarkstyle] (-0.08,0.08)--(0,0)--(-0.08,-0.08) (0.04,0.08)--(0.04,-0.08);}
},
rarr/.code={ % round arrow
\def\netmarkwidth{0.04}
\def\netmarkcode{\draw[netmarkstyle] (-0.04,-0.08)arc(90-180:90:0.08);}
},
dot/.code={ % dot
\def\netmarkwidth{0.08}
\def\netmarkcode{\draw[netmarkstyle] (0,0)circle(0.08);}
},
flag/.code={ % flag to the left
\def\netmarkwidth{0.06}
\def\netmarkcode{\draw[netmarkstyle] (-0.06,0)--(0,0.09)--(0.06,0)--cycle;}
},
darr/.code={ % "double arrow": half-arrow on the left
\def\netmarkwidth{0.08}
\def\netmarkcode{\draw[netmarkstyle] (-0.04,0)--(0.04,0)--(-0.04,0.08)--cycle;}
},
circ/.code={
\def\netmarkwidth{0.1}
\def\netmarkcode{\draw[netmarkstyle] (0,0) circle (0.1);}
},
hcirc/.code={ % half-circle on the left
\def\netmarkwidth{0.1}
\def\netmarkcode{\draw[netmarkstyle] (-0.1,0) arc (180:0:0.1);}
},
diamond/.code={
\def\netmarkwidth{0.1}
\def\netmarkcode{\draw[netmarkstyle] (-0.1,0)--(0,-0.1)--(0.1,0)--(0,0.1)--cycle;}
},
bar/.code={ % tick, short line perpendicular to line
\def\netmarkwidth{0.05}
\def\netmarkcode{
\draw[netmarkstyle] (0,-0.08cm-0.5*\pgflinewidth)--(0,0.08cm+0.5*\pgflinewidth);
}
},
dbar/.code={ % double tick
\def\netmarkwidth{0.13}
\def\netmarkcode{
\draw[netmarkstyle] (-0.04cm,-0.08cm-0.5*\pgflinewidth)--(-0.04cm,0.08cm+0.5*\pgflinewidth) (0.04cm,-0.08cm-0.5*\pgflinewidth)--(0.04cm,0.08cm+0.5*\pgflinewidth);
}
},
hbar/.code={ % tick, short perpendicular line towards the left
\def\netmarkwidth{0.05}
\def\netmarkcode{
\draw[netmarkstyle] (0, 0.5*\pgflinewidth)--++(0,0.12);
}
},
mill/.code={
\def\netmarkwidth{0.16}
\def\netmarkcode{
\draw[netmarkstyle] (0,-0.5*\pgflinewidth)--++(-0.08,-0.08)--++(0,0.08);
\draw[netmarkstyle] (0,0.5*\pgflinewidth)--++(0.08,0.08)--++(0,-0.08);
}
},
three dots/.code={
\def\netmarkwidth{0.2}
\def\netmarkcode{
\fill (-0.12,0) circle (0.5*0.05) (0,0) circle (0.5*0.05) (0.12,0) circle (0.5*0.05);
}
},
}

\tikzset{wid/.style={minimum width=#1cm}}
\tikzset{hei/.style={minimum height=#1cm}}

\tikzset{sx/.style={xshift=#1cm}}
\tikzset{sy/.style={yshift=#1cm}}

\tikzset{box/.style={draw,rectangle}}
\tikzset{fbox/.style={draw,rectangle, line width=1.1}}
\tikzset{roundbox/.style={draw,rectangle,rounded corners}}
\tikzset{froundbox/.style={draw,rectangle, rounded corners, line width=1.1}}
\tikzset{rounddiamond/.style={draw,diamond,rounded corners}}
\tikzset{dot/.style={draw, shape=circle, fill=black, scale=0.5}}

\tikzset{
netbox/.code={
\node[draw,netbdstyle] (\atomname) at (0,0) {#1};
\coordinate (\atomname-r) at (\atomname.east);
\coordinate (\atomname-l) at (\atomname.west);
\coordinate (\atomname-t) at (\atomname.north);
\coordinate (\atomname-b) at (\atomname.south);
\coordinate (\atomname-tr) at (\atomname.north east);
\coordinate (\atomname-br) at (\atomname.south east);
\coordinate (\atomname-tl) at (\atomname.north west);
\coordinate (\atomname-bl) at (\atomname.south west);
},
}

\tikzset{bdlw/.code={\tikzset{mybdstyle/.style={draw, line width=#1}}}}
\tikzset{bdcol/.code={\tikzset{mybdstyle/.append style={#1}}}}

\newcommand\setelements[1]{
\pgfkeys{/network/atom/.cd,#1}
}

\newcommand\atoms[2]{
\foreach \name/\keys in {#2}{
\expandafter\atom\expandafter{\keys,#1}{\name}
}
}

\newcommand\atom[2]{
\def\atomname{#2}
\tikzset{
nettrafo/.style={},
netatompos/.style={},
netdeco/.style={},
netpostdeco/.style={},
}

\pgfkeys{/network/atom/.cd,#1}

\begin{scope}[netatompos] % shift to atom position
\begin{scope}[nettrafo] % rotate, flip and scale
\netshapecoords % set the anchor coordinates
\fill[netbackstyle] \netshapepath;
\clip \netshapepath;
\tikzset{netdeco}
\draw[netbdstyle] \netshapepath;
\end{scope}
\tikzset{netpostdeco} % draw post-decorations, not rotated, flipped, or scaled
\end{scope}

}

\pgfkeys{
/network/atom/.cd,
square/.code={

\def\netshapepath{(-\tempsize,-\tempsize)rectangle (\tempsize,\tempsize)}
\def\netshapecoords{
\node[rectangle,wid=2*\tempsize,hei=2*\tempsize,inner sep=0,transform shape](\atomname)at(0,0){};
\coordinate(\atomname-c) at (0,0);
\coordinate(\atomname-r) at (\tempsize,0);
\coordinate(\atomname-l) at (-\tempsize,0);
\coordinate(\atomname-t) at (0,\tempsize);
\coordinate(\atomname-b) at (0,-\tempsize);
\coordinate(\atomname-br) at (\tempsize,-\tempsize);
\coordinate(\atomname-tr) at (\tempsize,\tempsize);
\coordinate(\atomname-bl) at (-\tempsize,-\tempsize);
\coordinate(\atomname-tl) at (-\tempsize,\tempsize);
}},
circ/.code={

\def\netshapepath{(0,0)circle(\tempsize)}
\def\netshapecoords{
\node[circle,wid=2*\tempsize,hei=2*\tempsize,inner sep=0,transform shape](\atomname)at(0,0){};
\coordinate(\atomname-c) at (0,0);
\coordinate(\atomname-r) at (\tempsize,0);
\coordinate(\atomname-l) at (-\tempsize,0);
\coordinate(\atomname-t) at (0,\tempsize);
\coordinate(\atomname-b) at (0,-\tempsize);
}},
triang/.code={

\def\netshapepath{(-30:\tempsize)--(90:\tempsize)--(-150:\tempsize)--cycle}
\def\netshapecoords{
\node[regular polygon,regular polygon sides=3,wid=2*\tempsize,inner sep=0,transform shape](\atomname)at(0,0){};
\coordinate(\atomname-c) at (0,0);
\coordinate(\atomname-cr) at (-30:\tempsize);
\coordinate(\atomname-cl) at (-150:\tempsize);
\coordinate(\atomname-ct) at (90:\tempsize);
\coordinate(\atomname-mb) at (-90:0.5*\tempsize);
\coordinate(\atomname-mr) at (30:0.5*\tempsize);
\coordinate(\atomname-ml) at (150:0.5*\tempsize);
}},
diamond/.code={

\def\netshapepath{(0,-\tempsize)--(\tempsize,0)--(0,\tempsize)--(-\tempsize,0)--cycle}
\def\netshapecoords{
\node[rotate=45,rectangle,wid=sqrt(2)*\tempsize,hei=sqrt(2)*\tempsize,inner sep=0,transform shape](\atomname)at(0,0){};
\coordinate(\atomname-c) at (0,0);
\coordinate(\atomname-r) at (\tempsize,0);
\coordinate(\atomname-l) at (-\tempsize,0);
\coordinate(\atomname-t) at (0,\tempsize);
\coordinate(\atomname-b) at (0,-\tempsize);
}},
pentagon/.code={

\def\netshapepath{(-126:\tempsize)--(-54:\tempsize)--(18:\tempsize)--(90:\tempsize)--(162:\tempsize)--cycle}
\def\netshapecoords{
\node[regular polygon,regular polygon sides=5,wid=2*\tempsize,inner sep=0,transform shape](\atomname)at(0,0){};
\coordinate(\atomname-c) at (0,0);
\coordinate (\atomname-mb)at(-90:{\tempsize*cos(36)});
\coordinate (\atomname-mbr)at(-18:{\tempsize*cos(36)});
\coordinate (\atomname-mtr)at(54:{\tempsize*cos(36)});
\coordinate (\atomname-mtl)at(126:{\tempsize*cos(36)});
\coordinate (\atomname-mbl)at(-162:{\tempsize*cos(36)});
\coordinate (\atomname-cbr)at(-54:\tempsize);
\coordinate (\atomname-cr)at(18:\tempsize);
\coordinate (\atomname-ct)at(90:\tempsize);
\coordinate (\atomname-cl)at(162:\tempsize);
\coordinate (\atomname-cbl)at(-126:\tempsize);
}},
halfcirc/.code={

\def\netshapepath{(\tempsize,0)arc(0:180:\tempsize)--++(0,-0.04)-|cycle}
\def\netshapecoords{
\node[circle,wid=2*\tempsize,hei=2*\tempsize,inner sep=0,transform shape](\atomname)at(0,0){};
\coordinate(\atomname-c) at (0,0);
\coordinate(\atomname-r) at (\tempsize,0);
\coordinate(\atomname-l) at (-\tempsize,0);
\coordinate(\atomname-t) at (0,\tempsize);
\coordinate(\atomname-b) at (0,0);
}},
void/.code={
\def\netshapepath{}
\def\netshapecoords{
\coordinate(\atomname) at (0,0);
\coordinate(\atomname-c) at (0,0);
}},
labbox/.code={
\def\netshapepath{(0,0)}
\def\netshapecoords{}
\tikzset{netpostdeco/.append style={netbox=#1}}
},
}

\pgfkeys{
/network/atom/.cd,
p/.code={\tikzset{netatompos/.append style={shift={(#1)}}}},
xscale/.code={\tikzset{nettrafo/.append style={xscale=#1}}},
yscale/.code={\tikzset{nettrafo/.append style={yscale=#1}}},
scale/.code={\tikzset{nettrafo/.append style={scale=#1}}},
rot/.code={\tikzset{nettrafo/.append style={rotate=#1}}},
hflip/.style={xscale=-1},
vflip/.style={yscale=-1},
big/.style={scale=3/2},
small/.style={scale=2/3},
huge/.style={scale=9/4},
tiny/.style={scale=4/9},
flat/.style={yscale=2/3},
fflat/.style={yscale=4/9},
}

\tikzset{
netbdstyle/.style={line width=0.15em}, % changed from pt(default)
netdecstyle/.style={},
netpostdecstyle/.style={},
netbackstyle/.style={white},
}
\pgfkeys{
/network/atom/.cd,
bdstyle/.code={\tikzset{netbdstyle/.style={#1}}},
bdastyle/.code={\tikzset{netbdstyle/.append style={#1}}},
decstyle/.code={\tikzset{netdecstyle/.style={#1}}},
decastyle/.code={\tikzset{netdecstyle/.append style={#1}}},
postdecstyle/.code={\tikzset{netpostdecstyle/.style={#1}}},
postdecastyle/.code={\tikzset{netpostdecstyle/.append style={#1}}},
backstyle/.code={\tikzset{netbackstyle/.style={#1}}},
backastyle/.code={\tikzset{netbackstyle/.append style={#1}}},
style/.style={bdstyle=#1, decstyle=#1, postdecstyle=#1},
astyle/.style={bdastyle=#1, decastyle=#1, postdecastyle=#1},
whiteblack/.style={decstyle=white,backstyle=black},
nobd/.style={bdstyle={line width=0}},
}

\tikzset{
netbscope/.code={\begin{scope}[#1]},
netescope/.code={\end{scope}},
}

\pgfkeys{
/network/atom/.cd,
bscope/.code={\tikzset{netdeco/.append style={netbscope={#1}}}},
escope/.code={\tikzset{netdeco/.append style={netescope}}},
dec/.style 2 args={bscope={#1},#2,escope}, % for applying transformation keys to certain decorations only
vbar/.style={dec={rotate=90}{hbar}},
dcross/.style={dec={rotate=45}{hbar},dec={rotate=-45}{hbar}},
cross/.style={hbar,vbar},
sect6/.style={sect=60},
sect8/.style={sect=45},
sect10/.style={sect=36},
}

\def\regdec#1{\pgfkeys{/network/atom/.cd,#1/.code={\tikzset{netdeco/.append style={net#1}}}}}
\regdec{all}\regdec{rhalf}\regdec{rquart}\regdec{brquart}\regdec{dot}\regdec{spiral}\regdec{swirl}\regdec{hstripe}\regdec{hbar}\regdec{rrey}
\pgfkeys{/network/atom/.cd,sect/.code={\tikzset{netdeco/.append style={netsect=#1}}}}

\tikzset{
netall/.code={\fill[netdecstyle] (-0.3,-0.3)rectangle (0.3,0.3);}, % fill all
netrhalf/.code={\fill[netdecstyle] (0,-0.3)rectangle (0.3,0.3);}, % right half
netrquart/.code={\fill[netdecstyle] (0.075,-0.3)rectangle (0.3,0.3);}, % right quarter
netbrquart/.code={\fill[netdecstyle] (0,0)rectangle (0.3,-0.3);}, % bottom right quarter
netsect/.code={\fill[netdecstyle] (0,0)--(0,-0.3)arc(-90:-90+#1:0.3)--cycle;}, % section of angle #1 starting from -90
netdot/.code={\fill[netdecstyle] (0,0)circle(0.07);}, % dot in the middle
netspiral/.code={\draw[netdecstyle] plot [variable=\t,domain=0:4] ({0.075*\t*cos(pi*(\t-0.5) r)},{0.075*\t*sin(pi*(\t-0.5) r)});}, % spiral
netswirl/.code={\fill[netdecstyle] plot [variable=\t,domain=0:2] ({0.15*\t*cos(pi*(\t-0.5) r)},{0.15*\t*sin(pi*(\t-0.5) r)}) arc(-90:-450:0.3)--cycle;}, % filled swirl
nethstripe/.code={\fill[netdecstyle] (-0.3,-0.05)rectangle(0.3,0.05);}, % horizontal stripe
nethbar/.code={\draw[netdecstyle] (-0.3,0)--(0.3,0);}, % horizontal line
netrrey/.code={\draw[netdecstyle] (0,0)--(0.3,0);} % line from the middle to the right
}

\pgfkeys{
/network/atom/.cd,
postbscope/.code={\tikzset{netpostdeco/.append style={netbscope={#1}}}},
postescope/.code={\tikzset{netpostdeco/.append style={netescope}}},
postdec/.style 2 args={postbscope={#1},#2,postescope}, % for applying transformation keys to certain decorations only
arc/.code={\tikzset{netpostdeco/.append style={netarc=#1}}},
lab/.code={\tikzset{netpostdeco/.append style={netlab={#1}}}},
shadecirc/.code={\tikzset{netpostdeco/.append style=netshadecirc}},
shaderect/.code={\tikzset{netpostdeco/.append style={netshaderect={#1}}}},
debug/.code={\tikzset{netpostdeco/.append style=netdebug}},
markline/.code 2 args={\tikzset{netpostdeco/.append style={netmarkline={#1}{#2}}}},
postcirc/.code={\tikzset{netpostdeco/.append style=netpostcirc}},
}

\tikzset{
netlab/.code={
\pgfkeys{/network/atom/lab/.cd,#1}
\node[netpostdecstyle] at (\ifdefined\netlabpos\netlabpos\else\netlabang:\netlabdist\fi) {\netlabwrap{\netlabtext}};
},
netarc/.code args={#1:#2:#3}{
\draw[netpostdecstyle] (#1:#3) arc (#1:#2:#3);
},
netshadecirc/.code= {
\fill[opacity=0.4,netpostdecstyle] (0,0)circle(0.4);
},
netpostcirc/.code= {
\draw[netpostdecstyle] (0,0)circle(0.15);
},
netshaderect/.code= {
\fill[rc,opacity=0.4,netpostdecstyle] ($-1*(#1)$) rectangle (#1);
},
netdebug/.code= {
\node[red] at (0,0){\atomname};
},
netmarkline/.code 2 args= {
\draw (\atomname)edge[mark={#2}]++(#1);
},
}

\def\netlabwrap#1{#1}
\pgfkeys{
/network/atom/lab/.cd,
t/.code={\def\netlabtext{#1}}, % label text
p/.code={\def\netlabpos{#1}}, % position of the label
ang/.code={\def\netlabang{#1}},
dist/.code={\def\netlabdist{#1}},
wrap/.code={\def\netlabwrap##1{#1}}, % store in doesn't work
}

\usepackage{amsmath, amssymb, mathtools, braket, tabularx}

\usepackage{diagbox}%,slashbox}
\usepackage{tikz}
\usepackage{mathtools}
\usepackage{amsthm}
\usepackage{mdframed}
\usepackage{parskip}
\theoremstyle{definition}
\newmdtheoremenv[linecolor=blue, linewidth=1pt]{mydef}{Definition}
\newmdtheoremenv[linecolor=blue, linewidth=1pt]{myprop}{Proposition}
\usetikzlibrary{matrix}

\newcommand\myparagraph[1]{\noindent\textbf{#1 ---}}

\newcommand{\fixedtilde}[2]{\mathrlap{\raisebox{-#1ex}{$\overset{\sim}{\phantom{#2}}$}}#2}
\DeclarePairedDelimiter{\floor}{\lfloor}{\rfloor}
\def\zz{\mathbb Z}
\def\rr{\mathbb R}
\def\cc{\mathbb C}
\def\mmod{\operatorname{mod}}
\def\idop{\operatorname{id}}
\def\ovl{\overline}
\def\ker{\operatorname{kernel}}
\def\img{\operatorname{image}}
\def\qinv{q^{\text{inv}}}

\def\eval{\operatorname{ev}}

\def\imag{\operatorname{Imag}}
\def\real{\operatorname{Real}}
\def\pf{\operatorname{Pf}}
\def\htild{\fixedtilde{0.4}{H}}
\def\homtild{\fixedtilde{0.8}{\operatorname{hom}}}
\def\omegatild{\fixedtilde{0.4}{\Omega}}
\def\bfpi{\boldsymbol{\pi}}
\def\circtild{\;\tilde\circ\;}

\newcommand{\mpm}[1]{\begin{pmatrix}#1\end{pmatrix}}

\usepackage[table]{xcolor}
\usepackage{colortbl}
\usepackage{graphicx}
\usepackage{bbold}
\usepackage{hyperref}
\usepackage{braket}
\usepackage[export]{adjustbox}

\makeatletter
\newcommand{\vphantomnl}[1]{%
  \vphantom{\let\label\@gobble #1}%
}
\makeatother
\ExplSyntaxOn
\NewDocumentEnvironment{multiequation}{b}
 {
  \seq_set_split:Nnn \l_tmpa_seq { \NEXT } {#1}
  \seq_map_inline:Nn \l_tmpa_seq
   {
    \begin{minipage}{\dim_eval:n {\displaywidth/\seq_count:N \l_tmpa_seq}}
    \begin{equation}
      ##1
    \end{equation}
    \end{minipage}
   }
}{}
\ExplSyntaxOff

\tikzset{
ind/.style={mark={lab=#1,a}}, % normal open index label
startind/.style={mark={lab=#1,b}}, % normal open index label
ar/.style={mark={arr,f,#1}},
/network/mark/.cd,
rlab/.style = {slab=$\scriptstyle\textcolor{red}{#1}$,sideoff=-0.06cm},
/tikz/.cd,
}

\setelements{
alg/.style={small,circ},
coalg/.style={small,circ,all},
antipode/.style={square,small,fflat},
linform/.style={square,small},
linfunc/.style={square,small},
}

\begin{document}
\title{Quadratic tensors as a unification of Clifford, Gaussian, and free-fermion physics}
\author{Andreas Bauer}
\email{andib@mit.edu}
\affiliation{\footnotesize Department of Mechanical Engineering, Massachusetts Institute of Technology, Cambridge, MA 02139, USA}
\author{Seth Lloyd}
\affiliation{\footnotesize Department of Mechanical Engineering, Massachusetts Institute of Technology, Cambridge, MA 02139, USA}

\begin{abstract}
Certain families of quantum mechanical models can be described and solved efficiently on a classical computer, including qubit or qudit Clifford circuits and stabilizer codes, free-boson or free-fermion models, and certain rotor and GKP codes.
We show that all of these families can be described as instances of the same algebraic structure, namely \emph{quadratic functions} over abelian groups, or more generally over (super) Hopf algebras.
Different kinds of degrees of freedom correspond to different ``elementary'' abelian groups or Hopf algebras:
$\zz_2$ for qubits, $\zz_d$ for qudits, $\rr$ for continuous variables, both $\zz$ and $\rr/\zz$ for rotors, and a super Hopf algebra $\mathcal F$ for fermionic modes.
Objects such as states, operators, superoperators, or projection-operator valued measures, etc, are tensors.
For the solvable models above, these tensors are \emph{quadratic tensors} based on quadratic functions.
Quadratic tensors with $n$ degrees of freedom are fully specified by only $O(n^2)$ coefficients.
Tensor networks of quadratic tensors can be contracted efficiently on the level of these coefficients, using an operation reminiscent of the \emph{Schur complement}.
Our formalism naturally includes models with mixed degrees of freedom, such as qudits of different dimensions.
We also use quadratic functions to define generalized stabilizer codes and Clifford gates for arbitrary abelian groups.
Finally, we give a generalization from quadratic (or 2nd order) to $i$th order tensors, which are specified by $O(n^i)$ coefficients but cannot be contracted efficiently in general.
\end{abstract}
\maketitle
\tableofcontents

\section{Introduction}
\myparagraph{Motivation}
General quantum mechanical models with many degrees of freedom cannot in general be simulated efficiently on classical machines:
Storing the coefficients of a many-body quantum state requires an exponential amount of memory in the number of degrees of freedom, just like storing a classical many-body probability distribution.
We can map a quantum state onto a probability distribution in order to sample from it, but in general we end up with some negative probabilities leading to a potential sign problem.
More heuristic methods, like approximate tensor-network contraction, are successful in many cases, but in general many (especially dynamical) models remain hard to solve.

For this reason, much of physics revolves around certain families of models that can be solved exactly and efficiently:
In both condensed-matter and high-energy physics, ``free'', ``single-particle'', ``quadratic'', or ``non-interacting'' models of fermions or bosons are of central importance.
Such models are described by a Hamiltonian that is quadratic in the annihilation and creation operators.
Often, higher-order terms are treated as perturbations on top of a quadratic theory.
In AMO physics and quantum information, free theories are also known as ``(fermionic) linear optics'' or ``Gaussian'' models, states, or circuits.
Circuits of Gaussian operations can be simulated efficiently on a classical computer~\cite{Bartlett2001,Weedbrook2011,Adesso2014}.

In quantum information, Clifford circuits~\cite{Gottesman1997a} over qubits or qudits~\cite{Gottesman1998a,Beaudrap2011,Tolar2018} play a central role.
Often, general quantum algorithms are compiled into circuits of Clifford unitaries, together with some ``magic'' non-Clifford gates such as the $T=\operatorname{diag}(1,e^{i\pi/4})$ gate.
Closely related to Clifford circuits are stabilizer codes~\cite{Gottesman1997}, which are the dominant language used in quantum error correction.
Clifford circuits can be efficiently emulated by tracking the stabilizer group of the current state, using the Gottesman-Knill theorem~\cite{Gottesman1998,Aaronson2004,Gidney2021}.

\myparagraph{Results}
In this work, we show that all families of models above are instances of the same algebraic structure, namely quadratic functions over abelian groups or more generally (co)commutative Hopf algebras.
The most general way in which quadratic functions can be used to represent exactly solvable models is as quadratic tensor networks:
Almost all objects in quantum mechanics, such as unitary evolution operators, pure states, density matrices, observables, quantum channels, POVMs, or measurement instruments, can be represented as tensors.
Almost all computations such as applying an operator to a state, applying a superoperator to a partial density matrix, or tracing over degrees of freedom, can be represented as tensor-network contractions.
In other words, simulating quantum many-body physics can be phrased as tensor-network contraction.
For exactly solvable models, all tensors are \emph{quadratic tensors} which can be stored and contracted efficiently, see Section\ref{sec:clifford_tensors}.

Each index $i$ of an $n$-index quadratic tensor takes values in an abelian group $G_i$, such that the associated vector space is $\cc^{G_i}$.
A tensor $T$ is an array of numbers, one for each index configuration $g=(g_0,\ldots,g_{n-1})\in G\coloneqq\bigtimes_i G_i$, that is, $g_i\in G_i$.
Alternatively, we can view $T: G\rightarrow \cc$ as a function that assigns the tensor entry $T(g)\in\cc$ to each index configuration $g$.
Roughly, a quadratic tensor is one where $T$ is a \emph{quadratic function}, that is,
\begin{equation}
b(g,h)\coloneqq T(g+h)T(g)^{-1}T(h)^{-1}
\end{equation}
is a bilinear form $b:G\times G\rightarrow \cc^\times$, where $\cc^\times$ is the multiplicative group of non-zero complex numbers.
Quadratic functions are a natural generalization of second-order polynomials $\rr\rightarrow\rr$ to arbitrary groups $G$.
In the literature, $T$ is also called a \emph{quadratic refinement} of $b$.

There can be different ways to decompose $G$ into factors $G_i$ corresponding to tensor indices, and it makes sense to decompose it into the smallest possible factors each of which corresponds to a single ``degree of freedom''.
For physically relevant abelian groups, these smallest ``elementary'' factors are given by $G_i\in\{\zz_k,\zz,\rr/\zz,\rr\}$.
$G_i=\zz_2$ corresponds to the computational basis vectors $\ket0$ and $\ket1$ of a qubit, and $\zz_k$ corresponds to a $k$-dimensional qudit.
$G_i=\rr$ corresponds to the position basis states $\ket x$ of a continuous variable/bosonic mode.
$G_i=\rr/\zz\simeq U(1)$ corresponds to the position/angle eigenstates $\ket\phi$ of a rotor, and $G_i=\zz$ belongs to the angular momentum eigenstates $\ket k$ of a rotor.

If $G$ consists of $n$ elementary factors $G_i$, then a bilinear form $b$ can be represented by an $n\times n$ coefficient matrix, and its quadratic refinement $T$ is fully specified by an additional length-$n$ vector.
That is, a quadratic tensor is specified by $O(n^2)$ coefficients, and can thus be efficiently represented in computer memory.
We also show that the contraction of a tensor network of quadratic tensors is again a quadratic tensor, and that its $O(n^2)$ representing coefficients can be computed efficiently in polynomial time.
Central to the contraction is a generalization of the \emph{Schur complement} of a matrix from $\rr$ or $\cc$ coefficients to arbitrary abelian-group coefficients.

In Section~\ref{sec:stabilizer_codes}, we also use quadratic functions to give natural definitions of stabilizer codes and Clifford operations for arbitrary abelian groups.
We show that objects such as code states, code space projectors, encoding maps, or Clifford unitaries are quadratic tensors.
We give concrete procedures to compute the $O(n^2)$ representing coefficients for these quadratic tensors.
Based on this, we give a long list of examples of quadratic tensors in Section~\ref{sec:quadratic_examples}.

In order to also represent free-fermion models as quadratic tensor networks (see Section~\ref{sec:free_fermions}), we need to generalize to \emph{fermionic tensor networks}~\cite{Kraus2009,Barthel2009,Corboz2009,tensor_type,Mortier2024}.
A fermionic tensor is one with additional \emph{odd-parity} configurations that experience $-1$ factors during an index transposition, reflecting the anti-commutativity of fermionic creation and annihilation operators.
We also need to (super)linearize the algebraic structure of abelian groups and quadratic functions.
That is, we replace sets, functions, and cartesian products by (super) vector spaces, (super) linear maps, and (super) tensor products.
The most systematic way to do this is to express the axioms of abelian groups and quadratic functions purely diagrammatically.
The result of super-linearization is given by quadratic functions over \emph{commutative and cocommutative (super) Hopf algebras}.
After generalizing to Hopf algebras we find one more elementary factor, namely a 2-dimensional super Hopf algebra $\mathcal F$ corresponding to free-fermion modes.
Quadratic functions in the general case (including $\mathcal F$ factors) can still be represented by $O(n^2)$ many coefficients.

Our formalism allows us to arbitrarily mix different elementary degrees of freedom.
That is, a single quadratic tensor can have indices corresponding to qudits of different dimensions, rotors, continuous variables, and free-fermion modes.
This way, we can represent objects such as the GKP encoding map from a qudit into a continuous variable, or interactions between a qubit and a four-dimensional qudit.
We note, however, that some pairs of degrees of freedom can only interact trivially, such as qudits with coprime dimensions, or free-fermion modes with any non-fermionic degrees of freedom.

Finally, we discuss the generalization from quadratic to higher-order tensors in Section~\ref{sec:ilinear}.
$i$th order tensors are based on $i$th order functions, which are a generalization of $i$th order polynomials to arbitrary groups $G$.
$i$th order tensors can be efficiently represented by $O(n^i)$ many coefficients.
However, tensor networks of $i$th order tensors can no longer be contracted efficiently.
An example of $i$th order tensors are diagonal operators in the $i$th level of the Clifford hierarchy, though we do not know if a similar correspondence holds for the non-diagonal operators.

\myparagraph{Literature review}
Due to their ubiquity in physics, results on ``free'' of ``Clifford'' models are scattered throughout many decades of literature in disciplines ranging from quantum algebra, high-energy physics, and condensed matter physics to AMO physics, quantum information, and computer science.
Below is our best effort to find references related to our approach across the different disciplines, though it is difficult to be exhaustive.

It is common knowledge that there are analogies between free/Gaussian/quadratic/non-interacting/linear-optics models of bosons and fermions, and Clifford/stabilizer models of qubits and qudits.
For example, the efficient simulability of Gaussian circuits~\cite{Bartlett2001} is also known as ``continuous-variable Gottesman-Knill''.
Further, Clifford circuits and free-fermion circuits have similar notions of ``magic'', ``stabilizer rank'', or Wigner-function positivity~\cite{Gross2006,Cudby2023,Bu2023,Bu2023a}.
The perhaps most quantitative connection in the literature between discrete-variable Cliffords and continuous-variable Gaussian operations is via Gottesman-Kitaev-Preskill (GKP) codes~\cite{Gottesman2000,Conrad2024}.
These are codes that encode a logical qubit (or qudit) into one (or many) continuous variables/bosonic modes.
After applying the encoding map of a GKP code, Clifford operations on qudits become Gaussian operations on continuous variables~\cite{Singh2025,Hahn2024}.
In this work, we find an even more direct connection between free-boson, free-fermion, and Clifford/stabilizer models, by showing that they are instances of the exact same algebraic structure, namely quadratic functions over (co)commutative Hopf algebras.

It is common in the literature to use ``quadratic forms'' or second-degree polynomials to describe Gaussian or stabilizer models.
The perhaps simplest manifestation of this is the bosonic Gaussian (squeezed-state) wave function, which is the exponential of a second-degree polynomial. 
There are also many places in the literature where second-degree polynomials have been used to describe stabilizer models for qubits~\cite{Dehaene2003,Gross2007,Beaudrap2008,Bravyi2016,Beaudrap2021,Yashin2025} and for qudits~\cite{Hostens2004,Nest2012,Bermejo2012,Bermejo2014,Huang2018,Moses2024}.
In particular, the \emph{standard form} of a qubit stabilizer state in terms of an affine subspace and a quadratic form is a special case of a quadratic tensor.
However, in most cases the definitions used for second-degree polynomials are more ad hoc than our algebraic notion of a quadratic function.
For example, the standard form in Appendix B of Ref.~\cite{Bravyi2016} has computational-basis entries given by $e^{2\pi i\frac18 q(\vec x)}$, where $q$ is a second-degree polynomial,
\begin{equation}
q:\zz_2^k\rightarrow \zz_8\;,\qquad
q(\vec x) = \vec x^T J \vec x + D \vec x + Q\;.
\end{equation}
Here, $J$ is restricted to values $\{0,4\}\subset \zz_8$, $D$ is restricted to $\{0,2,4,6\}\subset \zz_8$, and $Q\in\zz_8$.
In our approach, the choice of $\zz_8$ and the different restrictions for $J$ and $D$ do not need to be postulated; they follow directly from the basic algebraic definition of a quadratic function.
This algebraic definition also naturally explains the differences between odd and even-dimensional qudits.
A few works that adopt a notion of quadratic functions very similar to ours are Refs.~\cite{Nest2012,Bermejo2012,Bermejo2014,Moses2024}, but this notion appears to not have caught on in the literature so far.
Quadratic functions between abelian groups are also used in other contexts in mathematical physics.
For example, quadratic refinements have been used to study spin structures in fermionic models~\cite{Gaiotto2015}, or to define actions in topological gauge theories~\cite{Kapustin2013}.
Further, quadratic forms (which are quadratic refinements $q$ with the extra condition that $q(ng)=n^2q(g)$) describe abelian anyon models~\cite{Wang2020}.

We are not aware of any place in the literature that uses our simple definitions of generalized Clifford gates and generalized stabilizer codes using quadratic functions from Section~\ref{sec:stabilizer_codes}.
The \emph{normalizer gates} introduced in Ref.~\cite{Nest2012} (see also Refs.~\cite{Bermejo2012,Bermejo2014,Moses2024}) consist of three families of gates:
The Fourier transform, automorphisms of $G$, and quadratic phase gates.
Normalizer gates generate the same set of generalized Clifford gates.
Instead of three separate families, we use quadratic functions to define a single unified family containing all generalized Clifford gates, and we show that they can be efficiently composed in terms of the underlying coefficients.

An efficiently simulable class of planar qubit quantum circuits are so-called \emph{matchgate circuits}~\cite{Valiant2002,Cai2007,Bravyi2008,Jahn2017}.
Matchgate circuits are equivalent to planar free-fermion circuits under a Jordan-Wigner transformation~\cite{Terhal2001}, which provides a neat physical explanation for their efficient simulability.
The \emph{quon language}~\cite{Kang2025} is a recent approach to unify Clifford and matchgate circuits, but is quite different from ours:
Our approach addresses actual free-fermion circuits instead of matchgate circuits on qubits, and it is not restricted to planar connectivity.

Our work is not the only one that describes free-fermion, free-boson, or Clifford/stabilizer models in terms of tensor networks.
In particular, \emph{stabilizer tensor networks} are an established notion, which are tensor networks where each tensor is the unique code state of a stabilizer code~\cite{Nezami2016,Yang2021}.
\footnote{Note that in recent quantum simulation literature~\cite{Masot2024}, the term ``stabilizer tensor network'' is used for a related but different structure.}
However, these tensor networks are usually described via the stabilizer tableau formalism instead of quadratic functions as we do.
In the free-fermion case, \emph{matchgate tensor networks} are a well-known way to generalize matchgate circuits~\cite{Jahn2017}.
These are planar tensor networks of \emph{matchgate tensors}, and can be efficiently contracted.
Free-fermionic tensor networks on the other hand can be efficiently contracted independent of dimension and connectivity~\cite{tensor_type,Ren2025}.

The (super) Hopf algebra $\mathcal F$ that we use to describe free fermionic modes is known under many names in the mathematical physics literature:
It is equivalent to the \emph{exterior algebra} or \emph{Grassmann algebra} for a 1-dimensional vector space $\cc$, though the latter is not usually presented as a super Hopf algebra in the physics literature.
Alternatively, it is the \emph{universal enveloping algebra} of the super Lie algebra with a single odd generator and zero Lie bracket.
Universal enveloping algebras are known to be super Hopf algebras~\cite{Majid1995}.
However, the way we define $\mathcal F$ is quite different from the usual way the exterior algebra is defined via a quotient of the tensor algebra.
Note that the structure coefficients of $\mathcal F$ are also equal to the ``$W$ node'' in the ZW calculus~\cite{Felice2018,Poor2023}.

We are not aware of any physics literature that uses higher-order polynomial functions between groups as we introduce in Section~\ref{sec:ilinear}.
They are also a rather niche topic in mathematics, but appear in the context of cohomology theory.
Polynomial maps between groups can be found in Definition 2.4 of Ref.~\cite{Passi1968}, Section 27 (Chapter 4) of Ref.~\cite{Babakhanian1972}, Section 0.2 of Ref.~\cite{Leibman1999}, or Definition 5.1 of Ref.~\cite{Hartl2012}.
The definition in Ref.~\cite{Leibman1999} is directly equivalent to ours, whereas Refs.~\cite{Passi1968, Babakhanian1972, Hartl2012} define polynomial functions to vanish on powers of the \emph{augmentation ideal} of the group ring -- these definitions seem to become equivalent to ours if we choose $\zz$ as the base ring.
Refs.~\cite{Cui2016,Rengaswamy2019} show that diagonal gates in the Clifford hierarchy over $p$-dimensional qudits (for $p$ prime) are related to an ad-hoc notion of polynomials.
We generalize and clarify these results in Proposition~\ref{prop:diagonal_hierarchy} by showing that operators in the $i$th level of the Clifford hierarchy are precisely determined by $i$th order functions as defined in Section~\ref{sec:ilinear}.

\section{Tutorial}
\label{sec:tutorial}
In this tutorial-style section, we illustrate some key concepts informally using a few simple concrete examples.
This section is not a prerequisite for the more formal remainder of the text, but is intended to provide some helpful intuition.
Note that the core results of this paper are in the formal definitions and propositions of the remainder of the text, whereas this section only touches on the new results.

One of the goals of the paper is to show that common mathematical objects describing Clifford, free-boson, or free-fermion models are \emph{quadratic tensors}.
Roughly, a quadratic tensor is one whose entries are given by the exponential of a second-order polynomial in the index configuration.
It is clear what this means if the index configuration is an element of $\rr^n$, which is the case for systems of $n$ continuous variables.
However, there is a way to define second-order polynomials for arbitrary abelian groups and more generally (co)commutative Hopf algebras, which allows us to describe qudit Clifford and free-fermion systems as well.
Finally, we also allow quadratic tensors to be zero outside of an affine subspace, and determined by the second-order polynomial inside the subspace.

\myparagraph{Quadratic tensors for continuous variables}
Let's start with the simplest example, namely that of continuous variables.
The systems described by quadratic tensors have many names, such as free bosons, linear quantum optics, Gaussian states/operations/models, single-particle models, or quadratic Hamiltonians.
As a first example for a quadratic tensor consider a Gaussian pure state, or more precisely a squeezed state.
This is a pure state wave function $\ket\psi$ on a continuous variable specified by a variance $\sigma$ and a displacement $s$,
\begin{equation}
\label{eq:squeezed_state}
\braket{x=g|\psi} \propto e^{\frac12 \sigma^{-1} g^2 + sg}\;,
\end{equation}
where $\ket{x=g}$ denotes the position eigenstate with eigenvalue $g\in \rr$.
$\ket\psi$ is a function $\rr\rightarrow \cc$, or in other words a ``vector'' or ``tensor'' $\phi_g$ with one $\rr$-valued index.
The tensor entries are exponentials of a second-degree polynomial in $g$, so the tensor is a \emph{quadratic tensor}.
Let us next consider squeezed states on two continuous variables, determined by a $2\times 2$ covariance matrix $\sigma$ and length-2 displacement vector $s$.
The wave function is given by
\begin{equation}
\label{eq:2mode_quadratic}
\braket{x_0=g_0,x_1=g_1|\psi} \propto e^{\frac12 (\sigma^{-1})_{00} g_0^2 + \frac12(\sigma^{-1})_{11} g_1^2 + (\sigma^{-1})_{01}g_0g_1 + s_0g_0+s_1g_1}= e^{\frac12 g^T\sigma^{-1}g+sg}\;,
\end{equation}
Now, $\psi_{g_0g_1}$ is a tensor with two $\rr$-valued indices, and again the tensor entries are a second-degree polynomial in these real values.
As a next example, let us consider the time evolution under the Hamiltonian of a harmonic oscillator,
\begin{equation}
H=\frac12(\hat x^2+\hat p^2)\;, \quad U=e^{itH}\;.
\end{equation}
$U_{g_og_i}=\bra{g_o}U\ket{g_i}$ is a matrix or a tensor with two $\rr$-valued indices.
This matrix is also known as \emph{propagator} or \emph{kernel}, and is given by~\cite{Moriconi2004}
\begin{equation}
\bra{g_o} U\ket{g_i}
\propto e^{\frac{i}{2\sin(t)}(\cos(t)(g_o^2+g_i^2)-2g_og_i)}\;.
\end{equation}
Again, we see that the time-evolution unitary is a quadratic tensor.
We will show later in Section~\ref{sec:continuous_variable_examples} that arbitrary time evolutions under quadratic Hamiltonians on $n$ modes are examples of quadratic tensors.

\myparagraph{Quadratic tensors for qubits}
Next, let's try and extend our notion of quadratic tensors, such that also all stabilizer states and Clifford operators over qubits are quadratic tensors.
The computational basis for qubits is labeled by $\{0,1\}$ instead of $\rr$ for continuous variables.
Thus, it makes sense to identify $\{0,1\}$ with the finite field $\mathbb F_2$, and consider polynomials $p$ with coefficients in $\mathbb F_2$ instead.
Further, the exponential $e^{2\pi \bullet}$ of the continuous-variable case is a homomorphism from the additive group $\cc$ to the multiplicative group $\cc^\times$.
So in the qubit case, we should consider the exponential $(-1)^\bullet$ instead, as this is a homomorphism from the additive group $\mathbb F_2^+\simeq \zz_2$ to $\cc^\times$.
The tensor entries in the qubit case are then given by $T_g=(-1)^{p(g)}$, where $p$ is the second-degree polynomial.
This definition does indeed capture many examples of stabilizer states and Clifford operators:
The $\ket-$ state and the Hadamard gate are 1 and 2-index quadratic tensors, respectively,
\begin{equation}
\ket-=\mpm{1&-1}\;,\quad
\braket{g|-}=(-1)^g\;,\qquad
H\propto
\begin{pmatrix}
1&1\\1&-1
\end{pmatrix}\;,\quad
\bra{g_o} H\ket{g_i} = (-1)^{g_ig_o}\;.
\end{equation}
Indeed, both $g$ and $g_ig_o$ denote second-degree polynomials with $\mathbb F_2$ coefficients and variables.

Unfortunately, the above definition of a quadratic tensor fails to describe all stabilizer states and Clifford operations.
The simplest example that is not covered is a stabilizer state on one qubit, namely the $+1$ eigenstate of the Pauli-$Y$ operator,
\begin{equation}
\label{eq:intro_ystate}
\ket Y \propto \begin{pmatrix}1\\i\end{pmatrix}\;, \qquad \braket{g|Y} = (-1)^{\frac12 g^2}\;.
\end{equation}
The exponent $\frac12 g^2$ is not a second-degree polynomial in $\mathbb F_2$, as $\frac12\notin \mathbb F_2$.
The exponent would be a polynomial if we chose $i$ instead of $(-1)$ as the basis for the exponential, and wrote $\braket{g|Y}=i^{g^2}$.
However, this choice of basis is unjustified, as $i^{\bullet}$ is not a homomorphism from $\mathbb F_2^+$ to $\cc^\times$.

Fortunately, there is a better definition of a quadratic tensor that also captures the $\ket Y$ state:
This better definition allows us to define quadratic tensors for any abelian group $G$.
It includes quadratic tensors for continuous variables where $G$ consists of copies of $\rr$, and quadratic tensors for qubits where $G$ consists of copies of $\zz_2$.
A quadratic tensor $T$ is a tensor whose index configurations are labeled by elements $g\in G$, and whose entries of the form
\begin{equation}
\label{eq:quadratic_tensor_entry_tutorial}
T_g=\exp(q(g))\;.
\end{equation}
Here, $\exp: A\rightarrow \cc^\times$ is a group homomorphism, and $q$ is a \emph{quadratic function} valued in some abelian group $A$.
We define a quadratic function as a function $q:G\rightarrow A$ such that there exists a \emph{symmetric bilinear form} $q^{(2)}$ such that
\begin{equation}
\label{eq:invite_quadratic_axioms}
q(g+h) = q(g)+q(h)+q^{(2)}(g,h)\;,\quad\forall g,h\in G\;.
\end{equation}
Here, a symmetric bilinear form is a function $q^{(2)}:G\times G\rightarrow A$ such that
\begin{equation}
\label{eq:tutorial_bilinear}
q^{(2)}(g,h) = q^{(2)}(h,g)
\;,\quad q^{(2)}(g,h+h') = q^{(2)}(g,h)+q^{(2)}(g,h')\;,\quad
\forall g,h,h'\in G\;.
\end{equation}
Functions $q$ fulfilling Eq.~\eqref{eq:invite_quadratic_axioms} are known in the literature as \emph{quadratic refinements} of $q^{(2)}$, but we'll stick with the term quadratic function as this allows us to talk about $q$ without referencing $q^{(2)}$.

Let's now check that our new definition of quadratic tensors indeed captures the $\ket Y$ state, where the group $G$ is equal to $\zz_2$.
The group $A$ is given by $A=\rr/\zz$, the additive group of real numbers modulo 1, which is also known as the circle group and isomorphic to the multiplicative group $U(1)$.
The homomorphism $\exp: \rr/\zz\rightarrow\cc^\times$ is given by $\exp(q)=e^{2\pi i q}$.
The quadratic function $q: \zz_2\rightarrow \rr/\zz$ is given by $q(g)=\frac14 g^2$.
All that is left is to check that $q$ is indeed a quadratic function.
The corresponding bilinear form is given by
\begin{equation}
q^{(2)}(g,h) = \frac12 gh\;.
\end{equation}
It is not hard to see that this fulfills Eq.~\eqref{eq:tutorial_bilinear}; the key observation is that this equation only needs to hold $\mmod 1$ as it is valued in $\rr/\zz$.
Eq.~\eqref{eq:invite_quadratic_axioms} becomes
\begin{equation}
q(g+h)=\frac14 (g+h \mmod 2)^2 = \frac14 g^2 + \frac14 h^2 + \frac12 gh = q(g) + q(h) + q^{(2)}(g,h)\;.
\end{equation}
which again needs to hold $\mmod 1$.
Recall that the ``$\mmod 2$'' is there because ``$g+h$'' in Eq.~\eqref{eq:invite_quadratic_axioms} is an addition in $\zz_2$.
It is easy to see that the equation holds if $g=0$ or $h=0$, because we can ignore the ``$\mmod 2$'' in this case.
If $g=h=1$, then the equality in the middle becomes
\begin{equation}
\frac14(1+1\mmod 2)^2=0=\frac14 1^2+\frac14 1^2+\frac12 1\cdot 1\;,
\end{equation}
which is indeed true $\mmod 1$.

We should also briefly check that the new definition still covers our earlier examples.
For the Hadamard gate, we have $G=\zz_2\times \zz_2$, with $A=\rr/\zz$ and $\exp(q)=2^{2\pi iq}$ as before.
As $G$ consists of two factors, the quadratic function $q$ is a function in two arguments,
\begin{equation}
q(g_i,g_o)=\frac12 g_i g_o\;.
\end{equation}
We leave it to the reader to convince themselves that this is a quadratic function corresponding to Eqs.~\eqref{eq:tutorial_bilinear} and \eqref{eq:invite_quadratic_axioms}.
The corresponding bilinear form is
\begin{equation}
q^{(2)}((g_i,g_o),(g_i',g_o')) = \frac12 (g_ig_o'+g_og_i')\;.
\end{equation}
For the squeezed state $\ket\psi$, we have $G=\rr$, and $A=\cc$ and $\exp(a)=e^{2\pi a}$.
\begin{equation}
q(g)=\frac12 g^T \sigma^{-1} g + sg\;,\quad
q^{(2)}(g,g')=g^T \sigma^{-1} g'\;.
\end{equation}

Note that the new definition does not require the basis configurations to be valued in a field such as $\mathbb F_2$, but merely in an abelian group such as $\zz_2$.
Quadratic functions generalize second-degree polynomials, but they only need addition and no multiplication.
In this sense, our new definition is not only more general, it is also algebraically simpler.

To get a better intuition for quadratic tensors, let's look at the set of all possible quadratic tensors with a single qubit index, that is, with $G=\zz_2$.
There's four different such quadratic tensors.
Namely, there are two bilinear forms $q^{(2)}: \zz_2\times \zz_2\rightarrow \rr/\zz$, and each of them gives rise to two quadratic functions $q: \zz_2\rightarrow \rr/\zz$ and corresponding quadratic tensors $T$:
\begin{equation}
\label{eq:z2_quadratic_functions_list}
\begin{aligned}
q^{(2)}(g_0,g_1)=0:\qquad &q\in \{(0,0),\quad (0,\frac12)\}\;, \qquad T\in \{(1,1),\quad (1,-1)\}\;,\\
q^{(2)}(g_0,g_1)=\frac12 g_0 g_1:\qquad &q\in \{(0,\frac14),\quad (0,\frac34)\}\;, \qquad T\in \{(1,i),\quad (1,-i)\}\;.
\end{aligned}
\end{equation}
We observe that quadratic functions for the trivial bilinear form $q^{(2)}=0$ are simply group homomorphisms $\zz_2\rightarrow \rr/\zz$, and the corresponding tensors are group characters $\zz_2\rightarrow \cc^\times$.
To is a general feature for any group $G$; just set $q^{(2)}=0$ in Eq.~\eqref{eq:invite_quadratic_axioms}.
We also observe that even for the non-trivial bilinear form, the different quadratic functions differ by group homomorphisms.
Remember that we wanted to describe all stabilizer states as quadratic tensors.
There are 6 stabilizer states on one qubit, namely the eigenstates of $\pm X$, $\pm Y$, and $\pm Z$.
Eq.~\eqref{eq:z2_quadratic_functions_list} contains the eigenstates of $\pm X$ and $\pm Y$, but these of $\pm Z$ are still missing.
We will address this issue later.

\myparagraph{Quadratic tensors for qudits and rotors}
We've defined quadratic tensors such they work for arbitrary abelian groups $G$.
Apart from $\rr$ and $\zz_2$, there are a couple more abelian groups that are common in physics.
The group $\zz_d$ corresponds to $d$-dimensional qudits.
One might think that quadratic tensors for any $d$ behave just like these for $d=2$, but there turns out to be a qualitative difference between even and odd $d$.
Let's illustrate this by looking at the set of all quadratic tensors with a single qutrit index, corresponding to quadratic functions $q:\zz_3\rightarrow \rr/\zz$.
In total, there's 9 different quadratic functions and associated quadratic tensors.
Namely, there are 3 different symmetric bilinear forms $q^{(2)}$, and three quadratic functions for each of these,
\begin{equation}
\begin{aligned}
q^{(2)}(g_0,g_1)=0:\qquad &q\in \{(0,0,0),\quad (0,\frac13,\frac23),\quad (0,\frac23,\frac13)\}\;, &T\in \{(1,1,1),\quad (1,\omega,\ovl\omega),\quad (1,\ovl\omega,\omega)\}\;,\\
q^{(2)}(g_0,g_1)=\frac13 g_0 g_1:\qquad &q\in \{(0,0,\frac13),\quad (0,\frac13,0),\quad (0,\frac23,\frac23)\}\;, &T\in \{(1,1,\omega),\quad (1,\omega,1),\quad (1,\ovl\omega,\ovl\omega)\}\;,\\
q^{(2)}(g_0,g_1)=\frac23 g_0 g_1:\qquad &q\in \{(0,0,\frac23),\quad (0,\frac13,\frac13),\quad (0,\frac23, 0)\}\;, &T\in \{(1,1,\ovl\omega),\quad (1,\omega,\omega),\quad (1,\ovl\omega, 1)\}\;,
\end{aligned}
\end{equation}
with $\omega\coloneqq e^{2\pi i/3}$ and $\ovl\omega\coloneqq e^{2\pi i 2/3}$.
We notice the qualitative difference to the quadratic functions over $\zz_2$ shown in Eq.~\eqref{eq:z2_quadratic_functions_list}:
For $\zz_2$, the entries of quadratic functions are multiples of $\frac{1}{2\cdot 2}$, whereas for $\zz_3$ they are multiples of $\frac13$.
In general, we will find that for $\zz_d$ with $d$ even the entries are multiples of $\frac{1}{2d}$, whereas the $d$ odd they are multiples of $\frac{1}{d}$.
Note that it does not matter whether $d$ is a prime number or not.

There are two more abelian groups that are common in physics, namely the circle group $(\rr/\zz,+)\simeq (U(1),\times)$ and the integers $\zz$.
Both describe rotor degrees of freedom; $\rr/\zz$ corresponds to the position/angle basis, and $\zz$ correspond the angular momentum basis.
Quadratic tensors with $G=\rr/\zz$ are somewhat trivial, as there are no (non-trivial) symmetric bilinear forms $q^{(2)}:\rr/\zz\times \rr/\zz\rightarrow \rr/\zz$.
As discussed around Eq.~\eqref{eq:z2_quadratic_functions_list}, this means that all quadratic functions are homomorphisms $\rr/\zz\rightarrow \rr/\zz$.
Physically, the corresponding quadratic tensors are angular momentum eigenstates for different $k$.
For $G=\zz$, quadratic tensors just look like continuous-variable ones in Eq.~\eqref{eq:squeezed_state}.
The only difference is that we restrict $g\in G=\zz$ to integer values, whereas $\sigma$ and $s$ still can be complex-valued.

Note that the groups $\zz_d$ and $\rr$ are isomorphic to their own character groups, i.e., they are self-dual.
$\rr/\zz$ and $\zz$ are not duals of themselves, but duals of each other.
This has some important consequences when it comes to contracting large tensor networks of quadratic tensors:
Whereas contraction is always efficient in the self-dual case, it is not always efficient if there are $\zz$ subgroups.
Even though part of the appeal of quadratic tensors is lost for $\rr/\zz$ or $\zz$ factors, there are still many insightful examples including these factors.

\myparagraph{Mixing dimensions}
So far, we have only seen quadratic tensors with one or two indices.
For a 2-index tensor where each index corresponds to a degree of freedom with abelian group $C$, the over all group is $G=C^2$.
If there are $n$ indices, the group is $G=C^n$.
One of the nice features of quadratic tensors is that they work for any group $G$, including ones of the form $G=C_0\times C_1\times C_2\times \ldots$.
These are $n$-index tensors where each index corresponds to a different degree of freedom.
For example, $G=\zz_2\times \zz_4$ corresponds to a quadratic tensor with one qubit index and one 4-dimensional qudit index, possibly describing a joint pure state.
An example for a quadratic tensor that couples the qubit and qudit is as follows:
\begin{equation}
q(g_0,g_1)= \frac12 g_0g_1 + \frac18 g_1^2
=\mpm{0&\frac18&\frac12&\frac18\\0&\frac58&\frac12&\frac58}\;,\qquad
T=\mpm{1&e^{2\pi i\frac18}&-1&e^{2\pi i\frac18}\\1&e^{2\pi i\frac58}&-1&e^{2\pi i\frac58}}\;.
\end{equation}
The corresponding symmetric bilinear form is given by
\begin{equation}
q^{(2)}((g_0,g_1),(g_0',g_1')) = \frac12 (g_0g_1'+g_1g_0') + \frac14 g_1g_1'\;.
\end{equation}
We should note though that if we couple two qudits of dimensions $d$ and $d'$ that are coprime, $\gcd(d,d')=1$, then the coupling turns out to be always trivial.
That is, each such quadratic tensor is a tensor product of two independent tensors, one supported only on the $d$-qudits and one supported only on the $d'$-qudits.
Similarly, the coupling between $\rr$ or $\zz_d$ factors must be trivial.
However, it is possible to couple $\rr$ and $\zz_d$ factors indirectly via $\zz$ factors.

\myparagraph{Efficient representation}
In order to represent an arbitrary tensor $G\rightarrow \cc^\times$, we need $|G|$ numbers.
If $G=C_0\times C_1\times \ldots\times C_{n-1}$, then $|G|=|C_0|\cdot|C_1|\cdot\ldots\cdot|C_{n-1}|$, which grows exponentially with the number $n$ of factors.
If one of the factors is infinite, such as $C_i=\rr$, then we cannot exactly represent an arbitrary tensor with a finite amount of numbers, and even in the approximate case we might need a large amount of numbers for just a single factor to reach a sufficient accuracy.
One of the main reasons to consider quadratic tensors at all is that they can be represented efficiently with a number of coefficients that scales only quadratically with $n$.
This is most easily seen in the case of $n$ continuous variables, where $G=\rr^n$.
All quadratic tensors look like Eq.~\eqref{eq:2mode_quadratic}, and are determined by a $n\times n$ symmetric complex matrix $\sigma^{-1}$ and a length-$n$ vector $s$:
\begin{equation}
\braket{\vec x=\vec g|\psi} \propto e^{\frac12 \vec g^T\sigma^{-1}\vec g+s\vec g}\;.
\end{equation}
The situation is analogous if we replace $\rr$ by other elementary groups like $\zz_d$, $\zz$, or $\rr/\zz$:
Symmetric bilinear forms over $n$ elementary factors can be parameterized by a $n\times n$ matrix $q^{(\tilde2)}$, just that the matrix entries are not complex numbers but possibly different coefficients.
For example, if both $i$ and $j$ correspond to $\zz_2$ factors, we have seen in Eq.~\eqref{eq:z2_quadratic_functions_list} that there are only two bilinear forms $\zz_2\times \zz_2\rightarrow \rr/\zz$, so $q^{(\tilde2)}_{ij}$ is a $\zz_2$ coefficient.
For a fixed bilinear form there can be multiple quadratic functions.
As also discussed around Eq.~\eqref{eq:z2_quadratic_functions_list}, we can fix one particular ``standard'' quadratic function, and then all other quadratic functions are obtained by adding homomorphisms $G\rightarrow A$.
So the additional information needed to specify $q$ given $q^{(2)}$ is a homomorphism $G\rightarrow A$, and such a homomorphism can always be represented as a length-$n$ vector of coefficients.
In the continuous-variable case, bilinear forms look like $q^{(2)}(\vec g, \vec g')= \vec g^T\sigma^{-1}\vec g'$, and an obvious choice for the standard quadratic function is $q(\vec g) =\frac12 \vec g^T\sigma^{-1}\vec g$.
For other groups such as $\zz_2$, there's no obvious choice for the standard function, which leads to some subtleties.

\myparagraph{Quadratic functions on subspaces}
Remember that when we listed all quadratic tensors with a single qubit index in Eq.~\eqref{eq:z2_quadratic_functions_list}, we got the eigenstates of the Pauli $X$ and $Y$ operators, but unfortunately the following $Z$ eigenstates were missing:
\begin{equation}
T=(1,0)\;,\quad T=(0,1)\;.
\end{equation}
The simple reason for this is the $0$ entry, as $0$ is not an exponential of any number.
Fortunately, we can generalize our definition of a quadratic tensor such that the $Z$ eigenstates are quadratic tensors as well.
In the generalized definition, a quadratic tensor is specified by (1) an affine subspace of $G$, and (2) a quadratic function on this subspace.
More precisely, ``affine subspace'' here means a subgroup shifted by a constant, and the quadratic function is defined on this subgroup.
The quadratic tensor is then $0$ everywhere outside of the affine subspace and given by the exponential of the quadratic function inside the subspace.
For both $Z$ eigenstates above, the subspace/subgroup consists of only the identity element, and the affine shift is either by $0$ or by $1$.

There are many more examples of 2-qubit stabilizer states or single-qubit Clifford operations which have $0$ entries, and thus need the more general definition of a quadratic tensor.
One such example is the Pauli-$X$ operator,
\begin{equation}
X=
\begin{pmatrix}
0&1\\1&0
\end{pmatrix}\;,\qquad
\bra{g_o} X\ket{g_i} = \delta_{g_o=g_i+1\mmod 2}\;.
\end{equation}
The subgroup on which the tensor is non-zero is given by $\{(g,g)|g\in \zz_2\}\subset \zz_2^2$, the affine shift is $(0,1)\in \zz_2^2$, and the quadratic function is zero.
Note that the shift is not unique -- we could have chosen $(1,0)$ instead.
Another example is the $S$ gate,
\begin{equation}
\label{eq:s_matrix}
S=
\begin{pmatrix}
1&0\\0&i
\end{pmatrix}\;,\qquad
\bra{g_o} S\ket{g_i} = \delta_{g_o=g_i} e^{2\pi i \frac14 g_i^2}\;.
\end{equation}
The subgroup is the same as for the $X$ operator, but the affine shift is by $(0,0)$, and the quadratic function is the one same as for the $\ket Y$ state in Eq.~\eqref{eq:intro_ystate}.
In Section~\ref{sec:stabilizer_codes}, we show that our generalized notion of quadratic tensors does indeed capture all stabilizer states, as well as all Clifford operations.
In fact, we generalize stabilizer codes and Clifford operations to arbitrary abelian groups, and show that the relation holds on this more general level.

Now consider the generalized definition of a quadratic tensor in the case of continuous variables.
One new tensor we get is a position eigenstate with eigenvalue $h$, whose state-vector entries are given by a $\delta$-distribution/Dirac $\delta$-function,
\begin{equation}
\braket{x=g|x=h} = \delta_{g=h}\;.
\end{equation}
The subgroup of $G=\rr$ is the trivial one, shifted by the constant $h$.
One might argue that in the case of continuous variables, the generalized notion of a quadratic tensor only captures ``fringe'' cases which are unphysical.
However, such unphysical states very often occur in analytic computations.
Furthermore, the generalized notion is necessary to describe the most general Gaussian operators, which are certainly physical.
The most obvious example for this is the identity operator
\begin{equation}
\bra{x=g_o}\mathbb 1\ket{x=g_i}=\delta_{g_o=g_i}\;,
\end{equation}
where the subgroup is $\{(g,g)|g\in \rr\}\subset \rr\times\rr$ with zero shift and zero quadratic function.
Another example is the squeezing operation
\begin{equation}
H=\frac12(\hat a^2+(\hat a^\dagger)^2) = \frac12(\hat x\hat p+\hat p\hat x)\;,\quad U=e^{itH}\;,\quad\bra{x=g_o}U\ket{x=g_i}=\frac1{e^t} \delta_{g_o=e^t\cdot g_i}\;.
\end{equation}
The subgroup is $\{(e^t\cdot g,g)|g\in \rr\}\subset \rr\times\rr$ with zero shift and zero quadratic function.
Finally, we'd like to point out that the affine subgroup can also mix different kinds of abelian groups.
For example, the $\zz_2$ subgroup of $\zz_4$ gives rise to the following quadratic tensor with $G=\zz_4$,
\begin{equation}
\mpm{1&0&1&0}\;.
\end{equation}
Another example for this is the logical $0$ state for a single-mode square-lattice \emph{Gottesman-Kitaev-Preskill} (GKP) code,
\begin{equation}
\phi(x)=\sum_{y\in\zz} \delta_{x=y} .
\end{equation}
Here, the affine subgroup is $\zz\subset\rr$, the subgroup of integers in $\rr$.

\myparagraph{Tensor-network contraction}
Basic operations in quantum many-body physics, such as computing time evolutions or expectation values, can be phrased as tensor-network contractions.
A tensor network is a computation that takes a set of tensors as input and a single tensor as output, which consists of two operations:
(1) The tensor product of two tensors, and (2) the self-contraction of two indices of a tensor.
In general, contracting a tensor network with $n$ tensors is a hard problem, and the naive method requires runtime exponential in $n$.
One of the most important features of quadratic tensors is that tensor networks can be contracted efficiently, on the level of the representing $n\times n$ matrices and length-$n$ vectors.

Let us start by showing that we can efficiently perform the tensor product.
The tensor product of two tensors $T$ and $T'$ with groups $G$ and $G'$ corresponds to a ``tensor sum'' of the underlying quadratic functions $q$ and $q'$,
\begin{equation}
T''_{g,g'}=T_g T'_{g'}\;,\qquad q''(g,g')=q(g)+q'(g')\;.
\end{equation}
Now assume that $G$ ($G'$) has $n$ ($n'$) factors, and is described by a $n\times n$ ($n'\times n'$) matrix $q^{(\tilde2)}$ ($(q')^{(\tilde2)}$) and a length-$n$ (length-$n'$) vector $q^{(\tilde1)}$ ($(q')^{(\tilde1)}$).
Then the $(n+n')\times (n+n')$ matrix $(q'')^{(\tilde2)}$ is given by the direct sum of matrices, and the length-$(n+n')$ vector $(q'')^{(\tilde1)}$ is given by the concatenation of vectors,
\begin{equation}
(q'')^{(\tilde2)}=\mpm{q^{(\tilde2)}&0\\0&(q')^{(\tilde2)}}\;,\qquad
(q'')^{(\tilde1)}=\mpm{q^{(\tilde1)}&(q')^{(\tilde1)}}\;.
\end{equation}
If $T$ or $T'$ have non-trivial affine subgroups, then the affine subgroup of $T''$ is the product of these of $T$ and $T'$.

Next, let's show that we can efficiently perform the self-contraction.
The self-contraction is defined for a tensor $T$ with three indices corresponding to groups $H$, $C$, and again $C$, so that the total group is $G=H\times C\times C$.
The contraction over the two $C$-valued indices is then a tensor $T''$ with a single $H$ index,
\begin{equation}
T''_h \coloneqq \sum_{c\in C} T_{hcc}\;.
\end{equation}
Let us illustrate how the contraction works by looking at the simplest possible case where each index of $T$ corresponds to a single continuous variable, $H=\rr$, $C=\rr$, and $G=\rr^3$.
We further assume for simplicity that the length-$3$ vector (the displacement $s$) is zero, and that the affine subspace is all of $\rr^3$.
Such a quadratic tensor is fully specified by the symmetric $3\times 3$ matrix $q^{(\tilde2)}=\sigma^{-1}$:
\begin{equation}
\label{eq:tutorial_cv_before_contraction}
q^{(\tilde2)}=
\begin{pmatrix}
q_{00} & q_{01} & q_{02}\\
q_{01} & q_{11} & q_{12}\\
q_{02} & q_{12} & q_{22}
\end{pmatrix}
\;.
\end{equation}
The first step is force the contracted indices to take the same value, reducing the 3-index tensor $T$ to a 2-index tensor $T'_{hc}\coloneqq T_{hcc}$.
$T'$ is again a quadratic tensor, and the $2\times 2$ matrix $(q')^{(\tilde2)}$ given by adding both the second and third row and column of $q^{(\tilde2)}$:
\begin{equation}
\label{eq:tutorial_constraction_first_step}
(q')^{(\tilde2)}=
\begin{pmatrix}
q'_{00} & q'_{01}\\
q'_{01} & q_{11}'
\end{pmatrix}
=
\begin{pmatrix}
q_{00} & q_{01}+q_{02}\\
q_{01}+q_{02} & q_{11}+2q_{12}+q_{22}
\end{pmatrix}
\;.
\end{equation}
The second step is to perform the summation over $c$, which is actually an integration as $c$ is $\rr$-valued.
Assuming that $q_{11}'=q_{11}+2q_{12}+q_{22}\neq 0$, we can explicitly calculate
\begin{equation}
\label{eq:gaussian_integral}
\begin{gathered}
T''_h
= \int_c T_{hcc}
= \int_c T'_{hc}
= \int_c e^{2\pi (\frac12 q'_{00} h^2 + \frac12 q'_{11} c^2 + q'_{01}hc)}
= \int_c e^{2\pi (\frac12 q'_{11} (c+\frac{q'_{01}}{q'_{11}} h)^2 - \frac12 \frac{(q'_{01})^2}{q'_{11}} h^2  + \frac12 q'_{00} h^2)}\\
= e^{2\pi  \frac12(q'_{00}- \frac{(q'_{01})^2}{q'_{11}}) h^2} \int_c e^{2\pi (\frac12 q'_{11} (c+\frac{q'_{01}}{q'_{11}} h)^2)}
= e^{2\pi  \frac12(q'_{00}- \frac{(q'_{01})^2}{q'_{11}}) h^2} \int_c e^{2\pi \frac12 q'_{11} c^2}
= \frac1{\sqrt{q'_{11}}} e^{2\pi  \frac12(q'_{00}- \frac{(q'_{01})^2}{q'_{11}}) h^2}\;.
\end{gathered}
\end{equation}
In the calculation, we have used a quadratic completion, as well as a change of integration variable.
So we see that, up to a global prefactor, the result is again a quadratic tensor with $1\times 1$ matrix
\begin{equation}
(q'')^{(\tilde2)}=
\mpm{q'_{00}- \frac{(q'_{01})^2}{q'_{11}}}
\;.
\end{equation}
Now consider a more general case where $H$ consists of $n$, and $C$ of $m$ continuous variables.
$q^{(\tilde2)}$ can still be written as in Eq.~\eqref{eq:tutorial_cv_before_contraction}, just that the entries are now $m\times n$, $n\times n$, or $m\times m$ matrix blocks, and some transposes are missing.
The first step of the contraction in Eq.~\eqref{eq:tutorial_constraction_first_step} again looks identical, just that the entries are matrix blocks.
Next, we can still perform the calculation in Eq.~\eqref{eq:gaussian_integral} in the case of matrix blocks.
The only thing that changes is the the product of matrix blocks is non-commutative.
If we carefully keep track of the order in the products, we obtain the following expression for $(q'')^{(\tilde2)}$,
\begin{equation}
\label{eq:tutorial_schur_complement}
(q'')^{(\tilde2)}=q'_{00}-q'_{01} (q'_{11})^{-1} q'_{10}\;.
\end{equation}
$q_{01} (q_{11})^{-1} q_{10}$ denotes the product of an $n\times m$, $m\times m$, and a $m\times n$ matrix.
The expression above is know as the \emph{Schur complement} of the $2\times 2$ block matrix $(q')^{(\tilde2)}$.
Note that we can only perform the Schur complement if the matrix $q_{11}$ is invertible.

It is possible to generalize the calculation in Eq.~\eqref{eq:gaussian_integral} to the case of arbitrary abelian groups $H$ and $C$.
Instead of the quadratic expansion $\frac12(g+h)^2=\frac12g^2+\frac12h^2+gh$, we have to use the defining equation for a quadratic function in Eq.~\eqref{eq:invite_quadratic_axioms}.
The shift of integration (or summation) variable readily works for arbitrary abelian groups $H$ and $C$.
The result of this generalized calculation looks like Eq.~\eqref{eq:tutorial_schur_complement}, we just have to interpret it in a generalized sense.
$q_{11}:C\times C\rightarrow A$ is a bilinear map, so for every fixed $c\in C$ in the first component, $q_{11}(c,\bullet): C\rightarrow A$ is a homomorphism.
Denoting the group of homomorphisms $C\rightarrow A$ (equivalent to the character group) by $C^*$, we can view $q_{11}$ as a group homomorphism $C\rightarrow C^*$, mapping $c\in C$ to $q_{11}(c,\bullet)\in C^*$.
$(q_{11})^{-1}: C^*\rightarrow C$ in Eq.~\eqref{eq:tutorial_schur_complement} is then an inverse of this group homomorphism.
Similarly, $q_{10}$ can be viewed as a group homomorphism $H\rightarrow C^*$, and $q_{01}$ as a homomorphism $C\rightarrow H^*$.
$q_{01} (q_{11})^{-1} q_{10}$ is thus a composition of three homomorphisms $H\rightarrow C^*\rightarrow C\rightarrow H^*$, which in turn can be interpreted as a bilinear map $H\times H\rightarrow A$.

As a very simple example, consider a quadratic tensor with three qubit indices where we contract the second and third index.
Assume that after the first step of the contraction, the tensor $T'$ is given by
\begin{equation}
T'_{hc}=\mpm{1&i\\1&-i}\;.
\end{equation}
The associated bilinear form is
\begin{equation}
(q')^{(2)}((h,c),(h',c'))=\frac12\big(hc'+h'c+cc'\big)\;,\qquad
(q')^{(\tilde2)}=\mpm{0&1\\1&1}\;.
\end{equation}
We find that the result of the contraction is the $\ket{-Y}$ state,
\begin{equation}
T''_h=\sum_c T'_{hc}=\mpm{1+i\\1-i}\propto \mpm{1\\-i}\;,
\end{equation}
with associated bilinear form
\begin{equation}
(q'')^{(2)}(h,h')=\frac12hh'\;,\qquad (q'')^{(\tilde2)} = \mpm{1}\;.
\end{equation}
So Eq.~\eqref{eq:tutorial_schur_complement} indeed holds with
\begin{equation}
1= 0-1\cdot 1^{-1}\cdot 1\;.
\end{equation}
Here, the coefficients $q'_{01}=q'_{11}=(q'_{11})^{-1}=1=-1$ all represent the identity homomorphism on $\zz_2$.
Note that for the full contraction, we also need to compute $(q'')^{(\tilde1)}$, which is a bit more involved.
We discuss the contraction in all formality and generality in Section~\ref{sec:reduction}.

\myparagraph{Free fermions}
As free fermions are quite similar to free bosons, it is natural to try to also incorporate the former into our formalism.
There is indeed a type of tensors describing free-fermionic systems, known as \emph{matchgate tensors}.
A matchgate tensor defined on $n$ fermionic modes is fully specified by an anti-symmetric complex $n\times n$ matrix $q^{(\tilde2)}$.
Let $\vec x$ be a length-$n$ bitstring corresponding to a configuration of the $n$ modes.
The tensor entry of a matchgate tensor $T$ for the index configuration $\vec x$ is given by 
\begin{equation}
\label{eq:pfaffian_tensor_entries}
T(\vec x) = \pf(q^{(\tilde2)}|_{\vec x})\;.
\end{equation}
Here, $\pf$ denotes the \emph{Pfaffian} of a matrix, and $q^{(\tilde2)}|_{\vec x}$ denotes the matrix $q^{(\tilde2)}$ restricted to all rows and columns $i$ for which $\vec x_i=1$.
As an example for a 4-mode matchgate tensor, consider a ``fermionic beam splitter'', or the time evolution under a fermionic hopping operator on $m=2$ fermionic modes:
\begin{equation}
H=c_0c_1^\dagger+c_1c_0^\dagger = \mpm{0&0&0&0\\0&0&1&0\\0&1&0&0\\0&0&0&0},\quad U=e^{-itH}
=\mpm{1&0&0&0\\0&\cos(t)&i\sin(t)&0\\0&i\sin(t)&\cos(t)&0\\0&0&0&1}\;.
\end{equation}
The matrices are $2^m\times 2^m=4\times 4$ matrices, and the four rows or columns correspond to the computational basis states $\ket{00}$, $\ket{01}$, $\ket{10}$, and $\ket{11}$, respectively.
We find that this is indeed a matchgate tensor with $n=2m=4$ indices whose $2m\times 2m=4\times 4$ matrix $q^{(\tilde2)}$ is given by
\begin{equation}
q^{(\tilde2)}=\mpm{0&0&\cos(t)&i\sin(t)\\0&0&i\sin(t)&\cos(t)\\-\cos(t)&-i\sin(t)&0&0\\-\cos(t)&-i\sin(t)&0&0}
\;.
\end{equation}
Here, the four rows or columns correspond to the first input mode, second input mode, first output mode, and second output mode of $U$, respectively.

At a first glance, the definition of a matchgate tensor in Eq.~\eqref{eq:pfaffian_tensor_entries} looks very different from that of a quadratic tensor in Eq.~\eqref{eq:quadratic_tensor_entry_tutorial}.
However, in Section~\ref{sec:hopf_quadratic}, we show that matchgate tensors are precisely another instance of quadratic tensors, after we \emph{super-linearize} the algebraic structures ``group'' and ``quadratic tensor''.
Let's briefly outline how this works.
In general, an algebraic structure is given by a collection of sets, maps between these sets, and equations between different compositions of these maps.
In particular, an abelian group is defined by a single set $G$, a group multiplication map $+: G\times G\rightarrow G$ (which we denote by $+$ as it is abelian), and the associativity and commutativity axioms
\begin{equation}
\label{eq:tutorial_associativity_commutativity}
+\circ(+\times \idop) = +\circ(\idop\times+)\;,\qquad
+=+\circ\operatorname{Swap}\;.
\end{equation}
Here, $\times$ denotes the cartesian product of functions, $\circ$ denotes function composition, $\idop$ is the identity function, and
\begin{equation}
\operatorname{Swap}:G\times G\rightarrow G\times G\;,\qquad \operatorname{Swap}(g,h)=(h,g)
\end{equation}
is the function exchanging two arguments.
To super-linearize an algebraic structure means to replace every set by a \emph{super vector space}, every map by a \emph{super linear map}, every cartesian product by a ``super tensor product'', and every swap by a ``super swap''.
A super vector space $V$ is the direct sum of two complex vector spaces, $V=V_e\oplus V_o$, where $V_e$ is called the ``even'' and $V_o$ the ``odd'' part.
Physically, $V_o$ contains the configurations that have a fermionic charge, and $V_e$ contains ones that don't.
The tensor product $V\otimes_s V'$ of super vector spaces $V$ and $V'$ is given by
\begin{equation}
V''=V\otimes_s V'= V''_e\oplus V''_o\;,\quad
V''_e=V_e\otimes V'_e\oplus V_o\otimes V'_o\;,\quad
V''_o=V_e\otimes V'_o\oplus V_o\otimes V'_e\;.
\end{equation}
That is, even and odd parts mix in a $\zz_2$-graded fashion.
A super tensor is a tensor with super vector space indices that is only supported on the globally even sector -- so it is only non-zero if an even number of individual indices are in an odd configuration.
Whenever we use tensors to describe fermions we should use super tensors, whether we are describing free theories or not.
A super linear map is a super tensor with two indices.
Finally, the ``super swap'' of super vector spaces is given by
\begin{equation}
\operatorname{Swap}_s=\mpm{\operatorname{Swap}&\operatorname{Swap}\\\operatorname{Swap}&-\operatorname{Swap}}\;,
\end{equation}
where the four blocks correspond to $\operatorname{Swap}_s$ acting between $V_e$ and $V_o$ with $V'_e$ and $V'_o$, respectively.
We see that swapping $V_o$ and $V'_o$ yields an additional minus sign, corresponding to the exchange statistics of fermions.

For a group, super-linearization means to replace the set $G$ by a complex super vector space $V$, and the group multiplication $+$ by a super linear map $\cdot: V\otimes_s V\rightarrow V$.
The axioms for $\cdot$ are the super linear equations given by Eq.~\eqref{eq:tutorial_associativity_commutativity}, after replacing $+$ by $\cdot$, and $\times$ by $\otimes_s$.
The resulting algebraic structure is known as \emph{super associative and commutative algebras}.
However, there is some additional structure in a group we have neglected so far:
One can copy group elements via a copy map
\begin{equation}
\Delta: G\rightarrow G\times G\;,\qquad \Delta(g)=(g,g)\;.
\end{equation}
The copy map $\Delta$ fulfills a few simple identities.
For example, when making three copies, it does not matter whether we make the third copy from the first or from the second copy.
Or, swapping the two elements of $(g,g)$ has no effect.
Or, first adding two group elements $g+h$ and then copying the result is the same as first making copies $(g,g)$ and $(h,h)$, and then adding the pairs entry-wise: Both yield $(g+h,g+h)$.
These identities can be written as equations between function compositions,
\begin{equation}
\label{eq:tutorial_hopf_axioms}
\begin{gathered}
(\Delta\times \idop)\circ \Delta = (\idop\times \Delta)\circ \Delta\;,\qquad
\operatorname{Swap}\circ \Delta = \Delta\;,\\
\Delta\circ\cdot = (+\times+)\circ (\idop\times \operatorname{Swap}\times \idop)\circ (\Delta\times \Delta)\;.
\end{gathered}
\end{equation}
While a copy map automatically exists for any group, there is no canonical copy super linear map for a super algebra.
So we should explicitly add a super linear map $\Delta: V\rightarrow V\otimes V$, fulfilling Eq.~\eqref{eq:tutorial_hopf_axioms} where we replace $+$, $\times$, and $\operatorname{Swap}$ with $\cdot$, $\otimes_s$ and $\operatorname{Swap}_s$.
The resulting algebraic structure is similar to what is known as a \emph{super commutative and cocommutative Hopf algebra} in the math literature, where $\Delta$ is known as the \emph{coalgebra}.
So we should regard ``super (co)commutative Hopf algebra'' as the true super-linearization of ``abelian group''.

Next, we need to generalize quadratic tensors from groups to Hopf algebras.
Apply the replacement
\begin{equation}
\label{eq:group_to_hopf_quadratic_tensor}
T_g\coloneqq \exp(q(g))\;,\qquad
B_{g,h}\coloneqq \exp(q^{(2)}(g,h))
\end{equation}
to Eqs.~\eqref{eq:invite_quadratic_axioms} and \eqref{eq:tutorial_bilinear}.
Then a quadratic tensor $T$ is one for which there exists $B$ such that
\begin{equation}
\label{eq:super_quadratic}
T_{g+h}=T_g T_h B_{gh}\;,\qquad
B_{g,h+h'}=B_{g,h} B_{g,h'}\;.
\end{equation}
We can view the tensor $T$ as a super linear map $T:V\rightarrow \cc$ and $B$ as a super linear map $B: V\otimes_s V\rightarrow \cc$.
To get the super-linearized version, we should replace the function $+$ with the super linear map $\cdot$.
The right-hand sides of both equations contain multiple copies of $g$ or $h$, which we need to implement using the copy map $\Delta$.
After doing this, Eq.~\eqref{eq:super_quadratic} becomes
\begin{equation}
\label{eq:tutorial_hopf_quadratic}
T\circ \cdot = (T\otimes_s B\otimes_s T)\circ (\Delta\otimes_s\Delta)\;,\qquad
B\circ (\idop\otimes_s \cdot) = (B\otimes_s B)\circ (\idop\otimes_s \operatorname{Swap}_s\otimes_s\idop)\circ (\Delta\otimes_s \idop\otimes_s\idop)\;.
\end{equation}
That is, a quadratic tensor $T$ over a super Hopf algebra is one such that there exists $B$ fulfilling the above super linear equations.

After super-linearizing groups and their quadratic tensors, let's argue that the super-linearization is really a generalization.
Indeed, every abelian group $G$ yields a super (co)commutative Hopf algebra, which is known as the \emph{group Hopf algebra} of $G$:
The odd vector space is trivial, and the even vector space is given by $V=\cc^G$, such that there is a basis $\{e_g\}_{g\in G}$ labeled by the group elements.
The multiplication and copy maps are given by
\begin{equation}
e_g\cdot e_h=e_{g+h}\;,\qquad
\Delta(e_g)=e_g\otimes e_g\;.
\end{equation}
Further, any quadratic tensor over a group is also a quadratic tensor over the group Hopf algebra via Eq.~\eqref{eq:group_to_hopf_quadratic_tensor}.

Vice versa, we can ask: Are there super (co-)commutative Hopf algebras that are not equal (or at least isomorphic) to group Hopf algebras?
The answer is ``yes, but not too many'':
Essentially there one such super Hopf algebra, and it precisely corresponds to a free-fermion mode.
Both the even and the odd part of the vector space are one-dimensional, with basis vectors $e_0$ and $e_1$, respectively.
Physically, $e_0$ and $e_1$ correspond to the unoccupied and the occupied state of the mode.
The multiplication is given by
\begin{equation}
\label{eq:fermion_algebra_multiplication}
e_0\cdot e_0=e_0\;,\qquad e_0\cdot e_1 = e_1\cdot e_0 = e_1\;,\qquad e_1\cdot e_1 = 0\;,
\end{equation}
and the coalgebra by
\begin{equation}
\label{eq:fermion_algebra_comultiplication}
\Delta(e_0)=e_0\otimes e_0\;,\quad \Delta(e_1)=e_0\otimes e_1+e_1\otimes e_0\;.
\end{equation}
Note that for the last axiom in Eq.~\eqref{eq:tutorial_hopf_axioms} to hold, it is essential that we use the super version of the swap,
\begin{equation}
\operatorname{Swap}_s(e_1\otimes e_1) = -e_1\otimes e_1\;.
\end{equation}

Let's provide some justification for why the above super Hopf algebra corresponds to free fermions.
Consider the coefficient tensor of the group multiplication for the group $\zz_2$ in the $\{e_0,e_1\}$ basis,
\begin{equation}
\braket{e_j|e_i\cdot e_x}
=
\begin{tikzpicture}[baseline=-0.5ex]
\node (c) {
$\mpm{
\begin{tikzpicture}
\node[inner sep=0] (c0) {$\mpm{1&0\\0&1}$};
\draw ($(c0)+(-0.5,0.3)$)edge[->,mark={slab=$\scriptstyle{j}$,r}] ($(c0)+(-0.5,-0.3)$);
\path[use as bounding box] (current bounding box.south west)rectangle(current bounding box.north east);
\draw ($(c0)+(-0.3,0.5)$)edge[->,mark={slab=$\scriptstyle{i}$}] ($(c0)+(0.3,0.5)$);
\end{tikzpicture}
&
\begin{tikzpicture}
\node[inner sep=0] (c1) {$\mpm{0&1\\1&0}$};
\end{tikzpicture}
}$};
\draw ($(c)+(-0.9,0.75)$)edge[->,mark={slab=$\scriptstyle{x}$}] ($(c)+(1,0.75)$);
\end{tikzpicture}
=\mpm{\mathbb1, X}\;.
\end{equation}
As indicated, the value $x$ labels the block, whereas $i$ and $j$ label the column or row inside a block.
This shows that we can view the $\zz_2$ group algebra as ``the algebra formed by the operators $\mathbb1$ and Pauli-$X$''.
Similarly, consider the coefficient tensor of the coalgebra $\Delta$,
\begin{equation}
\begin{gathered}
\braket{e_j\otimes e_x|\Delta(e_i)}
=
\begin{tikzpicture}[baseline=-0.5ex]
\node (c) {
$\mpm{
\begin{tikzpicture}
\node[inner sep=0] (c0) {$\mpm{1&0\\0&0}$};
\draw ($(c0)+(-0.5,0.3)$)edge[->,mark={slab=$\scriptstyle{j}$,r}] ($(c0)+(-0.5,-0.3)$);
\path[use as bounding box] (current bounding box.south west)rectangle(current bounding box.north east);
\draw ($(c0)+(-0.3,0.5)$)edge[->,mark={slab=$\scriptstyle{i}$}] ($(c0)+(0.3,0.5)$);
\end{tikzpicture}
&
\begin{tikzpicture}
\node[inner sep=0] (c1) {$\mpm{0&0\\0&1}$};
\end{tikzpicture}
}$};
\draw ($(c)+(-0.9,0.75)$)edge[->,mark={slab=$\scriptstyle{x}$}] ($(c)+(1,0.75)$);
\end{tikzpicture}\;,\\
\mpm{1&0\\0&0}+\mpm{0&0\\0&1}=\mpm{1&0\\0&1}=\mathbb1\;,\qquad
\mpm{1&0\\0&0}-\mpm{0&0\\0&1}=\mpm{1&0\\0&-1}=Z\;.
\end{gathered}
\end{equation}
So we see that the coalgebra is equivalent to the coalgebra formed by the operators $\mathbb1$ and Pauli-$Z$.

Now let's look at the same tensor coefficients for the super Hopf algebra in Eqs.~\eqref{eq:fermion_algebra_multiplication} and \eqref{eq:fermion_algebra_comultiplication}.
We find
\begin{equation}
\braket{e_j|e_i\cdot e_x}
=
\begin{tikzpicture}[baseline=-0.5ex]
\node (c) {
$\mpm{
\begin{tikzpicture}
\node[inner sep=0] (c0) {$\mpm{1&0\\0&1}$};
\draw ($(c0)+(-0.5,0.3)$)edge[->,mark={slab=$\scriptstyle{j}$,r}] ($(c0)+(-0.5,-0.3)$);
\path[use as bounding box] (current bounding box.south west)rectangle(current bounding box.north east);
\draw ($(c0)+(-0.3,0.5)$)edge[->,mark={slab=$\scriptstyle{i}$}] ($(c0)+(0.3,0.5)$);
\end{tikzpicture}
&
\begin{tikzpicture}
\node[inner sep=0] (c1) {$\mpm{0&0\\1&0}$};
\end{tikzpicture}
}$};
\draw ($(c)+(-0.9,0.75)$)edge[->,mark={slab=$\scriptstyle{x}$}] ($(c)+(1,0.75)$);
\end{tikzpicture}
=\mpm{\mathbb1, c^\dagger}\;.
\end{equation}
So the algebra is equal to that formed by the identity and the fermionic creation operator $c^\dagger$.
Similarly, we find
\begin{equation}
\braket{e_j\otimes e_x|\Delta(e_i)}
=
\begin{tikzpicture}[baseline=-0.5ex]
\node (c) {
$\mpm{
\begin{tikzpicture}
\node[inner sep=0] (c0) {$\mpm{1&0\\0&1}$};
\draw ($(c0)+(-0.5,0.3)$)edge[->,mark={slab=$\scriptstyle{j}$,r}] ($(c0)+(-0.5,-0.3)$);
\path[use as bounding box] (current bounding box.south west)rectangle(current bounding box.north east);
\draw ($(c0)+(-0.3,0.5)$)edge[->,mark={slab=$\scriptstyle{i}$}] ($(c0)+(0.3,0.5)$);
\end{tikzpicture}
&
\begin{tikzpicture}
\node[inner sep=0] (c1) {$\mpm{0&1\\0&0}$};
\end{tikzpicture}
}$};
\draw ($(c)+(-0.9,0.75)$)edge[->,mark={slab=$\scriptstyle{x}$}] ($(c)+(1,0.75)$);
\end{tikzpicture}
=\mpm{\mathbb1, c}\;.
\end{equation}
So the coalgebra is generated by the identity and the annihilation operator $c$.
We can now see the analogy between the qubit and the free-fermion algebras:
The qubit Hopf algebra is the one of Pauli $X$ and $Z$ operators, whereas the free-fermion super Hopf algebra is the one formed by the creation and annihilation operators.
Finally, we note that the creation and annihilation operator are not proper super linear operators:
They map between the even and the odd sector, and thus have total odd parity.
Thus, swapping two annihilation or creation operators yields a minus sign,
\begin{equation}
\operatorname{Swap}_s\circ (c\otimes_s c)\circ \operatorname{Swap}_s = -(c\otimes_s c)\;.
\end{equation}
So creation and annihilation operators acting on different modes anti-commute as they should.

A quadratic tensor with $n$ free-fermion indices is defined as in Eq.~\eqref{eq:tutorial_hopf_quadratic}, where the Hopf algebra is given by the $n$-fold tensor product of the free-fermion Hopf algebra defined in Eq.~\eqref{eq:fermion_algebra_multiplication} and \eqref{eq:fermion_algebra_comultiplication}.
It remains to show that such quadratic tensors are precisely matchgate tensors of the form Eq.~\eqref{eq:pfaffian_tensor_entries}.
It is not quite trivial to see this, and we refer the reader to Section~\ref{sec:hopf_quadratic} for details.

\myparagraph{Higher-order tensors}
Finally, we also introduce a generalization of quadratic tensors to higher-order tensors, whose entries correspond to higher-order polynomial functions $G\rightarrow \cc^\times$.
For example, a cubic (or third order) function $q$ is one such that there exists a permutation-invariant trilinear form $q^{(3)}$ such that
\begin{equation}
q(g+h+i)=q(g+h)+q(g+i)+q(h+i)-q(g)-q(h)-q(i)+q^{(3)}(g,h,i)\;.
\end{equation}
An example for a cubic function $\zz_2\rightarrow \rr/\zz$ and the associated tensor is given by
\begin{equation}
q=\mpm{0&\frac18}\;,\quad T=\mpm{1&e^{2\pi i\frac18}}\;.
\end{equation}
This tensor corresponds to the so-called $\ket T$ state, which is the most-studied example of a \emph{magic state}.
The associated trilinear form is
\begin{equation}
q^{(3)}(g,h,i)=\frac12 ghi\;.
\end{equation}
We discuss higher-order tensors in Section~\ref{sec:ilinear}.

\section{Quadratic tensors}
\label{sec:clifford_tensors}
In this section, we introduce quadratic tensors.
We briefly review how many computations in physics can be phrased as tensor-network contractions in Section~\ref{sec:general_tensor_networks}.
Then we proceed with the definition of quadratic tensors in terms of quadratic and linear functions between abelian groups in Section~\ref{sec:clifford_definition}.
Next, we show how to efficiently store quadratic tensors using a quadratic number of coefficients in Sections~\ref{sec:factor_decomposition} and \ref{sec:coefficient_groups}.
Finally, in Sections~\ref{sec:homomorphism_composition} and \ref{sec:reduction} we show how to efficiently contract tensor networks of quadratic tensors.

\subsection{Tensor networks in physics}
\label{sec:general_tensor_networks}
Let us review tensors and tensor networks and how they can be used in quantum physics.
Roughly, a tensor is a multi-dimensional array of complex numbers.
\begin{mydef}
A \emph{tensor} $T$ with $n$ indices is defined with respect to a collection of sets $\{B_i\}_{0\leq i<n}$.
$B_i$ is called \emph{basis set} at \emph{index position} $0\leq i<n$.
$T$ is defined as a map
\begin{equation}
\label{eq:tensor_map}
T: \bigtimes_{0\leq i<n} B_i \rightarrow \cc\;.
\end{equation}
\end{mydef}
In other words, $T$ is the map which assigns a tensor entry to each \emph{index configuration} $(b_0,\ldots,b_{n-1})\in B_0\times \ldots\times B_{n-1}$.
It is common to write the arguments of the map $T$ as subscripts, $T(b_0,\ldots,b_{n-1})=T_{b_0,\ldots,b_{n-1}}$.
For the usual notion of a tensor, the sets $B_i$ are of the form $B_i=\{0,\ldots,d_i-1\}$, where $d_i$ is known as the \emph{bond dimension} of the index.
However, we will also consider basis sets like $B_i=\zz$, $B_i=\rr/\zz$, or $B_i=\rr$, in which case a tensor is a function over variables valued in a continuous or infinite set rather than a finite collection of complex numbers.
\begin{mydef}
\label{def:tensor_network}
A \emph{tensor network} is a specific form of computation which takes as input a set of tensors, and computes a single tensor as output, and consists of the following operations:
\begin{itemize}
\item Taking copies of a tensor.
\item The \emph{tensor product} of an $n$-index tensor $T$ and an $n'$-index tensor $T'$ is the $n+n'$-index tensor $T''$ given by
\begin{equation}
B''_i=\begin{cases} B_i & \text{if } 0\leq i<n\\ B'_{i-n} &\text{if } n\leq i<n+n'\end{cases}\;,\qquad
T''(b_0,\ldots,b_{n+n'-1}) = T(b_0,\ldots,b_{n-1}) \cdot T'(b_n,\ldots,b_{n+n'-1})\;.
\end{equation}
\item The \emph{index contraction} (or \emph{Einstein summation}) between the index positions $0\leq i<j<n$ of an $n$-index tensor $T$ is defined if $B_i=B_j$.
It given by the following $n-2$-index tensor $T'$ with sequence of basis sets $(B_0,\ldots,B_{i-1},B_{i+1},\ldots,B_{j-1},B_{j+1},\ldots,B_{n-1})$,
\begin{equation}
T'(b_0,\ldots,b_{n-3})
= \sum_{x\in B_i} T(b_0,\ldots,b_{i-1},x,b_i,\ldots,b_{j-2},x,b_{j-1},\ldots,b_{n-3})\;.
\end{equation}
\item Another operation that is not necessary for the definition of a tensor network but sometimes occurs in transformations of tensor networks is the blocking of indices.
The \emph{index blocking} at position $0\leq i<n-1$ of a tensor $T$ is a $n-1$-index tensor $T'$ with basis set sequence $(B_0,\ldots,B_{i-1},B_i\times B_{i+1},\ldots,B_{n-1})$,
\begin{equation}
T'(b_0,\ldots,b_{i-1},(b_i,b_{i+1}),\ldots,b_{n-1})=T(b_0,\ldots,b_{n-1})\;.
\end{equation}
\end{itemize}
\end{mydef}
As an example, consider the tensor network
\begin{equation}
\label{eq:example_tensor_network}
O_{ij}\coloneqq \sum_{xyz} T_{iyx} T_{xzj} M_{yz}\;.
\end{equation}
It has input tensors, a 3-index tensor $T$ and a 2-index tensor $M$, and results in a 2-index output tensor $O$.
The tensor network consists in (1) taking two copies of $T$ and one copy of $M$, (2) taking the tensor product of the three tensor copies, and (3) performing index contractions over three index pairs by summing over $x,y,z$.

Tensor networks can be drawn as diagrams, which is sometimes called \emph{Penrose notation} or simply \emph{tensor-network notation}.
To this end, we assign a distinct shape to each input tensor.
Sometimes we will also include a decoration, label, or color with the shape to distinguish different input tensors.
Then we draw one copy of the shape for each copy of the input tensor in the tensor network.
Finally, for each contraction between two indices of two tensor copies, we draw a line connecting the corresponding shapes.
Indices that are not summed over are part of the output tensor, and are drawn as ``open'' lines originating from the corresponding shape.
Additionally, we put markings near the place where a line connects to a shape, to indicate which of the index positions it corresponds to.
For example, the tensor network in Eq.~\eqref{eq:example_tensor_network} can be drawn as follows:
\begin{equation}
\label{eq:example_network_diagram}
\begin{tikzpicture}
\atoms{square,small}{{0/lab={p=0:0.25,t=$T$}}, {1/p={1.6,0},lab={p=180:0.25,t=$T$}}}
\atoms{circ,small}{{m/p={0.8,0.5},lab={p=90:0.25,t=$M$}}}
\draw (0)edge[ar=e,out=-90,in=-90](1) (0)edge[mark={bar,s},ar=e,out=90,in=180](m) (1)edge[mark={bar,s},out=90,in=0](m) (0)edge[ar=s,ind=$i$]++(180:0.6) (1)edge[ind=$j$]++(0:0.6);
\end{tikzpicture}
\;.
\end{equation}
$T$ is represented by a square shape, whereas $M$ is represented by a circle shape.
The first index of $T$ is marked by a triangle arrow, the second index is marked by a tick, and the third index is unmarked.
The first index of $M$ is distinguished from the second index by a triangle arrow.
Note that it is possible to omit the labels ``$M$'' and ``$T$'' in the diagram, since the different tensors are already distinguished by their different shapes.

\begin{figure}[h]
\begin{tikzpicture}
\node[inner sep=0] (x0) at (0,0){
\begin{tikzpicture}
\atoms{circ,small,lab={t=$\psi$,p=-90:0.25}}{0/}
\draw (0)--++(90:0.5);
\end{tikzpicture}
};
\node[inner sep=0] (x1) at (2,0){
\begin{tikzpicture}
\atoms{circ,small,lab={t=$U$,p=0:0.25}}{0/}
\draw (0)--++(90:0.5) (0)--++(-90:0.5);
\end{tikzpicture}
};
\node[inner sep=0] (x2) at (4,0){
\begin{tikzpicture}
\atoms{square,xscale=1.5,lab={t=$\rho$,p=-90:0.35}}{0/}
\draw ([sx=0.15]0.north)--++(90:0.4) ([sx=-0.15]0.north)--++(90:0.4);
\end{tikzpicture}
};
\node[inner sep=0] (x3) at (6,0){
\begin{tikzpicture}
\atoms{square,xscale=1.5,lab={t=$C$,p=0:0.45}}{0/}
\draw ([sx=0.15]0.north)--++(90:0.4) ([sx=-0.15]0.north)--++(90:0.4);
\draw ([sx=0.15]0.south)--++(-90:0.4) ([sx=-0.15]0.south)--++(-90:0.4);
\end{tikzpicture}
};
\node[inner sep=0] (x4) at (1,-2){
$
\begin{tikzpicture}
\atoms{square,xscale=1.5,lab={t=$C$,p=0:0.45}}{0/}
\draw ([sx=0.15]0.north)--++(90:0.4) ([sx=-0.15]0.north)--++(90:0.4);
\draw ([sx=0.15]0.south)--++(-90:0.4) ([sx=-0.15]0.south)--++(-90:0.4);
\end{tikzpicture}
=
\begin{tikzpicture}
\atoms{circ,small}{{0/p={-0.3,0},lab={t=$A$,p=180:0.25}}, {1/p={0.3,0},lab={t=$A^*$,p=0:0.3}}}
\draw (0)--(1) (0)--++(90:0.5) (1)--++(90:0.5) (0)--++(-90:0.5) (1)--++(-90:0.5);
\end{tikzpicture}
$
};
\node[inner sep=0] (x5) at (4,-2){
\begin{tikzpicture}
\atoms{square,xscale=1.5,lab={t=$\rho$,p=0:0.45}}{0/}
\atoms{square,xscale=1.5,lab={t=$O$,p=0:0.45}}{o/p={0,0.8}}
\draw ([sx=0.15]0.north)--([sx=0.15]o.south) ([sx=-0.15]0.north)--([sx=-0.15]o.south);
\end{tikzpicture}
};
\node[inner sep=0] (x6) at (6,-2){
$
\begin{tikzpicture}
\atoms{circ,small}{{0/lab={t=$U$,p=0:0.25}}, {1/p={0,0.6},lab={t=$U^*$,p=0:0.35}}}
\draw (0)--(1) (0)--++(-90:0.4) (1)--++(90:0.4);
\end{tikzpicture}
=
\begin{tikzpicture}
\draw (0,0)--++(90:0.8);
\end{tikzpicture}
$
};
\node[inner sep=0] (x7) at (9,-1){
\begin{tikzpicture}
\atoms{square,xscale=4,lab={t=$\rho$,p=0:0.85}}{r/}
\atoms{circ,small}{{u0/p={-0.6,0.6},lab={t=$U$,p={180:0.25}}}, {u1/p={-0.2,0.6},lab={t=$U^*$,p={0:0.35}}}}
\atoms{square,xscale=1.5,lab={t=$P$,p=180:0.45}}{p/p={-0.4,1.2}}
\atoms{circ,small}{{c0/p={-0.6,1.6},lab={t=$C$,p={180:0.25}}}, {c1/p={-0.2,1.9},lab={t=$C^*$,p={45:0.35}}}}
\atoms{square,xscale=1.5,lab={t=$O$,p=180:0.45}}{o/p={-0.4,2.4}}
\atoms{circ,all,tiny}{delta/p={0.4,1.6}}
\draw ([sx=-0.6]r.north)--(u0) ([sx=-0.2]r.north)--(u1) ([sx=-0.2]p.south)--(u0) ([sx=0.2]p.south)--(u1) ([sx=-0.2]p.north)--(c0) ([sx=0.2]p.north)--(c1) ([sx=-0.2]o.south)--(c0) ([sx=0.2]o.south)--(c1);
\draw[rc] (p.east)-|(delta) (delta)--++(c0) (delta)|-(c1) ([sx=0.2]r.north)--++(90:0.3)-|([sx=0.6]r.north);
\end{tikzpicture}
};
\node[anchor=east] at (x0.west){(a)};
\node[anchor=east] at (x1.west){(b)};
\node[anchor=east] at (x2.west){(c)};
\node[anchor=east] at (x3.west){(d)};
\node[anchor=east] at (x4.west){(e)};
\node[anchor=east] at (x5.west){(f)};
\node[anchor=east] at (x6.west){(g)};
\node[anchor=east] at (x7.west){(h)};
\end{tikzpicture}
\caption{
Examples of tensors and tensor networks representing objects and computations in quantum physics.
(a) A pure state $\psi$ is a vector, a tensor with one index.
(b) A unitary time-evolution operator is a matrix, a tensor with two indices.
(c) A density matrix is a tensor with two indices.
(d) A quantum channel is a superoperator acting on density matrices, and has four indices (two input and two output indices).
(e) Krauss decomposition of a channel, where $A_i$ are the Krauss operators.
(f) Expectation value of an observable (operator) $O$ yields a tensor-network diagram with no open indices (the result is a number).
(g) Tensor-network equation which states that $U$ is a unitary, $UU^\dagger=1$.
The line on the right-hand side represents the identity operator.
(h) Shows a more intricate example of a computation:
We start with a density matrix $\rho$ over two degrees of freedom (first degree of freedom corresponding to the left two indices), trace out the second degree of freedom, apply a unitary $U$ to the first degree of freedom, perform a measurement instrument $P$ whose right index represents the classical measurement outcome, apply a classically controlled unitary $C$ depending on this measurement outcome, and take the expectation value of the resulting state with respect to an observable $O$.
}
\label{fig:physics_tensors}
\end{figure}
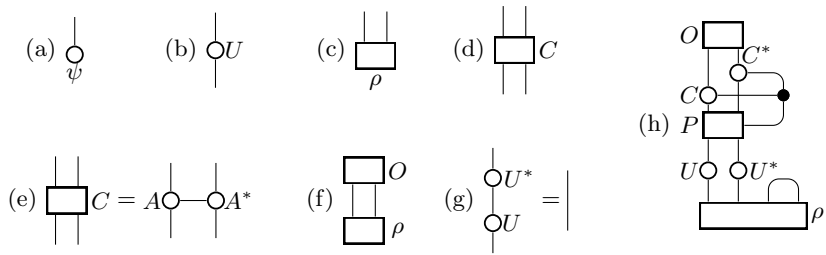

Many objects in quantum physics are tensors, and many computations are tensor networks.
Figure~\ref{fig:physics_tensors} shows a few examples for this.
In particular, quantum circuits consisting of unitaries, channels, or measurements are tensor networks, and circuit diagrams are basically a special case of tensor-network diagrams.
Tensor-network notation is particularly useful if there are multiple degrees of freedom involved.

One important example for an operation that is not directly equal to a tensor network is taking the exponential of a Hamiltonian $H$ to obtain a time evolution operator $U=e^{itH}$.
This is not fundamentally a problem, since we can alternatively describe the time evolution $U$ as a quantum circuit consisting of ``smaller'' unitaries.
Circuits composed of local unitaries and Hamiltonians composed of local terms are two valid kinds of models in many-body physics -- both be used to describe the same physical phenomena.
Further, continuous Hamiltonian evolution can be approximated by a tensor network or circuit using \emph{Trotterization}.

\subsection{Definition}
\label{sec:clifford_definition}
Having reviewed tensors and tensor networks in general, let us now define quadratic tensors, a special type of tensors.

\myparagraph{Quadratic functions}
The essential ingredient of quadratic tensors are 2nd order functions, or in short, quadratic functions.
Quadratic functions are a generalization of 2nd order polynomial functions $q:\rr\rightarrow \rr$, where each instance of $\rr$ is replaced with an arbitrary abelian group.%
\footnote{So when using the term ``quadratic function'', we should keep in mind that we do allow linear and constant terms.}
2nd order polynomials can be characterized by the property that their third derivative vanishes, so we start by defining a general-group analogue of the derivative of functions $q:\rr\rightarrow \rr$.
\begin{mydef}
Consider two abelian groups $G$ and $A$, and a function $q: G\rightarrow A$.
Define the \emph{$0$th derivative} of $q$ as
\begin{equation}
\label{eq:zeroth_derivative}
q^{(0)}\in A:\qquad q^{(0)}\coloneqq q(0)\;,
\end{equation}
the \emph{$1$st derivative} as
\begin{equation}
\label{eq:first_derivative}
\begin{aligned}
q^{(1)}: G &\rightarrow A\\
q^{(1)}(g) &\coloneqq q(g) -q^{(0)} = q(g)-q(0)\;,
\end{aligned}
\end{equation}
the \emph{$2$nd derivative} as
\begin{equation}
\label{eq:second_derivative}
\begin{aligned}
q^{(2)}: G\times G &\rightarrow A\\
q^{(2)}(g_0,g_1) &\coloneqq q^{(1)}(g_0+g_1)-q^{(1)}(g_0)-q^{(1)}(g_1) = q(g_0+g_1)-q(g_0)-q(g_1)+q(0)\;,
\end{aligned}
\end{equation}
and the \emph{$3$rd derivative} as
\begin{equation}
\begin{aligned}
q^{(3)}: G\times G\times G &\rightarrow A\\
q^{(3)}(g_0,g_1,g_2) &= q^{(2)}(g_0+g_1,g_2)-q^{(2)}(g_0,g_2)-q^{(2)}(g_1,g_2)\;.
\end{aligned}
\end{equation}
\end{mydef}
We note that each derivative is symmetric in its arguments.

After introducing derivatives, we define quadratic functions as in the $\rr$-valued case:
A quadratic function is one whose 3rd derivative vanishes.
We also define linear functions, which are the ones whose 2nd derivative vanishes.
\begin{mydef}
We will call a function $q: G\rightarrow A$ \emph{1st order}, or \emph{linear}
\footnote{Again, ``linear'' here is used in the sense of ``affine linear'', as we do allow for constant terms.}
function if
\begin{equation}
\label{eq:1linear_definition}
q^{(2)}(g_0,g_1)=0\quad\forall g_0,g_1\in G\;,
\end{equation}
and \emph{2nd order}, or \emph{quadratic} if
\begin{equation}
\label{eq:2linear_definition}
q^{(3)}(g_0,g_1,g_2)=0\quad\forall g_0,g_1,g_2\in G\;.
\end{equation}
We will write $F_1[G|A]$ and $F_2[G|A]$ for the set of linear and quadratic functions, respectively.
\end{mydef}
$q^{(2)}=0$ iff $q^{(1)}$ is a group homomorphism from $G\rightarrow A$.
So a linear function $q=q^{(1)}+q^{(0)}$ is the same as an affine group homomorphism.
$q$ is quadratic iff $q^{(2)}$ is a bilinear form (see Definition~\ref{def:bilinear_form_definition}).
In this context, quadratic functions $q$ are known in the literature as \emph{quadratic refinements} of the bilinear form $q^{(2)}$~\cite{Quadratic_refinement_nlab}.

\myparagraph{Quadratic tensors}
Using quadratic and linear functions, we can now define quadratic tensors.
We first describe the ``data'' that is necessary to fully specify a quadratic tensor, and then show how the tensor entries can be obtained from this data.
\begin{mydef}
\label{def:quadratic_tensor}
Let $A$ be some abelian group.
A \emph{quadratic tensor data} over an abelian group $G$ is either a triple $(E,\epsilon,q)$, where
\begin{itemize}
\item $E$ is an abelian group,
\item $\epsilon$ is a linear function $\epsilon\in F_1[E|G]$ called the \emph{embedding},
\item $q$ a quadratic function $q\in F_2[E|A]$,%
\end{itemize}
or it is equal to $0$.
Let $\exp\in \hom[A|\cc^\times]$ be a homomorphism from $A$ to the multiplicative group of complex numbers.
The \emph{quadratic tensor} associated with a quadratic tensor data $(E,\epsilon,q)$ is a vector in the vector space of complex functions over the abelian group $G$,
\begin{equation}
\label{eq:quadratic_tensor_definition}
\mathcal T[E,\epsilon,q] : G\rightarrow \cc:\qquad
\mathcal T[E,\epsilon,q](g)\coloneqq \sum_{e\in E: \epsilon(e)=g} \exp(q(e))\;.
\end{equation}
The quadratic tensor associated to $0$ is $\mathcal T[0](g)=0\forall g$.
Two quadratic tensor data $(E,\epsilon,q)$ and $(E',\epsilon',q')$ are considered \emph{equivalent} if they have the same associated quadratic tensor,
\begin{equation}
\label{eq:tensordata_equivalence}
\mathcal T[E',\epsilon',q'] = \mathcal T[E,\epsilon,q]\;.
\end{equation}
\end{mydef}
If not mentioned otherwise, $A$ will be given by its canonical choice $A=\rr\times \rr/\zz$, as this is just the additive version of $\cc^\times$.
The canonical choice for $\exp$ is the isomorphism%
\footnote{For the canonical choice, we could instead use a quadratic function $q\in F_2[E|\cc^\times]$ that directly results in $\cc^\times$ values, and get rid of $A$ and the homomorphism $\exp$.
One of the main reasons for using $A$ is that we prefer additive notation for abelian groups, and it would be confusing to use ``$+$'' for the multiplication in $\cc^\times$.
Another reason is that we will consider $A=\rr/\zz$ with $\exp(\phi)=e^{2\pi i\phi}$ at some point in Section~\ref{sec:reduction}.}
\begin{equation}
\exp((a,\phi))=e^{2\pi(a+i\phi)}\;.
\end{equation}
This is a canonical choice since we do not gain anything by choosing a ``larger'' $A$.
For the canonical choice, we can split $q=(q_a,q_\phi)$ into $q_a\in F_2[E|\rr]$ and $q_\phi\in F_2[E|\rr/\zz]$.%
\footnote{More generally, we may replace $\cc$ by an arbitrary ring $R$, and $\exp$ by an arbitrary homomorphism $A\rightarrow R^\times$, where $R^\times$ is the multiplicative group of $R$.
Most of the mathematical statements in this paper hold for this more general case, but $R=\cc$ with $A=\rr\times \rr/\zz$ is the choice relevant for quantum mechanics.}

\myparagraph{Physics use case}
A quadratic tensor with $n$ indices is one where $G$ is a cartesian product of $n$ factors, $G=\bigtimes_{0\leq i<n} G_i$, even though the definition of a quadratic tensor is agnostic towards this decomposition.
For the physics use case, we restrict ourselves to factors that we call \emph{elementary abelian groups},
\footnote{Equivalently, we could demand that $G$ is a finite-dimensional abelian Lie group.}
\begin{equation}
\label{eq:physical_groups}
G_i\in\{\zz_k\}_{k\in\zz}\cup\{\rr,\zz,\rr/\zz\}\;.
\end{equation}
Each of these elementary abelian groups corresponds to an elementary physical degree of freedom:
\begin{itemize}
\item $G_i=\zz_2$ corresponds to a qubit,
\item $G_i=\zz_k$ corresponds to a $k$-dimensional qudit,
\item $G_i=\rr$ corresponds to a continuous variable, or equivalently, an oscillator, or a bosonic mode,
\item and both $G_i=\rr/\zz$ and $G_i=\zz$ correspond to a rotor ($\rr/\zz$ for the position basis, and $\zz$ for the momentum basis).
\end{itemize}
Analogously, we consider $E$ equal to a product of $m$ elementary abelian groups.

\myparagraph{Injective or surjective $\epsilon$}
If $E=G$ and $\epsilon=\idop$, then $\mathcal T[E,\epsilon,q]$ is just the element-wise exponential of the quadratic function $q$.
If $\epsilon$ is not surjective, then there are elements of $G$ that are not in the image of $\epsilon$, and the corresponding tensor entries are zero.
If $\epsilon$ is not surjective but is injective, then $\mathcal T[E,\epsilon,q]$ is given by a quadratic function on some affine subspace of $G$, and zero outside of the subspace.
If $\epsilon$ is not injective, then there are elements of $G$ whose preimage consists of multiple elements of $E$, so the corresponding tensor entry is given by an exponential sum of $q$ over the preimage.
In fact, we will see in Section~\ref{sec:reduction} that if $E$ only contains $\zz_d$ or $\rr$ factors, every quadratic tensor data is equivalent to one with injective $\epsilon$.
However, for $\rr/\zz$ or $\zz$ factors, the non-injective case can give rise to more general quadratic tensors.

\myparagraph{Analytic considerations}
If $G$ or $E$ contain any infinite factors $\zz$, $\rr/\zz$, or $\rr$, we should make sure that the summation over $e\in E$ in Eq.~\eqref{eq:quadratic_tensor_definition} converges, or describe what to do if it does not.
We will briefly address these analytic aspects here.
In the rest of the paper, we focus on the algebraic properties of quadratic tensors, and may not always be fully explicit about analytic aspects.
First of all, if $\ker(\epsilon^{(1)})$ contains $\rr$ or $\rr/\zz$ factors, then the summation in Eq.~\eqref{eq:quadratic_tensor_definition} is replaced by an integration on these factors.
If $\ker(\epsilon^{(1)})$ contains $\rr$ or $\zz$ factors, then the summation or integration may not converge.
Consider $q_a^{(2)}|_{\ker(\epsilon^{(1)})}$, the second derivative of the quadratic function $q_a$ restricted to the kernel of $\epsilon^{(1)}$.
As we will see in Section~\ref{sec:coefficient_groups}, $q_a^{(2)}|_{\ker(\epsilon^{(1)})}$ is only non-zero on the $\rr$ and $\zz$ factors of $\ker(\epsilon^{(1)})$, where it is described by an $\rr$-valued matrix.
Now, the summation or integration is a Gaussian sum or integral if $q_a^{(2)}|_{\ker(\epsilon^{(1)})}<0$, and thus always converges in this case.
However, we will also consider quadratic tensor data where $q_a^{(2)}|_{\ker(\epsilon^{(1)})}\leq 0$, even if the summation or integration in Eq.~\eqref{eq:quadratic_tensor_definition} does not converge.
Even though these can be considered fringe cases, they occur quite often in analytical calculations in physics.
In these cases, the entries of $\mathcal T[E,\epsilon,q]$ are not given by a function $G\rightarrow \cc$, but a $\delta$-like distribution.
For example, consider the following quadratic tensor data over $G=\rr$,
\begin{equation}
E=\rr\times \rr\;,\quad
\epsilon(e_0,e_1)=e_1\;,\quad
q_a(e_0,e_1)=0\;,\quad
q_\phi(e_0,e_1) =e_0e_1\mmod 1\;.
\end{equation}
Even though integration in Eq.~\eqref{eq:quadratic_tensor_definition} does not converge, the associated tensor can be formally defined as a Dirac $\delta$-distribution,
\begin{equation}
\mathcal T[E,\epsilon,q](g)
=\sum_{e_0,e_1: \epsilon(e_0,e_1)=g} e^{2\pi(q_a(e_0,e_1)+iq_\phi(e_0,e_1))}
= \int_{e_0} e^{2\pi i e_0g}
= \delta_g\;.
\end{equation}
In physics, such a tensor describes a position eigenstate $\ket{x=0}$.
Even though such a state is unphysical, it often shows up in analytical calculations.

Apart from being ill-defined if $\ker(\epsilon^{(1)})$ contains $\rr$ or $\zz$ factors, we would also like to remark that the tensors resulting from Eq.~\eqref{eq:quadratic_tensor_definition} may not always be normalizable in 2-norm.
In particular, the norm is zero if the coimage of $\epsilon$ contains $\rr$ or $\rr/\zz$ factors.
As a minimal example, consider the following quadratic tensor data over $G=\rr$,
\begin{equation}
E=0\;,\quad
\epsilon=0\;,\quad
q_a=0\;,\quad
q_\phi=0\;.
\end{equation}
Here, by $E=0$, we mean the trivial 1-element group.
According to Eq.~\eqref{eq:quadratic_tensor_definition}, the associated tensor entries are given by a function with zero norm,
\begin{equation}
\label{eq:zero_norm_function}
\mathcal T[E,\epsilon,q](g)= \sum_{e\in 0: 0=g} =
\begin{cases}
1&\text{if } g=0\\0&\text{otherwise}
\end{cases}
.
\end{equation}
Further, the norm can be infinite if $q_a^{(2)}\leq 0$ is singular.
As a simple example, consider the following quadratic tensor data over $G=\rr$ with infinite 2-norm,
\begin{equation}
\label{eq:constant_function_example}
E=\rr,\quad
\epsilon=\idop,\quad
q_a=0,\quad
q_\phi=0
\qquad\Rightarrow\quad
\mathcal T[E,\epsilon,q](g)=1\forall g\;.
\end{equation}
In physics, such a tensor corresponds to an eigenstate of the momentum operator, $\ket{p=0}$.
Again, this state as such is unphysical, but occurs quite often in analytical calculations.

Finally, we would like to note that quadratic tensors are always normalizable (in 2-norm) and well defined if $q_a^{(2)}<0$.
Thus, one way to define everything rigorously is to replace each quadratic function $q$ with singular $q_a^{(2)}\leq 0$ by a converging sequence $q_i\xrightarrow{i\rightarrow\infty} q$ with $(q_i)_a^{(2)}<0$.
Then we apply all calculations to the elements in the sequence, and take the limit only at the end of the calculation (e.g., the tensor-network contraction).
The practiced physicist can instead informally keep track of the different infinities occurring in the calculation and cancel them against each other.
For example, the integral over the function in Eq.~\eqref{eq:zero_norm_function} is $1/\infty$, the integral over Eq.~\eqref{eq:constant_function_example} is $\infty$, and their product is $1$, even though this may be hard to justify formally.
Some of the calculations in the paper will involve $|G|$ which is the order of the group $G$ if $G$ is discrete, and the integration measure of $G$ if it is continuous (i.e., $|\rr/\zz|=1$).
For $G=\rr$ or $G=\zz$, the integration measure is $\infty$.

\subsection{Factor decomposition of linear and quadratic functions}
\label{sec:factor_decomposition}
In this section, we show that if $G$ is a direct product of $n$ factors,
\begin{equation}
\label{eq:product_group}
G=\bigtimes_{0\leq i<n} G_i\;,
\end{equation}
then linear or quadratic functions over $G$ can be decomposed into $O(n)$ or $O(n^2)$ smaller components related to individual factors.

\myparagraph{Factor decomposition of linear functions}
Let us start with linear functions.
\begin{myprop}
Let $G$ be a product as in Eq.~\eqref{eq:product_group}.
Then there is a bijection between (1) linear functions $q\in F_1[G|A]$, and (2) collections of
\begin{itemize}
\item a constant $q^{(0)}\in A$, and
\item for each $0\leq i<n$, a homomorphism $q^{(1)}_i\in \hom[G_i|A]$.
\end{itemize}
The collection can be obtained from $q$ as
\begin{equation}
\label{eq:linear_factor_decompositions}
q^{(0)}\coloneqq q^{(0)}\;,\quad
q^{(1)}_i(g)\coloneqq q^{(1)}(g|_i)\;,
\end{equation}
using the notation
\begin{equation}
\label{eq:argument_position_notation}
g|_i \coloneqq (\underbrace{0,\ldots,0}_{0\ldots i-1},\underbrace{g}_i,\underbrace{0,\ldots,0}_{i+1\ldots n-1})\in G\;,\qquad g\in G_i\;.
\end{equation}
Vice versa, $q$ can be obtained from the components above as
\begin{equation}
\label{eq:linear_func_reconstruction}
q(g) = q^{(0)}+\sum_{0\leq i<n} q^{(1)}_i(g_i)\;,
\end{equation}
where $g_i$ denotes the $i$th component of $g$ such that
\begin{equation}
g=(g_0,\ldots,g_{n-1})\;.
\end{equation}
\end{myprop}
\begin{proof}
It is easy to see that $q^{(1)}_i$ as defined in Eq.~\eqref{eq:linear_factor_decompositions} is indeed a homomorphism:
$q^{(1)}$ is a homomorphism since $q$ is linear, and $q^{(1)}_i$ is just a restriction of $q^{(1)}$ to the subgroup
\begin{equation}
\underbrace{0\times\ldots}_{0\ldots i-1}\times \underbrace{G_i}_i\times \underbrace{0\times \ldots}_{i+1\ldots n-1}\subset G\;.
\end{equation}
It is also easy to see that $q$ as defined in Eq.~\eqref{eq:linear_func_reconstruction} is indeed a linear function:
It is a sum of constants and homomorphisms evaluated on subgroups.
Finally we verify Eqs.~\eqref{eq:linear_func_reconstruction} and \eqref{eq:linear_factor_decompositions} are inverses:
\begin{equation}
q^{(0)}+\sum_{0\leq i<n} q^{(1)}_i(g_i)
=q^{(0)}+\sum_{0\leq i<n} q^{(1)}(g_i|_i)
\overset{q^{(2)}=0}{=}q^{(0)}+q^{(1)}(\sum_{0\leq i<n} g_i|_i)
=q^{(0)}+q^{(1)}(g)
=q(g)\;.
\end{equation}
\end{proof}

The embedding $\epsilon$ of a quadratic tensor is a linear function, where $E$ and $G$ play the role of $G$ and $A$ above.
Like $G$, $E$ also may be a product of $m$ factors
\begin{equation}
\label{eq:embedding_product}
E=\bigtimes_{0\leq j<m} E_j\;.
\end{equation}
We can slightly generalize the above proposition to also incorporate the decomposition of $E$.
\begin{myprop}
Let $G$ and $E$ be products as in Eqs.~\eqref{eq:product_group} and \eqref{eq:embedding_product}.
Then there is a bijection between (1) linear functions $\epsilon\in F_1[E|G]$ and (2) collections of
\begin{itemize}
\item for each $0\leq i<n$, a constant $\epsilon_i^{(0)}\in G_i$,
\item for each $0\leq i<n$ and $0\leq j<m$, a homomorphism $\epsilon^{(1)}_{ij}\in \hom[E_j|G_i]$.
\end{itemize}
The collection is obtained from $\epsilon$ as
\begin{equation}
\epsilon^{(0)}_i\coloneqq \epsilon^{(0)}_i\;,\quad
\epsilon^{(1)}_{ij}(e)\coloneqq \epsilon^{(1)}(e|_j)_i\;.
\end{equation}
Vice versa, $\epsilon$ can be obtained from the components above via
\begin{equation}
\epsilon(e)_i = \epsilon^{(0)}_i+\sum_{0\leq j<m} \epsilon^{(1)}_{ij}(e_j)\;.
\end{equation}
\end{myprop}

\myparagraph{Factor decomposition of quadratic functions}
Let us consider quadratic functions next.
The components will be normalized quadratic functions and bilinear forms as we define in the following.
\begin{mydef}
\label{def:bilinear_form_definition}
We call a function $q: G\rightarrow A$ \emph{normalized} if $q(0)=0$.
Normalized linear functions are nothing but group homomorphisms $G\rightarrow A$, so we denote their set by $\hom[G|A]$.
Inspired by this, we denote the set of normalized quadratic functions by $\hom_2[G|A]$.
We call a function $b:G_0\times G_1\rightarrow A$ a \emph{bilinear form} if it is a homomorphism in each component,
\begin{equation}
\begin{gathered}
b(g_0,g_1+g_1')=b(g_0,g_1)+b(g_0,g_1')\;,\quad b(g_0,0)=0\;,\\
b(g_0+g_0',g_1)=b(g_0,g_1)+b(g_0',g_1)\;,\quad b(0,g_1)=0\;.
\end{gathered}
\end{equation}
We denote the set of bilinear forms by $\hom^2[G_0,G_1|A]$.
\end{mydef}
We note that the sets $F_1[G|A]$, $F_2[G|A]$, $\hom[G|A]$, $\hom_2[G|A]$, and $\hom^2[G_0,G_1|A]$ form abelian groups under element-wise addition.

\begin{myprop}
\label{prop:quadratic_function_decomposition}
Let $G$ be a product as in Eq.~\eqref{eq:product_group}.
Then there is a bijection between (1) quadratic functions $q\in F_2[G|A]$ and (2) collections of
\begin{itemize}
\item a constant $q^{(0)}\in A$,
\item for every $0\leq k<n$, a normalized quadratic function $q^{(1)}_k\in \hom_2[G_k|A]$,
\item for every pair $0\leq k<l<n$, a bilinear form $q^{(2)}_{kl}\in \hom^2[G_k,G_l|A]$.
\end{itemize}
The components can be obtained from $q$ via
\begin{equation}
\label{eq:2linear_decomposition}
\begin{gathered}
q^{(0)}\coloneqq q^{(0)}\;,\quad
q^{(1)}_k(g) \coloneqq q^{(1)}(g|_k)\;,\quad
q^{(2)}_{kl}(g,h) \coloneqq q^{(2)}(g|_k,h|_l)\;.
\end{gathered}
\end{equation}
Vice versa, $q$ can be reconstructed from the components as
\begin{equation}
\label{eq:2linear_composition}
q(g) = q^{(0)} + \sum_{0\leq k<n} q^{(1)}_k(g_k) + \sum_{0\leq k<l<n} q^{(2)}_{kl}(g_k,g_l)\;.
\end{equation}
\end{myprop}
\begin{proof}
It is easy to see that $q^{(1)}_k$ as defined in Eq.~\eqref{eq:2linear_decomposition} is indeed a normalized quadratic function:
The first derivative $q^{(1)}$ of $q$ is a normalized quadratic function, and $q^{(1)}_k$ is the restriction to a subgroup.
Similarly, $q^{(2)}_{kl}$ is a bilinear form as it is a restriction of the bilinear form $q^{(2)}\in\hom^2[G,G|A]$ to a subgroup in each of the two arguments.
Further, we need to show that $q$ as defined in Eq.~\eqref{eq:2linear_composition} is a quadratic function.
To this end, we note that every bilinear form $b\in \hom^2[X,Y|Z]$ is also a (normalized) quadratic function $b\in \hom_2[X\times Y|Z]$, whose bilinear form $b^{(2)}$ is in turn given by
\begin{equation}
b^{(2)}((x,y),(x',y'))
=b(x+x',y+y')-b(x,y)-b(x',y')
=b(x,y')+b(x',y)\;.
\end{equation}
So Eq.~\eqref{eq:2linear_composition} is a sum over constants, quadratic functions, and bilinears which are also quadratic functions.
Finally, we show that Eqs.~\eqref{eq:2linear_decomposition} and Eq.~\eqref{eq:2linear_composition} are indeed inverses of another,
\begin{equation}
\begin{gathered}
q^{(0)} +\sum_{0\leq k<n} q_k^{(1)}(g_k) + \sum_{0\leq k<l<n} q_{kl}^{(2)}(g_k,g_l)
\overset{\eqref{eq:2linear_decomposition}}{=} q^{(0)} +\sum_{0\leq k<n} q^{(1)}(g_k|_k) + \sum_{0\leq k<l<n} q^{(2)}(g_k|_k,g_l|_l)\\
\overset{q^{(3)}=0}{=} q^{(0)} +\sum_{0\leq k<n} q^{(1)}(g_k|_k) + \sum_{0\leq k<n} q^{(2)}_k(g_k|_k,\sum_{k<l<n} g_l|_l)
\overset{\eqref{eq:second_derivative}}{=} q^{(0)} +q^{(1)}(\sum_{0\leq k<n} g_k|_k)
\overset{\eqref{eq:first_derivative}}{=} q(\sum_{0\leq k<n} g_k|_k)
= q(g)\;.
\end{gathered}
\end{equation}
\end{proof}
Note that for some applications it is useful to also define $q^{(2)}_{kl}$ for $k>l$ or $k=l$.
These components are determined by the components in Proposition~\ref{prop:quadratic_function_decomposition}, via $q^{(2)}_{kl}(g,h)=q^{(2)}(g|_k,h|_l)=q^{(2)}(h|_l,g|_k)=q^{(2)}_{lk}(h,g)$ for $k>l$, and $q^{(2)}_{ll}(g,h)=q^{(2)}(g|_l,h|_l)=(q^{(1)}_l)^{(2)}(g,h)$ for $k=l$.

\subsection{Coefficient groups}
\label{sec:coefficient_groups}
In the previous Section~\ref{sec:factor_decomposition}, we have shown how to decompose linear and quadratic functions into smaller components related to individual product factors.
The next step is to represent each of these components by an explicit coefficient, in the case where all factors are elementary abelian groups.
That is, for each choice of elementary abelian groups $G$, $G_0$, $G_1$, and $A$, we find elementary abelian groups that we call \emph{coefficient groups},
\begin{equation}
\homtild[G|A],\quad\homtild^2[G_0,G_1|A],\quad\homtild_2[G|A]\;,
\end{equation}
and which are isomorphic to the groups $\hom[G|A]$, $\hom^2[G_0,G_1|A]$, and $\hom_2[G|A]$ via \emph{coefficient isomorphisms}
\begin{equation}
\begin{aligned}
\htild[G|A]&: \homtild[G|A]\rightarrow \hom[G|A]\;,\\
\htild^2[G_0,G_1|A]&: \homtild^2[G_0,G_1|A]\rightarrow \hom^2[G_0,G_1|A]\;,\\
\htild_2[G|A]&: \homtild_2[G|A]\rightarrow \hom_2[G|A]\;.
\end{aligned}
\end{equation}
Often, $G$, $G_0$, $G_1$, and $A$ are clear from the context, in which case we may omit the arguments in square brackets.

\myparagraph{Coefficient group $\homtild$}
Homomorphisms between elementary abelian groups are widely known.
The corresponding coefficient groups $\homtild[G|A]$ are given by
\begin{equation}
\label{eq:elementary_homomorphism_groups}
\begin{tabular}{r|c|c|c|c}
\diagbox{$\scriptstyle{G}$}{$\scriptstyle{A}$}  & $\zz_l$ & $\zz$ & $\rr/\zz$ & $\rr$\\
\hline
$\zz_k$ & $\zz_{\operatorname{gcd}(k,l)}$ & $0$ & $\zz_k$ & $0$\\
\hline
$\zz$ & $\zz_l$ & $\zz$ & $\rr/\zz$ & $\rr$\\
\hline
$\rr/\zz$ & $0$ & $0$ & $\zz$ & $0$\\
\hline
$\rr$ & $0$ & $0$ & $\rr$ & $\rr$
\end{tabular}
\;,
\end{equation}
where $\gcd(k,l)$ denotes the greatest common divisor of two integers $k$ and $l$.
The following table lists the corresponding coefficient isomorphisms $\htild[G|A](h)(g)$ for $h\in \homtild[G|A]$ and $g\in G$,
\begin{equation}
\label{eq:elementary_homomorphisms}
\begin{tabular}{r|c|c|c|c}
\diagbox{$\scriptstyle{G}$}{$\scriptstyle{A}$}  & $\zz_l$ & $\zz$ & $\rr/\zz$ & $\rr$\\
\hline
$\zz_k$ & $\frac{l}{\operatorname{gcd}(k,l)}\ovl h \ovl g\mmod l$ & $0$ & $\frac{1}{k}\ovl h \ovl g\mmod 1$ & $0$\\
\hline
$\zz$ & $\ovl hg\mmod l$ & $hg$ & $\ovl h g \mmod 1$ & $hg$\\
\hline
$\rr/\zz$ & $0$ & $0$ & $h \ovl g\mmod 1$ & $0$\\
\hline
$\rr$ & $0$ & $0$ & $hg\mmod 1$ & $hg$
\end{tabular}
\;.
\end{equation}
Here, $\ovl{\bullet}$ denotes either the embedding of the set $\zz_n$ into $\zz$ as $\{0,\ldots,n-1\}$, or the embedding of $\rr/\zz$ into $\rr$ as $[0,1)$.
All products are either products in $\zz$ or $\rr$, all multiplicative inverses are in $\rr$, and the standard embedding of $\zz$ into $\rr$ as integers is implicit.

\myparagraph{Coefficient group $\homtild^2$}
In order to find the coefficient group for bilinear forms $b\in \hom^2[G_0,G_1|A]$, we make use of an isomorphism
\begin{equation}
\label{eq:currying}
\begin{aligned}
\hom^2[G_0,G_1|A]&\simeq \hom[G_0|\hom[G_1|A]]\;,\\
b(g_0,g_1) &\simeq b(g_0)(g_1)\;.
\end{aligned}
\end{equation}
In other words, for every $g_0\in G_0$, the partial evaluation $b(g_0,\bullet)$ is a homomorphism from $G_1$ to $A$.%
\footnote{This isomorphism exists for arbitrary functions $G_0\times G_1\rightarrow A$, and is known as \emph{currying} in the context of functional programming.}
Using this, we can set
\begin{equation}
\label{eq:bilinear_from_homomorphism}
\homtild^2[G_0,G_1|A] = \homtild[G_0|\homtild[G_1|A]]\;,
\end{equation}
and
\begin{equation}
\label{eq:bilinear_from_homomorphism_iso}
\htild^2[G_0,G_1|A](h)(g_0,g_1)
= \htild[G_1|A]\big(\htild[G_0|\homtild[G_1|A]](h)(g_0)\big)(g_1)
\;,
\end{equation}
perhaps more readable without square brackets as
\begin{equation}
\htild^2(h)(g_0,g_1)
= \htild(\htild(h)(g_0))(g_1)
\;.
\end{equation}
With this, the coefficient groups $\homtild^2[G_0,G_1|A]$ are given by
\begin{equation}
\label{eq:2hom_rr_table}
\begin{tabular}{r||c|c|c|c||c|c|c|c||c|c|c|c||c|c|c|c}
$A$ & \multicolumn{4}{c||}{$\zz_m$}&\multicolumn{4}{c||}{$\zz$} & \multicolumn{4}{c||}{$\rr/\zz$}&\multicolumn{4}{c}{$\rr$}\\
\hline
\diagbox{$\scriptstyle{G_0}$}{$\scriptstyle{G_1}$} &
$\zz_l$ & $\zz$ & $\rr/\zz$ & $\rr$ &
$\zz_l$ & $\zz$ & $\rr/\zz$ & $\rr$ &
$\zz_l$ & $\zz$ & $\rr/\zz$ & $\rr$ &
$\zz_l$ & $\zz$ & $\rr/\zz$ & $\rr$\\
\hline
$\zz_k$ &
$\zz_{\gcd(k,l,m)}$ & $\zz_{\gcd(k,m)}$ & $0$ & $0$ &
$0$ & $0$ & $0$ & $0$ &
$\zz_{\gcd(k,l)}$ & $\zz_k$ & $0$ & $0$ &
$0$ & $0$ & $0$ & $0$\\
\hline
\rowcolor[gray]{.9}[\tabcolsep]
$\zz$ &
$\zz_{\gcd(l,m)}$ & $\zz_m$ & $0$ & $0$ &
$0$ & $\zz$ & $0$ & $0$ &
$\zz_l$ & $\rr/\zz$ & $\zz$ & $\rr$ &
$0$ & $\rr$ & $0$ & $\rr$\\
\hline
$\rr/\zz$ &
$0$ & $0$ & $0$ & $0$ &
$0$ & $0$ & $0$ & $0$ &
$0$ & $\zz$ & $0$ & $0$ &
$0$ & $0$ & $0$ & $0$\\
\hline
\rowcolor[gray]{.9}[\tabcolsep]
$\rr$ &
$0$ & $0$ & $0$ & $0$ &
$0$ & $0$ & $0$ & $0$ &
$0$ & $\rr$ & $0$ & $\rr$ &
$0$ & $\rr$ & $0$ & $\rr$
\end{tabular}
\;,
\end{equation}
and the corresponding coefficient isomorphisms $\htild^2[G_0,G_1|A](h)(g_0,g_1)$ are
\begin{multline}
\label{eq:2hom_iso_rr_table}
\begin{tabular}{r||c|c|c|c||c|c|c|c||l}
$A$ & \multicolumn{4}{c||}{$\zz_m$}&\multicolumn{4}{c||}{$\zz$}&$\ldots$\\
\hline
\diagbox{$\scriptstyle{G_0}$}{$\scriptstyle{G_1}$} &
$\zz_l$ & $\zz$ & $\rr/\zz$ & $\rr$ &
$\zz_l$ & $\zz$ & $\rr/\zz$ & $\rr$ &$\ldots$ \\
\hline
$\zz_k$ &
$\ovl h \frac{m}{\gcd(k,l,m)} \ovl g_0\ovl g_1$ & $\ovl h \frac{m}{\gcd(k,m)} \ovl g_0 g_1$ & $0$ & $0$ &
$0$ & $0$ & $0$ & $0$ &$\ldots$\\
\hline
\rowcolor[gray]{.9}[\tabcolsep]
$\zz$ &
$\ovl h \frac{m}{\gcd(l,m)} g_0\ovl g_1$ & $\ovl h g_0g_1$ & $0$ & $0$ &
$0$ & $hg_0g_1$ & $0$ & $0$ &$\ldots$\\
\hline
$\rr/\zz$ &
$0$ & $0$ & $0$ & $0$ &
$0$ & $0$ & $0$ & $0$&$\ldots$\\
\hline
\rowcolor[gray]{.9}[\tabcolsep]
$\rr$ &
$0$ & $0$ & $0$ & $0$ &
$0$ & $0$ & $0$ & $0$&$\ldots$
\end{tabular}\\
\begin{tabular}{r||c|c|c|c||c|c|c|c|}
$\ldots$ & \multicolumn{4}{c||}{$\rr/\zz$}&\multicolumn{4}{c|}{$\rr$}\\
\hline
$\ldots$&
$\zz_l$ & $\zz$ & $\rr/\zz$ & $\rr$ &
$\zz_l$ & $\zz$ & $\rr/\zz$ & $\rr$\\
\hline
$\ldots$&
$\ovl h\frac{1}{\gcd(k,l)} \ovl g_0 \ovl g_1$ & $\ovl h \frac1k \ovl g_0 g_1$ & $0$ & $0$ &
$0$ & $0$ & $0$ & $0$\\
\hline
\rowcolor[gray]{.9}[\tabcolsep]
$\ldots$&
$\ovl h \frac1l g_0 \ovl g_1$ & $\ovl h g_0 g_1$ & $h g_0 \ovl g_1$ & $h g_0 g_1$ &
$0$ & $h g_0 g_1$ & $0$ & $h g_0 g_1$\\
\hline
$\ldots$&
$0$ & $h \ovl g_0 g_1$ & $0$ & $0$ &
$0$ & $0$ & $0$ & $0$\\
\hline
\rowcolor[gray]{.9}[\tabcolsep]
$\ldots$&
$0$ & $h g_0 g_1$ & $0$ & $h g_0 g_1$ &
$0$ & $h g_0 g_1$ & $0$ & $h g_0 g_1$
\end{tabular}
\;.
\end{multline}
Note that for brevity, we have omitted a ``$\mmod m$'' for every entry of the $A=\zz_m$ block, and a ``$\mmod 1$'' for every entry in the $A=\rr/\zz$ block.
Also note that only the blocks $A=\rr/\zz$ and $A=\rr$ are directly relevant for the purpose of this section, but the other ones occur as coefficients for the embedding $\epsilon$ of 3rd order tensors which we discuss in Section~\ref{sec:ilinear}.
It is easily confirmed that Eqs.~\eqref{eq:2hom_rr_table} and \eqref{eq:2hom_iso_rr_table} result from twice applying Eqs.~\eqref{eq:elementary_homomorphism_groups} and \eqref{eq:elementary_homomorphisms} according to Eqs.~\eqref{eq:bilinear_from_homomorphism} and \eqref{eq:bilinear_from_homomorphism_iso}.

\myparagraph{Coefficient group $\homtild_2$}
To determine the coefficient groups for normalized quadratic functions in $\hom_2[G|A]$, we make use of the group homomorphism
\begin{equation}
\bullet^{(2)}: \hom_2[G|A]\rightarrow \hom^2[G,G|A]\;.
\end{equation}
We start by considering $\img(\bullet^{(2)})$, that is, the subgroup of bilinears $b\in\hom^2[G,G|A]$ for which there exists $q\in \hom_2[G|A]$ such that $q^{(2)}=b$.
It turns out that $\img(\bullet^{(2)})=\hom^2[G,G|A]$ for almost all choices of elementary abelian groups $G$ and $A$.
The only case where $\img(\bullet^{(2)})$ is a proper subgroup is when $G=\zz_k$ and $A=\zz_l$ with both $k$ and $l$ even, and $k$ has at least as many prime factors of $2$ as $l$.

For any two $q_0,q_1\in \hom_2[G|A]$ such that $q_0^{(2)}=q_1^{(2)}=b$, we have
\begin{equation}
(q_0-q_1)^{(2)}=b-b=0\qquad q_0-q_1\in\hom[G|A]\;.
\end{equation}
Vice versa, for every $h\in \hom[G|A]$, $q_0+h$ is a quadratic function with $(q_0+h)^{(2)}=b$.
So if we choose a standard $Q_0(b)\in\hom[G|A]$ such that $Q_0(b)^{(2)}=b$, this identifies the preimage $(\bullet^{(2)})^{-1}(b)$ with $\hom[G|A]$.
Thus, as a set, $\hom_2[G|A]$ can be identified with the cartesian product $\img(\bullet^{(2)})\times \hom[G|A]$.

As a group, $\hom_2[G|A]$ is not in general isomorphic to the direct product above, but rather to some non-trivial group extension of $\img(\bullet^{(2)})$ by $\hom[G|A]$.
We can obtain the group structure as follows.
We find a minimal subgroup $X\subset\hom_2[G|A]$ such that $\bullet^{(2)}$ is surjective when restricted to $X$.
For example, if $\img(\bullet^{(2)})$ is discrete and generated by a single element $b$, and take $X\subset \hom_2[G|A]$ to be the subgroup generated by $Q_0(b)$.
This minimal subgroup $X$ is isomorphic to a single elementary abelian group.
Then we consider the intersection of $X$ with the subgroup $\hom[G|A]\subset\hom_2[G|A]$, $Y\coloneqq X\cap \hom[G|A]\subset\hom[G|A]$.
The quotient $\hom[G|A]/Y$ is again isomorphic to an elementary abelian group.
After this, $\hom_2[G|A]$ can be identified with a direct product of two elementary abelian groups, $\hom_2[G|A]\simeq X\times \hom[G|A]/Y$.
It turns out that in most cases we have $Y=0$, such that $\hom_2[G|A]\simeq \img(\bullet^{(2)})\times \hom[G|A]$ is actually a direct product.
However, for some cases we find $Y=\zz_2$, and the group structure schematically looks like ``$\hom_2[G|A]\simeq 2\img(\bullet^{(2)})\times \frac12\hom[G|A]$''.
For example, for $G=\zz_k$ with $k$ even and $A=\rr/\zz$, we can pick the generating bilinear form $b(g_0,g_1)=\frac1k \ovl g_0\ovl g_1\mmod 1$ with corresponding quadratic function $Q_0(b)(g)=\frac1{2k} \ovl g^2\mmod 1$.
The subgroup $X\subset \hom_2[G|A]$ generated by $Q_0(b)$ is isomorphic to $\zz_{2k}$, and it intersects with $\hom[G|A]$ on a subgroup $Y\simeq \zz_2$:
\begin{equation}
(\underbrace{Q_0(b)+Q_0(b)+\ldots}_{k\text{ times}})(g)=k\frac1{2k}\ovl g^2\mmod 1 = \frac12 \ovl g^2 \mmod 1 = \frac12 \ovl g \mmod 1 \in \hom[G|A]\;.
\end{equation}
So we have $\homtild_2[\zz_k\text{ even}|\rr/\zz]=\zz_{2k}\times \zz_k/\zz_2$.

All in all, the coefficient groups $\homtild_2[G|A]$ are listed in the following table,
\begin{equation}
\label{eq:hom2_groups}
\begin{tabular}{r|c|c|c|c|c}
\diagbox{$\scriptstyle{G}$}{$\scriptstyle{A}$}
 & $\zz_l$ even & $\zz_l$ odd & $\zz$ & $\rr/\zz$ & $\rr$\\
\hline
$\zz_k$ even &$\zz_{2\gcd(k,\frac{l}{2})}\times \zz_{\frac12\gcd(k,l)}$ & $\zz_{\gcd(k,l)}\times \zz_{\gcd(k,l)}$ & $0$ & $\zz_{2k}\times \zz_{\frac{k}2}$ & $0$\\
\hline
\rowcolor[gray]{.9}[\tabcolsep]
$\zz_k$ odd & $\zz_{\gcd(k,l)}\times \zz_{\gcd(k,l)}$ & $\zz_{\gcd(k,l)}\times \zz_{\gcd(k,l)}$ & $0$ & $\zz_k\times \zz_k$ & $0$\\
\hline
$\zz$ & $\zz_l\times \zz_l$ & $\zz_l\times \zz_l$ & $\zz\times\zz$ & $\rr/\zz\times \rr/\zz$ &  $\rr\times \rr$\\
\hline
\rowcolor[gray]{.9}[\tabcolsep]
$\rr/\zz$ & $0$ & $0$ & $0$ & $0\times \zz$ & $0$\\
\hline
$\rr$ & $0$ & $0$ & $0$ & $\rr\times \rr$ & $\rr\times \rr$
\end{tabular}\;.
\end{equation}
The corresponding coefficient isomorphisms $\htild_2[G|A]((h_2,h_1))(g)$ for $(h_2,h_1)=h\in \homtild_2[G|A]$ and $g\in G$ are given by
\begin{equation}
\label{eq:coefficient_iso_2}
\begin{tabular}{r|c|c|c|c|c}
\diagbox{$\scriptstyle{G}$}{$\scriptstyle{A}$}
 & $\zz_l$ even & $\zz_l$ odd & $\zz$ & $\rr/\zz$ & $\rr$\\
\hline
$\zz_k$ even & $\scriptstyle{\frac{l}{\gcd(k,\frac{l}{2})}\frac12 \ovl h_2 \ovl g^2 -\frac{l}{\gcd(k,l)} \ovl h_1 \ovl g^2 + \frac{l}{\gcd(k,l)}\ovl h_1 \ovl g}$ & $\scriptstyle{\frac{l}{\gcd(k,l)}\big(\ovl{(h_2/2)}\ovl g^2 + \ovl{h_1}\ovl g\big)}$ & $0$ & $\scriptstyle{\frac1k(\frac12 \ovl h_2 -\ovl h_1) \ovl g^2 + \frac1k \ovl h_1 \ovl g}$ & 0\\
\hline
\rowcolor[gray]{.9}[\tabcolsep]
$\zz_k$ odd & $\scriptstyle{\frac{l}{\gcd(k,l)}\big(\ovl{(h_2/2)}\ovl g^2 + \ovl{h_1}\ovl g\big)}$ & $\scriptstyle{\frac{l}{\gcd(k,l)}\big(\ovl{(h_2/2)}\ovl g^2 + \ovl{h_1}\ovl g\big)}$ & $0$ & $\scriptstyle{\frac1k\big(\ovl{(h_2/2)}\ovl g^2 + \ovl{h_1}\ovl g\big)}$ & 0\\
\hline
$\zz$ & $\scriptstyle{\frac12 \ovl{h_2} g(g+1) + \ovl{h_1}g}$ & $\scriptstyle{\ovl{(h_2/2)} g^2 + \ovl h_1g}$ & $\scriptstyle{\frac12 h_2 g(g+1) +h_1g}$ & $\scriptstyle{(\ovl{h_2}-\frac12\ovl h_1) g^2 + \frac12\ovl h_1g}$ & $\scriptstyle{\frac12 h_2 g^2+h_1g}$\\
\hline
\rowcolor[gray]{.9}[\tabcolsep]
$\rr/\zz$ & $0$ & $0$ & $0$ & $\scriptstyle{h_1\ovl g}$ & 0\\
\hline
$\rr$ & $0$ & $0$ & $0$ & $\scriptstyle{\frac12 h_2g^2+h_1g}$ & $\scriptstyle{\frac12 h_2 g^2+h_1g}$
\end{tabular}\;.
\end{equation}
To save space we have omitted ``$\mmod l$'' in every entry of the first and second column, and ``$\mmod 1$'' in the fourth column.
$h/2$ denotes the unique element of $\zz_k$ ($k$ odd) such that $h/2+h/2=h$.
It is straightforward but tedious to verify that the above table entries are indeed quadratic functions in the argument $g$, and we do so in Appendix~\ref{sec:2cocycle}.
In computations, we will sometimes need to find the bilinear form $h^{(2)}$ for a normalized quadratic function $h$ on the level of coefficient groups.
The following table lists
\begin{equation}
\htild^2[G,G|A]^{-1}(\htild_2[G|A](h_2,h_1)^{(2)})
\end{equation}
for $(h_2,h_1)\in \homtild_2[G|A]$,
\begin{equation}
\label{eq:coefficient_form_from_function}
\begin{tabular}{r|c|c|c|c|c}
\diagbox{$\scriptstyle{G}$}{$\scriptstyle{A}$}
 & $\zz_l$ even & $\zz_l$ odd & $\zz$ & $\rr/\zz$ & $\rr$\\
\hline
$\zz_k$ even & $(\frac{\gcd(k,l)}{\gcd(k,\frac{l}2)}\ovl h_2-2\ovl h_1) \mmod \gcd(k,l)$ & $h_2$ & $0$ & $(\ovl h_2 -2\ovl h_1) \mmod k$ & 0\\
\hline
\rowcolor[gray]{.9}[\tabcolsep]
$\zz_k$ odd & $h_2$ & $h_2$ & $0$ & $h_2$ & 0\\
\hline
$\zz$ & $h_2$ & $h_2$ & $h_2$ & $(2\ovl h_2-\ovl h_1) \mmod 1$ & $h_2$\\
\hline
\rowcolor[gray]{.9}[\tabcolsep]
$\rr/\zz$ & $0$ & $0$ & $0$ & $0$ & 0\\
\hline
$\rr$ & $0$ & $0$ & $0$ & $h_2$ & $h_2$
\end{tabular}\;.
\end{equation}
It is also often handy to find the inclusion $\hom[G|A]\subset\hom_2[G|A]$ in terms of coefficient groups:
The following table lists
\begin{equation}
\htild_2[G|A]^{-1}(\htild[G|A](h))
\end{equation}
for $h\in\homtild[G|A]$,
\begin{equation}
\label{eq:1in2_inclusion_coefficients}
\begin{tabular}{r|rl|c|c|rl|c}
\diagbox{$\scriptstyle{G}$}{$\scriptstyle{A}$}
 & \multicolumn{2}{c|}{$\zz_l$ even} & $\zz_l$ odd & $\zz$ & \multicolumn{2}{c|}{$\rr/\zz$} & $\rr$\\
\hline
$\zz_k$ even & $\frac{2\gcd(k,\frac{l}2)}{\gcd(k,l)} \ovl h \mmod 2\gcd(k,\frac{l}2),$ & $\ovl h \mmod \frac12\gcd(k,l)$ & $0,h$ & $0$ & $2\ovl h\mmod 2k,$ & $\ovl h\mmod \frac{k}2$ & 0\\
\hline
\rowcolor[gray]{.9}[\tabcolsep]
$\zz_k$ odd & $0,$ & $h$ & $0,h$ & $0$ & $0,$ & $h$ & 0\\
\hline
$\zz$ & $0,$ & $h$ & $0,h$ & $0,h$ & $h,$ & $2\ovl h\mmod 1$ & $0,h$\\
\hline
\rowcolor[gray]{.9}[\tabcolsep]
$\rr/\zz$ & \multicolumn{2}{c|}{$0$} & $0$ & $0$ & $0,$ & $h$ & 0\\
\hline
$\rr$ & \multicolumn{2}{c|}{$0$} & $0$ & $0$ & $0,$ & $h$ & $0,h$
\end{tabular}\;.
\end{equation}
Eqs.~\eqref{eq:coefficient_form_from_function} and \eqref{eq:1in2_inclusion_coefficients} can be used to show that the coefficient groups in Eq.~\eqref{eq:hom2_groups} do contain all normalized quadratic functions.
In Appendix~\ref{sec:2cocycle}, we explicitly list a choice of the standard function $Q_0(b)$, and we describe $\hom_2[G|A]$ as a product of $\img(\bullet^{(2)})$ and $\hom[G|A]$ twisted by a group 2-cocycle $\Omega$.

\myparagraph{Putting it all together}
Let us now summarize the results from this and the previous section and spell out explicitly what coefficients are necessary to specify $q$ and $\epsilon$ of a quadratic tensor data where $G$ is a product of $n$ elementary abelian groups $G_i$ and $E$ is a product of $m$ elementary abelian groups $E_j$.

The quadratic function $q\in F_2[E|\rr\times\rr/\zz]$ is fully specified by
\begin{itemize}
\item $q_a^{(\tilde0)}\coloneqq q_a^{(0)}\in \rr$ and $q_\phi^{(\tilde0)}\coloneqq q_\phi^{(0)}\in \rr/\zz$.
\item $q^{(\tilde1)}_{a,i}\coloneqq \htild_2^{-1}((q_a)^{(1)}_i) \in \homtild_2[E_i|\rr]$ and $q^{(\tilde1)}_{\phi,i}\coloneqq \htild_2^{-1}((q_\phi)^{(1)}_i) \in \homtild_2[E_i|\rr/\zz]$ for every $0\leq i<m$, each of which consists of two elementary abelian coefficients in general.
In practice, we arrange these coefficients into length-$m$ vectors which we call $q_a^{(\tilde1)}$ and $q_\phi^{(\tilde1)}$.
\item $q^{(\tilde2)}_{a,ij}\coloneqq (\htild^2)^{-1}((q_a)^{(2)}_{ij})\in \homtild^2[E_i,E_j|\rr]$ and $q^{(\tilde2)}_{\phi,ij}\coloneqq (\htild^2)^{-1}((q_\phi)^{(2)}_{ij})\in \homtild^2[E_i,E_j|\rr/\zz]$ for every $0\leq i<j<m$.
In practice, we arrange these coefficients into strictly upper triangular $m\times m$ matrices which we call $q_a^{(\tilde2)}$ and $q_\phi^{(\tilde2)}$.
We omit the entries on and below the diagonal, such that the matrix looks like a non-strictly-upper triangular $(m-1)\times (m-1)$ matrix.
\end{itemize}
Note that we can turn the upper triangular matrix $q^{(\tilde2)}$ into a full symmetric matrix.
The diagonal entries $q^{(\tilde2)}_{ii}$ are determined by $q^{(\tilde1)}_i$ via Eq.~\eqref{eq:coefficient_form_from_function}.

The linear function $\epsilon\in F_1[E|G]$ is fully specified by
\begin{itemize}
\item $\epsilon^{(\tilde0)}_i\coloneqq \epsilon^{(0)}_i\in G_i$ for every $0<i<n$, which we arrange into a length-$n$ vector $\epsilon^{(\tilde0)}$.
\item $\epsilon^{(\tilde1)}_{ij}\coloneqq \htild^{-1}(\epsilon^{(1)}_{ij})\in \homtild[E_j|G_i]$, which we arrange into an $n\times m$ matrix $\epsilon^{(\tilde1)}$.
\end{itemize}

\subsection{Homomorphism composition}
\label{sec:homomorphism_composition}
In this section, we show how to compute the composition of a quadratic function with a homomorphism in terms of the underlying coefficients.
This will be necessary for computing reductions as in Section~\ref{sec:reduction}.

\begin{myprop}
\label{prop:homomorphism_composition_quadratic}
Consider a quadratic function $q\in F_2[G|A]$, and a homomorphism $\gamma\in \hom[H|G]$.
The precomposition of $q$ with $\gamma$,
\begin{equation}
q\circ\gamma: H\rightarrow A\;,\quad (q\circ\gamma)(h)\coloneqq q(\gamma(h))\;,
\end{equation}
is again a quadratic function, $q\circ\gamma\in F_2[H|A]$.
\end{myprop}
\begin{proof}
We compute
\begin{equation}
\label{eq:precomposition_is_quadratic}
\begin{gathered}
(q\circ\gamma)^{(2)}(h,h')
= q(\gamma(h+h'))-q(\gamma(h))-q(\gamma(h'))+q(0)
= q(\gamma(h)+\gamma(h'))-q(\gamma(h))-q(\gamma(h'))+q(0)\\
=q^{(2)}(\gamma(h),\gamma(h'))
=(q^{(2)}\circ(\gamma\times\gamma))(h,h')\;.
\end{gathered}
\end{equation}
So $q\circ\gamma$ is indeed a quadratic function with associated bilinear form $q^{(2)}\circ(\gamma\times \gamma)$.
\end{proof}

Let us now assume that $H$ and $G$ are products of many factors, and $q$ and $\gamma$ are given in terms of their components as in Section~\ref{sec:factor_decomposition}.
We note that $\gamma$ is a homomorphism, which is a linear function with $\gamma^{(0)}=0$, so $\gamma$ is specified by the components $\gamma^{(1)}_{ij}$ alone, which we simply denote by $\gamma_{ij}$.
\begin{myprop}
The components of $q\circ\gamma$ are given by
\begin{equation}
\label{eq:precomposition_factoring}
\begin{gathered}
(q\circ\gamma)^{(2)}_{ij}
= \sum_{k,l} q^{(2)}_{kl} \circ (\gamma_{ki}\times \gamma_{lj})\;,\\
(q\circ\gamma)^{(1)}_i(h)
=\sum_k (q^{(1)}_k\circ\gamma_{ki})(h) + \sum_{l<k} q^{(2)}_{kl}\circ (\gamma_{ki}\times \gamma_{li})(h,h)\;,\\
(q\circ\gamma)^{(0)}
=q^{(0)}\;.
\end{gathered}
\end{equation}
\end{myprop}
\begin{proof}
For the first equation, we find
\begin{equation}
\begin{gathered}
(q\circ\gamma)^{(2)}_{ij}(h,h')
=(q\circ\gamma)^{(2)}(h|_i,h'|_j)
=q^{(2)}(\gamma(h|_i),\gamma(h'|_j))\\
=\sum_{k,l} q^{(2)}(\gamma_{ki}(h)|_k,\gamma_{lj}(h')|_l)
= \sum_{k,l} q^{(2)}_{kl} \circ (\gamma_{ki}\times \gamma_{lj})(h,h')\;.
\end{gathered}
\end{equation}
For the second equation, we find
\begin{equation}
\begin{gathered}
(q\circ\gamma)^{(1)}_i(h)
=(q\circ\gamma)^{(1)}(h|_i)
=q^{(1)}(\gamma(h|_i))
=q^{(1)}(\sum_k \gamma_{ki}(h)|_k)\\
=\sum_k q^{(1)}(\gamma_{ki}(h)|_k) + \sum_{l<k} q^{(2)}(\gamma_{ki}(h)|_k, \gamma_{li}(h)|_l)
=\sum_k q^{(1)}_k(\gamma_{ki}(h)) + \sum_{l<k} q^{(2)}_{kl}(\gamma_{ki}(h), \gamma_{li}(h))\;.
\end{gathered}
\end{equation}
The third equation is trivial,
\begin{equation}
(q\circ\gamma)^{(0)}=(q\circ\gamma)(0)=q(\gamma(0))=q(0)=q^{(0)}\;.
\end{equation}
\end{proof}

Finally, let us assume that all the factors of $H$ and $G$ are elementary abelian groups, and that all the components of $q$ and $\gamma$ are given as explicit coefficients as described in Section~\ref{sec:coefficient_groups}.
Note that we will write $\tilde\gamma_{ij}$ instead of $\gamma^{(\tilde1)}_{ij}$, as $\gamma^{(\tilde0)}_{ij}$ is trivial.
We will need the following notion of a dual.
\begin{mydef}
\label{def:duals}
For each abelian group $G$, define the \emph{dual} with respect to an abelian group $A$ as%
\footnote{If $A=\rr/\zz$, this is known as the \emph{Pontryagin dual} and we have $G^{**}\simeq G$.
For general $A$, we may have $G^{**}\not\simeq G$, so the name ``dual'' is maybe not entirely justified.
Instead, $G$ can be identified with a subgroup of $G^{**}$, $G\subset G^{**}$ via the injective homomorphism
\begin{equation*}
\eval\in\hom[G|G^{**}]:\qquad
(\eval(g))(g')=g'(g)\quad \forall g'\in G^*,\quad g\in G\;.
\end{equation*}
}
\begin{equation}
G^*\coloneqq \hom[G|A]\;.
\end{equation}
For each homomorphism $\gamma\in\hom[H|G]$, define the \emph{dual} as
\begin{equation}
\label{eq:dual_definition}
\gamma^*\in\hom[G^*|H^*]:\qquad
\gamma^*(g')(h) = g'(\gamma(h))\quad\forall g'\in G^*,\quad h\in H\;.
\end{equation}
\end{mydef}
\begin{myprop}
Let $H$ and $G$ be products of elementary abelian groups, $A$ an elementary abelian group, $\gamma\in\hom[H|G]$, and $q\in F_2[G|A]$.
Then the coefficients of $q\circ\gamma$ can be calculated from these of $q$ as follows
\begin{equation}
\label{eq:precomposition_coefficients}
\begin{gathered}
(q\circ\gamma)^{(\tilde2)}_{ij}
= \sum_{k,l} \tilde\gamma_{ki}^{\tilde*} \circtild q^{(\tilde2)}_{kl} \circtild \tilde\gamma_{lj}\;,\\
(q\circ\gamma)^{(\tilde1)}_i
=\sum_k \Phi(q^{(\tilde1)}_k,\tilde\gamma_{ki}) + \sum_{l<k} \Lambda(\tilde\gamma_{ki}^{\tilde*} \circtild q^{(\tilde2)}_{kl}\circtild \tilde\gamma_{li})\;,\\
(q\circ\gamma)^{(\tilde0)}
=q^{(\tilde0)}\;.
\end{gathered}
\end{equation}
Here, $h'\circtild h$ for $h\in\homtild[G_0|G_1]$ and $h'\in \homtild[G_1|G_2]$ for elementary abelian groups $G_0$, $G_1$, and $G_2$ denotes the composition of homomorphisms conjugated by the coefficient isomorphism $\htild$,
\begin{equation}
\label{eq:coefficient_composition}
h\circtild h'\coloneqq  \htild[G_0|G_2]^{-1}\big(\htild[G_1|G_2](h')\circ \htild[G_0|G_1](h)\big)\in\homtild[G_0|G_2]\;.
\end{equation}
It is listed for different $G_0$, $G_1$, and $G_2$ below in Eq.~\eqref{eq:composition_table}.
In order to apply this, we note that by definition in Eq.~\eqref{eq:bilinear_from_homomorphism}, we have
\begin{equation}
\label{eq:homomorphism_composition_curry}
q^{(\tilde2)}_{kl}\in \homtild^2[G_k,G_l|A] =\homtild[G_k|\homtild[G_l|A]]\;.
\end{equation}
$\gamma^{\tilde*}$ for $\gamma\in\homtild[H_0|G_0]$ denotes the dual of a homomorphism conjugated with the coefficient isomorphism $\htild$,
\begin{equation}
\gamma^{\tilde*}=\htild^{-1}[\homtild[G_0|A]|\homtild[H_0|A]]\big(\htild[H_0|G_0](\gamma)^*\big)\;.
\end{equation}
It is listed below in Eq.~\eqref{eq:dual_coefficient_table}.
$\Phi(q,\gamma)$ for $q\in \hom_2[G_0|A]$ and $\gamma\in \hom[H_0|G_0]$ and elementary groups $G_0$, $H_0$ and $A$ is given by the composition of $q$ with $\gamma$, conjugated by the coefficient isomorphisms $\htild$ and $\htild_2$,
\begin{equation}
\label{eq:precomposition_phi}
\Phi[G_0,H_0|A](q,\gamma) \coloneqq \htild_2[H_0|A]^{-1}\big(\htild_2[G_0|A](q)\circ \htild[H_0|G_0](\gamma)\big)\in \homtild_2[H_0|A]\;.
\end{equation}
It is listed for different $G_0$, $H_0$, and $A$ in Eqs.~\eqref{eq:hom21_precomposition} and \eqref{eq:hom21_precomposition_rr}.
Last, we can turn a bilinear form $b\in\hom^2[G_0,G_0|A]$ into a quadratic function $q\in\hom_2[G_0|A]$ by taking twice the same argument, $q(g)=b(g,g)=b\circ\operatorname{Copy}(g)$ or $q=b\circ\operatorname{Copy}$.
$\Lambda(h)$ for $h\in \homtild^2[G_0,G_0|A]$ denotes that same operation conjugated by the coefficient isomorphisms,
\begin{equation}
\label{eq:lambda_definition}
\Lambda[G_0|A](h)\coloneqq \htild_2[G_0|A]^{-1}\big(\htild^2[G_0,G_0|A](h)\circ\operatorname{Copy}\big)\;.
\end{equation}
$\Lambda$ is listed for different $G_0$ and $A$ in Eq.~\eqref{eq:precomposition_lambda_table}.
\end{myprop}
\begin{proof}
This follows immediately from conjugating both sides of Eq.~\eqref{eq:precomposition_factoring} by the coefficient isomorphisms $\htild$, $\htild^2$, and $\htild_2$.
Thereby, we note that we can rewrite
\begin{equation}
q^{(2)}_{kl} \circ (\gamma_{ki}\times \gamma_{lj})
=\gamma_{lj}^* \circ q^{(2)}_{kl} \circ \gamma_{ki}\;,
\end{equation}
using the identification in Eq.~\eqref{eq:homomorphism_composition_curry}.
\end{proof}
We now give the explicit formulas for $\circtild$, $\bullet^{\tilde*}$, $\Phi$, and $\Lambda$ used in the above proposition.
We start with the composition $\circtild$ from Eq.~\eqref{eq:coefficient_composition}, which we list for all triples $G_0$, $G_1$, and $G_2$ such that all of $\hom[G_0|G_1]$, $\hom[G_1|G_2]$, and $\hom[G_0|G_2]$ are non-zero:
\begin{equation}
\label{eq:composition_table}
\begin{tabular}{l|l|l|l|l|l|l}
$G_0$ & $G_1$ & $G_2$ & $\homtild[G_0|G_1]$ & $\homtild[G_1|G_2]$ & $\homtild[G_0|G_2]$ & $h\cdot h'$\\
\hline
$\zz$ & $\zz$ & $\zz$ & $\zz$ & $\zz$ & $\zz$ & $hh'$\\\hline
\rowcolor[gray]{.9}[\tabcolsep]
$\zz$ & $\zz$ & $\rr/\zz$ & $\zz$ & $\rr/\zz$ & $ \rr/\zz$ & $h\ovl h'\mmod 1$\\\hline
$\zz$ & $\zz$ & $\zz_k$ & $\zz$ & $\zz_k$ & $\zz_k$ & $h\ovl h'\mmod k$\\\hline
\rowcolor[gray]{.9}[\tabcolsep]
$\zz$ & $\zz$ & $\rr$ & $\zz$ & $\rr$ & $\rr$ & $hh'$\\\hline
$\zz$ & $\rr/\zz$ & $\rr/\zz$ & $\rr/\zz$ & $\zz$ & $\rr/\zz$ & $\ovl hh'\mmod 1$\\\hline
\rowcolor[gray]{.9}[\tabcolsep]
$\zz$ & $\zz_k$ & $\zz_l$ & $\zz_k$ & $\zz_{\gcd(k,l)}$ & $\zz_l$ & $\ovl h\ovl h'\frac{l}{\gcd(k,l)}\mmod l$\\\hline
$\zz$ & $\rr$ & $\rr$ & $\rr$ & $\rr$ & $\rr$ & $hh'$\\\hline
\rowcolor[gray]{.9}[\tabcolsep]
$\zz$ & $\zz_k$ & $\rr/\zz$ & $\zz_k$ & $\zz_k$ & $\rr/\zz$ & $\frac1k \ovl h\ovl h'\mmod 1$\\\hline
$\zz$ & $\rr$ & $\rr/\zz$ & $\rr$ & $\rr$ & $\rr/\zz$ & $h h'\mmod 1$\\\hline
\rowcolor[gray]{.9}[\tabcolsep]
$\zz_k$ & $\zz_l$ & $\zz_m$ & $\zz_{\gcd(k,l)}$ & $\zz_{\gcd(l,m)}$ & $\zz_{\gcd(k,m)}$ & $\ovl h\ovl h' \frac{l\gcd(k,m)}{\gcd(k,l)\gcd(l,m)} \mmod\gcd(k,m)$\\\hline
$\zz_k$ & $\zz_l$ & $\rr/\zz$ & $\zz_{\gcd(k,l)}$ & $\zz_l$ & $\zz_k$ & $\ovl h\ovl h'\frac{k}{\gcd(k,l)}\mmod k$\\\hline
\rowcolor[gray]{.9}[\tabcolsep]
$\zz_k$ & $\rr/\zz$ & $\rr/\zz$ & $\zz_k$ & $\zz$ & $\zz_k$ & $\ovl hh'\mmod k$\\\hline
$\rr$ & $\rr$ & $\rr$ & $\rr$ & $\rr$ & $\rr$ & $hh'$\\\hline
\rowcolor[gray]{.9}[\tabcolsep]
$\rr$ & $\rr$ & $\rr/\zz$ & $\rr$ & $\rr$ & $\rr$ & $hh'$\\\hline
$\rr$ & $\rr/\zz$ & $\rr/\zz$ & $\rr$ & $\zz$ & $\rr$ & $hh'$\\\hline
\rowcolor[gray]{.9}[\tabcolsep]
$\rr/\zz$ & $\rr/\zz$ & $\rr/\zz$ & $\zz$ & $\zz$ & $\zz$ & $hh'$
\end{tabular}
\;.
\end{equation}
Next, we give $\gamma^{\tilde*}$ for different $G=G_0$, $H=H_0$, and $A$, where $\gamma\in\homtild[H|G]$.
The following table lists entries of the form $\homtild[H|G]\rightarrow \gamma^{\tilde*}\in\homtild[\homtild[G|A]|\homtild[H|A]]$:
\begin{multline}
\label{eq:dual_coefficient_table}
\begin{tabular}{r||c|c|c|c||c|c|c|c|l}
$A$ & \multicolumn{4}{c|}{$\zz_m$} & \multicolumn{4}{c|}{$\zz$} & $\ldots$\\
\hline
\diagbox{$\scriptstyle{H}$}{$\scriptstyle{G}$} &
$\zz_l$ & $\zz$ & $\rr/\zz$ & $\rr$ &
$\zz_l$ & $\zz$ & $\rr/\zz$ & $\rr$ &$\ldots$ \\
\hline
$\zz_k$ &
$\scriptstyle{\zz_{\gcd(k,l)}\rightarrow\ovl\gamma \mmod \gcd(k,l,m)\in \zz_{\gcd(k,l,m)}}$ & $\scriptstyle{0\rightarrow 0\in \zz_{\gcd(k,m)}}$ & $\scriptstyle{\zz_k\rightarrow 0\in 0}$ & $\scriptstyle{0\rightarrow 0\in 0}$ &
$\scriptstyle{\zz_{\gcd(k,l)}\rightarrow 0\in 0}$ & $\scriptstyle{0\rightarrow 0\in 0}$ & $\scriptstyle{\zz_k\rightarrow 0\in 0}$ & $\scriptstyle{0\rightarrow 0\in 0}$ & $\ldots$\\
\hline
\rowcolor[gray]{.9}[\tabcolsep]
$\zz$ &
$\scriptstyle{\zz_l\rightarrow \ovl\gamma\mmod \gcd(l,m) \in \zz_{\gcd(l,m)}}$ & $\scriptstyle{\zz\rightarrow \gamma\mmod m\in \zz_m}$ & $\scriptstyle{\rr/\zz\rightarrow 0\in 0}$ & $\scriptstyle{\rr\rightarrow0\in 0}$ &
$\scriptstyle{\zz_l\rightarrow 0\in 0}$ & $\scriptstyle{\zz\rightarrow \gamma\in\zz}$ & $\scriptstyle{\rr/\zz\rightarrow 0\in 0}$ & $\scriptstyle{\rr\rightarrow 0\in 0}$ & $\ldots$\\
\hline
$\rr/\zz$ &
$\scriptstyle{0\rightarrow 0\in 0}$ & $\scriptstyle{0\rightarrow 0\in 0}$ & $\scriptstyle{\zz\rightarrow 0\in 0}$ & $\scriptstyle{0\rightarrow 0\in 0}$ &
$\scriptstyle{0\rightarrow 0\in 0}$ & $\scriptstyle{0\rightarrow 0\in 0}$ & $\scriptstyle{\zz\rightarrow 0\in 0}$ & $\scriptstyle{0\rightarrow 0\in 0}$ & $\ldots$\\
\hline
\rowcolor[gray]{.9}[\tabcolsep]
$\rr$ &
$\scriptstyle{0\rightarrow 0\in 0}$ & $\scriptstyle{0\rightarrow 0\in 0}$ & $\scriptstyle{\rr\rightarrow 0\in 0}$ & $\scriptstyle{\rr\rightarrow 0\in 0}$ &
$\scriptstyle{0\rightarrow 0\in 0}$ & $\scriptstyle{0\rightarrow 0\in 0}$ & $\scriptstyle{\rr\rightarrow 0\in 0}$ & $\scriptstyle{\rr\rightarrow 0\in 0}$ & $\ldots$
\end{tabular}\\
\begin{tabular}{r|c|c|c|c||c|c|c|c}
$\ldots$ & \multicolumn{4}{c||}{$\rr/\zz$}&\multicolumn{4}{c}{$\rr$}\\
\hline
\vphantom{\diagbox{$\scriptstyle{H}$}{$\scriptstyle{G}$}}$\ldots$&
$\zz_l$ & $\zz$ & $\rr/\zz$ & $\rr$ &
$\zz_l$ & $\zz$ & $\rr/\zz$ & $\rr$\\
\hline
$\ldots$ &
$\scriptstyle{\zz_{\gcd(k,l)}\rightarrow\gamma\in\zz_{\gcd(k,l)}}$ & $\scriptstyle{0\rightarrow 0\in 0}$ & $\scriptstyle{\zz_k\rightarrow \gamma\in \zz_k}$ & $\scriptstyle{0\rightarrow 0\in 0}$ &
$\scriptstyle{\zz_{\gcd(k,l)}\rightarrow 0\in 0}$ & $\scriptstyle{0\rightarrow 0\in 0}$ & $\scriptstyle{\zz_k\rightarrow 0\in 0}$ & $\scriptstyle{0\rightarrow 0\in 0}$\\
\hline
\rowcolor[gray]{.9}[\tabcolsep]
$\ldots$ &
$\scriptstyle{\zz_l\rightarrow \gamma\in \zz_l}$ & $\scriptstyle{\zz\rightarrow \gamma\in \zz}$ & $\scriptstyle{\rr/\zz\rightarrow \gamma\in\rr/\zz}$ & $\scriptstyle{\rr\rightarrow\gamma\in\rr}$ &
$\scriptstyle{\zz_l\rightarrow 0\in 0}$ & $\scriptstyle{\zz\rightarrow\gamma\in\rr}$ & $\scriptstyle{\rr/\zz\rightarrow 0\in 0}$ & $\scriptstyle{\rr\rightarrow\gamma\in\rr}$\\
\hline
$\ldots$ &
$\scriptstyle{0\rightarrow 0\in 0}$ & $\scriptstyle{0\rightarrow 0\in 0}$ & $\scriptstyle{\zz\rightarrow \gamma\in\zz}$ & $\scriptstyle{0\rightarrow 0\in 0}$ &
$\scriptstyle{0\rightarrow 0\in 0}$ & $\scriptstyle{0\rightarrow 0\in 0}$ & $\scriptstyle{\zz\rightarrow 0\in 0}$ & $\scriptstyle{0\rightarrow 0\in 0}$\\
\hline
\rowcolor[gray]{.9}[\tabcolsep]
$\ldots$ &
$\scriptstyle{0\rightarrow 0\in 0}$ & $\scriptstyle{0\rightarrow 0\in 0}$ & $\scriptstyle{\rr\rightarrow \gamma\in\rr}$ & $\scriptstyle{\rr\rightarrow \gamma\in\rr}$ &
$\scriptstyle{0\rightarrow 0\in 0}$ & $\scriptstyle{0\rightarrow 0\in \rr}$ & $\scriptstyle{\rr\rightarrow 0\in 0}$ & $\scriptstyle{\rr\rightarrow\gamma\in\rr}$
\end{tabular}
\;.
\end{multline}
Next, we give $\Phi[G,H|A]((h_2,h_1),\gamma)$ from Eq.~\eqref{eq:precomposition_phi} for different elementary abelian groups $H$, $G$, and $A$, where $(h_2,h_1)\in\homtild_2[G|A]$ and $\gamma\in \homtild[H|G]$.
We restrict to $A=\rr/\zz$ and $A=\rr$, which are cases directly relevant for the purposes of this section.
For $A=\rr/\zz$, we find
\begin{multline}
\label{eq:hom21_precomposition}
\begin{tabular}{r|rl|rl|l}
\diagbox{$\scriptstyle{H}$}{$\scriptstyle{G}$}
 & \multicolumn{2}{c|}{$\zz_k$ even} & \multicolumn{2}{c|}{$\zz_k$ odd} & $\ldots$\\
\hline
$\zz_m$ even &$\scriptstyle{(\ovl h_2-2\ovl h_1)\ovl\gamma^2 \frac{mk}{\gcd(m,k)^2} + 2\ovl h_1\ovl\gamma\frac{m}{\gcd(m,k)},}$ & $\scriptstyle{\ovl h_1\ovl\gamma \frac{m}{\gcd(m,k)}}$ & $\scriptstyle{2\ovl{(h_2/2)}\ovl\gamma^2 \frac{mk}{\gcd(m,k)^2} + 2\ovl h_1\ovl\gamma\frac{m}{\gcd(m,k)},}$ & $\scriptstyle{\ovl h_1\ovl\gamma \frac{m}{\gcd(m,k)}}$ & $\ldots$\\
\hline
\rowcolor[gray]{.9}[\tabcolsep]
$\zz_m$ odd & $\scriptstyle{(\ovl h_2-2\ovl h_1)\ovl\gamma^2 \frac{mk}{\gcd(m,k)^2},}$ & $\scriptstyle{\ovl h_1\ovl\gamma \frac{m}{\gcd(m,k)}}$ & $\scriptstyle{2\ovl{(h_2/2)}\ovl\gamma^2 \frac{mk}{\gcd(m,k)^2},}$ & $\scriptstyle{\ovl h_1\ovl\gamma \frac{m}{\gcd(m,k)}}$ & $\ldots$\\
\hline
$\zz$ & $\scriptstyle{\frac1k(\frac12 \ovl h_2-\ovl h_1) \ovl\gamma^2+\ovl h_1\ovl\gamma,}$ & $\scriptstyle{\frac2k \ovl h_1 \ovl\gamma}$ & $\scriptstyle{\frac1k\big(\ovl{(h_2/2)}\ovl\gamma^2 + \ovl h_1\ovl\gamma\big),}$ & $\scriptstyle{\frac2k\ovl h_1\ovl\gamma}$ & $\ldots$\\
\hline
\rowcolor[gray]{.9}[\tabcolsep]
$\rr/\zz$ & \multicolumn{2}{c|}{$0$} & \multicolumn{2}{c|}{$0$} & $\ldots$\\
\hline
$\rr$ & \multicolumn{2}{c|}{$0$} & \multicolumn{2}{c|}{$0$} & $\ldots$\\
\end{tabular}\\
\begin{tabular}{r|rl|rl|rl}
\vphantom{\diagbox{$\scriptstyle{H}$}{$\scriptstyle{G}$}}$\ldots$
 & \multicolumn{2}{c|}{$\zz$} & \multicolumn{2}{c|}{$\rr/\zz$} & \multicolumn{2}{c}{$\rr$}\\
\hline
$\ldots$ & \multicolumn{2}{c|}{$0$} & $\scriptstyle{\frac2m h_1 \ovl\gamma,}$ & $\scriptstyle{\frac1m h_1 \ovl\gamma}$ & \multicolumn{2}{c}{$0$}\\
\hline
\rowcolor[gray]{.9}[\tabcolsep]
$\ldots$ & \multicolumn{2}{c|}{$0$} & $\scriptstyle{0,}$ & $\scriptstyle{\frac1m h_1 \ovl\gamma}$ & \multicolumn{2}{c}{$0$}\\
\hline
$\ldots$ & $\scriptstyle{(\ovl h_2-\frac12 h_1)\gamma^2 +\frac12\ovl h_1\gamma,}$ & $\scriptstyle{\ovl h_1 \gamma}$ & $\scriptstyle{h_1\ovl\gamma,}$ & $\scriptstyle{2 h_1 \ovl\gamma}$ &  $\scriptstyle{\frac12 h_2 \gamma^2+h_1 \gamma,}$ & $\scriptstyle{2h_1\gamma}$\\
\hline
\rowcolor[gray]{.9}[\tabcolsep]
$\ldots$ & \multicolumn{2}{c|}{$0$} & $\scriptstyle{0,}$ & $\scriptstyle{h_1\gamma}$ & \multicolumn{2}{c}{$0$}\\
\hline
$\ldots$ & \multicolumn{2}{c|}{$0$} & $\scriptstyle{0,}$ & $\scriptstyle{h_1\gamma}$ & $\scriptstyle{h_2 \gamma^2,}$ & $\scriptstyle{h_1\gamma}$
\end{tabular}\;.
\end{multline}
For $A=\rr$, we have
\begin{equation}
\label{eq:hom21_precomposition_rr}
\begin{tabular}{r|c|c|rl|c|rl}
\diagbox{$\scriptstyle{H}$}{$\scriptstyle{G}$}
 & $\zz_l$ even & $\zz_l$ odd & \multicolumn{2}{c|}{$\zz$} & $\rr/\zz$ & \multicolumn{2}{c}{$\rr$}\\
\hline
$\zz_k$ even &$0$ & $0$ & \multicolumn{2}{c|}{$0$} & $0$ & \multicolumn{2}{c}{$0$}\\
\hline
\rowcolor[gray]{.9}[\tabcolsep]
$\zz_k$ odd &$0$ & $0$ & \multicolumn{2}{c|}{$0$} & $0$ & \multicolumn{2}{c}{$0$}\\
\hline
$\zz$ & $0$ & $0$ & $h_2 \gamma^2,$ & $h_1\gamma$ & $0$ &  $h_2 \gamma^2,$ & $h_1\gamma$\\
\hline
\rowcolor[gray]{.9}[\tabcolsep]
$\rr/\zz$ & $0$ & $0$ & \multicolumn{2}{c|}{$0$} & $0$ & \multicolumn{2}{c}{$0$}\\
\hline
$\rr$ & $0$ & $0$ & $h_2 \gamma^2,$ & $h_1\gamma$ & $0$ & $h_2 \gamma^2,$ & $h_1\gamma$
\end{tabular}\;.
\end{equation}
We confirm both tables above in Appendix~\ref{sec:precompositon_psi_derivation}.
Finally, we give $\Lambda[G|A](h)$ for different abelian groups $G$ and $A$, where $h\in \homtild^2[G,G|A]$:
\begin{equation}
\label{eq:precomposition_lambda_table}
\begin{tabular}{r|rl|rl|c|rl|c}
\diagbox{$\scriptstyle{G}$}{$\scriptstyle{A}$} & \multicolumn{2}{c|}{$\zz_l$ even} & \multicolumn{2}{c|}{$\zz_l$ odd} & $\zz$ & \multicolumn{2}{c|}{$\rr/\zz$} & $\rr$\\
\hline
$\zz_k$ even & $\frac{2\gcd(k,\frac{l}{2})}{\gcd(k,l)}\ovl h\mmod 2\gcd(k,\frac{l}{2}),$ & $0$& $2\ovl h\mmod \gcd(k,l),$&$0$ & $0$ & $2\ovl h\mmod 2k,$&$0$ & $0$\\
\hline
\rowcolor[gray]{.9}[\tabcolsep]
$\zz_k$ odd & $2\ovl h\mmod \gcd(k,l),$&$0$ & $2\ovl h\mmod \gcd(k,l),$&$0$ & $0$ & $2\ovl h \mmod k,$&$0$ & $0$\\
\hline
$\zz$ & $2\ovl h\mmod l,$&$-h$ & $2\ovl h\mmod l,$&$0$ & $2h,-h$ & $h,$&$0$ & $2h,0$\\
\hline
\rowcolor[gray]{.9}[\tabcolsep]
$\rr/\zz$ & \multicolumn{2}{c|}{$0$} & \multicolumn{2}{c|}{$0$} & $0$& \multicolumn{2}{c|}{$0$}& $0$\\
\hline
$\rr$ & \multicolumn{2}{c|}{$0$} & \multicolumn{2}{c|}{$0$} & $0$ & $2h,$&$0$ & $2h,0$
\end{tabular}
\;.
\end{equation}
We confirm this table in Appendix~\ref{sec:lambda_verification}.

\subsection{Contraction and reduction}
\label{sec:reduction}
In this section, we discuss how to contract tensor networks of quadratic tensors directly in terms of the underlying coefficients $E$, $\epsilon^{(\tilde1)}$, $\epsilon^{(\tilde0)}$, $q^{(\tilde2)}$, $q^{(\tilde1)}$, and $q^{(\tilde0)}$.

\myparagraph{Tensor product and self-contraction}
We start by showing that tensor networks of quadratic tensors result again in quadratic tensors.

\begin{myprop}
Consider a quadratic tensor data $(E,\epsilon,q)$ over $G$, and another quadratic tensor data $(E',\epsilon',q')$ over $G'$.
Then the tensor product of the associated quadratic tensors is again a quadratic tensor, with data given given as follows:
\begin{equation}
\mathcal T[E,\epsilon,q]\otimes \mathcal T[E',\epsilon',q']
= \mathcal T[E'',\epsilon'',q'']\;,\qquad
E''=E\times E',\quad \epsilon''=\epsilon\times \epsilon',\quad q''((e,e'))= q(e)+q'(e')\;.
\end{equation}
If one of the quadratic tensor data is $0$, the tensor product is $0$.
\end{myprop}
\begin{proof}
\begin{equation}
\begin{gathered}
(\mathcal T[E,\epsilon,q]\otimes \mathcal T[E',\epsilon',q'])(g,g')
= \mathcal T[E,\epsilon,q](g)\cdot \mathcal T[E',\epsilon',q'](g')\\
=\sum_{e\in E:\epsilon(e)=g} \exp(q(e))\cdot \sum_{e'\in E':\epsilon'(e')=g'} \exp(q'(e'))
= \sum_{(e,e')\in E\times E': (\epsilon\times \epsilon')(e,e')=(g,g')} \exp(q(e)+q'(e'))\\
= \sum_{e\in E'': \epsilon''(e)=(g,g')} \exp(q''(e))
=\mathcal T[E'',\epsilon'',q''](g,g')
\;.
\end{gathered}
\end{equation}
\end{proof}
\begin{myprop}
Let $G_r$ and $G_c$ be abelian groups.
Consider a quadratic tensor data $(E,\epsilon,q)$ over $G=G_r\times G_c\times G_c$, and write $\epsilon=\epsilon_0\times \epsilon_1\times \epsilon_2$.
Then, the self-contraction of the two $G_c$ indices is again a quadratic tensor over $G_r$ with data given by
\begin{equation}
\label{eq:quadratic_self_contraction}
\begin{gathered}
\sum_{g_c\in G_c} \mathcal T[E,\epsilon,q](g_r,g_c,g_c) =
\begin{cases}
\mathcal T[E',\epsilon',q'](g_r)&\text{if } \epsilon_2^{(0)}-\epsilon_1^{(0)}\in \img(\epsilon_1^{(1)}-\epsilon_2^{(1)})\\
0 &\text{otherwise}
\end{cases}\\
E' = \ker(\epsilon_1^{(1)}-\epsilon_2^{(1)})\;,\quad
\epsilon' = \epsilon_0\circ (\kappa+\tilde e)\;,\quad
q' = q\circ (\kappa+\tilde e)\;.
\end{gathered}
\end{equation}
Here, $\kappa\in \hom[E'|E]$ is the homomorphism that identifies $E'$ as a subgroup of $E$, and $\tilde e\in E$ is chosen such that $(\epsilon_1^{(1)}-\epsilon_2^{(1)})\tilde e = \epsilon_2^{(0)}-\epsilon_1^{(0)}$, or equivalently $\epsilon_1(\tilde e)=\epsilon_2(\tilde e)$.
\end{myprop}
\begin{proof}
The second case in Eq.~\eqref{eq:quadratic_self_contraction} is obvious.
For the first case, we find
\begin{equation}
\begin{gathered}
\sum_{g_c\in G_c} \mathcal T[E,\epsilon,q](g_r,g_c,g_c)
=\sum_{g_c\in G_c, e\in E: \epsilon_0(e)=g_r, \epsilon_1(e)=g_c,\epsilon_2(e)=g_c} \exp(q(e))
=\sum_{e\in E: \epsilon_0(e)=g_r, \epsilon_1(e)=\epsilon_2(e)} \exp(q(e))\\
=\sum_{e\in E': \epsilon_0(\kappa e+\tilde e)=g_r} \exp(q(\kappa e+\tilde e))
=\sum_{e\in E': \epsilon'(e)=g_r} \exp(q'(e))
=\mathcal T[E',\epsilon',q'](g_r)\;.
\end{gathered}
\end{equation}
\end{proof}
An arbitrary tensor network can be contracted by first performing pairwise tensor products until there is a single tensor, and then performing self-contractions of all the index pairs.
So we see that it is in principle possible to evaluate arbitrary tensor networks of quadratic tensors in terms of the underlying quadratic tensor data.
However, this evaluation leads to a growing group $E$.
Roughly speaking, if we evaluate a tensor network with $n$ $G_c$ contractions, then the resulting $E$ is by up to $n$ $G_c$ factors larger than the resulting $G$.
So we merely shifted the computational cost from the contraction to the calculation of the resulting tensor entries $\mathcal T[E,\epsilon,q]$ in Eq.~\eqref{eq:quadratic_tensor_definition}.
To implement contraction in a practically useful way, we thus also need a way to reduce the size of $E$, which we do in the following.

\myparagraph{Reduction via invertible subgroups}
We will now show two ways to \emph{reduce} a quadratic tensor data $(E,\epsilon,q)$ to an equivalent quadratic tensor data $(E',\epsilon',q')$, such that $E'$ is smaller than $E$, in the sense that it consists of fewer elementary factors.
Roughly, the first type of reduction can be applied if the bilinear form $q^{(2)}$ is non-degenerate on some subgroup of $\ker(\epsilon)$, and the second reduction can be applied if $q^{(2)}$ is zero on some subgroup of $\ker(\epsilon)$.
The following two propositions are two variants of the first type of reduction.
We will use the notation for duals from Definition~\ref{def:duals}, and use the isomorphism from Eq.~\eqref{eq:currying} implicitly in the notation.
We denote composition of homomorphisms by just writing them in sequence, saving space by omitting the usual $\circ$ symbol.
Combining the above notations, we can for example write
\begin{equation}
q^{(2)}(\gamma(h),g) = q^{(2)}\gamma(h,g)\;,\qquad q^{(2)}(g,\gamma h) = \gamma^* q^{(2)}(g,h)
\end{equation}
for $h\in H$, $g\in G$, $q^{(2)}\in \hom^2[G,G|A]\simeq \hom[G|G^*]$, $\gamma\in\hom[H|G]$.

For reasons that will come clear later, we will consider the notion of quadratic tensors for some general abelian $A$, instead of the more specific physics use case $A=\rr\times\rr/\zz$.
\begin{myprop}
\label{prop:invert_reduction}
Let $(E,\epsilon,q)$ be a quadratic tensor data over $G$.
Let $\rho\in\hom[R|E]$ be an injective homomorphism that identifies $R$ as a subgroup of $E$, such that the following holds.
\begin{itemize}
\item $E$ is a subgroup of the kernel of $\epsilon^{(1)}$,
\begin{equation}
\label{eq:invertiblered_derivation2}
\epsilon^{(1)}\rho=0\;.
\end{equation}
\item $q^{(2)}$ is non-degenerate on $R$, that is,
\begin{equation}
\rho^*q^{(2)}\rho\in\hom[R|R^*]
\end{equation}
has an inverse
\begin{equation}
\label{eq:invertiblered_derivation4}
\qinv\in \hom[R^*|R]:\qquad
\rho^*q^{(2)}\rho \qinv = 1_{R^*},\quad \qinv \rho^*q^{(2)}\rho=1_R
\;.
\end{equation}
\item The exponential sum of $q$ over $R$ is itself an exponential (and in particular non-zero):
\begin{equation}
\label{eq:invertiblered_derivation5}
\sum_{r\in R} \exp(q(\rho(r))) = \exp(q^{\sum}),\quad q^{\sum}\in A\;.
\end{equation}
\end{itemize}
Then the following quadratic tensor data $(E',\epsilon',q')$ is equivalent to $(E,\epsilon,q)$:
\begin{equation}
\label{eq:invertiblered}
\begin{aligned}
E'&=E/R\;,\\
\epsilon'(r^\perp)&=\epsilon(\rho^\perp(r^\perp))\;,\\
q'(r^\perp)&= \widetilde q(\rho^{\perp}(r^\perp)),\quad \widetilde q(e) \coloneqq q(e)-q(\gamma(e))+q^{\sum}\;,
\end{aligned}
\end{equation}
where
\begin{equation}
\label{eq:invertiblered_gamma_def}
\gamma\coloneqq \rho\qinv \rho^* q^{(2)}\in \hom[E|E]\;,
\end{equation}
and
\begin{equation}
\label{eq:reduction_lift}
\rho^{\perp}: E/R\rightarrow E: \qquad (\rho^{\perp}(r^\perp))/R = r^\perp
\end{equation}
is any left inverse of the quotient map from $E$ to $E/R$.
\end{myprop}
\begin{proof}
Note that $\widetilde q\in F_2(E,A)$ is a quadratic function since it is obtained from $q$ via sums, homomorphism composition (see Proposition~\ref{prop:homomorphism_composition_quadratic}), and constants.
Further, we find that $\widetilde q$ is invariant under adding elements of $R$ to its argument:
\begin{equation}
\label{eq:qtilde_lift_independent}
\begin{gathered}
\widetilde q(e+\rho(r))
\overset{\eqref{eq:invertiblered}}{=} q(e+\rho(r)) - q(\gamma (e+\rho(r))) + q^{\sum}\\
\overset{\eqref{eq:invertiblered_derivation6}}{=} q(e+\rho(r)) - q(\gamma(e) + \rho (r))) + q^{\sum}\\
\overset{\eqref{eq:second_derivative}}{=} q(e)+q(\rho(r)) + q^{(2)}(\rho(r),e) - q(\gamma(e)) - q(\rho(r))- q^{(2)} (\rho(r),\gamma(e)) + q^{\sum}\\
\overset{\eqref{eq:invertiblered_derivation7}}{=} q(e) - q(\gamma(e)) + q^{\sum} = \widetilde q(e)\;.
\end{gathered}
\end{equation}
Here, we have used that
\begin{gather}
\gamma\rho=\rho\;,\quad
q^{(2)}\gamma=\gamma^* q^{(2)}\;,
\label{eq:invertiblered_derivation6}\\
\Rightarrow q^{(2)}(\rho(r),e) = q^{(2)}\rho(r,e)=q^{(2)}\gamma\rho(r,e)=\gamma^*q^{(2)}\rho(r,e)=q^{(2)}(\rho(r),\gamma(e))\;.
\label{eq:invertiblered_derivation7}
\end{gather}
Thus, we can reduce $\widetilde q$ to a quadratic function $q'$ on $E/R$, which is independent of the choice of $\rho^\perp$.
Let us now confirm Eq.~\eqref{eq:tensordata_equivalence}.
First, we find that
\begin{equation}
\label{eq:invertiblered_derivation3}
\begin{gathered}
\sum_{r\in R}\exp(q(e+\rho(r)))
\overset{\eqref{eq:second_derivative}}{=} \sum_{r\in R} \exp\big(q(e)+q(\rho(r))+q^{(2)}(\rho(r),e)-q(0)\big)\\
\overset{\eqref{eq:invertiblered_derivation7}}{=} \sum_{r\in R} \exp\big(q(e)+q(\rho(r))+q^{(2)}(\rho(r),\gamma(e))-q(0)\big)\\
\overset{\eqref{eq:second_derivative}}= \sum_{r\in R} \exp\big(q(e)+q\big(\rho(r)+\gamma(e)\big)-q(\gamma(e))\big)\\
\overset{\eqref{eq:invertiblered_gamma_def}}{=} \exp\big(q(e)-q(\gamma(e))\big) \sum_{r\in R} \exp\big(q(\rho(r+\qinv \rho^*q^{(2)}(e)))\big)\\
\overset{*}{=} \exp\big(q(e)-q(\gamma(e))\big)\sum_{r\in R} \exp(q(\rho(r)))
\overset{\eqref{eq:invertiblered_derivation5}}{=} \exp\big(q(e)-q(\gamma(e))+q^{\sum}\big)
\overset{\eqref{eq:invertiblered}}{=} \exp(\widetilde q(e))\;.
\end{gathered}
\end{equation}
For the equality labeled $*$, we have performed a change in summation variable $r\rightarrow r-\qinv \rho^*q^{(2)}(e)$.
Thus, we have
\begin{equation}
\label{eq:invertiblered_derivation_final}
\begin{gathered}
\mathcal T[E,e,q](g)
\overset{\eqref{eq:quadratic_tensor_definition}}{=} \sum_{e\in E: \epsilon(e)=g} \exp(q(e))
\overset{\eqref{eq:invertiblered_derivation1}, \eqref{eq:invertiblered_derivation2}}{=} \sum_{r\in R, r^\perp\in E/R: \epsilon(\rho^\perp(r^\perp))=g} \exp(q(\rho^\perp(r^\perp)+\rho(r)))\\
\overset{\eqref{eq:invertiblered_derivation3}}{=} \sum_{r^\perp \in E/R: \epsilon(\rho^\perp(r^\perp))=g} \exp(\widetilde q(\rho^{\perp}(r^\perp)))
\overset{\eqref{eq:invertiblered}}{=} \sum_{r^\perp\in E': \epsilon'(r^\perp)=g} \exp(q'(r^\perp))
\overset{\eqref{eq:quadratic_tensor_definition}}{=} \mathcal T[E',\epsilon',q'](g)\;.
\end{gathered}
\end{equation}
Here, we have used that for every $e\in E$, there are unique $r\in R$ and $r^\perp\in E/R$, such that
\begin{equation}
\label{eq:invertiblered_derivation1}
e=\rho^\perp(r^\perp)+\rho(r)\;.
\end{equation}
\end{proof}
The reduction above holds for any choice of $A$ and $\rho$.
However, if $A=\rr\times \rr/\zz$, then we can never find a non-trivial subgroup such that Eq.~\eqref{eq:invertiblered_derivation4} holds.
So we can only really apply the reduction above if $A=\rr/\zz$, which is equivalent to saying that $q_a=0$.
In fact, it suffices if $q_a$ is zero restricted to the subgroup $R$, which is automatic if $R$ only consists of $\zz_k$ or $\rr/\zz$ factors.

In Section~\ref{sec:tutorial}, we have seen that we can also perform reduction over (continuous-variable) $\rr^n$ subgroups, where both $q_\phi$ and $q_a$ are non-zero.
In this contraction, we treat $q_a^{(2)}+iq_\phi^{(2)}$ as a complex number, and we perform the inverse $q_{11}^{-1}$ in the field of complex numbers.
If we want to define such a contraction in a more general case where only the subgroup $R$ is equal to $\rr^n$, but the full group $E$ can be arbitrary, we have to keep real and imaginary parts such as $q_a$ and $q_\phi$ separate.
The result is the following, slightly lengthy reduction.

In the following, we use the notation $X^*=\hom[X|\rr/\zz]$, and $X^+=\hom[X|\rr]$ for any abelian group $X$, and the analogous notation $\rho^*\in \hom[Y^*|X^*]$ and $\rho^+\in\hom[Y^+|X^+]$ for a homomorphism $\rho\in \hom[X|Y]$.
We also make use of an isomorphism
\begin{equation}
\chi: (\rr^n)^+\rightarrow (\rr^n)^*\;:\qquad
\chi(x)(\vec r) = x(\vec r)\mmod 1\;.
\end{equation}
\begin{myprop}
\label{prop:complex_invert_reduction}
Let $(E,\epsilon,q)$ be a quadratic tensor data over $G$.
Let $\rho\in \hom[\rr^n|E]$ be an injective homomorphism that identifies $\rr^n$ as a subgroup of $E$, such that the following holds.
\begin{itemize}
\item Eq.~\eqref{eq:invertiblered_derivation2} holds.
\item There exist
\begin{equation}
\qinv_a\in \hom[(\rr^n)^+|\rr^n]\;,\qquad \qinv_\phi\in \hom[(\rr^n)^*|\rr^n]
\end{equation}
such that
\begin{equation}
\label{eq:invertiblered_complex_derivation3}
\begin{gathered}
\chi\rho^+ q_a^{(2)}\rho \qinv_a\chi^{-1}-\rho^* q_\phi^{(2)}\rho \qinv_\phi=1,\qquad
\chi \rho^+ q_a^{(2)}\rho \qinv_\phi+\rho^* q_\phi^{(2)}\rho \qinv_a\chi^{-1}=0\;,\\
(\qinv_a \rho^+ q_a^{(2)}-\qinv_\phi\rho^* q_\phi^{(2)})\rho=1,\qquad
(\qinv_a\chi^{-1}\rho^* q_\phi^{(2)}+\qinv_\phi\chi\rho^+ q_a^{(2)})\rho=0\;.
\end{gathered}
\end{equation}
\item $q$ is integrable over the $\rr^n$ subgroup,
\begin{equation}
\int_{r\in \rr^n} \exp\Big(q_a(\rho(r))),\quad q_\phi(\rho(r))\Big) = \exp(q_a^{\sum},q_\phi^{\sum})\;.
\end{equation}
\end{itemize}
Define $\beta_a,\beta_\phi\in \hom[E|\rr^n]$, and $\gamma_a,\gamma_\phi\in \hom[E|E]$ as
\begin{equation}
\begin{gathered}
\beta_a\coloneqq \qinv_a \rho^+ q_a^{(2)}- \qinv_\phi \rho^* q_\phi^{(2)}\;,\quad
\beta_\phi\coloneqq \qinv_a\chi^{-1} \rho^* q_\phi^{(2)}+ \qinv_\phi\chi \rho^+ q_a^{(2)}\;,\quad
\gamma_a\coloneqq \rho \beta_a\;,\quad
\gamma_\phi\coloneqq \rho \beta_\phi\;.
\end{gathered}
\end{equation}
Then the following quadratic tensor data $(E',\epsilon',q')$ is equivalent to $(E,\epsilon,q)$:
\begin{equation}
\label{eq:invertiblered_complex}
\begin{aligned}
E'&=E/R\;,\\
\epsilon'(r^\perp)&=\epsilon(\rho^\perp(r^\perp))\;,\\
q_a'(r^\perp)&= \widetilde q_a(\rho^{\perp}(r^\perp)),\quad \widetilde q_a(e) = q_a(e)-q_a(\gamma_a(e))+q_a(\gamma_\phi(e))+\chi^{-1}\rho^*q^{(2)}_\phi (\gamma_\phi(e),\beta_a(e))+q_a^{\sum}\\
q_\phi'(r^\perp)&= \widetilde q_\phi(\rho^{\perp}(r^\perp)),\quad \widetilde q_\phi(e) = q_\phi(e)-q_\phi(\gamma_a(e))+q_\phi(\gamma_\phi(e))-\chi\rho^+q^{(2)}_a (\gamma_\phi(e),\beta_a(e))+q_\phi^{\sum}\;.
\end{aligned}
\end{equation}
\end{myprop}
\begin{proof}
As a first step, we show that $\widetilde q_a$ and $\widetilde q_\phi$ are independent of choice of $\rho^\perp$, analogous to Eq.~\eqref{eq:qtilde_lift_independent}:
\begin{equation}
\begin{gathered}
\widetilde q_a(e+\rho(r))
\overset{\eqref{eq:invertiblered_complex}}{=} q_a(e+\rho(r)) - q_a(\gamma_a(e+\rho(r))) + q_a(\gamma_\phi(e+\rho(r))) + \chi^{-1}\rho^* q^{(2)}_\phi(\gamma_\phi(e+\rho(r)),\beta_a(e+\rho(r))) + q_a^{\sum}\\
\overset{\eqref{eq:invertiblered_complex_derivation0},\eqref{eq:invertiblered_complex_derivation1}}{=} q_a(e+\rho(r)) - q_a(\gamma_a(e)+\rho(r)) + q_a(\gamma_\phi(e)) + \chi^{-1}\rho^*q_\phi^{(2)}(\gamma_\phi(e), r) + \chi^{-1}\rho^* q^{(2)}_\phi(\gamma_\phi(e),\beta_a(e)) + q_a^{\sum}\\
\overset{\eqref{eq:second_derivative}}{=} q_a(e)+q_a(\rho(r))+q_a^{(2)}(e,\rho(r)) - q_a(\gamma_a(e))- q_a(\rho(r)) -q^{(2)}_a(\gamma_a(e),\rho(r))\\
+ q_a(\gamma_\phi(e)) + \chi^{-1}\rho^*q_\phi^{(2)}(\gamma_\phi(e),r) + \chi^{-1}\rho^*q^{(2)}_\phi(\gamma_\phi(e),\beta_a(e)) + q_a^{\sum}\\
\overset{\eqref{eq:invertiblered_complex_derivation2}}{=}q_a(e)-q_a(\gamma_a(e))+q_a(\gamma_\phi(e))+\chi^{-1}\rho^*q^{(2)}_\phi (\gamma_\phi(e),\beta_a(e))+q_a^{\sum}
=\widetilde q_a(e)\;.
\end{gathered}
\end{equation}
Here, we have used that
\begin{gather}
\label{eq:invertiblered_complex_derivation0}
\beta_a\rho \overset{\eqref{eq:invertiblered_complex_derivation3}}{=}1\;,\\
\label{eq:invertiblered_complex_derivation1}
\beta_\phi\rho \overset{\eqref{eq:invertiblered_complex_derivation3}}{=}0\;,\\
\label{eq:invertiblered_complex_derivation2}
\begin{multlined}
\rho^+ q^{(2)}_a\gamma_a-\chi^{-1}\rho^* q^{(2)}_\phi \gamma_\phi
= \rho^+q_a^{(2)} \rho (\qinv_a \rho^+ q_a^{(2)}- \qinv_\phi \rho^* q_\phi^{(2)})-\chi^{-1}\rho^* q^{(2)}_\phi\rho (\qinv_a \chi^{-1}\rho^* q_\phi^{(2)}+ \qinv_\phi \chi\rho^+ q_a^{(2)})\\
= (\rho^+q_a^{(2)} \rho \qinv_a - \chi^{-1}\rho^* q^{(2)}_\phi\rho \qinv_\phi\chi) \rho^+ q_a^{(2)} - (\rho^+q_a^{(2)} \rho \qinv_\phi +\chi^{-1}\rho^* q^{(2)}_\phi\rho \qinv_a\chi^{-1}) \rho^* q_\phi^{(2)}
= \rho^+q^{(2)}_a\;,\\
\Rightarrow
q_a^{(2)}(e,\rho(r))
=q_a^{(2)}(\gamma_a(e),\rho(r))-\chi^{-1}\rho^*q_\phi^{(2)}(\gamma_\phi(e),r)\;,
\end{multlined}\\
\label{eq:invertiblered_complex_derivation4}
\rho^*q_\phi^{(2)}\gamma_a + \chi\rho^+q_a^{(2)}\gamma_\phi = \rho^*q_\phi^{(2)}
\quad\Rightarrow\quad
q_\phi^{(2)}(e,\rho(r)) = q_\phi^{(2)}(\gamma_a(e),\rho(r)) + \chi\rho^+q_a^{(2)}(\gamma_\phi(e),r)\;.
\end{gather}
We omit the same computation for $\widetilde q_\phi$ as it is very similar.
Analogous to Eq.~\eqref{eq:invertiblered_derivation3} we find:
\begin{equation}
\label{eq:invertiblered_complex_derivation3x}
\begin{gathered}
\int_{r\in \rr^n}\exp\Big(q_a(e+\rho(r)),\quad q_\phi(e+\rho(r))\Big)\\
\overset{\eqref{eq:second_derivative}}{=} \int_{r\in \rr^n} \exp\Big(q_a(e)+q_a(\rho(r))+q_a^{(2)}(e,\rho(r))-q_a(0),\quad q_\phi(e)+q_\phi(\rho(r))+q_\phi^{(2)}(e,\rho(r))-q_\phi(0)\Big)\\
\overset{\eqref{eq:invertiblered_complex_derivation2},\eqref{eq:invertiblered_complex_derivation4}}{=} \int_{r\in \rr^n} \exp\Big(q_a(e)+q_a(\rho(r))+q_a^{(2)}(\gamma_a(e),\rho(r)) - \chi^{-1}\rho^* q_\phi^{(2)}(\gamma_\phi(e),r)-q_a(0),\\
q_\phi(e)+q_\phi(\rho(r))+q_\phi^{(2)}(\gamma_a(e),\rho(r)) + \chi\rho^+q_a^{(2)}(\gamma_\phi(e),r)-q_\phi(0)\Big)\\
\overset{\eqref{eq:second_derivative}}= \int_{r\in \rr^n} \exp\Big(q_a(e)+q_a(\gamma_a(e)+\rho(r))-q_a(\gamma_a(e)) - \chi^{-1}\rho^*q_\phi^{(2)}(\gamma_\phi(e),r),\\
q_\phi(e)+q_\phi(\gamma_a(e)+\rho(r))-q_\phi(\gamma_a(e)) + \chi\rho^+q_a^{(2)}(\gamma_\phi(e),r)\Big)\\
\begin{multlined}
= \exp\Big(q_a(e)-q_a(\gamma_a(e))+q_a(\gamma_\phi(e))+\chi^{-1}\rho^* q^{(2)}_\phi (\gamma_\phi(e),\beta_a(e)),\\
q_\phi(e)-q_\phi(\gamma_a(e))+q_\phi(\gamma_\phi(e))-\chi\rho^+q^{(2)}_a (\gamma_\phi(e),\beta_a(e))\Big)\\
\int_{r\in \rr^n} \exp\Big(q_a(\rho(r+\beta_a(e)))-q_a(\rho(\beta_\phi(e))) -\chi^{-1}\rho^*q_\phi^{(2)}(\gamma_\phi(e),r+\beta_a(e)),\\
q_\phi(\rho(r+\beta_a(e)))-q_\phi(\rho(\beta_\phi(e)))+\chi\rho^+q_a^{(2)}(\gamma_\phi(e),r+\beta_a(e))\Big)
\end{multlined}\\
\overset{*}{=}\exp\Big(q_a(e)-q_a(\gamma_a(e))+q_a(\gamma_\phi(e))+\chi^{-1}\rho^*q^{(2)}_\phi (\gamma_\phi(e),\beta_a(e)),\\
q_\phi(e)-q_\phi(\gamma_a(e))+q_\phi(\gamma_\phi(e))-\chi\rho^+q^{(2)}_a (\gamma_\phi(e),\beta_a(e))\Big)\\
\int_{r\in \rr^n} \exp\Big(q_a(\rho(r))),\quad q_\phi(\rho(r))\Big)\\
\overset{\eqref{eq:invertiblered_complex}}{=}\exp\Big(\tilde q_a(e),\quad \tilde q_\phi(e)\Big)\;.
\end{gathered}
\end{equation}
For the equality labeled $*$, we have used that $q_a(\rho(\bullet))$ and $q_\phi(\rho(\bullet))$ are quadratic functions $\rr^n$, and take the following simple form,
\begin{equation}
q_a(\rho(r))=\frac12 r^TA_a r+d_a r+c_a\;,\quad
q_\phi(\rho(r))=\frac12 r^TA_\phi r+d_\phi r+c_\phi \mod 1\;.
\end{equation}
Using this concrete form, we can calculate the integral,
\begin{equation}
\begin{gathered}
\begin{multlined}
\int_{r\in \rr^n}\exp\Big(q_a(\rho(r+\beta_a(e)))-q_a(\rho(\beta_\phi(e))) -\chi^{-1}\rho^*q_\phi^{(2)}(\rho(r+\beta_a(e)),\beta_\phi(e)),\\
q_\phi(\rho(r+\beta_a(e)))-q_\phi(\rho(\beta_\phi(e)))+\chi\rho^+q_a^{(2)}(\rho(r+\beta_a(e)),\beta_\phi(e))\Big)
\end{multlined}\\
\begin{multlined}
=\int_{r\in \rr^n}\exp\Big(\frac12 (r+\beta_a(e))^TA_a(r+\beta_a(e)) + d_a(r+\beta_a(e)) + c_a\\ - \beta_\phi(e)^T A_a \beta_\phi(e) - d_a \beta_\phi(e) - c_a - (r+\beta_a(e))^TA_\phi \beta_\phi(e),\\
\frac12 (r+\beta_a(e))^TA_\phi(r+\beta_a(e)) + d_\phi(r+\beta_a(e)) + c_\phi\\ - \beta_\phi(e)^T A_\phi \beta_\phi(e) - d_\phi \beta_\phi(e) - c_\phi - (r+\beta_a(e))^TA_a \beta_\phi(e)\Big)
\end{multlined}\\
=\int_{r\in \rr^n}e^{2\pi\Big(\frac12(r+\beta_a(e)+i\beta_\phi(e))^T(A_a+iA_\phi)(r+\beta_a(e)+i\beta_\phi(e)) + (d_a+id_\phi)(r+\beta_a(e)+i\beta_\phi(e)) + c_a+c_\phi\big)}\\
=\int_{r\in \rr^n}e^{2\pi\Big(\frac12 r^T(A_a+iA_\phi)r + (d_a+id_\phi)r + c_a+c_\phi\big)}\\
\int_{r\in \rr^n} \exp\Big(q_a(\rho(r))),\quad q_\phi(\rho(r))\Big)\;.
\end{gathered}
\end{equation}
Here, we have used that $\exp$ is an analytic function that vanishes at infinity, and so we can perform a change of variable $r\rightarrow r+\beta_a(e)+i\beta_\phi(e)$ shifting the real variable $r$ by a complex value $\beta_e(e)+i\beta_\phi(e)$.
With this, the last step of the proof, showing that $(E,\epsilon,q)$ and $(E',\epsilon',q')$ are equivalent, is the same as in Eq.~\eqref{eq:invertiblered_derivation_final}.
\end{proof}

The reductions above only apply if the quadratic form $q^{(2)}$ is invertible on the subgroup $R$.
The second kind of reduction, introduced below, applies in the opposite case, namely when $q^{(2)}$ is zero on $R$.
\begin{myprop}
\label{prop:zero_reduction}
Let $(E,\epsilon,q)$ be a quadratic tensor data over a group $G$.
Let $\rho: R\rightarrow E$ be an injective homomorphism that identifies $R$ as a subgroup of $E$, such that the following holds.
\begin{itemize}
\item Eq.~\eqref{eq:invertiblered_derivation2} holds.
\item The bilinear form $q^{(2)}$ is zero on $R$,
\begin{equation}
\label{eq:zeroreduction_derivation1}
\rho^*q^{(2)}\rho = 0\;.
\end{equation}
\end{itemize}
Then the quadratic tensor data $(E',\epsilon',q')$ defined in the following is equivalent to $(E,\epsilon,q)$.
Let $\rho^\perp:E/R\rightarrow E$ be a left inverse of the quotient map as in Eq.~\eqref{eq:reduction_lift}.
Consider the homomorphism
\begin{equation}
\label{eq:zeroreduction_alphadef}
\alpha: E/R\rightarrow R^*:\qquad
\alpha(r^\perp) = \rho^* q^{(2)}(\rho^\perp(r^\perp))\;.
\end{equation}
If there exists $r^\perp_0\in E/R$ such that
\begin{equation}
\label{eq:zeroreduction_offset}
\alpha(r^\perp_0)+ q^{(1)}\circ\rho = 0\;,
\end{equation}
then define
\begin{equation}
\label{eq:zeroreduction}
\begin{aligned}
E' &=\ker(\alpha)\subset E/R\;,\\
\epsilon'(r^\perp) &=\epsilon(\rho^\perp (r^\perp+r^\perp_0))\;,\\
q'(r^\perp) &=q(\rho^\perp (r^\perp+r^\perp_0)) + \log(|R|)\;.
\end{aligned}
\end{equation}
Otherwise, $(E,\epsilon,q)$ is equivalent to $0$.
Here, $\log$ denotes the inverse of $\exp$ (which has an additional $2\pi$ compared to the usual logarithm).
$|R|$ denotes the order of the finite-group part of $R$, ignoring factors of $\rr/\zz$, $\zz$, or $\rr$.
\end{myprop}
\begin{proof}
Let us start by confirming that $E'$, $\epsilon'$ and $q'$ are well defined.
First, we note that in Eq.~\eqref{eq:zeroreduction_offset}, the composition of functions $q^{(1)}\circ\rho$ is an element of $R^*$:
\begin{equation}
(q^{(1)}\circ\rho)(r_0+r_1)
\overset{\eqref{eq:second_derivative}}{=} q^{(1)}(\rho(r_0))+q^{(1)}(\rho(r_1)) + \rho^*q^{(2)}\rho(r_0,r_1)
\overset{\eqref{eq:zeroreduction_derivation1}}{=} (q^{(1)}\circ\rho)(r_0)+(q^{(1)}\circ\rho)(r_1)\;.
\end{equation}
We also need to check that $\alpha$ itself defines a valid homomorphism:
\begin{equation}
\begin{gathered}
\alpha(r^\perp_0 + r^\perp_1)
\overset{\eqref{eq:zeroreduction_alphadef}}{=} \rho^*q^{(2)}(\rho^\perp (r^\perp_0+r^\perp_1))
= \rho^*q^{(2)}(\rho^\perp (r^\perp_0)+\rho^\perp(r^\perp_1)+\rho(r))\\
= \rho^* q^{(2)}(\rho^\perp (r^\perp_0)) + \rho^* q^{(2)}(\rho^\perp (r^\perp_1)) + \rho^* q^{(2)} (\rho(r))
\overset{\eqref{eq:zeroreduction_derivation1}}{=} \alpha(r^\perp_0) + \alpha(r^\perp_1)\;.
\end{gathered}
\end{equation}
Next, we show that $\alpha$ (and thus $E'$), $\epsilon'$, and $q'$ do not depend on the choice of $\rho^\perp$.
That is, they do not change if we replace $\widetilde\rho^\perp(r^\perp)\coloneqq \rho^\perp(r^\perp)+\rho(r)$ for some fixed $r\in R$:
\begin{equation}
\widetilde\alpha(r^\perp)
= \rho^*q^{(2)}(\widetilde\rho^\perp(r^\perp))
= \rho^*q^{(2)}(\rho^\perp(r^\perp)+\rho(r))
\overset{\eqref{eq:zeroreduction_derivation1}}{=} \rho^*q^{(2)}(\rho^\perp(r^\perp))
= \alpha(r^\perp)\;,
\end{equation}
\begin{equation}
\widetilde\epsilon'(r^\perp)
= \epsilon(\widetilde \rho^\perp(r^\perp+r_0^\perp))
= \epsilon(\rho^\perp(r^\perp+r_0^\perp)+\rho(r))
= \epsilon(\rho^\perp(r^\perp+r_0^\perp)) + \epsilon^{(1)}\rho(r)
\overset{\eqref{eq:invertiblered_derivation2}}{=} \epsilon(\rho^\perp(r^\perp+r_0^\perp))
= \epsilon'(r^\perp)\;,
\end{equation}
\begin{equation}
\begin{gathered}
\widetilde q'(r^\perp)
= q(\widetilde \rho^\perp(r^\perp+r_0^\perp)) + \log(|R|)
= q(\rho^\perp(r^\perp+r_0^\perp)+\rho(r)) + \log(|R|)\\
\overset{\eqref{eq:second_derivative}}{=} q(\rho^\perp(r^\perp+r_0^\perp)) +q(\rho(r)) -q(0) + q^{(2)}(\rho^\perp(r^\perp+r_0^\perp), \rho(r)) + \log(|R|)\\
\overset{\eqref{eq:zeroreduction},\eqref{eq:first_derivative}}{=} q'(r^\perp) +(q^{(1)}\circ\rho)(r) + \rho^* q^{(2)}(\rho^\perp(r^\perp+r_0^\perp), r)\\
\overset{\eqref{eq:zeroreduction_alphadef}}{=} q'(r^\perp) +(q^{(1)}\circ\rho)(r) + \alpha(r^\perp_0)(r)+ \alpha(r^\perp)(r)
\overset{*,\eqref{eq:zeroreduction_offset}}{=} q'(r^\perp)\;.
\end{gathered}
\end{equation}
For the equality labeled $*$, we have used that by assumption $r^\perp\in\ker(\alpha)$.
Finally, we confirm Eq.~\eqref{eq:tensordata_equivalence} by direct calculation:
\begin{equation}
\begin{gathered}
\mathcal T[E,\epsilon,q](g)
\overset{\eqref{eq:quadratic_tensor_definition}}{=} \sum_{e\in E: \epsilon(e)=g} \exp(q(e))
\overset{\eqref{eq:invertiblered_derivation1},\eqref{eq:invertiblered_derivation2}}{=} \sum_{r^\perp\in E/R: \epsilon(\rho^\perp(r^\perp))=g} \sum_{r\in R} \exp(q(\rho^\perp(r^\perp)+\rho(r)))\\
\overset{\eqref{eq:second_derivative}}{=} \sum_{r^\perp\in E/R: \epsilon(\rho^\perp(r^\perp))=g} \sum_{r\in R} \exp\big(q(\rho^\perp(r^\perp))+q(\rho(r))+  q^{(2)}(\rho^\perp(r^\perp),\rho(r)) - q(0)\big)\\
\overset{\eqref{eq:first_derivative}}{=} \sum_{r^\perp\in E/R: \epsilon(\rho^\perp(r^\perp))=g} \exp(q(\rho^\perp(r^\perp))) \sum_{r\in R} \exp(q^{(1)}(\rho(r))+\rho^* q^{(2)}(\rho^\perp(r^\perp),r))\\
\overset{\eqref{eq:zeroreduction_alphadef}}{=} \sum_{r^\perp\in E/R: \epsilon(\rho^\perp(r^\perp))=g} \exp(q(\rho^\perp(r^\perp))) \sum_{r\in R} \exp((\alpha(r^\perp)+q^{(1)}\circ\rho)(r))\\
\overset{\eqref{eq:group_character_sum}}{=} \sum_{r^\perp\in E/R: \epsilon(\rho^\perp(r^\perp))=g} \exp(q(\rho^\perp(r^\perp))) |R| \delta_{\alpha(r^\perp)+q^{(1)}\circ\rho=0}\\
= \sum_{r^\perp\in E/R: \alpha(r^\perp)+q^{(1)}\circ\rho=0, \epsilon(\rho^\perp(r^\perp))=g} \exp(q(\rho^\perp(r^\perp))+\log(|R|))\\
\overset{*,\eqref{eq:zeroreduction_offset}}{=} \sum_{r^\perp\in \ker(\alpha): \epsilon(\rho^\perp (r^\perp+r^\perp_0))=g} \exp(q(\rho^\perp(r^\perp+r_0^\perp))+\log(|R|))\\
\overset{\eqref{eq:zeroreduction}}{=} \sum_{e\in E': \epsilon'(e)=g} \exp(q'(e))
\overset{\eqref{eq:quadratic_tensor_definition}}{=} \mathcal T[E',\epsilon',q'](g)\;.
\end{gathered}
\end{equation}
For the equality labeled $*$, we have performed a change of summation variable from $r^\perp$ to $r^\perp+r^\perp_0$.
We have used the exponential sum over group characters,
\begin{equation}
\label{eq:group_character_sum}
\sum_{r\in R} \exp(\chi(r))=|R|\delta_{\chi=0}\;,
\end{equation}
for any abelian group $R$ and $\chi\in R^*$.
Note that if $|R|$ contains $\zz$ or $\rr$ factors, then we are missing factors of ``$\infty$''.
This means that the resulting quadratic tensor is a $\delta$-distribution rather than the characteristic function of a subspace of zero measure.
\end{proof}

\myparagraph{Reducibility}
The two types of reductions above reduce $E$ (and also $\ker(\epsilon^{(1)})$) to some quotient group.
Let us now study to which extent $\ker(\epsilon^{(1)})$ can be reduced by reductions.
We find that we can apply reductions until $\ker(\epsilon^{(1)})$ only consists of $\zz$ factors.
\begin{myprop}
Assume that a quadratic tensor data cannot be further reduced using the three propositions above.
Then $\ker(\epsilon^{(1)})$ is isomorphic to $\zz^n$.
\end{myprop}
\begin{proof}
Assume $\ker(\epsilon^{(1)})$ contains an elementary abelian factor $A$ that is equal to either $\zz_k$, or $\rr$, or $\rr/\zz$.
Let $\alpha$ be the injective homomorphism that embeds $A$ into $\ker(\epsilon^{(1)})$.
We note that $\alpha^{+/*} q^{(2)}_{a/\phi} \alpha\in \hom[A|A^{+/*}]$ and write $x_{a/\phi}\coloneqq\htild^{-1}(\alpha^{+/*}q_{a/\phi}^{(2)}\alpha)=(\htild^2)^{-1}(q^{(2)}_{a/\phi}\circ (\alpha\times\alpha))$.
\begin{itemize}
\item If $A=\zz_k$ and $x_\phi\neq 0$, then let $R=\zz_{k/\gcd(k,x_\phi)}$ and $\rho=\alpha \htild(x_\phi\mmod k/\gcd(k,x_\phi))\in\hom[R|E]$.
Then $\htild^{-1}(\rho^* q_\phi^{(2)}\rho)=1$ is invertible, and we can apply the reduction from Proposition~\ref{prop:invert_reduction}.
\item If $A=\zz_k$ and $x_\phi=0$, we can apply the reduction from Proposition~\ref{prop:zero_reduction}.
\item If $A=\rr$ and $x_\phi\neq 0$ or $x_a\neq 0$, then we can apply the reduction from Proposition~\ref{prop:complex_invert_reduction} with $R=A$ and $\rho=\alpha$.
\item If $A=\rr$ and $x_\phi=0$ and $x_a= 0$, then we can apply the reduction from Proposition~\ref{prop:zero_reduction} with $R=A$ and $\rho=\alpha$.
\item If $A=\rr/\zz$ then $x_\phi=0$ and $x_a=0$ automatically, and we can apply the reduction from Proposition~\ref{prop:zero_reduction}.
\end{itemize}
\end{proof}
On the other hand, if $A=\zz$, then $\alpha^*q_\phi^{(2)}\alpha\in \hom[\zz|\rr/\zz]$ cannot have an inverse since there is only the trivial homomorphism from $\rr/\zz$ to $\zz$.
Thus the reduction in Proposition~\ref{prop:invert_reduction} can never be applied for $R=\zz$.
Further, the reduction in Proposition~\ref{prop:complex_invert_reduction} only applies to $R=\rr$ and not $R=\zz$.

Let us now provide some evidence that performing tensor network contractions is in general hard if $\ker(\epsilon^{(1)})$ contains a large number of $\zz$ factors.
First, consider the following quadratic tensor over the trivial group $G=0$, which represents a number
\begin{equation}
E=\zz^n\;,\quad \epsilon=0\;,\quad q_\phi(x)=(x+d)^T\operatorname{imag}(A)(x+d)\;,\quad q_a(x)=(x+d)^T\operatorname{real}(A)(x+d)\;.
\end{equation}
So the reduction computes the number
\begin{equation}
\mathcal T(E,\epsilon,q) = \Theta(A,d)=\sum_{x\in \zz^n} e^{2\pi (x+d)^TA(x+d)}\;,
\end{equation}
which is know as \emph{lattice theta function}.
There is no known polynomial-time algorithm that computes it for large $n$.
Let us also provide a second piece of evidence.
It is well-known that Clifford operations together with any non-Clifford operation are universal, meaning that they can be used to approximately compose any unitary as a quantum circuit.
Similarly, quadratic tensors over $G=\zz_2^n$ with injective $\epsilon^{(1)}$ (or ``stabilizer tensors'') become universal as soon as we add any non-stabilizer tensor.
Now, consider the following quadratic tensor data over $G=\zz_2$,
\begin{equation}
\begin{gathered}
E=\zz\;,\quad \epsilon^{(\tilde1)}=1\;,\quad \epsilon^{(\tilde0)}=0\;,\quad q_\phi^{(\tilde2)}= q_a^{(\tilde2)}=\varnothing\;,\\
q_\phi^{(\tilde1)}=\mpm{(\imag(a),\imag(b))}\;,\quad q_a^{(\tilde1)}=\mpm{(\real(a),\real(b))}\;,\quad q^{(\tilde0)}=0\;.
\end{gathered}
\end{equation}
whose quadratic tensor is given by
\begin{equation}
\mathcal T[E,\epsilon,q]=\Big(\sum_{k\in\zz} e^{2\pi (a (2k)^2+b)}, \sum_{k\in\zz} e^{2\pi (a (2k+1)^2+b)}\Big)\;.
\end{equation}
For different choices of $a$ and $b$ we get at least one 2-element vector that is not equal to an eigenstate of $\pm X$, $\pm Y$, or $\pm Z$.
So if we allow for $\zz$ factors inside $\epsilon^{(1)}$ then we get a ``universal'' set of tensors, and contracting tensor networks from such a universal set is computationally hard.

Note that even though it is in general hard to reduce quadratic tensor data with $\zz$ factors, it is still possible to apply the reduction from Proposition~\ref{prop:invert_reduction} to remove $\zz$ factors from $\ker(\epsilon^{(1)})$ in some cases.
To this end, we have to use an additional $\rr/\zz$ factor in $\ker(\epsilon^{(1)})$.
Assume we have an injective homomorphism $\rho\in \hom[R,\ker(\epsilon^{(1)})]$ for $R=\zz\times \rr/\zz$ with $\rho^+q_a^{(2)}\rho=0$ and
\begin{equation}
\htild^{-1}(\rho^*q_\phi^{(2)}\rho)
=\mpm{b_{00}&1\\1& 0}
\in \mpm{\rr/\zz&\zz\\\zz&0}\;.
\end{equation}
This matrix in invertible with
\begin{equation}
\qinv=\mpm{0 &1\\1& -b_{00}}\in \mpm{0&\zz\\\zz&\rr/\zz}\;,
\end{equation}
so we can apply the reduction from Proposition~\ref{prop:invert_reduction}.
Equivalently, we can apply the reduction in Proposition~\ref{prop:zero_reduction} for the $\rr/\zz$ factor alone, which also removes the $\zz$ factor.

\myparagraph{Reduction in the direct product case}
Let us now give some more detail on how one would compute the reduction in Proposition~\ref{prop:invert_reduction} in practice.
We first note that we can efficiently compute $\widetilde q$ in terms of its coefficients, using the methods from Section~\ref{sec:homomorphism_composition}.
Namely, combining Eqs.~\eqref{eq:invertiblered} and \eqref{eq:precomposition_coefficients}, we find
\begin{equation}
\label{eq:invert_reduction_coefficients}
\begin{gathered}
\widetilde q^{(\tilde2)}_{ij}=q^{(\tilde2)}_{ij}-\sum_{k,l} \tilde\gamma_{ki}^{\tilde*} \circtild q^{(\tilde2)}_{kl} \circtild \tilde\gamma_{lj}\;,\\
\widetilde q^{(\tilde1)}_i=q^{(\tilde1)}_i-\sum_k \Phi(q^{(\tilde1)}_k,\tilde\gamma_{ki}) - \sum_{l<k} \Lambda(\tilde\gamma_{ki}^{\tilde*} \circtild q^{(\tilde2)}_{kl}\circtild \tilde\gamma_{li})\;,\\
\widetilde q^{(\tilde0)}=q^{(\tilde0)}+q^{\sum}\;.
\end{gathered}
\end{equation}
Going from $\widetilde q$ to $q'$ in the most general case requires to choose a lift for every homomorphism $\rho$, which we will not do explicitly in this paper.
Instead we discuss the simpler case of quadratic tensors over $G$, where
\begin{equation}
E=E_0\times R,\quad \epsilon=\mpm{\epsilon_0&0}\;,\quad \rho=\mpm{0&1}\;,
\end{equation}
for some $\epsilon_0\in \hom[E_0|G]$.
Note that when evaluating a tensor network where $\epsilon$ of each individual tensor is injective, all reductions are of this form.
In this case, we have $E'=E_0$, $\epsilon'=\epsilon_0$, and $(q')^{(\tilde2)}$ and $(q')^{(\tilde1)}$ are obtained by restricting $\widetilde q^{(\tilde2)}$ and $\widetilde q^{(\tilde1)}$ to the submatrix/subvector corresponding to $E_0$.

Note that we can rewrite the first line of Eq.~\eqref{eq:invert_reduction_coefficients} as
\begin{equation}
\widetilde q^{(2)}
=q^{(2)} - (q^{(2)})^* \rho(\qinv)^*\rho^* q^{(2)} \rho\qinv \rho^*q^{(2)}
=q^{(2)} - ((q^{(2)})^* \rho) \qinv (\rho^*q^{(2)})\;.
\end{equation}
In this form, the reduction on the level of the bilinear forms is reminiscent of the \emph{Schur complement}, as already mentioned in the context of partial Gaussian integrals in Section~\ref{sec:tutorial}.

\myparagraph{Evaluating the quadratic sum}
For the first type of reduction, we have to calculate the value of $q^{\sum}$.
In practice, it suffices to be able to evaluate the sum $\sum_{g\in G} \exp(q(g))$ for all elementary abelian groups $G$ and all normalized quadratic functions $q$ such that $q^{(2)}$ is non-degenerate.
We will not do this here explicitly, but make two general comments.
The first comment is that, if $G$ is finite, the absolute value of the exponential sum is $\sqrt{|G|}$:
\begin{equation}
\begin{gathered}
|\sum_{g\in G}\exp(q(g))|^2
= \sum_{g\in G} \exp(-q(g)) \sum_{g'\in G}\exp(q(g'))
\overset{*}{=} \sum_{g\in G} \exp(-q(g)) \sum_{g'\in G}\exp(q(g+g'))\\
= \sum_{g,g'\in G}\exp(q(g+g')-q(g))
\overset{\eqref{eq:second_derivative}}{=} \sum_{g,g'\in G} \exp(q(g')+q^{(2)}(g',g))
= \sum_{g'\in G}\exp(q(g')) \sum_{g\in G}\exp(q^{(2)}(g',g))\\
\overset{\eqref{eq:group_character_sum}}{=} \sum_{g'\in G}\exp(q(g'))\cdot |G|\delta_{q^{(2)}(g')=0}
\overset{**}{=} \sum_{g'\in G}\exp(q(g'))\cdot |G|\delta_{g'=0}
= |G|\;.
\end{gathered}
\end{equation}
For the equality labeled $*$, we have performed a change of summation variable $g'\rightarrow g'+g$, and for the equality labeled $**$ we have used that $q^{(2)}$ is invertible.
The second comment is that if we know the exponential sum for $q\in\hom_2[G|A]$, then it is easy to calculate the exponential sum for $q+h$ with $h\in\hom[G|A]$:
\begin{equation}
\sum_{g\in G} \exp(q(g)+h(g))
\overset{\eqref{eq:second_derivative}}{=}\sum_{g\in G} \exp\big(q(g+(q^{(2)})^{-1}(h)) - q((q^{(2)})^{-1}(h))\big)
\overset{*}{=}\exp(-q((q^{(2)})^{-1}(h))) \sum_{g\in G} \exp(q(g))\;.
\end{equation}
For the equality labeled $*$ we have performed a change of summation variable $g+(q^{(2)})^{-1}(h)\rightarrow g$, and the inverse $(q^{(2)})^{-1}$ exists per assumption that $q^{(2)}$ is non-degenerate.
In other words, if we know the exponential sum for one $q$ with $q^{(2)}=b$, we know it for all $q$ with $q^{(2)}=b$.

\section{Generalized stabilizer codes and states}
\label{sec:stabilizer_codes}
In this section, we show that various objects related to (Pauli) stabilizer codes are quadratic tensors.
To this end, we generalize the notion of a stabilizer code from copies of $\zz_2$ to arbitrary abelian groups, using quadratic functions.
Similarly, we show how quadratic functions appear in generalized Clifford operations, and how these operations are described by quadratic tensors.

\subsection{Generalized Pauli operators}
We start by defining Pauli operators for arbitrary abelian groups $H$.
We then show that these Pauli operators are quadratic tensors, and compute the corresponding quadratic tensor data.

\begin{mydef}
\label{def:generalized_pauli_operator}
For any abelian group $H$ and $h\in H, h'\in H^*$, define the \emph{generalized Pauli operator} $\rho(h,h')\in GL(\cc^H)$ as
\begin{equation}
\label{eq:generalized_pauli_operator_def}
(\rho(h,h')\phi)(h_0)=e^{2\pi i h'(h_0)} \phi(h_0-h)\quad \forall \phi\in \cc^H\;.
\end{equation}
Here, the dual $H^*$ (see Definition~\ref{def:duals}) is the Pontryagin dual with $A=\rr/\zz$, $H^*=\hom[H|\rr/\zz]$.
\end{mydef}
\begin{myprop}
The map
\begin{equation}
\rho: H\times H^*\rightarrow GL(\cc^H)
\end{equation}
defines a projective representation of $H\times H^*$.
That is, it fulfills
\begin{equation}
\label{eq:pauli_projective_rep}
\rho(h_1+h_2,h_1'+h_2')
=e^{2\pi i\omega((h_1,h_1'),(h_2,h_2'))}\rho(h_1,h_1')\rho(h_2,h_2')\;,
\end{equation}
with
\begin{equation}
\omega((h_1,h_1'),(h_2,h_2'))\coloneqq h_2'(h_1)\;.
\end{equation}
\end{myprop}
\begin{proof}
Direct calculation shows
\begin{equation}
\begin{gathered}
(\rho(h_1+h_1,h_1'+h_2')\phi)(h_0)
=e^{2\pi i (h_1'+h_2')(h_0)}\phi(h_0-(h_1+h_2))\\
=e^{2\pi i (h_2'(h_1)+h_1'(h_0)+h_2'(h_0-h_1))}\phi(h_0-h_1-h_2)
=e^{2\pi i (h_2'(h_1)+h_1'(h_0))} (\rho(h_2,h_2')  \phi)(h_0-h_1)\\
=e^{2\pi i h_2'(h_1)} (\rho(h_1,h_1')\rho(h_2,h_2')\phi)(h_0)
=e^{2\pi i\omega((h_1,h_1'),(h_2,h_2'))}(\rho(h_1,h_1')\rho(h_2,h_2')\phi)(h_0)\;.
\end{gathered}
\end{equation}
\end{proof}
Mathematically, $\omega$ is a group 2-cocycle $\omega\in H^2(B(H\times H^*),\rr/\zz)$, or equivalently, $e^{2\pi i \omega}\in H^2(B(H\times H^*),U(1))$.
$\omega$ also is a bilinear form, $\omega\in \hom^2[H\times H^*,H\times H^*|\rr/\zz]\simeq \hom[H\times H^*|H^*\times H]$, which as a matrix could be written as
\begin{equation}
\omega=\begin{pmatrix}0&1\\0&0\end{pmatrix}\;.
\end{equation}
Note that since $\omega$ is in a non-trivial group cohomology class, it is not possible to add prefactors to $\rho$ such that it becomes a proper linear representation.

If we want the generalized Pauli operators to form a group under multiplication, the generalized Pauli group $\mathcal P_H$, we need to consider operators of the form $e^{2\pi i \alpha}\rho(h,h')$ for some $\alpha\in\rr/\zz$.
$H\times H^*$ is then the generalized Pauli group modulo prefactors.
$\mathcal P_H$ is given by the group extension
\begin{equation}
\rr/\zz\rightarrow \mathcal P_H \rightarrow H\times H^*
\end{equation}
defined by the 2-cocycle $\omega$.
That is, as a set, we have $\mathcal P_H=\rr/\zz\times H\times H^*$, and the group multiplication given by
\begin{equation}
(\phi_0,h_0,h_0')\cdot (\phi_1,h_1,h_1') = \big(\phi_0+\phi_1+\omega((h_0,h_0'),(h_1,h_1')), h_0+h_1,h_0'+h_1'\big)\;.
\end{equation}
Note that in the $n$-qubit case, $H=\zz_2^n$, Pauli operators are usually considered only with special prefactors in $\{1,i,-1,-i\}\simeq \zz_4$, and the extension above is by $\zz_4$ instead of $\rr/\zz$, however, these special prefactors do not appear to have a natural generalization to arbitrary groups.

\begin{myprop}
\label{prop:pauli_operator_tensor}
Each generalized Pauli operator is a quadratic tensor over $H\times H$, $\bra{h_o}e^{2\pi i \alpha} \rho(h,h')\ket{h_i}=\mathcal T[E,\epsilon,q](h_o,h_i)$, with
\begin{equation}
E=H,\quad
\epsilon^{(1)}=\mpm{1\\1}\;,\quad
\epsilon^{(0)}=\mpm{h\\0}\;,\quad
q_\phi^{(2)}=\varnothing\;,\quad
q_\phi^{(1)}=\mpm{h'},\quad
q_\phi^{(0)}=\alpha,\quad
q_a=0\;.
\end{equation}
Here, $\varnothing$ denotes the fact that there are no entries in $q_\phi^{(2)}$ since $E$ consists of only one factor.
\end{myprop}
\begin{proof}
We can see this by direct computation,
\begin{equation}
\begin{gathered}
\bra{h_o} e^{2\pi i\alpha} \rho(h,h') \ket{h_i}
= e^{2\pi i (\alpha+h'(h_i))} \braket{h_o|h_i+h}
= e^{2\pi i (\alpha+ h'(h_i))} \delta_{h_o=h_i+h}
= \sum_{e\in H: (h_o,h_i)=(e+h,e)} e^{2\pi i (\alpha+ h'(e))}\;.
\end{gathered}
\end{equation}
\end{proof}

Not only the individual generalized Pauli operators are quadratic tensors, but also the collection of all Pauli operators is one single quadratic tensor.
\begin{myprop}
The projective representation $\rho$ is a quadratic tensor over $H\times H\times H\times H^*$, $\bra{h_o}\rho(h,h')\ket{h_i}=\mathcal T[E,\epsilon,q](h_o,h_i,h,h')$, where
\begin{equation}
E=H\times H\times H^*\;,\quad
\epsilon^{(1)}=\mpm{1&1&0\\1&0&0\\0&1&0\\0&0&1}\;,\quad
\epsilon^{(0)}=\mpm{0\\0\\0\\0}\;,\quad
q_\phi^{(2)}=\mpm{0&1\\&0}\;,\quad
q_\phi^{(1)}=\mpm{0&0&0}\;,\quad
q_\phi^{(0)}=q_a=0\;.
\end{equation}
\end{myprop}
\begin{proof}
The tensor entries are given by $\bra{h_o}\rho(h,h')\ket{h_i}=\delta_{h_o=h_i+h} e^{2\pi i h'(h_i)}$.
Elements of $E$ correspond to 3-tuples $(h_i,h,h')$, such that the first row of $\epsilon^{(1)}$ represents $h_o=h_i+h$.
The ``$1$'' entry of $q_\phi^{(2)}$ denotes the canonical pairing $h'(h_i)$ between $h'\in H^*$ and $h_i\in H$, which is the same as the identity homomorphism $1\in\hom[H|H^{**}]=\hom[H|H]$.
\end{proof}
In fact, this quadratic tensor can also be viewed as a map from the space of operators $\cc^{H\times H}$ to the ``phase space'' $\cc^{H\times H^*}$.
Up to some normalization factors, this is (a generalization of) the map that transforms an operator in its matrix representation to its \emph{characteristic function}.
The \emph{Wigner function} can be obtained from the characteristic function by a Fourier transform, so the map from the matrix representation to the Wigner function is again a quadratic tensor.

\subsection{Generalized stabilizer codes}
\label{sec:stabilizer_projector}
In this section, we introduce generalized (Pauli) stabilizer codes in a language compatible with Section~\ref{sec:clifford_tensors}.
For simplicity, we start with the special case of generalized Calderbank-Shor-Steane (CSS) codes~\cite{Calderbank1995,Steane1996}.
We start by defining the data needed to specify a generalized CSS code.
\begin{mydef}
A \emph{CSS stabilizer tableau data} over a group $H$ is a tuple $(S_x,S_z,\sigma_x,\sigma_z)$, where
\begin{itemize}
\item $S_x$ and $S_y$ are abelian groups,
\item $\sigma_x$ and $\sigma_y$ are injective homomorphisms
\begin{equation}
\sigma_x\in \hom[S_x|H],\quad \sigma_z \in \hom[S_z|H^*]\;,
\end{equation}
such that
\begin{equation}
\label{eq:css_stabilizer_tableau_condition}
\sigma_z^*\sigma_x=0\;.
\end{equation}
\end{itemize}
\end{mydef}
Next, we show how a generalized CSS code can be constructed from its data.

\begin{myprop}
For a CSS stabilizer tableau data $(S_x,S_z,\sigma_x,\sigma_z)$ over $H$, consider the following linear map $R[S_x,S_z,\sigma_x,\sigma_z]: S_x\times S_z\rightarrow GL(\cc^H)$:
\begin{equation}
\label{eq:css_stabilizer_representation}
R[S_x,S_z,\sigma_x,\sigma_z](s_x,s_z)\coloneqq\rho(\sigma_xs_x,\sigma_zs_z)\;.
\end{equation}
$R[S_x,S_z,\sigma_x,\sigma_z]$ defines a linear representation $R[S_x,S_z,\sigma_x,\sigma_z] \in \hom[S_x\times S_z|GL(\cc^H)]$ of $S_x\times S_z$ on $\cc^H$.
\end{myprop}

\begin{mydef}
For a CSS stabilizer tableau data $(S_x,S_z,\sigma_x,\sigma_z)$, the \emph{associated quantum code} is the invariant subspace of $R[S_x,S_z,\sigma_x,\sigma_z]$.
$\cc^H$ is called the \emph{physical Hilbert space}, and $S_x\times S_z$ is called the \emph{stabilizer group}.
\end{mydef}

\begin{mydef}
We say that a CSS stabilizer tableau data $(S_x,S_z,\sigma_x,\sigma_z)$ over $H$ is \emph{complete} if
\begin{equation}
\forall h\in H :\quad \sigma_z^*h=0\Rightarrow\exists s_x\in S_x: h=\sigma_x s_x\;.
\end{equation}
For a complete CSS stabilizer tableau data, the \emph{associated stabilizer state} $\phi[S_x,S_z,\sigma_x,\sigma_z]$ is the unique common eigenstate of all stabilizer operators,
\begin{equation}
\rho(\sigma_xs_x,\sigma_zs_z) \phi[S_x,S_z,\sigma_x,\sigma_z]=\phi[S_x,S_z,\sigma_x,\sigma_z] \quad\forall s_x\in S_x,\quad s_z\in S_z\;.
\end{equation}
\end{mydef}
Note that $\phi$ is only unique up to a global prefactor.
Being complete is equivalent to the condition that the length-2 chain complex
\begin{equation}
S_x\xrightarrow{\sigma_x} H \xrightarrow{\sigma_z^*} S_z^*
\end{equation}
has trivial 1-homology, or in other words, that the above defines a short exact sequence.

Next, we continue with general (non-CSS) stabilizer codes.
Again, we start by defining the data needed to specify a generalized stabilizer code.
\begin{mydef}
A \emph{stabilizer tableau data} over an abelian group $H$ is a tuple $(S,\sigma,p)$, where
\begin{itemize}
\item $S$ is an abelian group,
\item $\sigma$ is an injective homomorphism
\begin{equation}
\sigma\in \hom[S|H\times H^*]\;,
\end{equation}
\item and $p$ is a normalized quadratic function $p\in \hom_2[S|\rr/\zz]$ with
\begin{equation}
\label{eq:stabilizer_tableau_condition}
p^{(2)} = \sigma^*\omega \sigma\;.
\end{equation}
\end{itemize}
\end{mydef}
Note that $p^{(2)}$ must be symmetric, and thus
\begin{equation}
\label{eq:code_jconjugation_zero}
0=p^{(2)}-(p^{(2)})^* = \sigma^*(\omega-\omega^*) \sigma = \sigma^*J \sigma\;,
\end{equation}
where
\begin{equation}
J=\omega-\omega^*=\mpm{0&1\\-1&0}
\in \hom[H\times H^*|H^*\times H]
\end{equation}
is a generalization of the symplectic form to arbitrary abelian groups.
Also note that if we write
\begin{equation}
\sigma=
\begin{pmatrix}
\sigma_x\\\sigma_z
\end{pmatrix}
\;,
\end{equation}
then Eq.~\eqref{eq:stabilizer_tableau_condition} becomes
\begin{equation}
\label{eq:stabilizer_tableau_condition_rewritten}
p^{(2)}=\sigma_x^*\sigma_z = \sigma_z^* \sigma_x\;.
\end{equation}

Next, we define the stabilizer code from its data.
\begin{myprop}
Given a stabilizer tableau data $(S,\sigma,p)$ over $H$,
\begin{equation}
\label{eq:stabilizer_group_representation}
R(s)\coloneqq e^{-2\pi i p(s)} \rho(\sigma s)\;.
\end{equation}
defines a linear representation of $S$ on $\cc^H$, $R \in \hom[S|\operatorname{GL}(\cc^H)]$.
\end{myprop}
\begin{proof}
Direct computation shows
\begin{equation}
\begin{gathered}
(R(s_0+s_1)\phi)(h)
\overset{\eqref{eq:stabilizer_group_representation}}{=} e^{-2\pi ip(s_0+s_1)}(\rho(\sigma(s_0+s_1))\phi)(h)\\
\overset{\eqref{eq:second_derivative},\eqref{eq:generalized_pauli_operator_def}}{=} e^{2\pi i\big(-p(s_0)-p(s_1)-p^{(2)}(s_0,s_1)+\sigma_z(s_0+s_1)(h)\big)} \phi(h-\sigma_x(s_0+s_1))\\
\overset{\eqref{eq:stabilizer_tableau_condition_rewritten}}{=} e^{2\pi i\big(-p(s_0)+\sigma_z (s_0)(h)\big)} e^{2\pi i\big(-p(s_1)+\sigma_z (s_1)(h-\sigma_x s_0)\big)} \phi(h-\sigma_xs_0-\sigma_xs_1)\\
\overset{\eqref{eq:stabilizer_group_representation}}{=} e^{2\pi i\big(-p(s_0)+\sigma_z (s_0)(h)\big)} (R(s_1) \phi)(h-\sigma_x s_0)
\overset{\eqref{eq:stabilizer_group_representation}}{=}(R(s_0)R(s_1) \phi)(h)\;.
\end{gathered}
\end{equation}
\end{proof}
Let us provide some insight into this computation:
Since $\rho$ is a projective representation with group 2-cocycle $\omega$ (or, more precisely $e^{2\pi i \omega}$), $\rho(\sigma\bullet)$ is a projective representation with 2-cocycle $\sigma^*\omega \sigma$.
Eq.~\eqref{eq:stabilizer_tableau_condition} implies that $p$ is a group 1-cochain whose coboundary is $\sigma^*\omega \sigma$, so $\sigma^*\omega \sigma$ is cohomologically trivial and $\rho(\sigma\bullet)$ can be turned into a proper representation by adding $p$ (or more precisely $e^{2\pi i p}$) as a prefactor, as in Eq.~\eqref{eq:stabilizer_group_representation}.

\begin{mydef}
For a stabilizer tableau data $(S,\sigma,p)$ over $H$, the \emph{associated quantum code} is the invariant subspace of the corresponding representation $R$.
In particular, the physical Hilbert space is $\cc^H$, and the stabilizer group is $S$.
\end{mydef}

\begin{mydef}
We say that a stabilizer tableau data $(S,\sigma,p)$ over $H$ is \emph{complete} if
\begin{equation}
\forall (h,h')\in H\times H^* :\quad \sigma^*J (h,h')=0\Rightarrow \exists s\in S: (h,h')=\sigma s\;.
\end{equation}
For a complete stabilizer tableau data, the \emph{associated stabilizer state} $\phi[S,\sigma,p]$ is the unique common eigenstate of all stabilizer operators,
\begin{equation}
\label{eq:stabilizer_state_condition}
e^{-2\pi i p(s)} \rho(\sigma_xs,\sigma_zs) \phi[S,\sigma,p]=\phi[S,\sigma,p] \quad\forall s\in S\;.
\end{equation}
\end{mydef}
In other words, a stabilizer tableau data is complete if the length-2 chain complex
\begin{equation}
S\xrightarrow{\sigma} H\times H^*\xrightarrow{\sigma^*J} S^*
\end{equation}
has trivial 1-homology, and defines a short exact sequence.

Finally, let us spell out explicitly how generalized CSS codes are a special case of generalized stabilizer codes.
\begin{myprop}
Any CSS stabilizer tableau data $(S_x,S_z,\sigma_x,\sigma_z)$ gives rise to a stabilizer tableau data $(S',\sigma',p)$ via
\begin{equation}
\label{eq:css_to_general_mapping}
S'=S_x\times S_z,\quad
\sigma'=
\begin{pmatrix}
\sigma_x&0\\
0& \sigma_z
\end{pmatrix},\quad
p=0\;.
\end{equation}
The associated quantum codes for $(S_x,S_z,\sigma_x,\sigma_z)$ and $(S',\sigma',p)$ are the same.
\end{myprop}
\begin{proof}
Eq.~\eqref{eq:css_to_general_mapping} indeed fulfills Eq.~\eqref{eq:stabilizer_tableau_condition}, since
\begin{equation}
p^{(2)}= 0 = \sigma_z^*\sigma_x = \sigma^*\omega \sigma\;.
\end{equation}
\end{proof}

\subsection{Code projector and state as quadratic tensors}
In this section, we show that the projector onto the code space for any generalized (CSS) stabilizer code is a quadratic tensor, and the same is true for the unique code state of a complete (CSS) stabilizer tableau data.
As usual, we start with the much simpler CSS case.

\begin{myprop}
For a CSS stabilizer tableau data $(S_x,S_z,\sigma_x,\sigma_z)$, let $P[S_x,S_z,\sigma_x,\sigma_z]$ denote the projector onto the code space of the associated quantum code.
$\bra{h_o}P[S_x,S_z,\sigma_x,\sigma_z]\ket{h_i}=\mathcal T[E,\epsilon,q](h_o,h_i)$ is a quadratic tensor over $H\times H$ with
\begin{equation}
\begin{gathered}
E=K_z\times S_x\;,\quad
\epsilon^{(1)}=\mpm{\kappa_z&\sigma_x\\\kappa_z&0}\;,\quad
\epsilon^{(0)}=\mpm{0\\0}\;,\quad
q_\phi^{(2)}=\mpm{0}\;,\quad
q_\phi^{(1)}=\mpm{0&0}\;,\quad
q_\phi^{(0)}=0\;,\\
q_a^{(0)}=-\log(|S_x|)\;.
\end{gathered}
\end{equation}
Here, $K_z$ is some abelian group isomorphic to the kernel of $\sigma_z^*$, $K_z\simeq\ker(\sigma_z^*)$, and $k_z\in \hom[K_z|H]$ is an injective homomorphism that maps $K_z$ onto $\ker(\sigma_z^*)$.
\end{myprop}
\begin{proof}
Direct computation yields
\begin{equation}
\label{eq:css_code_space_projector}
\begin{gathered}
\bra{h_o} P[S_x,S_z,\sigma_x,\sigma_z] \ket{h_i}
\overset{*,\eqref{eq:css_stabilizer_representation}}{=}\frac{1}{|S_x||S_z|} \sum_{s_x\in S_x,s_z\in S_z} \bra{h_o}\rho((\sigma_x s_x,\sigma_z s_z))\ket{h_i}\\
\overset{\eqref{eq:generalized_pauli_operator_def}}{=}\frac{1}{|S_x||S_z|} \sum_{s_x\in S_x,s_z\in S_z} e^{2\pi i (\sigma_z s_z)(h_i)} \delta_{h_o=h_i+\sigma_x s_x}
\overset{\eqref{eq:group_character_sum}}{=}\frac{1}{|S_x|}\sum_{s_x\in S_x} \delta_{\sigma_z^*(h_i)=0} \delta_{h_o=h_i+\sigma_x s_x}\\
=\frac{1}{|S_x|}\sum_{s_x\in S_x,k_z\in K_z} \delta_{h_i=\kappa_z k_z} \delta_{h_o=h_i+\sigma_x s_x}
\;.
\end{gathered}
\end{equation}
In the equation labeled $*$, we have used the fact that (up to normalization), $P[S_x,S_z,\sigma_x,\sigma_z]$ can be obtained from summing the representation $R[S_x,S_z,\sigma_x,\sigma_z]$ over all elements $(s_x,s_z)\in S_x \times S_z$.
\end{proof}

Next, we consider the code state for a complete CSS stabilizer tableau data.
\begin{myprop}
For a complete CSS stabilizer tableau data $(S_x,S_z,\sigma_x,\sigma_z)$, the associated code state $\phi[S_x,S_z,\sigma_x,\sigma_z](h)=\mathcal T[E,\epsilon,q](h)$ is a quadratic tensor over $H$ with
\begin{equation}
E=S_x\;,\quad
\epsilon^{(1)} = \mpm{\sigma_x}\;,\quad
\epsilon^{(0)} = \mpm{0}\;,\quad
q_\phi^{(2)}=\varnothing\;,\quad
q_\phi^{(1)}=\mpm{0}\;,\quad
q_\phi^{(0)}=0\;,\quad
q_a^{(0)}=-\frac12 \log(|S_x|)\;.
\end{equation}
\end{myprop}
\begin{proof}
We can show this by applying the code space projector in Eq.~\eqref{eq:css_code_space_projector} onto the $\ket0$ state (up to normalization),
\begin{equation}
\begin{gathered}
\phi[S_x,S_z,\sigma_x,\sigma_z](h)
\propto \bra{h}P[S_x,S_z,\sigma_x,\sigma_z]\ket0\\
=\frac{1}{|S_x|}\sum_{s_x\in S_x,k_z\in K_z} \delta_{0=\kappa_z k_z} \delta_{h=\sigma_x s_x}
=\frac{1}{|S_x|}\sum_{s_x\in S_x} \delta_{h=\sigma_x s_x}
\propto\mathcal T[E,\epsilon,q](h)
\;.
\end{gathered}
\end{equation}
\end{proof}

Let us now consider general (non-CSS) stabilizer codes.
\begin{myprop}
\label{prop:stabilizer_codestate_tensor}
Let $P[S,\sigma,p]$ denote the code space projector of a generalized stabilizer code $(S,\sigma,p)$.
$\bra{h_o}P[S,\sigma,p]\ket{h_i}=\mathcal T[E,\epsilon,q](h_o,h_i)$ is a quadratic tensor over $H\times H$ with $(E,\epsilon,q)$ constructed as follows.
Let $\kappa_x\in \hom[K_x|S]$ be a homomorphism that isomorphically identifies $\ker(\sigma_x)$ with some standalone group $K_x$.
In particular, $\kappa_x$ is injective and
\begin{equation}
\label{eq:stabilizer_data_kernel_condition}
\sigma_x\kappa_x=0\;.
\end{equation}
In addition, we choose a left inverse $\kappa_x^\perp: S/K_x\rightarrow S$ of the quotient map $S\rightarrow S/K_x$, similar to Eq.~\eqref{eq:reduction_lift}.
Further, choose $h_0\in H$ such that
\begin{equation}
\label{eq:h0_definition}
\kappa_x^*\sigma_z^*h_0 = p\circ \kappa_x\;.
\end{equation}
Finally, let $\epsilon_z\in \hom[E_z|H]$ be a homomorphism that identifies $\ker(\kappa_x^* \sigma_z^*)\subset H$ with some standalone group $E_z$.
In particular, $\epsilon_z$ is injective, and
\begin{equation}
\label{eq:stabilizer_data_kernel_condition2}
\kappa_x^*\sigma_z^*\epsilon_z=0\;.
\end{equation}
With this, the quadratic tensor data is given by
\begin{equation}
\label{eq:stabilizer_projector_tensor_data}
\begin{gathered}
E=E_z\times S/K_x,\quad
\epsilon^{(2)}=\mpm{\epsilon_z & \sigma_x\circ \kappa_x^\perp\\\epsilon_z&0}\;,\quad
\epsilon^{(1)}=\mpm{h_0\\h_0}\;,\quad
q_\phi^{(2)}=\mpm{\epsilon_z^*\sigma_z\circ\kappa_x^\perp}\;,\\
q_\phi^{(1)}=\mpm{0&(\sigma_z^*(h_0)-p)\circ \kappa_x^\perp}\;,\quad
q_\phi^{(0)}=0\;,\quad
q_a^{(0)}=\log(|K_x|)-\log(|S|)\;.
\end{gathered}
\end{equation}
\end{myprop}
\begin{proof}
First we should check that Eq.~\eqref{eq:h0_definition} makes sense and there always exists a choice of $h_0$.
We start by noting that the right-hand side is valued in $K_x^*$ like the left-hand side,
\begin{equation}
(p\circ \kappa_x)^{(2)}=\kappa_x^*p^{(2)}\kappa_x\overset{\eqref{eq:stabilizer_tableau_condition_rewritten}}{=}\kappa_x^* \sigma_z^*\sigma_x \kappa_x\overset{\eqref{eq:stabilizer_data_kernel_condition}}{=}0\;,\quad
(p\circ \kappa_x)^{(0)}=p(0)=0\qquad
\Rightarrow\quad 
p\circ\kappa_x \in \hom[K_x|\rr/\zz]= K_x^*\;.
\end{equation}
Now let us check that $h_0$ always exists:
Since $\kappa_x$ and $\sigma$ are injective, $\sigma\kappa_x$ is injective.
Since additionally $\sigma_x\kappa_x=0$, $\sigma_z\kappa_x$ is injective.
Thus, $(\sigma_x\kappa_z)^*=\kappa_x^*\sigma_z^*\in\hom[H|K_x^*]$ is surjective and $h_0$ always exists.
Next, $\kappa_x^\perp$ is in general not a homomorphism, so we should check that all three entries in Eq.~\eqref{eq:stabilizer_projector_tensor_data} involving a composition with $\kappa_x^\perp$ remain homomorphisms or quadratic functions after the composition.
We should also check that these entries do not depend on the choice of $\kappa_x^\perp$.
For both properties it suffices to show that we get zero if we replace $\kappa_x^\perp$ with the homomorphism $\kappa_x$ in each expression.
For the first expression $\epsilon^{(2)}_{01}=\sigma_x\circ\kappa_x^\perp: S/K_x\rightarrow H$, we indeed have $\sigma_x\kappa_x=0$ due to Eq.~\eqref{eq:stabilizer_data_kernel_condition}.
The second expression is $q^{(2)}_{\phi,01}=\epsilon_z^*\sigma_z\circ\kappa_x^\perp: S/K_x\rightarrow E_z^*$, and again we have $\epsilon_z^*\sigma_z\kappa_x=0$ due to Eq.~\eqref{eq:stabilizer_data_kernel_condition2}.
The third expression is $q^{(1)}_{\phi,1}=(\sigma_z^*(h_0)-p)\circ \kappa_x^\perp: S/K_x\rightarrow \rr/\zz$, and again we find
\begin{equation}
(\sigma_z^*(h_0)-p)\kappa_x
=\kappa_x^*\sigma_z^*h_0-p\circ\kappa_x
\overset{\eqref{eq:h0_definition}}{=}0\;.
\end{equation}
After these initial checks, the result follows from direct computation:
\begin{equation}
\label{eq:stabilizer_projector_data_derivation}
\begin{gathered}
\bra{h_o} P[S,\sigma,p] \ket{h_i}
\overset{*}{=} \frac{1}{|S|}\sum_{s\in S} \bra{h_o}R(s)\ket{h_i}
\overset{\eqref{eq:stabilizer_group_representation},\eqref{eq:generalized_pauli_operator_def}}{=} \frac{1}{|S|} \sum_{s\in S} e^{2\pi i \big(-p(s)+(\sigma_zs)(h_i)\big)} \delta_{h_o=h_i+\sigma_x s}\\
\overset{\sim\eqref{eq:invertiblered_derivation1}}{=} \frac{1}{|S|} \sum_{k\in K_x, r\in S/K_x} e^{2\pi i \big(-p(\kappa_x^\perp(r)+\kappa_x k)+(\sigma_z (\kappa_x^\perp(r)+\kappa_x k))(h_i)\big)} \delta_{h_o=h_i+\sigma_x (\kappa_x^\perp(r)+\kappa_x k)}\\
\overset{\eqref{eq:second_derivative},\eqref{eq:stabilizer_tableau_condition_rewritten},\eqref{eq:stabilizer_data_kernel_condition}}{=} \frac{1}{|S|} \sum_{k\in K_x, r\in S/K_x} e^{2\pi i \big(-p(\kappa_x^\perp(r))-p(\kappa_x k)-\kappa_x^\perp(r)(\sigma_z^*\sigma_x\kappa_x k) +(\sigma_z\circ\kappa_x^\perp(r))(h_i) +(\sigma_z\kappa_x k)(h_i)\big)} \delta_{h_o=h_i+\sigma_x (\kappa_x^\perp(r))}\\
\overset{\eqref{eq:stabilizer_data_kernel_condition}}{=} \frac{1}{|S|} \sum_{r\in S/K_x} e^{2\pi i \big(-p\circ \kappa_x^\perp(r) +(\sigma_z\circ \kappa_x^\perp(r))(h_i)\big)} \sum_{k\in K_x} e^{2\pi i(\kappa_x^*\sigma_z^* h_i-p\circ \kappa_x) (k)} \delta_{h_o=h_i+\sigma_x \circ\kappa_x^\perp(r)}\\
\overset{\eqref{eq:group_character_sum}}{=} \frac{|K_x|}{|S|} \sum_{r\in S/K_x} e^{2\pi i \big(-p\circ \kappa_x^\perp(r) +(\sigma_z\circ \kappa_x^\perp(r))(h_i)\big)} \delta_{\kappa_x^*\sigma_z^* h_i-p\circ \kappa_x=0} \delta_{h_o=h_i+\sigma_x \circ\kappa_x^\perp(r)}\\
\overset{\eqref{eq:code_projector_kernel_parameterization}}{=} \frac{|K_x|}{|S|} \sum_{r\in S/K_x} e^{2\pi i \big(-p\circ\kappa_x^\perp(r) +(\sigma_z\circ \kappa_x^\perp(r))(h_i)\big)} \sum_{e\in E_z} \delta_{h_i=h_0+\epsilon_z e} \delta_{h_o=h_i+\sigma_x\circ \kappa_x^\perp(r)}\\
=\frac{|K_x|}{|S|} \sum_{\substack{r\in S/K_x,e\in E_z\\h_i=h_0+\epsilon_z e\\h_o=h_0+\epsilon_ze+\sigma_x\circ \kappa_x^\perp(r)}} e^{2\pi i \big(-p\circ\kappa_x^\perp(r) +(\sigma_z\circ \kappa_x^\perp(r))(h_0+\epsilon_z e)\big)}\;.
\end{gathered}
\end{equation}
For the equality labeled $*$, we have again used that the sum over a representation is the projector onto its invariant subspace.
We have also used the explicit parametrization of the solution of a linear equation,
\begin{equation}
\label{eq:code_projector_kernel_parameterization}
\kappa_x^*\sigma_z^* h_i-p\circ \kappa_x=0
\quad \overset{\eqref{eq:stabilizer_data_kernel_condition2},\eqref{eq:h0_definition}}{\Leftrightarrow}\quad
\exists e\in E_z: h_i=h_0+\epsilon_z e\;.
\end{equation}
\end{proof}
Note that already the third term in Eq.~\eqref{eq:stabilizer_projector_data_derivation} has the form of a quadratic tensor with data
\begin{equation}
E=H\times S\;,\quad
\epsilon^{(1)}=\mpm{1&\sigma_x\\1&0}\;,\quad
\epsilon^{(0)}=\mpm{0\\0}\;,\quad
q_\phi^{(2)}=\mpm{\sigma_z}\;,\quad
q_\phi^{(1)}=\mpm{0&-p}\;,\quad
q_a^{(0)}=-\log(|S|)\;.
\end{equation}
However, this quadratic tensor data is not fully reduced.
Applying the reduction from Proposition~\ref{prop:zero_reduction} for $\rho=\kappa_x$ yields Eq.~\eqref{eq:stabilizer_projector_tensor_data}.

\begin{myprop}
\label{prop:stabilizer_tensor}
The associated stabilizer state of a complete stabilizer tableau data $(S,\sigma,p)$ is a quadratic tensor over $H$, $\phi[S,\sigma,p](h)=\mathcal T[E,\epsilon,q](h)$, with
\begin{equation}
\label{eq:stabilizer_tensor_data}
\begin{gathered}
E = S/K_x\;,\quad
\epsilon^{(1)} = \mpm{\sigma_x\circ\kappa_x^\perp}\;,\quad
\epsilon^{(0)} = \mpm{h_0}\;,\quad
q_\phi^{(2)}=\varnothing\;,\quad
q_\phi^{(1)}=\mpm{(\sigma_z^*(h_0)-p)\circ\kappa_x^\perp}\;,\\
q_\phi^{(0)}=0\;,\quad
q_a^{(0)}= \frac12 \log(|K_x|)-\frac12 \log(|S|)\;,
\end{gathered}
\end{equation}
using $K_x$, $\kappa_x^\perp$, and $h_0$ from Proposition~\ref{prop:stabilizer_codestate_tensor}.
\end{myprop}
\begin{proof}
This is obtained from applying the code space projector to the state $\ket{h_0}$, setting $h_i=h_0$ in the result of Eq.~\eqref{eq:stabilizer_projector_data_derivation}:
\begin{equation}
\begin{gathered}
\phi[S,\sigma,p](h)\propto \bra{h} P[S,\sigma,p] \ket{h_0}
=\frac{|K_x|}{|S|} \sum_{\substack{r\in S/K_x,e\in E_z\\h_0=h_0+\epsilon_z e\\h=h_0+\epsilon_ze+\sigma_x\circ \kappa_x^\perp(r)}} e^{2\pi i \big(-p\circ\kappa_x^\perp(r) +(\sigma_z\circ \kappa_x^\perp(r))(h_0+\epsilon_z e)\big)}\\
=\frac{|K_x|}{|S|} \sum_{\substack{r\in S/K_x:h=h_0+\sigma_x\circ \kappa_x^\perp(r)}} e^{2\pi i \big(-p\circ\kappa_x^\perp(r) +(\sigma_z\circ \kappa_x^\perp(r))(h_0)\big)}\;.
\end{gathered}
\end{equation}
\end{proof}

Let us now provide some intuition for the above propositions.
$K_x$ describes the space of pure $Z$-type stabilizers, which give rise to the zero entries of the tensor.
The action of the pure-$Z$ stabilizer $k\in K_x$ is the exponential of $(\sigma_z \kappa_x k)(h)-p(\kappa_x k)=(k_x^*\sigma_z^*h-p\circ\kappa_x)k$.
The subspace of $H$ outside of which all tensor entries are zero is thus given determined by $\{h\in H: \kappa_x^*\sigma_z^*h=p\circ\kappa_x\}$.
Finally, we note that in the case of a complete stabilizer tableau, $E_z$ is isomorphic to $S/K_x$.

Finally, we consider a special case for illustration.
\begin{myprop}
\label{eq:prop_stabilizertensor_zerokernel}
Consider a complete stabilizer tableau data with $\ker(\sigma_x)=0$.
Then the associated stabilizer state is a quadratic tensor over $H$ with data
\begin{equation}
E = S\;,\quad
\epsilon^{(1)} = \mpm{\sigma_x}\;,\quad
\epsilon^{(0)} = 0\;,\quad
q_\phi^{(2)}=\varnothing\;,\quad
q_\phi^{(1)}=-p^{(1)}\;,\quad
q_\phi^{(0)}=0\;,\quad
q_a^{(0)}=-\frac12\log(|S|)\;.
\end{equation}
\end{myprop}
\begin{proof}
This is a special case of Proposition~\ref{prop:stabilizer_tensor}, where we set $K_x=0$, $h_0=0$, $\kappa_x^\perp=1$.
\end{proof}

\subsection{Pauli measurement as quadratic tensor}
In this section, we show that generalized Pauli measurements are quadratic tensors.
More precisely, we show that the corresponding projection-operator valued measure (POVM) is a 3-index quadratic tensor, with one index corresponding to the ket, bra, and the classical outcome each.
The most general way to define a Pauli measurement for arbitrary abelian groups is as the stabilizer measurement of a stabilizer tableau data.

\begin{mydef}
For a stabilizer tableau data $(S,\sigma,p)$ over $H$, the \emph{associated Pauli measurement} is the POVM with the following projection operators $\{P_{s'}\}_{s'\in S^*}$,
\begin{equation}
P_{s'}=\frac{1}{|S|} \sum_{s\in S} e^{2\pi i (-p(s)+s'(s))} \rho(\sigma s)\;.
\end{equation}
\end{mydef}
That is, the POVM measures the syndrome of the stabilizer code defined by $(S,\sigma,p)$, which is valued in the dual $S^*$.
If we want to construct a POVM from a single Pauli operator $\rho(h,h')$, we can take $S$ to be the subgroup of $H\times H^*$ generated by $(h,h')$.
Note that even though we need the quadratic function $p$ to define the Pauli measurement, different $p$ yield equivalent measurements:
$p^{(2)}$ is fixed by Eq.~\eqref{eq:stabilizer_tableau_condition}, and adding a linear function $x\in \hom[S,\rr/\zz]=S^*$ to $p$, $p\rightarrow p+x$, simply corresponds to a relabeling of the measurement outcomes, $s'\rightarrow s'-x$.

\begin{myprop}
$\bra{h_o}P_{s'}\ket{h_i}=\mathcal T[E,\epsilon,q](h_o,h_i,s')$ is a quadratic tensor over $H\times H\times S^*$, whose data is constructed as follows.
Recall $K_x$, $\kappa_x$, and $\kappa_x^\perp$ from Proposition~\ref{prop:stabilizer_codestate_tensor}.
Further, choose $s_0'\in S^*$ such that
\begin{equation}
\label{eq:pauli_measurement_offset}
\kappa_x^*s_0'=p\circ\kappa_x\;,
\end{equation}
and let $\lambda\in \hom[L|S^*]$ be a homomorphism that identifies $\ker(\kappa_x^*)$ with some standalone group $L$, in particular,
\begin{equation}
\label{eq:pauli_measurement_kernel_condition}
\kappa_x^*\lambda=0\;.
\end{equation}
With this, the data is given by
\begin{equation}
\begin{gathered}
E=H\times S/K_x\times L,\quad
\epsilon^{(1)}=\mpm{1&\sigma_x\circ\kappa_x^\perp & 0\\1&0&0\\-\sigma_z^*&0&\lambda}\;,\quad
\epsilon^{(0)}=\mpm{0\\0\\s_0'}\;,\quad
q_\phi^{(2)}=\mpm{0&0\\&(\lambda^*\circ\kappa_x^\perp)^*}\\
q_\phi^{(1)}=\mpm{0&(s_0'-p)\circ\kappa_x^\perp & 0}\;,\quad
q_a^{(0)}=\log{|K_x|}-\log(|S|)\;.
\end{gathered}
\end{equation}
\end{myprop}
\begin{proof}
First we check that the entries which are a composition with $\kappa_x^\perp$ are still homomorphisms or quadratic functions.
$\sigma_x\circ\kappa_x^\perp$ was already an entry in Eq.~\eqref{eq:stabilizer_projector_tensor_data}.
$\lambda^*\circ\kappa_x^\perp$ is a homomorphism independent of $\kappa_x^\perp$ due to Eq.~\eqref{eq:pauli_measurement_kernel_condition}.
$(s_0'-p)\circ\kappa_x^\perp$ is a normalized quadratic function independent of $\kappa_x^\perp$ due to Eq.~\eqref{eq:pauli_measurement_offset}.
After these initial checks, the result follows from direct computation:
\begin{equation}
\begin{gathered}
\bra{h_o}P_{s'}\ket{h_i}
=\frac{1}{|S|} \sum_{s\in S} e^{2\pi i(-p(s)+ s'(s))} \bra{h_o}\rho(\sigma s)\ket{h_i}\\
=\frac{1}{|S|} \sum_{s\in S} e^{2\pi i (-p(s)+(\sigma_z s)(h_i)+s'(s))} \delta_{h_o=h_i+\sigma_x s}\\
\overset{\sim\eqref{eq:stabilizer_projector_data_derivation}}{=} \frac{1}{|S|} \sum_{r\in S/K_x} e^{2\pi i \big(-p\circ \kappa_x^\perp(r) +(\sigma_z\circ \kappa_x^\perp(r))(h_i)+s'\circ\kappa_x^\perp(r)\big)} \sum_{k\in K_x} e^{2\pi i(\kappa_x^*\sigma_z^* h_i-p\circ \kappa_x+\kappa_x^*(s')) (k)} \delta_{h_o=h_i+\sigma_x \circ\kappa_x^\perp(r)}\\
\overset{\eqref{eq:group_character_sum}}{=} \frac{|K_x|}{|S|} \sum_{r\in S/K_x} e^{2\pi i \big((s'+\sigma_z^*(h_i)-p)\circ \kappa_x^\perp(r)\big)} \delta_{\kappa_x^*(s'+\sigma_z^* h_i)-p\circ \kappa_x=0} \delta_{h_o=h_i+\sigma_x \circ\kappa_x^\perp(r)}\\
\overset{\eqref{eq:pauli_measurement_kernel_parameterization}}{=} \frac{|K_x|}{|S|} \sum_{r\in S/K_x} \sum_{l\in L} e^{2\pi i \big((s_0'+\lambda l-p)\circ \kappa_x^\perp(r)\big)} \delta_{s'=s_0'+\lambda l-\sigma_z^*h_i} \delta_{h_o=h_i+\sigma_x \circ\kappa_x^\perp(r)}\;.
\end{gathered}
\end{equation}
We have used that
\begin{equation}
\label{eq:pauli_measurement_kernel_parameterization}
\kappa_x^*(s'+\sigma_z^* h_i)-p\circ \kappa_x=0\quad\overset{\eqref{eq:pauli_measurement_kernel_condition},\eqref{eq:pauli_measurement_offset}}{\Leftrightarrow}\quad
\exists l\in L: s'+\sigma_z^*h_i = s_0'+\lambda l\;.
\end{equation}
\end{proof}

\subsection{Generalized Clifford operations}
\label{sec:generalized_clifford}
In this section, we generalize Clifford operations to arbitrary abelian groups, and show that the corresponding unitaries are quadratic tensors.
We start by defining the data needed to specify a generalized Clifford operation.
\begin{mydef}
A \emph{Clifford automorphism data} over a group $H$ is a pair $(\alpha,u)$, where
\begin{itemize}
\item $\alpha$ is a homomorphism, $\alpha\in \hom[H\times H^*|H\times H^*]$,
\item $u$ is a normalized quadratic function, $u\in \hom_2[H\times H^*|\rr/\zz]$ with
\begin{equation}
\label{eq:clifford_data_condition}
u^{(2)}=\alpha^*\omega \alpha-\omega\;.
\end{equation}
\end{itemize}
\end{mydef}
Note that for a given $\alpha$, $u$ can only exist if the bilinear form on the right-hand side of Eq.~\eqref{eq:clifford_data_condition} is symmetric, that is,
\begin{equation}
0=u^{(2)}-(u^{(2)})^* = \alpha^*(\omega-\omega^*)\alpha-(\omega-\omega^*) = \alpha^*J \alpha-J\;,
\end{equation}
or
\begin{equation}
\label{eq:symplectic_condition}
\alpha^*J \alpha=J\;.
\end{equation}
So the homomorphism $\alpha$ can be viewed as a symplectic map.
If we write $\alpha$ in components,
\begin{equation}
\label{eq:clifford_automorphism_decomposition}
\alpha=
\begin{pmatrix}
\alpha_{xx}&\alpha_{xz}\\
\alpha_{zx}&\alpha_{zz}
\end{pmatrix}
\;,
\end{equation}
then Eq.~\eqref{eq:clifford_data_condition} becomes
\begin{equation}
\label{eq:symplectic_clifford_explicit}
u^{(2)}=
\begin{pmatrix}
\alpha_{xx}^*\alpha_{zx} & \alpha_{xx}^* \alpha_{zz}-1\\
\alpha_{xz}^* \alpha_{zx} & \alpha_{xz}^* \alpha_{zz}
\end{pmatrix}
\;,
\end{equation}
and Eq.~\eqref{eq:symplectic_condition} becomes
\begin{equation}
\alpha_{xx}^*\alpha_{zx}=\alpha_{zx}^*\alpha_{xx}\;,\quad
\alpha_{xx}^* \alpha_{zz}-\alpha_{zx}^*\alpha_{xz}=1\;,\quad
\alpha_{xz}^* \alpha_{zz}=\alpha_{zz}^*\alpha_{xz}\;.
\end{equation}

As the name suggests, a Clifford automorphism data $(\alpha,u)$ gives rise to a (generalized) Clifford operation.
\begin{mydef}
\label{eq:associated_clifford}
For a Clifford automorphism data $(\alpha,u)$, the \emph{associated Clifford operation} $C[\alpha,u]\in GL(GL(\cc^H))$ is a superoperator that maps generalized Pauli operators to generalized Pauli operators as follows,
\begin{equation}
\label{eq:clifford_automorphism_definition}
C[\alpha,u](\rho(h,h')) \coloneqq e^{-2\pi i u((h,h'))} \rho(\alpha(h,h'))\;.
\end{equation}
\end{mydef}
\begin{myprop}
$C[\alpha,u]$ is an automorphism of $GL(\cc^H)$.
\end{myprop}
\begin{proof}
By direct calculation:
\begin{equation}
\begin{gathered}
C[\alpha,u] (\rho(h,h')\rho(g,g'))
\overset{\eqref{eq:pauli_projective_rep}}{=} e^{-2\pi i\omega((h,h'),(g,g'))} C[\alpha,u]\rho((h+g,h'+g'))\\
\overset{\eqref{eq:clifford_automorphism_definition}}{=} e^{-2\pi i \big(\omega((h,h'),(g,g'))+u((h+g,h'+g'))\big)} \rho(\alpha(h+g,h'+g'))\\
\overset{\eqref{eq:second_derivative},\eqref{eq:clifford_data_condition}}{=} e^{-2\pi i\big(u((h,h'))+u((g,g'))+(\alpha^*\omega \alpha)((h,h'),(g,g'))\big)}\rho(\alpha(h,h')+\alpha(g,g'))\\
\overset{\eqref{eq:pauli_projective_rep}}{=} e^{-2\pi i u((h,h'))}\rho(\alpha(h,h'))e^{-2\pi i u((g,g'))}\rho(\alpha(g,g'))\\
\overset{\eqref{eq:clifford_automorphism_definition}}{=} (C[\alpha,u]\rho(h,h'))(C[\alpha,u] \rho(g,g'))
\;.
\end{gathered}
\end{equation}
\end{proof}

\begin{mydef}
Like all automorphisms of $GL(\cc^H)$, $C[\alpha,u]$ is given by conjugation with a unitary.
Let $U[\alpha,u]$ denote this \emph{associated Clifford unitary},
\begin{equation}
\label{eq:unitary_conjugation}
C[\alpha,u](\rho(h,h'))= U[\alpha,u]\rho(h,h') U[\alpha,u]^\dagger\;.
\end{equation}
Note that $U[\alpha,u]$ is only determined up to a global phase factor.
\end{mydef}

As a very simple example, generalized Pauli operators are a special case of Clifford unitaries.
\begin{myprop}
The generalized Pauli operator $\rho(h,h')$ is the associated Clifford unitary for the following Clifford automorphism data:
\begin{equation}
\alpha=1,\quad u^{(2)}=0,\quad u^{(1)}=(h,h')J=(-h',h).
\end{equation}
\end{myprop}
\begin{proof}
Direct computation:
\begin{equation}
\begin{gathered}
\rho(h,h')\rho(g,g')\rho(h,h')^\dagger
\overset{\eqref{eq:pauli_projective_rep}}{=} e^{-2\pi ig'(h)} \rho(h+g,h'+g') e^{-2\pi i h'(h)} \rho(-h,-h')\\
\overset{\eqref{eq:pauli_projective_rep}}{=} e^{-2\pi i\big(g'(h)+h'(h)-h'(g+h)\big)} \rho(g+h-h,g'+h'-h')
= e^{-2\pi i\big(-h'(g)+g'(h)\big)} \rho(g,g')
= C[\alpha,u](\rho(g,g'))\;.
\end{gathered}
\end{equation}
\end{proof}

Generalized Clifford gates can be composed efficiently on the level of the underlying data.
\begin{myprop}
Let $(\alpha,u)$ and $(\alpha',u')$ be two Clifford automorphism data.
Then the product of the associated Clifford operations is again a Clifford operation with data $(\alpha'',u'')$,
\begin{equation}
C[\alpha',u'] C[\alpha,u] = C[\alpha'',u'']\;,
\end{equation}
where
\begin{equation}
\alpha''=\alpha'\alpha\;,\qquad u''=u+u'\circ\alpha\;.
\end{equation}
\end{myprop}
\begin{proof}
\begin{equation}
C[\alpha',u']C[\alpha,u]\rho(\xi)
= C[\alpha',u'] e^{-2\pi i u(\xi)} \rho(\alpha(\xi))
= e^{-2\pi i \big(u(\xi)+u'(\alpha(\xi))\big)} \rho(\alpha'(\alpha(\xi)))\;.
\end{equation}
\end{proof}
The coefficients of $\alpha''$ and $u''$ can be efficiently computed from these of $\alpha$, $\alpha'$, $u$, and $u'$ using the methods from Section~\ref{sec:homomorphism_composition}.

Finally, the following main proposition of this subsection calculates the quadratic tensor data of a Clifford unitary.
\begin{myprop}
\label{prop:clifford_to_quadratic}
For each Clifford automorphism data $(\alpha,u)$, the associated Clifford unitary is a quadratic tensor $\bra{h_o}U[\alpha,u]\ket{h_i}=\mathcal T[E,\epsilon,q](h_i,h_o)$, whose quadratic tensor data $(E,\epsilon,q)$ is constructed as follows.
Decompose $u\in \hom_2[H\times H^*|\rr/\zz]$ into its components $u^{(1)}_0\in\hom_2[H|\rr/\zz]$, $u^{(1)}_1\in\hom_2[H^*|\rr/\zz]$, and $u^{(2)}_{01}=\alpha_{zx}^*\alpha_{xz}\in \hom^2[H^*,H|\rr/\zz]= \hom[H^*|H^*]$.
Choose an injective homomorphism $\widetilde \kappa_x\in \hom[\widetilde K_x|H^*]$ that identifies $\ker(\alpha_{xz})$ with some standalone group $\widetilde K_x$, in particular,
\begin{equation}
\label{eq:clifford_to_quadratic_kernel}
\alpha_{xz}\widetilde \kappa_x=0\;.
\end{equation}
Also choose a left inverse $\widetilde\kappa_x^\perp: H^*/\widetilde K_x\rightarrow H^*$ of the quotient map.
Finally, choose $\widetilde h_0\in H$ such that
\begin{equation}
\label{eq:clifford_data_h0_definition}
\widetilde \kappa_x^* \alpha_{zz}^* \widetilde h_0=u\circ \mpm{0\\\widetilde\kappa_x}\;.
\end{equation}
With this, the data is given by
\begin{equation}
\label{eq:clifford_tensor_data}
\begin{gathered}
E = H\times (H^*/\widetilde K_x)\;,\quad
\epsilon^{(1)} = \mpm{1&0\\\alpha_{xx}&\alpha_{xz}\circ\widetilde\kappa_x^\perp}\;,\quad
\epsilon^{(0)} = \mpm{0\\\widetilde h_0}\;,\quad
q_\phi^{(2)}
=\mpm{\alpha_{zx}^*\alpha_{xz}\circ\widetilde\kappa_x^\perp}\;,\\
q_\phi^{(1)}=\mpm{\alpha_{zx}^*\widetilde h_0-u^{(1)}_0 & (\alpha_{zz}^*\widetilde h_0-u^{(1)}_1)\circ\widetilde \kappa_x^\perp}\;.
%q_a^{(0)}=\log(|\widetilde K_x|)-\log(|H|)-\log(|H^*|)\;.
\end{gathered}
\end{equation}
\end{myprop}
\begin{proof}
Even though implicit in the following proof, it is instructive to first check that all expressions in the proposition are well defined.
First, the right-hand side of Eq.~\eqref{eq:clifford_data_h0_definition} is an element of $\widetilde K_x^*$ just like the left-hand side,
\begin{equation}
\begin{gathered}
\big(u\circ\mpm{0\\\widetilde\kappa_x}\big)^{(2)}
= \widetilde \kappa_x^*u^{(2)}_{11}\widetilde \kappa_x
\overset{\eqref{eq:symplectic_clifford_explicit}}{=} \widetilde \kappa_x^*\alpha_{xz}^*\alpha_{zz}\widetilde \kappa_x
\overset{\eqref{eq:clifford_to_quadratic_kernel}}{=}0
\quad\Rightarrow\quad u\circ \mpm{0\\\widetilde \kappa_x}\in \hom[\widetilde K_x|\rr/\zz]=\widetilde K_x^*\;.
\end{gathered}
\end{equation}
Next, we should check that all entries involving a composition with $\widetilde\kappa_x^\perp$ are actually homomorphisms or normalized quadratic functions.
The key to this is that each such expression becomes $0$ if we replace $\widetilde\kappa_x^\perp$ with $\widetilde\kappa_x$.
For $\epsilon^{(1)}_{11}=\alpha_{xz}\circ\widetilde\kappa_x^\perp$ and $q^{(2)}_{\phi,01}=\alpha_{zx}^*\alpha_{xz}\circ\widetilde\kappa_x^\perp$ this is a direct consequence of Eq.~\eqref{eq:clifford_to_quadratic_kernel}.
For $q_{\phi,1}^{(1)}=(\alpha_{zz}^*\widetilde h_0-u^{(1)}_1)\circ\widetilde\kappa_x^\perp$, this is precisely Eq.~\eqref{eq:clifford_data_h0_definition}.

After the initial cross checks, we start by showing that $U[\alpha,u]$ is the associated stabilizer state of a complete stabilizer tableau data $(S,\sigma,p)$ over $H\times H$.
Re-arranging Eq.~\eqref{eq:unitary_conjugation} yields
\begin{equation}
C[\alpha,u](\rho(h,h')) U[\alpha,u] \rho(h,h')^\dagger = U[\alpha,u]\;.
\end{equation}
So the tensor $U[\alpha,u]$ is invariant under action of the following operators,
\begin{equation}
\begin{gathered}
\overline{\rho(h,h')}\otimes C[\alpha,u](\rho(h,h'))
= e^{-2\pi i u((h,h'))} \rho(h,-h')\otimes \rho(\alpha(h,h'))\\
\overset{\eqref{eq:clifford_automorphism_decomposition}}{=} e^{-2\pi i u((h,h'))} \rho_{H\times H}((h,\alpha_{xx}h+\alpha_{xz}h'),(-h',\alpha_{zx}h+\alpha_{zz}h'))
\;.
\end{gathered}
\end{equation}
Here, the first factor corresponds to the $h_i$ index of $U[\alpha,u]$, and the second component to the $h_o$ index.
$\ovl{\rho(h,h')}$ denotes the complex conjugate, and $\rho_{H\times H}$ denotes the projective representation $\rho$ with $H\times H$ instead of $H$.
From this, we can read off the complete stabilizer tableau data $(S,\sigma,p)$:
\begin{equation}
\label{eq:stabilizer_from_clifford}
\begin{gathered}
S=H\times H^*,\quad
\sigma_x =
\mpm{1 & 0\\\alpha_{xx} & \alpha_{xz}},\quad
\sigma_z =
\mpm{0 & -1\\\alpha_{zx} & \alpha_{zz}},\quad
p=u
\;.
\end{gathered}
\end{equation}
A little cross check confirms that $(S,\sigma,p)$ is a valid stabilizer tableau data,
\begin{equation}
\begin{gathered}
p^{(2)}
\overset{\eqref{eq:stabilizer_from_clifford}}{=} u^{(2)}
\overset{\eqref{eq:symplectic_clifford_explicit}}{=}
\begin{pmatrix}
\alpha_{xx}^*\alpha_{zx} & \alpha_{xx}^*\alpha_{zz}-1\\
\alpha_{xz}^* \alpha_{zx}& \alpha_{xz}^*\alpha_{zz}
\end{pmatrix}
=
\begin{pmatrix}
1 & \alpha_{xx}^*\\
0 & \alpha_{xz}^*
\end{pmatrix}
\begin{pmatrix}
0 & -1\\
\alpha_{zx} & \alpha_{zz}
\end{pmatrix}
\overset{\eqref{eq:stabilizer_from_clifford}}{=} \sigma_x^* \sigma_z
\;.
\end{gathered}
\end{equation}
Having obtained a complete stabilizer tableau data for $U[\alpha,u]$, we can now follow the general prescription from Proposition~\ref{prop:stabilizer_tensor} to turn it into a quadratic tensor data.
To this end, we need to determine the objects $K_x$, $\kappa_x$, $\kappa_x^\perp$, and $h_0$ from Propositions~\ref{prop:stabilizer_tensor} and \ref{prop:stabilizer_codestate_tensor}.
Since $\sigma_x$ has a $1$ entry at the top left, its kernel is only supported on the second factor, and we can set
\begin{equation}
\ker(\sigma_x)=
\begin{pmatrix}
0\\\ker(\alpha_{xz})
\end{pmatrix}
\quad\Rightarrow\quad
K_x=\widetilde K_x\;,\quad
\kappa_x =
\begin{pmatrix}
0\\\widetilde \kappa_x
\end{pmatrix}
\;.
\end{equation}
Further, we can set $S/K_x\simeq H\times (H^*/\widetilde K_x)$, and
\begin{equation}
\kappa_x^\perp=\mpm{1&0\\0&\widetilde\kappa_x^\perp}\;,
\end{equation}
Next, we can set
\begin{equation}
h_0=
\begin{pmatrix}
0\\\widetilde h_0
\end{pmatrix}
\;,
\end{equation}
as we indeed find
\begin{equation}
\kappa_x^*\sigma_z^*h_0=
\mpm{0 & \widetilde \kappa_x^*}
\mpm{0 & \alpha_{zx}^*\\-1 & \alpha_{zz}^*}
\mpm{0\\\widetilde h_0}
=\widetilde \kappa_x^* \alpha_{zz}^* \widetilde h_0
= u\circ \mpm{0\\\widetilde\kappa_x}
= p\circ\kappa_x\;.
\end{equation}
Plugging all these choices into Eq.~\eqref{eq:stabilizer_tensor_data} yields
\begin{equation}
\epsilon^{(1)}
=\sigma_x\circ\kappa_x^\perp
=\mpm{1 & 0\\\alpha_{xx} & \alpha_{xz}}\circ\mpm{1&0\\0&\widetilde\kappa_x^\perp}
=\mpm{1&0\\\alpha_{xx}&\alpha_{xz}\circ\widetilde\kappa_x^\perp}\;,
\end{equation}
and
\begin{equation}
q_\phi^{(2)}
=p^{(2)}\circ (\kappa_x^\perp\times \kappa_x^\perp)
=\mpm{\ldots&p^{(2)}_{01}\circ\widetilde\kappa_x^\perp\\\ldots&\ldots}
=\mpm{\ldots&\alpha_{zx}^*\alpha_{xz}\circ\widetilde\kappa_x^\perp\\\ldots&\ldots}
\;,
\end{equation}
and
\begin{equation}
\begin{gathered}
q_{\phi,0}^{(1)}=\alpha_{zx}^*\widetilde h_0-u^{(1)}_0\;,\\
q_{\phi,1}^{(1)}=(\alpha_{zz}^*\widetilde h_0-u^{(1)}_1)\circ\widetilde \kappa_x^\perp\;,
\end{gathered}
\end{equation}
which is Eq.~\eqref{eq:clifford_tensor_data}.
\end{proof}

After the general case, let us consider a simpler special case.
\begin{myprop}
\label{prop:clifford_to_quadratic_invertible}
Let $(\alpha,u)$ be a Clifford automorphism data such that $\alpha_{xz}$ is invertible.
Then the corresponding quadratic tensor data is
\begin{equation}
\begin{gathered}
E = H\times H^*\;,\quad
\epsilon^{(1)} = \mpm{1&0\\\alpha_{xx}&\alpha_{xz}}\;,\quad
\epsilon^{(0)} = \mpm{0\\0}\;,\quad
q_\phi=u\;,\quad
q_a^{(0)}=-\log(|H|)-\log(|H^*|)\;.
\end{gathered}
\end{equation}
\end{myprop}
\begin{proof}
Following Proposition~\ref{prop:clifford_to_quadratic}, we find
\begin{equation}
\widetilde K_x=0\;,\quad \widetilde \kappa_x=0\;,\quad \widetilde h_0=0\;,\quad \widetilde\kappa_x^\perp=1\;,
\end{equation}
leading to the above data.
Note that $\epsilon^{(1)}$ has an inverse,
\begin{equation}
(\epsilon^{(1)})^{-1}=
\mpm{1 & 0\\-\alpha_{xz}^{-1}\alpha_{xx} & \alpha_{xz}^{-1}}
\in\hom[H\times H|H\times H^*]\;.
\end{equation}
Applying this isomorphism to $E$, we get an equivalent quadratic tensor data $(E',\epsilon',q')$ with
\begin{equation}
E'=H\times H\;,\quad
\epsilon'=1_{H\times H}\;,\quad
q_\phi'=u\circ (\epsilon^{(1)})^{-1}\;.
\end{equation}
The corresponding bilinear form is then given by
\begin{equation}
\label{eq:cv_quadratic_hamiltonian_entries}
\begin{gathered}
(q_\phi')^{(2)} =
((\epsilon^{(1)})^{-1})^*u^{(2)}(\epsilon^{(1)})^{-1}=
\mpm{1 & 0\\-\alpha_{xz}^{-1}\alpha_{xx} & \alpha_{xz}^{-1}}^*
\mpm{\alpha_{xx}^*\alpha_{zx} & \alpha_{zx}^* \alpha_{xz}\\\alpha_{xz}^* \alpha_{zx} & \alpha_{xz}^* \alpha_{zz}}
\mpm{1 & 0\\-\alpha_{xz}^{-1}\alpha_{xx} & \alpha_{xz}^{-1}}\\
=
\mpm{1 & -\alpha_{xx}^* (\alpha_{xz}^{-1})^*\\0 & (\alpha_{xz}^{-1})^*}
\mpm{\alpha_{xx}^* \alpha_{zx} - \alpha_{zx}^*\alpha_{xx}& \alpha_{zx}^*\\\alpha_{xz}^*\alpha_{zx} - \alpha_{zz}^*\alpha_{xx}& \alpha_{zz}^*}
=
\mpm{\alpha_{xz}^{-1}\alpha_{xx}& \alpha_{zx}^* - \alpha_{xx}^*(\alpha_{xz}^{-1})^*\alpha_{zz}^*\\\alpha_{zx}-\alpha_{zz} \alpha_{xz}^{-1} \alpha_{xx} & \alpha_{zz} \alpha_{xz}^{-1}}
\;.
\end{gathered}
\end{equation}
\end{proof}

\section{Examples of quadratic tensors}
\label{sec:quadratic_examples}
In this section, we give a collection of examples spanning a wide range from general and abstract to specific and concrete.
We start with some simple abstract mathematical examples, continue with stabilizer/Clifford related examples on qubits, then go to continuous variables, and finish with some rotor and GKP examples.
When writing down quadratic tensor data, we adopt the convention to only write down non-zero entries.
\subsection{Abstract group examples}
In this section, we give some simple examples of quadratic tensors for arbitrary groups.
For any abelian group $G$ and any $h\in G$, the Dirac $\delta$-distribution $\delta_h$ is a quadratic tensor over $G$ with data
\begin{equation}
E=0\;,\quad
\epsilon^{(0)} = \mpm{h}\;.
\end{equation}
For any abelian group $G$, the \emph{Fourier transform} is a linear map $F:\cc^G\rightarrow \cc^{G^*}$ defined by $\bra{g'}F\ket{g}=e^{2\pi ig'(g)}$.
The Fourier transform is a quadratic tensor $\bra{g'}F\ket{g}=\mathcal T[E,\epsilon,q](g,g')$ over $G\times G^*$ with
\begin{equation}
E=G\times G^*\;,\quad
\epsilon^{(1)}=\mpm{1&0\\0&1}\;,\quad
q_\phi^{(2)} = \mpm{1}\;,\quad
q_a^{(0)} = -\frac12 \log(|G|)\;.
\end{equation}
For any abelian group $G$, the convolution defines a bilinear map $\cc^G\times \cc^G\rightarrow \cc^G$, or equivalently a linear map $C:\cc^{G\times G}\rightarrow \cc^{G}$ defined by $\bra{g_2}C\ket{g_0,g_1}=\delta_{g_0+g_1=g_2}$.
$C$ is a quadratic tensor $\bra{g_2}C\ket{g_0,g_1}=\mathcal T[E,\epsilon,q](g_0,g_1,g_2)$ over $G\times G\times G$, with data
\begin{equation}
E=G\times G\;,\quad
\epsilon^{(1)}=
\mpm{1&0\\0&1\\1&1}\;.
\end{equation}
Consider two abelian groups $G$ and $H$ and a homomorphism $\eta\in \hom[G|H]$.
$\eta$ can be linearized to a linear map $\eta:\cc^G\rightarrow \cc^H$ with $\bra{h}\eta\ket{g}=\delta_{h=\eta(g)}$.
$\eta$ is a quadratic tensor $\bra{h}\eta\ket{g}=\mathcal T[E,\epsilon,q](g,h)$ over $G\times H$ with
\begin{equation}
E=G\;,\quad
\epsilon^{(1)}=\mpm{1\\\eta}\;.
\end{equation}

As a last example we consider diagonal generalized Clifford operations.
Consider a quadratic function $\widetilde u\in F_2[H|\rr/\zz]$ on an abelian group $H$.
Then we can define a Clifford automorphism data over $H$ as follows:
\begin{equation}
\alpha=\mpm{1&0\\\widetilde u^{(2)}&1}\;,\quad
u^{(2)}=\mpm{0}\;,\quad
u^{(1)}=\mpm{\widetilde u&0}\;,
\end{equation}
noting that we indeed have
\begin{equation}
u^{(2)}
=\mpm{\widetilde u^{(2)}&0\\0&0}
=\mpm{1&\widetilde u^{(2)}\\0&1}\mpm{0&1\\0&0}\mpm{1&0\\\widetilde u^{(2)}&1}-\mpm{0&1\\0&0}
=\alpha^*\omega\alpha-\omega\;.
\end{equation}
Let us now derive the quadratic tensor data describing the associated Clifford unitary.
Following Proposition~\ref{prop:clifford_to_quadratic}, we find
\begin{equation}
K_x=H^*\;,\quad
\widetilde\kappa_x=1\;,\quad
\widetilde\kappa_x^\perp=0\;,\quad
\widetilde h_0=0\;.
\end{equation}
Plugging into Eq.~\eqref{eq:clifford_tensor_data} yields
\begin{equation}
E=H\;,\quad
\epsilon^{(1)}=\mpm{1\\1}\;,\quad
q_\phi^{(1)}=-\widetilde u^{(1)}\;.
\end{equation}
In this form it is indeed obvious that the Clifford unitary is diagonal in the computational basis.
This is no surprise, as it can be seen from $\alpha$ that it leaves $Z$-type Pauli operators invariant.

\subsection{Qubits}
Let us now give some examples of quadratic tensors for qubits, where all factors of $G$ and $E$ are equal to $\zz_2$.
We recall from Section~\ref{sec:coefficient_groups} that (as for all finite groups) the coefficients for $q_a^{(\tilde2)}$ and $q_a^{(\tilde1)}$ are trivial, the coefficients for $q_\phi^{(\tilde2)}$ are in $\homtild^2[\zz_2,\zz_2|\rr/\zz]=\zz_2$, and the coefficients for $q_\phi^{(\tilde1)}$ are in $\homtild_2[\zz_2|\rr/\zz]=\zz_4$.
The entries of $\epsilon^{(\tilde1)}$ are in $\homtild[\zz_2|\zz_2]=\zz_2$, and the entries of $\epsilon^{(\tilde0)}$ are in $\zz_2$.
The state $\ket+=\frac1{\sqrt2}(1,1)$ is a quadratic tensor over $\zz_2$ with data
\begin{equation}
\label{eq:plus_state}
E=\zz_2\;,\quad
\epsilon^{(\tilde1)}=\mpm{1}\;,\quad
q_a^{(\tilde0)}=-\frac12 \log(2)\;.
\end{equation}
This state is stabilized by the Pauli operator $X$, so the corresponding stabilizer tableau data over $\zz_2$ is
\begin{equation}
S=\zz_2,\quad
\tilde\sigma_x=\mpm{1},\quad
\tilde\sigma_z=\mpm{0},\quad
p^{(\tilde1)}=\mpm{0}\;.
\end{equation}
Following Proposition~\ref{prop:stabilizer_tensor}, we find
\begin{equation}
K_x=0\;,\quad
\kappa_x=\varnothing\;,\quad
\kappa_x^\perp=1\;,\quad
h_0=0\;,
\end{equation}
leading back to Eq.~\eqref{eq:plus_state}.

As a next example, the state $\ket-=\frac1{\sqrt2}(1,-1)$ is a quadratic tensor over $\zz_2$ with data
\begin{equation}
E=\zz_2\;,\quad
\epsilon^{(\tilde1)}=\mpm{1}\;,\quad
q_\phi^{(\tilde1)}=\mpm{2}\;,\quad
q_a^{(\tilde0)}=-\frac12 \log(2)\;.
\end{equation}
The state is stabilized by $-X$, so the corresponding stabilizer tableau data is
\begin{equation}
S=\zz_2\;,\quad
\tilde\sigma_x=\mpm{1}\;,\quad
\tilde\sigma_z=\mpm{0}\;,\quad
p^{(\tilde1)}=\mpm{2}\;.
\end{equation}

We already introduced the state $\ket Y=\frac1{\sqrt2}(1,i)$ in Section~\ref{sec:tutorial}.
It is a quadratic tensor over $\zz_2$ with data
\begin{equation}
E=\zz_2\;,\quad
\epsilon^{(\tilde1)}=\mpm{1}\;,\quad
q_\phi^{(\tilde1)}=\mpm{1}\;,\quad
q_a^{(\tilde0)}=-\frac12 \log(2)\;.
\end{equation}

Next, consider the state $\ket0=(1,0)$ with quadratic tensor data
\begin{equation}
E=0\;,\quad
\epsilon^{(\tilde0)}=\mpm{0}\;.
\end{equation}
and similarly for the $\ket1=(0,1)$ state,
\begin{equation}
E=0\;,\quad
\epsilon^{(\tilde0)}=\mpm{1}\;.
\end{equation}

The Hadamard gate $\bra{a}H\ket{b}=\frac1{\sqrt 2}(-1)^{ab}$ is a quadratic tensor over $\zz_2^2$,
\begin{equation}
E=\zz_2^2\;,\quad
\epsilon^{(\tilde1)}=\mpm{1&0\\0&1}\;,\quad
q_\phi^{(\tilde2)}=\mpm{1}\;,\quad
q_a^{(\tilde0)}=-\frac12 \log(2)\;.
\end{equation}
The Hadamard gate is a Clifford operation, acting on Pauli operators as
\begin{equation}
HXH^\dagger = Z,\quad HZH^\dagger = X.
\end{equation}
Thus, the corresponding Clifford automorphism data is given by
\begin{equation}
\tilde\alpha=\mpm{0&1\\1&0}\;,\quad
u^{(\tilde2)}=\mpm{1}\;,\quad
u^{(\tilde1)}=\mpm{0&0}\;.
\end{equation}
We can check explicitly that Eq.~\eqref{eq:clifford_data_condition} holds:
\begin{equation}
u^{(2)}
=\mpm{0&1\\1&0}
=\mpm{0&1\\1&0}\mpm{0&1\\0&0}\mpm{0&1\\1&0}-\mpm{0&1\\0&0}
=\alpha^*\omega\alpha-\omega\;.
\end{equation}
Note that the off-diagonal $1$ entry of $u^{(2)}$ corresponds to the fact that
\begin{equation}
H(XZ)H^\dagger=ZX=(-1)^1 XZ\;.
\end{equation}

Next, the $S$ matrix from Eq.~\eqref{eq:s_matrix} is a quadratic tensor over $\zz_2^2$ with data
\begin{equation}
\label{eq:smatrix_quadratic_data}
E=\zz_2\;,\quad
\epsilon^{(\tilde1)}=\mpm{1\\1}\;,\quad
q_\phi^{(\tilde1)}=\mpm{1}\;.
\end{equation}
The $S$ matrix is also a Clifford operation, acting on Pauli operators as
\begin{equation}
SXS^\dagger=-iZX,\quad SZS^\dagger=Z\;.
\end{equation}
So the corresponding Clifford automorphism data given by
\begin{equation}
\tilde\alpha=\mpm{1&0\\1&1}\;,\quad
u^{(\tilde2)}=\mpm{0}\;,\quad
u^{(\tilde1)}=\mpm{1&0}\;.
\end{equation}
Again, Eq.~\eqref{eq:clifford_data_condition} holds:
\begin{equation}
u^{(2)}
=\mpm{1&0\\0&0}
=\mpm{1&1\\0&1}\mpm{0&1\\0&0}\mpm{1&0\\1&1}-\mpm{0&1\\0&0}
=\alpha^*\omega\alpha-\omega\;.
\end{equation}
Following Proposition~\ref{prop:clifford_to_quadratic}, we find
\begin{equation}
\widetilde K_x=\zz_2^*\;,\quad
\widetilde \kappa_x = 1\;,\quad
\widetilde \kappa_x^\perp=\varnothing\;,\quad
\widetilde h_0=0\;,
\end{equation}
which leads again to Eq.~\eqref{eq:smatrix_quadratic_data}.

Next, the controlled-$Z$ ($CZ$) gate $\bra{g_{o0},g_{o1}}CZ\ket{g_{i0},g_{i1}}=\mathcal T[E,\epsilon,q](g_{i0},g_{i1},g_{o0},g_{o1})$ is a quadratic tensor over $\zz_2^4$ with data
\begin{equation}
E=\zz_2^2\;,\quad
\epsilon^{(\tilde1)}=\mpm{1&0\\0&1\\1&0\\0&1}\;,\quad
q_\phi^{(\tilde2)}=\mpm{1}\;,\quad
q_\phi^{(\tilde1)}=\mpm{0&0}\;.
\end{equation}
The controlled-$X$ ($CX$) gate  $\bra{g_{o0},g_{o1}}CX\ket{g_{i0},g_{i1}}=\mathcal T[E,\epsilon,q](g_{i0},g_{i1},g_{o0},g_{o1})$ is a quadratic tensor over $\zz_2^4$ with data
\begin{equation}
E=\zz_2^2\;,\quad
\epsilon^{(\tilde1)}=\mpm{1&0\\0&1\\1&0\\1&1}\;.
\end{equation}
The $CX$ gate is also a Clifford unitary with Clifford automorphism data given by
\begin{equation}
\tilde\alpha=
\mpm{
1&0&0&0\\
1&1&0&0\\
0&0&1&1\\
0&0&0&1}\;,\quad
u^{(\tilde2)}=\mpm{0&0&0\\&0&0\\&&0}\;,\quad
u^{(\tilde1)}=\mpm{0&0&0&0}\;.
\end{equation}
Eq.~\eqref{eq:clifford_data_condition} holds:
\begin{equation}
u^{(2)}
=
\mpm{
0&0&0&0\\
0&0&0&0\\
0&0&0&0\\
0&0&0&0}
=
\mpm{
1&1&0&0\\
0&1&0&0\\
0&0&1&0\\
0&0&1&1}
\mpm{
0&0&1&0\\
0&0&0&1\\
0&0&0&0\\
0&0&0&0}
\mpm{
1&0&0&0\\
1&1&0&0\\
0&0&1&1\\
0&0&0&1}
-
\mpm{
0&0&1&0\\
0&0&0&1\\
0&0&0&0\\
0&0&0&0}
=
\alpha^*\omega\alpha-\omega\;.
\end{equation}

As a more intricate example, consider the encoding map of the 5-qubit code~\cite{Bennett1996,Laflamme1996}.
It is a quadratic tensor over $\zz_2^6$, where the first 5 factors correspond to the physical qubits, and the last factor corresponds to the logical qubit.
The quadratic tensor data is given by
\begin{equation}
E=\zz_2^5\;,\quad
\epsilon^{(\tilde1)}=\mpm{1&0&0&0&0\\0&1&0&0&0\\0&0&1&0&0\\0&0&0&1&0\\0&0&0&0&1\\1&1&1&1&1}\;,\quad
q^{(\tilde2)}=\mpm{1&0&0&1\\&1&0&0\\&&1&0\\&&&1}\;.
\end{equation}

\subsection{Continuous variables}
\label{sec:continuous_variable_examples}
Let us now consider some continuous-variable examples, where both $G$ and $E$ consist of only $\rr$ factors.
The coefficients of $q_{a/\phi}^{(\tilde2)}$ are in $\homtild^2[\rr,\rr|\rr]=\rr$ and $\homtild^2[\rr,\rr|\rr/\zz]=\rr$.
The coefficients of $q_{a/\phi}^{(\tilde1)}$ are in $\homtild_2[\rr|\rr]=\rr\times\rr$ and $\homtild_2[\rr|\rr/\zz]=\rr\times\rr$.
The coefficients of $\epsilon^{(\tilde1)}$ are in $\homtild[\rr|\rr]=\rr$, and the coefficients of $\epsilon^{(\tilde0)}$ are in $\rr$.

\myparagraph{Squeezed states}
We have already seen in Section~\ref{sec:tutorial} that a squeezed state on $n$ continuous variables,
\begin{equation}
\braket{x=g|\phi}=e^{2\pi(\frac12 x^T a x+bx+c)}\;,
\end{equation}
is a quadratic tensor over $G=\rr^n$.
The corresponding quadratic tensor data is given by
\begin{equation}
\begin{gathered}
E=\rr^n\;,\quad
\epsilon^{(\tilde1)}=1_{n\times n}\;,\\
q_\phi^{(\tilde2)}=\imag(a)\;,\quad
q_\phi^{(\tilde1)}=(\operatorname{diag}(\imag(a)), \imag(b))\;,\quad
q_\phi^{(\tilde0)}=\imag(c)\;,\\
q_a^{(\tilde2)}=\real(a)\;,\quad
q_a^{(\tilde1)}=(\operatorname{diag}(\real(a)), \real(b))\;,\quad
q_a^{(\tilde0)}=\real(c)\;.
\end{gathered}
\end{equation}
We note that it suffices to give the upper triangular part of $q_{a/\phi}^{(\tilde2)}$ as the diagonal is contained in $q_{a/\phi}^{(\tilde1)}$ and the lower triangular part is symmetric.

\myparagraph{Displacement operators}
Next, let us show that the generalized Pauli operators from Definition~\ref{def:generalized_pauli_operator} for continuous variables ($H=\rr^n$) are equal to \emph{displacement operators} in quantum optics.
Proposition~\ref{prop:pauli_operator_tensor} then shows that these displacement operators are quadratic tensors over $\rr\times\rr$.
\begin{mydef}
For any pair $x\in \rr^n$ and $p\in \rr^n$, the \emph{displacement operator} $D(x,p)$ is given by
\begin{equation}
\label{eq:displacement_operator_def}
D(x,p)\coloneqq e^{i(p^T\hat x-x^T\hat p)}\;,
\end{equation}
where $\hat x$ and $\hat p$ are length-$n$ vectors containing the position and momentum operators for the $n$ different modes, respectively.
They satisfy the commutation relation,
\begin{equation}
\label{eq:xp_commutation}
[p^T\hat x,x^T\hat p]=ip^Tx\;,
\end{equation}
and act on wave functions $\phi:\rr^n\rightarrow\cc$ as
\begin{equation}
\label{eq:wave_function_action}
(e^{ip^T\hat x}\phi)(g)=e^{ip^Tg}\phi(g)\;,\quad
(e^{ix^T\hat p}\phi)(g)=\phi(g+x)\;.
\end{equation}
We can write Eq.~\eqref{eq:displacement_operator_def} in a more compact form,
\begin{equation}
\label{eq:displacement_operator_def_compact}
D(\xi) =  e^{i\xi^T \widetilde J \hat\xi}\;,\quad
\xi=\mpm{x\\p},\quad
\hat\xi=\mpm{\hat x\\\hat p},\quad
\widetilde J = \htild^{-1}(J)=\mpm{0&\mathbb1\\-\mathbb1&0}\;.
\end{equation}
\end{mydef}
\begin{myprop}
\label{prop:pauli_from_displacement}
The displacement operator $D(x,p)$ is equal to a generalized Pauli operator over $H=\rr^n$ from Definition~\ref{def:generalized_pauli_operator}, up to normalization,
\begin{equation}
\label{eq:pauli_from_displacement}
D(x,p) = e^{-i \frac12 p^T x} \rho(x,\frac{\htild(p)}{2\pi})\;.
\end{equation}
Here, $\htild$ is the coefficient isomorphism $\rr\rightarrow \hom[\rr|\rr/\zz]= \rr^*$ from Eq.~\eqref{eq:elementary_homomorphisms}, which we apply to the individual components of the vector $p$.
\end{myprop}
\begin{proof}
We verify Eq.~\eqref{eq:pauli_from_displacement} by showing that both sides act in the same way on a wave function $\phi: \rr^n\rightarrow \cc$.
\begin{equation}
\begin{gathered}
(D(x,p)\phi)(g)
\overset{\eqref{eq:displacement_operator_def}}{=} (e^{i(p^T\hat x-x^T\hat p)}\phi)(g)
\overset{\eqref{eq:xp_commutation}}{=} e^{-i\frac12 p^Tx} (e^{ip^T\hat x-ix^T\hat p + \frac12[ip^T\hat x,-ix^T\hat p]}\phi)(g)
\overset{*}{=} e^{-i\frac12 p^Tx} (e^{ip^T\hat x} e^{-ix^T\hat p}\phi)(g)\\
\overset{\eqref{eq:wave_function_action}}{=} e^{-i\frac12 p^Tx} e^{ip^Tg}\phi(g-x)
\overset{\eqref{eq:elementary_homomorphisms}}{=} e^{-i\frac12 p^Tx} e^{2\pi i\frac{
%\htild
\widetilde H
(p)(g)}{2\pi}}\phi(g-x)
\overset{\eqref{eq:generalized_pauli_operator_def}}{=} e^{-i\frac12 p^Tx} (\rho(x,\frac{\htild(p)}{2\pi})\phi)(g)\;.
\end{gathered}
\end{equation}
For the equality labeled $*$, we have used the Baker-Campbell-Hausdorff formula and the fact that all higher commutators vanish.
\end{proof}
Note that we can also write Eq.~\eqref{eq:pauli_from_displacement} in the compact form,
\begin{equation}
\label{eq:pauli_from_displacement_compact}
D(\xi) = e^{-i \frac12 \xi^T\widetilde\omega\xi} \rho(\htild(\ovl{\bfpi}\xi))\;,\quad
\widetilde\omega=\htild^{-1}(\omega)=\mpm{0&\mathbb1\\0&0}\;,\quad
\bfpi\coloneqq \mpm{1&0\\0&2\pi}\;,\quad
\ovl\bfpi\coloneqq \bfpi^{-1}\;.
\end{equation}

\begin{myprop}
According to Proposition~\ref{prop:pauli_operator_tensor}, $D(x,p)$ is a quadratic tensor over $\rr^{2n}$ with data
\begin{equation}
E=\rr^n\;,\quad
\epsilon^{(\tilde1)}=\mpm{1\\1}\;,\quad
\epsilon^{(\tilde0)}=\mpm{x\\0}\;,\quad
q_\phi^{(\tilde1)}=\frac{p}{2\pi}\;,\quad
q_\phi^{(\tilde0)}=-\frac{1}{4\pi} p^Tx\;.
\end{equation}
\end{myprop}

\myparagraph{Gaussian unitaries}
Let us next show that the generalized Clifford operators from Definition~\ref{eq:associated_clifford} for the case of continuous variables ($H=\rr^n$) are equal to Gaussian unitary operators.
It then follows from Proposition~\ref{prop:clifford_to_quadratic} that Gaussian unitaries are quadratic tensors over $\rr^n\times \rr^n$.
\begin{mydef}
A unitary $U$ is \emph{Gaussian} if it acts on the quadrature operators $\hat x$ and $\hat p$ as follows,
\begin{equation}
\label{eq:gaussian_quadrature_action}
U^\dagger \hat\xi U = L\hat \xi + d ,
\end{equation}
where $d\in \rr^{2n}$ is a length-$2n$ real ``displacement'' vector, and $L$ is a real $2n\times 2n$ symplectic matrix,
\begin{equation}
\label{eq:symplectic_matrix}
L^T\widetilde JL=\widetilde J\;.
\end{equation}
\end{mydef}

\begin{myprop}
A Gaussian unitary is a Clifford unitary from Definition~\ref{eq:associated_clifford} over the group $H=\rr^n$,
\begin{equation}
\label{eq:gaussian_is_clifford}
U=U[\alpha,u]\;,
\end{equation}
for the following Clifford automorphism data,
\begin{equation}
\label{eq:gaussian_is_clifford_data}
\alpha= \htild(L^{-1}_\pi)\;,\quad
u(g)=\frac12 \htild^{-1}(g)^T\big(L^{-T}_\pi\widetilde\omega L^{-1}_\pi-\widetilde\omega\big)\htild^{-1}(g) + d^T\ovl\bfpi\widetilde J \htild^{-1}(g)\;,\quad
L^{-1}_\pi\coloneqq \ovl\bfpi L^{-1}\bfpi\;.
\end{equation}
\end{myprop}
\begin{proof}
First, we verify that $(\alpha,u)$ defines a valid Clifford automorphism data and fulfills Eq.~\eqref{eq:clifford_data_condition}:
\begin{equation}
u^{(2)}
=\htild(L^{-T}_\pi\widetilde\omega L^{-1}_\pi-\widetilde\omega)
=\alpha^*\omega\alpha-\omega\;.
\end{equation}
Then we verify Eq.~\eqref{eq:gaussian_is_clifford} by showing that $U$ and $U[\alpha,u]$ act in the same way on all displacement operators $D(\xi)$:
\begin{equation}
\begin{gathered}
U^\dagger D(\xi) U
\overset{\eqref{eq:gaussian_quadrature_action},\eqref{eq:displacement_operator_def_compact}}{=}e^{i\xi^T \widetilde J(L\hat\xi+d)}
\overset{\eqref{eq:symplectic_matrix}}{=}e^{i\xi^T \widetilde J d} D(L^{-1}\xi)
\overset{\eqref{eq:pauli_from_displacement_compact}}{=}e^{i\xi^T \widetilde J d-i\frac12 \xi^TL^{-T}\widetilde\omega L^{-1}\xi} \rho(\htild(\ovl\bfpi L^{-1}\xi))\\
\overset{\eqref{eq:symplectic_pi_commutation}}{=}e^{-i\frac12 \xi^T\widetilde\omega \xi} e^{-2\pi i \big(d^T\ovl\bfpi \widetilde J \ovl\bfpi\xi + \frac12 (\ovl\bfpi\xi)^T(L^{-T}_\pi\widetilde\omega L^{-1}_\pi-\widetilde\omega)\ovl\bfpi\xi\big)} \rho(\htild(L^{-1}_\pi\ovl\bfpi\xi))
\overset{\eqref{eq:gaussian_is_clifford_data}}{=}e^{-i\frac12 \xi^T\widetilde\omega \xi} e^{-2\pi iu(
%\htild
\widetilde H
(\ovl\bfpi\xi))} \rho(\alpha(\htild(\ovl\bfpi\xi)))\\
\overset{\eqref{eq:clifford_automorphism_definition}}{=}e^{-i\frac12 \xi^T\widetilde\omega \xi} U[\alpha,u]^\dagger \rho(\htild(\ovl\bfpi\xi)) U[\alpha,u]
\overset{\eqref{eq:pauli_from_displacement_compact}}{=}U[\alpha,u]^\dagger D(\xi) U[\alpha,u]\;.
\end{gathered}
\end{equation}
Here we have used that
\begin{equation}
\label{eq:symplectic_pi_commutation}
\ovl\bfpi\widetilde\omega\ovl\bfpi = \frac{\widetilde\omega}{2\pi}\;,\quad
\ovl\bfpi\widetilde J\ovl\bfpi = \frac{\widetilde J}{2\pi}\;.
\end{equation}
\end{proof}

\begin{myprop}
Let $L$ be a symplectic matrix such that $L_{xz}$ is invertible.
Then following Proposition~\ref{prop:clifford_to_quadratic_invertible}, the Gaussian unitary determined by $L$ is a quadratic tensor over $\rr^n\times \rr^n$ with
\begin{equation}
q^{(\tilde2)}=
\mpm{
L_{xz}^{-1}L_{xx}& L_{zx}^T - L_{xx}^T(L_{xz}^{-1})^TL_{zz}^T\\
L_{zx}-L_{zz} L_{xz}^{-1} L_{xx} & L_{zz} L_{xz}^{-1}
}
\end{equation}
\end{myprop}

Another way of defining a Gaussian unitary is as the time evolution under a quadratic (Bogoliubov-de Gennes) Hamiltonian.
The relation to the above definition is textbook knowledge, but we give it for completeness:
\begin{mydef}
For any real symmetric $2n\times 2n$ matrix $M$, the associated \emph{quadratic Hamiltonian} $H$ and unit-time-evolution unitary $U$ are the following operators on $\rr^n$:
\begin{equation}
H=\frac12 \sum_{0\leq i,j<2n} M_{ij}  \hat\xi_i \hat\xi_j = \frac12 \hat \xi^T M \hat\xi\;,\quad
U = e^{iH}=e^{i \frac12 \hat\xi^T M \hat\xi}\;.
\end{equation}
\end{mydef}

Note that one could add a time parameter $t$ to the unitary $U$, but it can be absorbed into $M$.
\begin{myprop}
$U$ is a Gaussian unitary with
\begin{equation}
L= e^{\widetilde J M}\;,\quad
d=0\;.
\end{equation}
\end{myprop}
\begin{proof}
We start by using the commutation relations
\begin{equation}
[\hat\xi_i,\hat\xi_j]=i\widetilde J_{ij},\quad
[\hat\xi_i,\hat\xi_j\hat\xi_k]=i\widetilde J_{ij}\hat\xi_k + i\widetilde J_{ik}\hat\xi_j
\end{equation}
to calculate
\begin{equation}
\begin{gathered}
[\xi^T \hat\xi,\hat\xi^T M\hat\xi]
=[\sum_i \xi_i \hat\xi_i,\sum_{jk} M_{jk} \hat \xi_j \hat \xi_k]
=\sum_{ijk} \xi_i M_{jk} [\hat\xi_i,\hat\xi_j\hat\xi_k]
=i\sum_{ijk} \xi_i M_{jk} (\widetilde J_{ij}\hat\xi_k + \widetilde J_{ik}\hat\xi_j)\\
=i(\xi^T \widetilde J M\hat\xi+\xi^T \widetilde J M^T\hat\xi) = 2i\xi^T \widetilde J M\hat\xi\;.
\end{gathered}
\end{equation}
Given this, we can calculate how $U$ acts on the quadrature operators:
\begin{equation}
\begin{gathered}
e^{-i\frac12 \hat\xi^TM\hat\xi} \hat\xi e^{i\frac12 \hat\xi^TM\hat\xi}
\overset{*}{=}1+[\frac12 i\hat\xi^TM\hat\xi,\hat\xi] +\frac{1}{2!}[\frac12 i\hat\xi^TM\hat\xi,[\frac12 i\hat\xi^TM\hat\xi,\hat\xi]] + \ldots\\
=1+(\widetilde J M)\hat\xi + \frac{1}{2!}(\widetilde J M)^2 \hat\xi + \frac{1}{3!}(\widetilde J M)^3 \hat\xi +\ldots
= e^{\widetilde J M} \hat\xi
=L\hat\xi\;.
\end{gathered}
\end{equation}
In the equality labeled $*$, we have used the Taylor expansion for the conjugation by an exponential.
\end{proof}
Note that matrices $\widetilde JM$ for a symmetric matrix $M$ are known as \emph{Hamiltonian matrices}, and their exponentials are known to be a symplectic matrices.

Let us look again at a special case.
Consider the Hamiltonian for
\begin{equation}
M=\mpm{0&0\\0&M_{zz}}
\;,
\end{equation}
where $M_{zz}$ is some invertible matrix.
So this is a Hamiltonian that only involves the momentum operators, but not the position operators.
The unit-time evolution of this Hamiltonian is a Gaussian unitary with symplectic matrix
\begin{equation}
L = e^{\widetilde JM}
= \exp(\mpm{0&1\\-1&0}\mpm{0&0\\0&M_{zz}})
= \exp(\mpm{0&M_{zz}\\0&0})
=\mpm{1&M_{zz}\\0&1}\;.
\end{equation}
Plugging into Eq.~\eqref{eq:cv_quadratic_hamiltonian_entries} thus yields that this Gaussian unitary is a quadratic tensor over $\rr^n\times \rr^n$ with
\begin{equation}
E=\rr^n\times\rr^n\;,\quad
\epsilon^{(\tilde1)}=1\;,\quad
q^{(\tilde2)}=\mpm{M_{zz}^{-1} & -(M_{zz}^{-1})^*\\-M_{zz}^{-1}&M_{zz}^{-1}}\;.
\end{equation}

\subsection{Rotors and GKP codes}
In this section, we consider examples for tensors describing systems with rotors, or codes with discrete but infinite stabilizer groups.

\myparagraph{GKP codes}
We start by showing that generalized stabilizer codes over continuous variables whose stabilizer group consists of copies of $\zz$ are GKP codes.
Like in the continuous-variable case, all we need to do is identify $\rr$ with its dual $\rr^*$, and take care of some normalization in the definition of displacement operators.
\begin{mydef}
A \emph{Gottesman-Kitaev-Preskill} (GKP) code on $n$ modes is specified by a real $2n\times 2n$ matrix $L$ which is symplectic, fulfilling Eq.~\eqref{eq:symplectic_matrix}.
It is given by the subspace of $\rr^n$ that is invariant under all stabilizer operators of the form $D(L \xi)$, where $D$ is the displacement operator from Eq.~\ref{eq:displacement_operator_def}, and $\xi$ is an integer vector, $\xi\in \zz^{2n}$.
\end{mydef}
\begin{myprop}
A GKP code is a generalized stabilizer code over $H=\rr^n$ with stabilizer tableau data
\begin{equation}
\label{eq:gkp_stabilizer_tableau_data}
S=\zz^{2n}\;,\quad
\sigma = \htild(\ovl\bfpi L)\;,\quad
p(\xi)=\frac{1}{4\pi} \xi^T L^T\widetilde\omega L\xi\;.
\end{equation}
Here, $\htild$ denotes the application of the coefficient isomorphism to individual factors, which can either be $\htild:\rr\rightarrow\hom[\zz|\rr]$ or $\htild: \rr\rightarrow \hom[\zz|\rr^*]$.
\end{myprop}
\begin{proof}
First we check that the above indeed defines a valid stabilizer tableau data fulfilling Eq.~\eqref{eq:stabilizer_tableau_condition}:
\begin{equation}
p^{(2)}
=\frac{1}{2\pi}\htild(L^T\widetilde\omega L)
=\htild((\ovl\bfpi L)^T\widetilde\omega \ovl\bfpi L)
=\htild(\ovl\bfpi L)^*\htild(\widetilde\omega)\htild(\ovl\bfpi L)
=\sigma^*\omega\sigma\;.
\end{equation}
Next we verify that the stabilizers $D(L\xi)$ of the GKP code are indeed the same as the stabilizers of the generalized codes in Eq.~\eqref{eq:stabilizer_group_representation}:
\begin{equation}
\begin{gathered}
e^{-2\pi i p(\xi)} \rho(\sigma\xi)
\overset{\eqref{eq:gkp_stabilizer_tableau_data}}{=} e^{-2\pi i (\frac{1}{4\pi} \xi^TL^T\widetilde\omega L\xi)} \rho(\htild(\ovl\bfpi L)\xi)
= e^{-i\frac12 \xi^T L^T\widetilde\omega L\xi} \rho(\htild(\ovl\bfpi L\xi))
\overset{\eqref{eq:pauli_from_displacement_compact}}{=}D(L\xi)\;.
\end{gathered}
\end{equation}
\end{proof}

After showing that GKP codes are just a special case of generalized stabilizer codes, the results of Section~\ref{sec:stabilizer_codes} imply that their generalized code states, code space projectors, encoding maps, etc, are all quadratic tensors.
As an example consider the code state.
Consider a GKP code with a unique code state, which is the case when $\img(L)=\ker(\widetilde J L^T)$, and assume that $\ker(L_x)$ is trivial when writing $L=(L_x,L_z)^T$.
Then following Proposition~\ref{eq:prop_stabilizertensor_zerokernel}, the code state is a quadratic tensor over $\rr^n$ with
\begin{equation}
E=\zz^{2n}\;,\quad
\epsilon^{(\tilde1)}=L_x\;,\quad
q_\phi^{(\tilde2)} = \frac{1}{2\pi} L^T\widetilde\omega L\;,\quad
q_\phi^{(\tilde1)}=(\frac1{2\pi} \operatorname{diag}(L^T\widetilde\omega L), 0)\;.
\end{equation}

The exact GKP code states are unnormalizable and thus unphysical.
Luckily, approximate GKP states are also quadratic tensors.
Let us show this for a trivial single-mode square-lattice GKP code.
One notion for an approximate GKP state with a variance of $a^{-1}$ in $\hat x\mmod 1$ and a variance $b^{-1}$ in $\hat p\mmod 1$ is given by
\begin{equation}
\phi(x) = \sum_{i\in\zz} e^{-2\pi \frac12\big( a(x-i)^2+bx^2\big)}\;.
\end{equation}
This is a quadratic tensor over $G=\rr$ with
\begin{equation}
E=\rr\times \zz\;,\quad
\epsilon^{(\tilde1)}=\mpm{1\\0}\;,\quad
\epsilon^{(\tilde0)}=0\;,\quad
q_a^{(\tilde2)}=\mpm{-a-b&a\\a&-a}\;,\quad
q_a^{(\tilde1)}=(\operatorname{diag}(q^{(\tilde2)}),0)\;.
\end{equation}

We have seen that by using additional $\zz$ factors in $E$, it is possible to get quadratic tensors for continuous-variable wave functions that are not just Gaussians.
The simplest example that captures this behavior is the Jacobi theta-function over $\rr/\zz$,
\begin{equation}
\theta_\alpha: \rr/\zz\rightarrow \rr\;,\qquad \theta_\alpha(\phi) = \sum_{i\in \zz} e^{\alpha (\ovl\phi+i)^2}\;.
\end{equation}
This is a quadratic tensor over $G=\rr/\zz$ given by
\begin{equation}
E=\rr\;,\quad \epsilon(e)=e\mmod 1\;,\quad q(x)=\alpha x^2\;.
\end{equation}

\myparagraph{Rotor codes}
The rotor codes studied in Ref.~\cite{Vuillot2023} are examples of generalized CSS stabilizer codes.
In fact, they are exactly the generalized CSS codes in the case where $G=\rr/\zz^n$, $S_x=\rr/\zz^k$ and $S_z=\zz^l$.
The integer-valued check matrices $H_x$ and $H_z$ in Ref.~\cite{Vuillot2023} are the coefficient matrices of $\sigma_x$ and $\sigma_z$.
We can generalize Ref.~\cite{Vuillot2023} by considering non-CSS stabilizer codes with $G=\rr/\zz^n$ and $S=\rr/\zz^k\times \zz^l$.
The coefficient matrix of $\sigma$ for such a code is of the form
\begin{equation}
\htild^{-1}(\sigma)=\mpm{H_x&h_{xz}\\0&H_z}\;,
\end{equation}
where $h_{xz}\in \rr/\zz^{n\times l}$, and $H_x$ and $H_z$ are integer matrices as before.
The above $\sigma$ can be made into a valid code if Eq.~\eqref{eq:code_jconjugation_zero} holds,
\begin{equation}
\mpm{H_x^T&0\\h_{xz}^T&H_z^T}\widetilde J \mpm{H_x&h_{xz}\\0&H_z}
=\mpm{0&H_x^T\\-H_z^T&h_{xz}^T} \mpm{H_x&h_{xz}\\0&H_z}
=\mpm{0&H_x^TH_z\\-H_z^TH_x&h_{xz}^TH_z-H_z^T h_{xz}}
= 0\;.
\end{equation}
That is, it is a valid code of $H_x^TH_z=0$ like for the CSS version, and additionally $h_{xz}^TH_z$ is a symmetric matrix valued in $\rr/\zz$.

\section{Free-fermion quadratic tensors}
\label{sec:free_fermions}
In this section, we generalize quadratic tensors in such a way that they also incorporate free-fermion models.
This is done in two steps:
The first step (see Section~\ref{sec:fermionic_tensor_network}) is to realize that for describing any fermionic model (free or not), we need fermionic tensor networks.
The second step (see Sections~\ref{sec:hopf_algebras} and \ref{sec:hopf_quadratic}) is to abstractify the axioms of abelian groups and their quadratic functions by representing them in terms of string diagrams.
These string diagrams can then be interpreted in terms of fermionic tensor networks.
\subsection{Fermionic tensor networks}
\label{sec:fermionic_tensor_network}
When we describe fermionic systems with tensor networks, it is natural to use \emph{fermionic tensor networks}.
Just like an ordinary tensor network, a fermionic tensor network is a computation on tensor data using tensor products and contractions.
Fermionic tensor networks can therefore be denoted by the same diagrams as ordinary tensor networks.
The only difference is that ``tensor data'', ``tensor product'', and ``contraction'' are defined slightly differently.
Namely, they automatically keep track of the fermionic reordering sign:
Exchanging two fermions yields a factor of $-1$, which is most commonly stated as $c_i^\dagger c_j^\dagger = -c_j^\dagger c_i^\dagger$ for $i\neq j$, where $c^\dagger_i$ is the creation operator for the $i$th fermionic mode.

The mathematically most formal way to define fermionic tensors and tensor networks is as a \emph{compact closed category}, or alternatively as a \emph{tensor type}~\cite{tensor_type}.
Here, we will resort to a less formal definition.
\begin{mydef}
A \emph{fermionic tensor} $T$ with $n$ indices is defined as an ordinary tensor (see Eq.~\eqref{eq:tensor_map}) with the following constraints and additional structure.
For each index position $0\leq i<n$, there is a map
\begin{equation}
|\bullet|: B_i\rightarrow \zz_2\;.
\end{equation}
$T$ has to satisfy
\begin{equation}
\label{eq:z2_grading}
\sum_{0\leq i<n} |b_i| \neq 0
\quad\Rightarrow\quad
T(b_0,\ldots,b_{n-1}) = 0\;.
\end{equation}
Further, for each index position $i$, the $i$th index is either \emph{ingoing} or \emph{outgoing}.
\end{mydef}
That is, a fermionic tensor is a tensor where each basis set $B$ is divided \emph{even-parity} configurations $b\in B$ with $|b|=0$ and \emph{odd-parity} configurations with $|b|=1$.
The entries are only non-zero for index configurations with total even parity, where parities are added $\mmod 2$.
In other words, a fermionic tensor is a $\zz_2$-graded tensor. A fermionic tensor network is defined in a similar way to the non-fermionic case in Definition~\ref{def:tensor_network}.
\begin{mydef}
A \emph{fermionic tensor network} is a computation with a set of fermionic tensors as input and a single fermionic tensor as output, consisting of the following operations:
\begin{itemize}
\item The \emph{fermionic index transposition} at the $i$th index position of a tensor $T$ is a tensor $T'$ with sequence of basis sets $(B_0,\ldots,B_{i+1},B_i,\ldots,B_{n-1})$.
$T'$ is obtained by (1) exchanging the $i$th with the $i+1$th index, and (2) multiplying by $-1$ if they are both in an odd-parity configuration:
\begin{equation}
T'(b_0,\ldots,b_i,b_{i+1},\ldots,b_{n-1}) = (-1)^{|b_i||b_{i+1}|} \cdot T(b_0,\ldots,b_{i+1},b_i,\ldots,b_{n-1})\;.
\end{equation}
\item The \emph{fermionic index contraction} at the $i$th index position is defined if $B_i=B_{i+1}$ and the $i$th index is ingoing and the $i+1$th index is outgoing.
In this case, it is defined like an ordinary index contraction between the $i$th and $i+1$th index from Definition~\ref{def:tensor_network}.
\item The \emph{fermionic tensor product} is equal to the ordinary tensor product from Definition~\ref{def:tensor_network}.
\item \emph{Fermionic index blocking} at position $i$ is defined if the indices at positions $i$ and $i+1$ are either both ingoing or both outgoing.
If they are both outgoing, it is defined as in Definition~\ref{def:tensor_network}.
If they are both ingoing, it is defined with an additional factor of $(-1)^{|b_i||b_{i+1}|}$.
\end{itemize}
\end{mydef}
It is easy to see that all four operations preserve Eq.~\eqref{eq:z2_grading}.
In particular, in the contraction, both contracted indices are forced to the same configuration $x$, and the pair $(x,x)$ has even total parity.

Even though we only define fermionic index contraction between consecutive indices, we can emulate the contraction of an arbitrary index pair by first applying index transpositions.
The $-1$ factors collected during these index transpositions are the difference between evaluating an ordinary tensor network and evaluating a fermionic tensor network.
These signs are known in the physics literature as \emph{fermionic reordering signs}, and in the mathematics literature as \emph{Kozul signs}.
Let us give an example for how to compute the reordering signs.
Recall the tensor-network diagram given in Eq.~\eqref{eq:example_network_diagram}:
\begin{equation}
\label{eq:reordering_derivation}
\begin{tikzpicture}
\atoms{square,small}{{0/lab={p=0:0.25,t=$T$}}, {1/p={1.6,0},lab={p=180:0.25,t=$T$}}}
\atoms{circ,small}{{m/p={0.8,0.5},lab={p=90:0.25,t=$M$}}}
\draw (0)edge[ar=e,out=-90,in=-90,mark={rlab=z,sideoff=0.06cm,r}](1) (0)edge[mark={bar,s},ar=e,out=90,in=180,mark={rlab=x,sideoff=0.06cm}](m) (1)edge[mark={bar,s},out=90,in=0,mark={rlab=y,sideoff=0.06cm,r}](m) (0)edge[ar=s,ind=$i$]++(180:0.6) (1)edge[ind=$j$]++(0:0.6);
\end{tikzpicture}
\;,\qquad
(
\begin{tikzpicture}
\atoms{square,small}{{0/lab={p=-90:0.3,t=$T$}}}
\draw (0)edge[ind=$i$]++(180:0.5) (0)edge[ar=s,ind=$j$]++(90:0.5) (0)edge[mark={bar,s},ind=$k$]++(0:0.5);
\end{tikzpicture}
,
jk'i'
)
\;,\qquad
(
\begin{tikzpicture}
\atoms{circ,small}{{0/lab={p=-90:0.3,t=$M$}}}
\draw (0)edge[ar=s,ind=$i$]++(180:0.5) (0)edge[ind=$j$]++(00:0.5);
\end{tikzpicture}
,
ij
)\;.
\end{equation}
On the right, we indicate which lines in the diagram correspond to which index position, and whether the corresponding index is ingoing or outgoing.
We do this by labeling the lines originating from the tensor, and then writing the labels in sequence, adding a prime to each outgoing index.
When contracting the tensor network, we can start by taking the tensor product of all tensor copies.
After this, the index sequence is obtained by concatenating the index sequences of all individual tensor copies.
Note that the order of tensor product/concatenation does not matter since each tensor copy is individually parity even.
The labels we use in this sequence are either the open-index labels in the diagram in Eq.~\eqref{eq:reordering_derivation} or the small red labels corresponding to contractions.
So for the diagram in Eq.~\eqref{eq:reordering_derivation} we get
\begin{equation}
ix'z'\quad zy'j'\quad xy\;.
\end{equation}
In order to perform the contraction between an ingoing index with label $x$ and an outgoing index with label $x'$ we need to apply fermionic index transpositions such that they occur consecutively as $xx'$ in the sequence.
We record the $-1$ factors that occur during these fermionic index transpositions and write them in front of the current index sequence.
Once $xx'$ are in consecutive order, we can perform the contraction, which removes the pair from the index sequence.
So for the above sequence, we get
\begin{equation}
\begin{multlined}
ix'z' zy'j xy
\rightarrow
(-1)^{|z|}ix'zz'y'jxy
\rightarrow
(-1)^{|z|+|j||y|}ix'jxyy'\\
\rightarrow
(-1)^{|z|+|j||y|+|x|+|x||j|}ijxx'
\rightarrow
(-1)^{|z|+|j||y|+|x|+|x||j|}ij
\;.
\end{multlined}
\end{equation}
We can rewrite the reordering sign using the fact that each individual tensor copy has even parity, for example we have $|i|+|x|+|z|=0$ as one tensor copy has indices $i$, $x'$, and $z'$.
Note that the final result depends on a chosen ordering of the remaining open indices, $ij$ in the example above.
We can also consider the relative reordering sign between two sides of a tensor-network equation, which does not depend on the chosen ordering of the open indices, as long as we choose the same ordering on both sides.

Finally, we note that ordinary tensors (tensor networks) can be seen as a subset of fermionic tensors (tensor networks), where we set $|b_i|=0$ for every basis configuration $b_i\in B_i$.

\subsection{(Super) Hopf algebras}
\label{sec:hopf_algebras}
In order to capture free fermions as quadratic tensors, we have to generalize the notion of an abelian group to what is known as a (co)commutative (super) Hopf algebra.
\begin{mydef}
\label{def:hopf_algebra}
A \emph{commutative and cocommutative (super) Hopf algebra} is a set of (fermionic) tensors
\begin{equation}
\label{eq:hopf_algebra_tensors}
\begin{tikzpicture}
\atoms{alg}{0/}
\draw (0)edge[ind=$a$,ar=s]++(45:0.6) (0)edge[ind=$b$,ar=s]++(135:0.6) (0)edge[ind=$c$]++(-90:0.6);
\end{tikzpicture}
\;,\qquad
\begin{tikzpicture}
\atoms{alg}{0/}
\draw (0)edge[ind=$a$]++(-90:0.6);
\end{tikzpicture}\;,\qquad
\begin{tikzpicture}
\atoms{coalg}{0/}
\draw (0)edge[ind=$c$]++(-45:0.6) (0)edge[ind=$b$]++(-135:0.6) (0)edge[ind=$a$,ar=s]++(90:0.6);
\end{tikzpicture}
\;,\qquad
\begin{tikzpicture}
\atoms{coalg}{0/}
\draw (0)edge[ind=$a$,ar=s]++(90:0.6);
\end{tikzpicture}\;,\qquad
\begin{tikzpicture}
\atoms{antipode}{0/}
\draw (0)edge[ind=$a$,ar=s]++(90:0.6) (0)edge[ind=$b$]++(-90:0.6);
\end{tikzpicture}\;,
\end{equation}
where indices marked with an arrow are ingoing, satisfying the following set of (fermionic) tensor-network equations:
\begin{gather*}
\begin{multiequation}
\label{eq:hopf_commutative}
\begin{tikzpicture}
\atoms{alg}{0/}
\draw (0)edge[ind=$b$,ar=s]++(45:0.5) (0)edge[ind=$a$,ar=s]++(135:0.5) (0)edge[ind=$c$]++(-90:0.5);
\end{tikzpicture}
=
\begin{tikzpicture}
\atoms{alg}{0/}
\draw (0)edge[ind=$c$]++(-90:0.5);
\draw[rc,ind=$b$,ar=s] (0)--++(135:0.4)--++(30:0.7);
\draw[rc,ind=$a$,ar=s] (0)--++(45:0.4)--++(150:0.7);
\end{tikzpicture}
\NEXT
\label{eq:hopf_unital}
\begin{tikzpicture}
\atoms{alg}{0/, 1/p={45:0.6}}
\draw (0)edge[ar=s](1) (0)edge[ind=$a$,ar=s]++(90:0.6) (0)edge[ind=$b$]++(-90:0.6);
\end{tikzpicture}
=
\begin{tikzpicture}
\draw (0,0)edge[startind=$b$,ind=$a$]++(90:0.8);
\end{tikzpicture}
\NEXT
\label{eq:hopf_associative}
\begin{tikzpicture}
\atoms{alg}{0/, 1/p={45:0.6}}
\draw (0)edge[ar=s](1) (1)edge[ind=$c$,ar=s]++(45:0.6) (1)edge[ind=$b$,ar=s]++(135:0.6) (0)edge[ind=$a$,ar=s]++(135:0.6) (0)edge[ind=$d$]++(-90:0.6);
\end{tikzpicture}
=
\begin{tikzpicture}
\atoms{alg}{0/, 1/p={135:0.6}}
\draw (0)edge[ar=s](1) (1)edge[ind=$b$,ar=s]++(45:0.6) (1)edge[ind=$a$,ar=s]++(135:0.6) (0)edge[ind=$c$,ar=s]++(45:0.6) (0)edge[ind=$d$]++(-90:0.6);
\end{tikzpicture}
\end{multiequation}
\\
\begin{multiequation}
\label{eq:hopf_cocommutative}
\begin{tikzpicture}
\atoms{coalg}{0/}
\draw (0)edge[ind=$c$]++(-45:0.5) (0)edge[ind=$b$]++(-135:0.5) (0)edge[ind=$a$,ar=s]++(90:0.5);
\end{tikzpicture}
=
\begin{tikzpicture}
\atoms{coalg}{0/}
\draw (0)edge[ind=$a$,ar=s]++(90:0.5);
\draw[rc,ind=$c$] (0)--++(-135:0.4)--++(-30:0.7);
\draw[rc,ind=$b$] (0)--++(-45:0.4)--++(-150:0.7);
\end{tikzpicture}
\NEXT
\label{eq:hopf_counital}
\begin{tikzpicture}
\atoms{coalg}{0/, 1/p={-45:0.6}}
\draw (0)edge[ar=e](1) (0)edge[ind=$b$]++(-90:0.6) (0)edge[ind=$a$,ar=s]++(90:0.6);
\end{tikzpicture}
=
\begin{tikzpicture}
\draw (0,0)edge[startind=$b$,ind=$a$]++(90:0.8);
\end{tikzpicture}
\NEXT
\label{eq:hopf_coassociative}
\begin{tikzpicture}
\atoms{coalg}{0/, 1/p={-45:0.6}}
\draw (0)edge[ar=e](1) (1)edge[ind=$d$]++(-45:0.6) (1)edge[ind=$c$]++(-135:0.6) (0)edge[ind=$b$]++(-135:0.6) (0)edge[ind=$a$,ar=s]++(90:0.6);
\end{tikzpicture}
=
\begin{tikzpicture}
\atoms{coalg}{0/, 1/p={-135:0.6}}
\draw (0)edge[ar=e](1) (1)edge[ind=$c$]++(-45:0.6) (1)edge[ind=$b$]++(-135:0.6) (0)edge[ind=$d$]++(-45:0.6) (0)edge[ind=$a$,ar=s]++(90:0.6);
\end{tikzpicture}
\end{multiequation}
\\
\begin{multiequation}
\label{eq:hopf_bialg0}
\begin{tikzpicture}
\atoms{alg}{0/, 1/p={0.8,0}}
\atoms{coalg}{2/p={0,0.8}, 3/p={0.8,0.8}}
\draw (0)edge[ar=s](2) (0)edge[ar=s](3) (1)edge[ar=s](2) (1)edge[ar=s](3);
\draw (2)edge[ar=s]++(90:0.5) (3)edge[ar=s]++(90:0.5) (0)--++(-90:0.5) (1)--++(-90:0.5);
\end{tikzpicture}
=
\begin{tikzpicture}
\atoms{alg}{0/}
\atoms{coalg}{1/p={0,-0.8}}
\draw (1)edge[ar=s](0) (0)edge[ar=s]++(45:0.5) (0)edge[ar=s]++(135:0.5) (1)--++(-45:0.5) (1)--++(-135:0.5);
\end{tikzpicture}
\NEXT
\label{eq:hopf_bialg1}
\begin{tikzpicture}
\atoms{alg}{0/, 1/p={0.8,0}}
\draw (0)--++(-90:0.5) (1)--++(-90:0.5);
\end{tikzpicture}
=
\begin{tikzpicture}
\atoms{alg}{0/}
\atoms{coalg}{1/p={0,-0.8}}
\draw (1)edge[ar=s](0) (1)--++(-45:0.5) (1)--++(-135:0.5);
\end{tikzpicture}
\NEXT
\label{eq:hopf_bialg2}
\begin{tikzpicture}
\atoms{coalg}{2/p={0,0.8}, 3/p={0.8,0.8}}
\draw (2)edge[ar=s]++(90:0.5) (3)edge[ar=s]++(90:0.5);
\end{tikzpicture}
=
\begin{tikzpicture}
\atoms{alg}{0/}
\atoms{coalg}{1/p={0,-0.8}}
\draw (1)edge[ar=s](0) (0)edge[ar=s]++(45:0.5) (0)edge[ar=s]++(135:0.5);
\end{tikzpicture}
\NEXT
\label{eq:hopf_bialg3}
\varnothing
=
\begin{tikzpicture}
\atoms{alg}{0/}
\atoms{coalg}{1/p={0,-0.6}}
\draw (1)edge[ar=s](0);
\end{tikzpicture}
\end{multiequation}
\\
\begin{multiequation}
\label{eq:hopf_antipode0}
\begin{tikzpicture}
\atoms{alg}{0/}
\atoms{antipode}{1/p={-90:0.5}}
\draw (1)edge[ar=s](0) (0)edge[ind=$b$,ar=s]++(45:0.5) (0)edge[ind=$a$,ar=s]++(135:0.5) (1)edge[ind=$c$]++(-90:0.5);
\end{tikzpicture}
=
\begin{tikzpicture}
\atoms{alg}{0/}
\atoms{antipode}{1/p={-0.4,0.6}, 2/p={0.4,0.6}}
\draw (1)edge[ar=s,ind=$a$]++(90:0.5) (2)edge[ar=s,ind=$b$]++(90:0.5) (0)edge[ind=$c$]++(-90:0.5);
\draw[rc,ar=e] (1)--++(-90:0.3)--(0);
\draw[rc,ar=e] (2)--++(-90:0.3)--(0);
\end{tikzpicture}
\NEXT
\label{eq:hopf_antipode1}
\begin{tikzpicture}
\atoms{alg}{0/}
\atoms{antipode}{1/p={-90:0.5}}
\draw (1)edge[ar=s](0) (1)edge[ind=$a$]++(-90:0.5);
\end{tikzpicture}
=
\begin{tikzpicture}
\atoms{alg}{0/}
\draw (0)edge[ind=$a$]++(-90:0.5);
\end{tikzpicture}
\NEXT
\label{eq:hopf_antipode2}
\begin{tikzpicture}
\atoms{antipode}{0/, 1/p={0.8,0}}
\atoms{coalg}{2/p={0.4,0.6}}
\draw[ar=s,rc] (0)--++(90:0.3)--(2);
\draw[ar=s,rc] (1)--++(90:0.3)--(2);
\draw (2)edge[ar=s]++(90:0.5) (0)--++(-90:0.5) (1)--++(-90:0.5);
\end{tikzpicture}
=
\begin{tikzpicture}
\atoms{antipode}{0/}
\atoms{coalg}{1/p={0,-0.6}}
\draw (1)edge[ar=s](0) (0)edge[ar=s]++(90:0.5) (1)--++(-45:0.5) (1)--++(-135:0.5);
\end{tikzpicture}
\end{multiequation}
\\
\begin{multiequation}
\label{eq:hopf_antipode3}
\begin{tikzpicture}
\atoms{coalg}{2/p={0.4,0.6}}
\draw (2)edge[ar=s]++(90:0.5);
\end{tikzpicture}
=
\begin{tikzpicture}
\atoms{antipode}{0/}
\atoms{coalg}{1/p={0,-0.6}}
\draw (1)edge[ar=s](0) (0)edge[ar=s]++(90:0.5);
\end{tikzpicture}
\NEXT
\label{eq:hopf_antipode4}
\begin{tikzpicture}
\atoms{coalg}{0/}
\atoms{alg}{1/p={0,-1.2}}
\atoms{antipode}{a/p={-0.4,-0.6}}
\draw[rc,ar=s] (a)--++(90:0.3)--(0);
\draw[rc,ar=s] (1)--++(-0.4,0.4)--(a);
\draw[rc,ar=s] (1)--++(0.4,0.4)--++(90:0.4)--(0);
\draw (0)edge[ar=s,ind=$a$]++(90:0.4) (1)edge[ind=$b$]++(-90:0.4);
\end{tikzpicture}
=
\begin{tikzpicture}
\atoms{coalg}{0/}
\atoms{alg}{1/p={0,-0.6}}
\draw (0)edge[ar=s,ind=$a$]++(90:0.4) (1)edge[ind=$b$]++(-90:0.4);
\end{tikzpicture}
\NEXT
\label{eq:hopf_antipode5}
\begin{tikzpicture}
\atoms{antipode}{0/, 1/p={90:0.5}}
\draw (0)edge[ar=s](1) (1)edge[ar=s,ind=$a$]++(90:0.5) (0)edge[ind=$b$]++(-90:0.5);
\end{tikzpicture}
=
\begin{tikzpicture}
\draw[ind=$a$,startind=$b$](0,0)--++(90:0.8);
\end{tikzpicture}
\end{multiequation}
\end{gather*}
The tensors in Eq.~\eqref{eq:hopf_algebra_tensors} are called \emph{multiplication}, \emph{unit}, \emph{comultiplication}, \emph{counit}, and \emph{antipode}, respectively.
Multiplication and unit together form an \emph{(associative unital commutative) algebra}, and comultiplication and counit form a \emph{(coassociative counital cocommutative) coalgebra}.
The tensor-network equations are commonly called \emph{commutativity} (Eq.~\eqref{eq:hopf_commutative}), \emph{unitality} (Eq.~\eqref{eq:hopf_unital}), \emph{associativity} (Eq.~\eqref{eq:hopf_associative}), \emph{cocommutativity} (Eq.~\eqref{eq:hopf_cocommutative}), \emph{counitality} (Eq.~\eqref{eq:hopf_counital}), and \emph{coassociativity} (Eq.~\eqref{eq:hopf_coassociative}).
Eqs.~\eqref{eq:hopf_bialg0}, \eqref{eq:hopf_bialg1}, \eqref{eq:hopf_bialg2}, and \eqref{eq:hopf_bialg3} make the algebra and the coalgebra into a \emph{bialgebra}, and Eqs.~\eqref{eq:hopf_antipode0}, \eqref{eq:hopf_antipode1}, \eqref{eq:hopf_antipode2}, \eqref{eq:hopf_antipode3}, \eqref{eq:hopf_antipode4}, and \eqref{eq:hopf_antipode5} make the bialgebra into a Hopf algebra.
Note that the set of axioms and tensors is symmetric under (1) exchanging full and empty tensors, (2) flipping the ingoing and outgoing indices (including vertical reflection if we keep the flow of time from top to bottom).
So for a Hopf algebra $G$, exchanging full and empty tensors yields another Hopf algebra, which we call the \emph{dual} of $G$ and denote by $G^*$.
We do not claim that the axioms are fully independent:
At least Eq.~\eqref{eq:hopf_antipode5} can be derived diagrammatically from the remaining axioms.
We will often drop the adjectives ``commutative'' and ``cocommutative'', as they will hold for all the Hopf algebras we consider here.
Finally, it will sometimes be convenient to introduce algebra and coalgebra tensors with more indices, such as
\begin{equation}
\label{eq:algeb_coalg_more_indices}
\begin{tikzpicture}
\atoms{alg}{0/}
\draw (0)edge[ar=s]++(-0.7,0.7) (0)edge[ar=s]++(0,0.7) (0)edge[ar=s]++(0.7,0.7) (0)edge[]++(-90:0.6);
\end{tikzpicture}
\coloneqq
\begin{tikzpicture}
\atoms{alg}{0/, 1/p={45:0.6}}
\draw (0)edge[ar=s](1) (1)edge[ar=s]++(45:0.6) (1)edge[ar=s]++(135:0.6) (0)edge[ar=s]++(135:0.6) (0)edge[]++(-90:0.6);
\end{tikzpicture}
\;,\qquad
\begin{tikzpicture}
\atoms{coalg}{0/}
\draw (0)edge[]++(-0.7,-0.7) (0)edge[]++(0,-0.7) (0)edge[]++(0.7,-0.7) (0)edge[ar=s]++(90:0.6);
\end{tikzpicture}
\coloneqq
\begin{tikzpicture}
\atoms{coalg}{0/, 1/p={-45:0.6}}
\draw (0)edge[ar=e](1) (1)edge[]++(-45:0.6) (1)edge[]++(-135:0.6) (0)edge[]++(-135:0.6) (0)edge[ar=s]++(90:0.6);
\end{tikzpicture}
\;.
\end{equation}
\end{mydef}
More commonly, one would view the tensors as (multi-)linear maps from the top to the bottom indices.
Note that as any (non-fermionic) tensor is also a fermionic tensor, any (non-super) Hopf algebra is also a super Hopf algebra.
Next, we show that (co)commutative (super) Hopf algebras are indeed a generalization of abelian groups.
\begin{myprop}
Any abelian group $G$ yields an commutative and cocommutative Hopf algebra.
The basis set is $G$, and the tensors are given by
\begin{equation}
\label{eq:group_hopf_algebra}
\begin{tikzpicture}
\atoms{alg}{0/}
\draw (0)edge[ind=$a$,ar=s]++(45:0.6) (0)edge[ind=$b$,ar=s]++(135:0.6) (0)edge[ind=$c$]++(-90:0.6);
\end{tikzpicture}
=\delta_{a+b=c}\;,\qquad
\begin{tikzpicture}
\atoms{alg}{0/}
\draw (0)edge[ind=$a$]++(-90:0.6);
\end{tikzpicture}
=\delta_{a=0}
\;,\qquad
\begin{tikzpicture}
\atoms{coalg}{0/}
\draw (0)edge[ind=$c$]++(-45:0.6) (0)edge[ind=$b$]++(-135:0.6) (0)edge[ind=$a$,ar=s]++(90:0.6);
\end{tikzpicture}
= \delta_{a=b=c}
\;,\qquad
\begin{tikzpicture}
\atoms{coalg}{0/}
\draw (0)edge[ind=$a$,ar=s]++(90:0.6);
\end{tikzpicture}
=1\forall 1
\;,\qquad
\begin{tikzpicture}
\atoms{antipode}{0/}
\draw (0)edge[ind=$a$,ar=s]++(90:0.6) (0)edge[ind=$b$]++(-90:0.6);
\end{tikzpicture}
=\delta_{-a=b}
\;.
\end{equation}
Here, we have used a Kronecker-$\delta$ symbol, which is $1$ if the condition in its subscript is fulfilled, and $0$ otherwise.
We call this Hopf algebra the \emph{group Hopf algebra} of $G$.
\end{myprop}
\begin{proof}
We show this by realizing that the above tensors/linear maps are just ``linearizations'' of functions $f$ between copies of the set $G$.
A linearization of $f$ is a tensor $L(f)$ with entries
\begin{equation}
L(f)_{ab}=\delta_{f(a)=b}\;.
\end{equation}
Concretely, we have
\begin{equation}
\label{eq:group_functions}
\begin{gathered}
\begin{tikzpicture}
\atoms{alg}{0/}
\draw (0)edge[ar=s]++(45:0.4) (0)edge[ar=s]++(135:0.4) (0)edge[]++(-90:0.4);
\end{tikzpicture}
=L(+:G\times G\rightarrow G)
\;,\qquad
\begin{tikzpicture}
\atoms{alg}{0/}
\draw (0)edge[]++(-90:0.4);
\end{tikzpicture}
=L(0: \{\varnothing\}\rightarrow G)\;,\qquad
\begin{tikzpicture}
\atoms{coalg}{0/}
\draw (0)edge[]++(-45:0.4) (0)edge[]++(-135:0.4) (0)edge[ar=s]++(90:0.4);
\end{tikzpicture}
=L(\operatorname{Copy}: G\rightarrow G\times G)
\;,\\
\begin{tikzpicture}
\atoms{coalg}{0/}
\draw (0)edge[ar=s]++(90:0.4);
\end{tikzpicture}
=L(\operatorname{Erase}: G\rightarrow\{\varnothing\})
\;,\qquad
\begin{tikzpicture}
\atoms{antipode}{0/}
\draw (0)edge[ar=s]++(90:0.4) (0)edge[]++(-90:0.4);
\end{tikzpicture}
=L(-: G\rightarrow G)\;.
\end{gathered}
\end{equation}
Here, $+$, $0$, and $-$ are the group multiplication, unit, and inverse, and $\operatorname{Copy}$ and $\operatorname{Erase}$ are the following functions:
\begin{equation}
\operatorname{Copy}(g)=(g,g)\;,\qquad
\operatorname{Erase}(g)=\varnothing\;.
\end{equation}
Under this mapping, the tensor-network equations in Definition~\ref{def:hopf_algebra} become equations between different compositions of functions.
The corresponding diagrams are then ``circuit'' representations of these compositions.
Concretely, the equations are mapped to the following axioms for the functions:
\begin{equation}
\label{eq:group_algebra_axioms}
\begin{gathered}
\eqref{eq:hopf_commutative}:\quad
a+b = b+a\;,\qquad
\eqref{eq:hopf_unital}:\quad
a+0 = a\;,\qquad
\eqref{eq:hopf_associative}:\quad
a+(b+c)=(a+b)+c\;,\\
\eqref{eq:hopf_cocommutative}:\quad
\operatorname{Copy}(a) = \operatorname{Swap}(\operatorname{Copy}(a))\;,\qquad
\eqref{eq:hopf_counital}:\quad
(\idop\times \operatorname{Erase})(\operatorname{Copy}(a)) = a\;,\\
\eqref{eq:hopf_coassociative}:\quad
(\idop\times \operatorname{Copy})(\operatorname{Copy}(a)) = (\operatorname{Copy}\times \idop)(\operatorname{Copy}(a))\;,\\
\eqref{eq:hopf_antipode0}:\quad
-(a+b) = (-a)+(-b)\;,\qquad
\eqref{eq:hopf_antipode1}:\quad
-0=0\;,\qquad
\eqref{eq:hopf_antipode4}:\quad
a+(-a)=0\;,\qquad
\eqref{eq:hopf_antipode5}:\quad
-(-a) = a\;.
\end{gathered}
\end{equation}
Further, for any function $f$ we have the following equality:
On the left side we copy all inputs of $f$ and then apply $f$ to both sets of copies.
On the right side we apply $f$ to its input, and then copy all outputs.
The tensor-network equations in Eqs.~\eqref{eq:hopf_bialg0}, \eqref{eq:hopf_bialg1}, \eqref{eq:hopf_antipode2} map to this equality for $f$ equal to the multiplication, unit, and group inverse, respectively.
There is also the following trivial equality:
On the left side we erase all inputs of $f$, and on the right side we apply $f$ and then erase all of its outputs.
The tensor-network equations in Eqs.~\eqref{eq:hopf_bialg2}, \eqref{eq:hopf_bialg3}, \eqref{eq:hopf_antipode3} map to this equality for $f$ equal to the multiplication, unit, and group inverse, respectively.
All in all, we see that all the tensor-network equations of Hopf algebras are mapped to either the group axioms, or two trivial equalities involving the copy and erase functions.
\end{proof}
Mathematically, groups, Hopf algebras, and super Hopf algebras are all just Hopf algebras, in the \emph{symmetric monoidal category} of sets, vector spaces, and super vector spaces, respectively.
The diagrammatic axioms are \emph{string diagrams} of symmetric monoidal categories.
The group algebra can be obtained from applying a functor from the category of sets to the category of vector spaces.

A natural question is whether there are (super) Hopf algebras that are not group Hopf algebras.
If we restrict ourselves to proper, non-super, Hopf algebras, then the answer is no:
It is known that every cocommutative Hopf algebra is isomorphic to a group Hopf algebra as in Eq.~\eqref{eq:group_hopf_algebra}, and the group is abelian if the Hopf algebra is additionally commutative.
Thus, going from abelian groups to commutative and cocommutative Hopf algebras does not actually yield a more general family of efficiently solvable models.
It does, however, give us the freedom to express the Hopf-algebra tensors in a basis other than the one spanned by the group elements.
As an example, consider the group Hopf algebra for $G=\rr$.
A natural basis is that of eigenfunctions of the harmonic oscillator, in other words, the energy eigenbasis or the occupation-number basis, labeled by non-negative integers.
In this basis, the $\rr$-group Hopf algebra becomes the following Hopf algebra with basis set $B=\zz_{\geq 0}$, and tensors
\begin{equation}
\label{eq:free_boson_hopf_algebra}
\begin{gathered}
\begin{tikzpicture}
\atoms{alg}{0/}
\draw (0)edge[ind=$a$,ar=s]++(45:0.6) (0)edge[ind=$b$,ar=s]++(135:0.6) (0)edge[ind=$c$]++(-90:0.6);
\end{tikzpicture}
=
\sqrt{\frac{(a+b)!}{a!b!}}
\delta_{a+b=c}
\qquad
\begin{tikzpicture}
\atoms{alg}{0/}
\draw (0)edge[ind=$a$]++(-90:0.6);
\end{tikzpicture}
=
\delta_{a=0}\;,
\\
\begin{tikzpicture}
\atoms{coalg}{0/}
\draw (0)edge[ind=$a$]++(-45:0.6) (0)edge[ind=$b$]++(-135:0.6) (0)edge[ind=$c$,ar=s]++(90:0.6);
\end{tikzpicture}
=
\sqrt{\frac{(a+b)!}{a!b!}}
\delta_{a+b=c}
\qquad
\begin{tikzpicture}
\atoms{coalg}{0/}
\draw (0)edge[ind=$a$,ar=s]++(90:0.6);
\end{tikzpicture}
=
\delta_{a=0}\;,
\qquad
\begin{tikzpicture}
\atoms{antipode}{0/}
\draw (0)edge[ind=$a$,ar=s]++(90:0.6) (0)edge[ind=$b$]++(-90:0.6);
\end{tikzpicture}
=(-1)^a \delta_{a=b}
\;.
\end{gathered}
\end{equation}
This basis has the neat feature that it is completely symmetric, in the sense that the structure coefficients of the algebra and coalgebra are identical.
Note that this is the algebra generated by bosonic creation and annihilation operators.

In contrast to non-super Hopf algebras, there are super Hopf algebras that not isomorphic to group Hopf algebras.
Concretely, there is essentially just one such example.
\begin{myprop}
The following defines a super Hopf algebra, which we call $\mathcal F$.
The basis configurations are given by the set $B=\{0,1\}$ with $|0|=0$ and $|1|=1$.
The tensors are given by
\begin{equation}
\label{eq:fermion_hopf_algebra}
\begin{gathered}
\big(
\begin{tikzpicture}
\atoms{alg}{0/}
\draw (0)edge[ind=$a$,ar=s]++(45:0.6) (0)edge[ind=$b$,ar=s]++(135:0.6) (0)edge[ind=$c$]++(-90:0.6);
\end{tikzpicture}
,cab\big)
=
\delta_{a+b=c}
\simeq\mpm{\mpm{1&0\\0&0}&\mpm{0&1\\1&0}}
\;,
\qquad
\big(
\begin{tikzpicture}
\atoms{alg}{0/}
\draw (0)edge[ind=$a$]++(-90:0.6);
\end{tikzpicture}
,a\big)
=
\delta_{a=0}
\simeq
\begin{pmatrix}
1&0
\end{pmatrix}
\;,
\\
\big(
\begin{tikzpicture}
\atoms{coalg}{0/}
\draw (0)edge[ind=$a$]++(-45:0.6) (0)edge[ind=$b$]++(-135:0.6) (0)edge[ind=$c$,ar=s]++(90:0.6);
\end{tikzpicture}
,bac\big)
=
\delta_{a+b=c}
\simeq\mpm{\mpm{1&0\\0&1}&\mpm{0&1\\0&0}}
\;,
\qquad
\big(
\begin{tikzpicture}
\atoms{coalg}{0/}
\draw (0)edge[ind=$a$,ar=s]++(90:0.6);
\end{tikzpicture}
,a\big)
=
\delta_{a=0}
\simeq
\begin{pmatrix}
1&0
\end{pmatrix}
\;,\\
\big(
\begin{tikzpicture}
\atoms{antipode}{0/}
\draw (0)edge[ind=$a$,ar=s]++(90:0.6) (0)edge[ind=$b$]++(-90:0.6);
\end{tikzpicture}
,ba\big)
=(-1)^a \delta_{a=b}
\simeq
\begin{pmatrix}
1&0\\0&-1
\end{pmatrix}
\;.
\end{gathered}
\end{equation}
\end{myprop}
\begin{proof}
It is straightforward to verify that all the Hopf algebra axioms are fulfilled.
To this end, we need to compute the relative reordering sign of each tensor-network equation.
The algebraic structure we have defined so far has a flow of time, which we have chosen to draw from the top to bottom.
In this case, we can order the indices of each tensor as follows:
We start with the output indices from left to right, and continue with the input indices from right to left, as chosen in Eq.~\eqref{eq:fermion_hopf_algebra}.
We can choose the same ordering for the open indices of the resulting tensor.
With this, it is easy to read off the reordering sign directly from the diagram:
For every crossing of two lines with index values $x$ and $y$, we get a reordering sign of $(-1)^{|x||y|}$.

Using this simple rule, we can see that the reordering signs are only non-trivial for Eqs.~\eqref{eq:hopf_commutative}, \eqref{eq:hopf_cocommutative}, and \eqref{eq:hopf_bialg0}.
We start by checking that the equations with trivial reordering signs hold.
Eqs.~\eqref{eq:hopf_associative}, \eqref{eq:hopf_unital}, \eqref{eq:hopf_coassociative}, and \eqref{eq:hopf_counital} hold since the full tensors and the empty tensors in Eq.~\eqref{eq:fermion_hopf_algebra} each form a well-known unital and associative (non-super) algebra, with one non-trivial generator $e_1$ satisfying $e_1^2=0$.
Eqs.~\eqref{eq:hopf_antipode0}, \eqref{eq:hopf_antipode1}, \eqref{eq:hopf_antipode2}, \eqref{eq:hopf_antipode3} conjugate the algebra and coalgebra tensors by the antipode, which is a Pauli-$Z$ operator.
These equations hold since the algebra and coalgebra tensors are $\zz_2$-graded, as any fermionic tensor must be.
Eq.~\eqref{eq:hopf_antipode4} holds due to
\begin{equation}
\begin{tikzpicture}
\atoms{coalg}{0/}
\atoms{alg}{1/p={0,-1.2}}
\atoms{antipode}{a/p={-0.4,-0.6}}
\draw[rc,ar=e,mark={rlab=x,sideoff=0.06cm,r}] (0)--++(-0.4,-0.4)--(a);
\draw[rc,ar=s,mark={rlab=y}] (1)--++(-0.4,0.4)--(a);
\draw[rc,ar=s,mark={rlab=z,sideoff=0.06cm,r}] (1)--++(0.4,0.4)--++(90:0.4)--(0);
\draw (0)edge[ar=s,ind=$a$]++(90:0.4) (1)edge[ind=$b$]++(-90:0.4);
\end{tikzpicture}
=
\sum_{x,y,z\in\{0,1\}}\delta_{x+z=a}\delta_{y+z=b}(-1)^y\delta_{x=y}=\delta_{a=b}\sum_{\substack{x,z\in\{0,1\}\\x+z=a}}(-1)^x = \delta_{a=b}\delta_{a=0} = \delta_{a=0}\delta_{b=0}
=
\begin{tikzpicture}
\atoms{coalg}{0/}
\atoms{alg}{1/p={0,-0.6}}
\draw (0)edge[ar=s,ind=$a$]++(90:0.4) (1)edge[ind=$b$]++(-90:0.4);
\end{tikzpicture}
\;.
\end{equation}
Eq.~\eqref{eq:hopf_bialg1} holds as
\begin{equation}
\delta_{a=0}\delta_{b=0}=\delta_{a+b=0}\;,
\end{equation}
and similar for Eqs.~\eqref{eq:hopf_bialg2} and \eqref{eq:hopf_bialg3}.
The relative reordering sign for Eq.~\eqref{eq:hopf_commutative} $(-1)^{|a||b|}$, however, this reordering sign has no effect as the multiplication tensor in Eq.~\eqref{eq:fermion_hopf_algebra} has entry $0$ for $a=1,b=1$.
Thus the fact that the $e_1^2=0$ algebra is commutative also implies that the corresponding superalgebra is supercommutative.
Eq.~\eqref{eq:hopf_cocommutative} holds analogously.
Finally, Eq.~\eqref{eq:hopf_bialg0} holds due to
\begin{equation}
\begin{tikzpicture}
\atoms{alg}{0/, 1/p={0.8,0}}
\atoms{coalg}{2/p={0,0.8}, 3/p={0.8,0.8}}
\draw (0)edge[ar=s,mark={rlab=w}](2) (0)edge[ar=s,mark={rlab=x,p=0.8}](3) (1)edge[ar=s,mark={rlab=y,p=0.8,r,sideoff=-0.06cm}](2) (1)edge[ar=s,mark={rlab=z,r,sideoff=-0.06cm}](3);
\draw (2)edge[ar=s,ind=$a$]++(90:0.5) (3)edge[ar=s,ind=$b$]++(90:0.5) (0)edge[ind=$c$]++(-90:0.5) (1)edge[ind=$d$]++(-90:0.5);
\end{tikzpicture}
=
\sum_{w,x,y,z} (-1)^{xy} \delta_{w+x=c} \delta_{y+z=d} \delta_{w+y=a} \delta_{x+z=b}
=
\delta_{ab=0} \delta_{a+b=c+d}
=
\sum_x \delta_{a+b=x} \delta_{c+d=x}
=
\begin{tikzpicture}
\atoms{alg}{0/}
\atoms{coalg}{1/p={0,-0.8}}
\draw (1)edge[ar=s,mark={rlab=x}](0) (0)edge[ar=s,ind=$b$]++(45:0.5) (0)edge[ar=s,ind=$a$]++(135:0.5) (1)edge[ind=$d$]++(-45:0.5) (1)edge[ind=$c$]++(-135:0.5);
\end{tikzpicture}
\;.
\end{equation}
The relative reordering sign is $(-1)^{|x||y|}$, due to the line crossing on the left-hand side.
\end{proof}
Note that $\mathcal F$ looks different in different bases.
We have chosen a basis such that the structure coefficients for the algebra and coalgebra are identical, just like for the continuous-variable Hopf algebra in Eq.~\eqref{eq:free_boson_hopf_algebra}.
In a sense, the fermionic Hopf algebra $\mathcal F$ is the only example:
It is known~\cite{Konstant1977} that each super commutative and cocommutative Hopf algebra is isomorphic to a product of an abelian group Hopf algebra, and copies of the $\mathcal F$.
Taken together, the Hopf algebras we consider here are products of \emph{elementary Hopf algebras} $G\in \{\zz_k,\rr,\rr/\zz,\zz,\mathcal F\}$.

\subsection{Hopf quadratic tensors}
\label{sec:hopf_quadratic}
\myparagraph{Quadratic Hopf vectors}
After generalizing from groups to (super) Hopf algebras, we next need to generalize our definition of a quadratic function to the case of Hopf algebras.
This generalization is given by the notion of a quadratic Hopf vector below.
\begin{mydef}
Consider a (super) Hopf algebra $G$.
A \emph{Hopf vector} over $G$ is a vector $q$ with one $G$-index,
\begin{equation}
\label{eq:hopf_vector}
\begin{tikzpicture}
\atoms{linfunc,dot}{0/}
\draw (0)edge[ind=$g$,ar=s]++(180:0.5);
\end{tikzpicture}
\;.
\end{equation}
such that there exists an ``inverse'',
\begin{equation}
\label{eq:hopf_function_inverse}
\begin{tikzpicture}
\atoms{linfunc,dot}{0/lab={t=$-1$,p=90:0.3}}
\draw (0)edge[ind=$g$,ar=s]++(180:0.5);
\end{tikzpicture}
\;,
\end{equation}
such that
\begin{equation}
\label{eq:hopf_function_inverse_axiom}
\begin{tikzpicture}
\atoms{coalg}{0/}
\atoms{linfunc,dot}{1/p={-0.4,-0.6}, {2/p={0.4,-0.6},lab={t=$-1$,p=-90:0.3}}}
\draw[rc,ar=s] (1)--++(90:0.3)--(0);
\draw[rc,ar=s] (2)--++(90:0.3)--(0);
\draw (0)edge[ar=s,ind=$g$]++(90:0.5);
\end{tikzpicture}
=
\begin{tikzpicture}
\atoms{coalg}{0/}
\draw (0)edge[ar=s,ind=$g$]++(90:0.5);
\end{tikzpicture}
\;.
\end{equation}
We define the \emph{zeroth, first, second, and third derivative} of $q$ as
\begin{equation}
\label{eq:hopf_derivatives}
\begin{tikzpicture}
\atoms{linfunc}{0/}
\end{tikzpicture}
\coloneqq
\begin{tikzpicture}
\atoms{linfunc,dot}{0/}
\atoms{alg}{1/p={90:0.5}}
\draw (0)edge[ar=s](1);
\end{tikzpicture}\;,\qquad
\begin{tikzpicture}
\atoms{linfunc}{0/}
\draw (0)edge[ar=s]++(90:0.5);
\end{tikzpicture}
\coloneqq
\begin{tikzpicture}
\atoms{linfunc,dot}{0/}
\draw (0)edge[ar=s]++(90:0.5);
\atoms{linfunc}{{1/p={0.5,0},lab={t=$-1$,p=-90:0.3}}}
\end{tikzpicture}
\;,\qquad
\begin{tikzpicture}
\atoms{linfunc}{0/}
\draw (0)edge[ar=s]++(180:0.5) (0)edge[ar=s]++(0:0.5);
\end{tikzpicture}
\coloneqq
\begin{tikzpicture}
\atoms{alg}{0/}
\atoms{coalg}{1/p={-0.7,0}, 2/p={0.7,0}}
\atoms{linfunc}{a/p={0,-0.5}, {b/p={-0.7,-0.5},lab={t=$-1$,p=-90:0.3}}, {c/p={0.7,-0.5},lab={t=$-1$,p=-90:0.3}}}
\draw (0)edge[ar=e](a) (1)edge[ar=e](b) (2)edge[ar=e](c) (0)edge[ar=s](1) (0)edge[ar=s](2) (1)edge[ar=s]++(180:0.5) (2)edge[ar=s]++(0:0.5);
\end{tikzpicture}
\;,\qquad
\begin{tikzpicture}
\atoms{linfunc}{0/}
\draw (0)edge[ar=s]++(180:0.5) (0)edge[ar=s]++(0:0.5) (0)edge[ar=s]++(-90:0.5);
\end{tikzpicture}
\coloneqq
\begin{tikzpicture}
\atoms{alg}{0/}
\atoms{coalg}{1/p={-0.7,0}, 2/p={0.7,0}, 3/p={0,-1.3}}
\atoms{linfunc}{a/p={0,-0.5}, {b/p={-0.7,-0.5},lab={t=$-1$,p=180:0.4}}, {c/p={0.7,-0.5},lab={t=$-1$,p=0:0.4}}}
\draw (0)edge[ar=e](a) (1)edge[ar=e](b) (2)edge[ar=e](c) (0)edge[ar=s](1) (0)edge[ar=s](2) (1)edge[ar=s]++(180:0.5) (2)edge[ar=s]++(0:0.5);
\draw[rc,ar=s] (a)--++(-90:0.4)--(3);
\draw[rc,ar=s] (b)--++(-90:0.4)--(3);
\draw[rc,ar=s] (c)--++(-90:0.4)--(3);
\draw (3)edge[ar=s]++(-90:0.5);
\end{tikzpicture}
\;,
\end{equation}
respectively.
In text, we denote the derivatives as $q^{(0)}$, $q^{(1)}$, $q^{(2)}$, and $q^{(3)}$, respectively.
We call $q$ \emph{constant} if $q^{(1)}$ is trivial, \emph{linear} if $q^{(2)}$ is trivial, and \emph{quadratic} if $q^{(3)}$ is trivial, in the following sense, respectively,
\begin{equation}
\begin{tikzpicture}
\atoms{linfunc}{0/}
\draw (0)edge[ar=s]++(90:0.5);
\end{tikzpicture}
=
\begin{tikzpicture}
\atoms{coalg}{0/}
\draw (0)edge[ar=s]++(90:0.5);
\end{tikzpicture}
\;,\qquad
\begin{tikzpicture}
\atoms{linfunc}{0/}
\draw (0)edge[ar=s]++(180:0.5) (0)edge[ar=s]++(0:0.5);
\end{tikzpicture}
=
\begin{tikzpicture}
\atoms{coalg}{0/, 1/p={0.5,0}}
\draw (0)edge[ar=s]++(-90:0.5) (1)edge[ar=s]++(-90:0.5);
\end{tikzpicture}
\;,\qquad
\begin{tikzpicture}
\atoms{linfunc}{0/}
\draw (0)edge[ar=s]++(180:0.5) (0)edge[ar=s]++(0:0.5) (0)edge[ar=s]++(-90:0.5);
\end{tikzpicture}
=
\begin{tikzpicture}
\atoms{coalg}{0/, 1/p={0.5,0}, 2/p={1,0}}
\draw (0)edge[ar=s]++(-90:0.5) (1)edge[ar=s]++(-90:0.5) (2)edge[ar=s]++(-90:0.5);
\end{tikzpicture}
\;.
\end{equation}
We denote the set of constant, linear, and quadratic Hopf vectors by $F_0[G]$, $F_1[G]$, and $F_2[G]$, respectively.
We call $q$ \emph{normalized} if
\begin{equation}
\label{eq:hopf_vector_normalized}
\begin{tikzpicture}
\atoms{linfunc}{0/}
\end{tikzpicture}
=1\;.
\end{equation}
We denote the set of normalized constant, linear, and quadratic Hopf functions by $\hom_0[G]$, $\hom_1[G]$, and $\hom_2[G]$, respectively.
\end{mydef}
Note that the coefficient group $A$ of an ordinary quadratic function does not turn into a Hopf algebra.
Instead, $A$ directly corresponds to the $\cc^\times$-valued entries of the quadratic Hopf vector.
Next, we should check that Hopf quadratic vectors are indeed a generalization of ordinary quadratic functions.
\begin{myprop}
Consider a quadratic function $q\in F_2[G|\rr\times\rr/\zz]$.
Then $q$ gives rise to a quadratic Hopf vector over the group Hopf algebra of $G$ as follows,
\begin{equation}
\label{eq:hopf_quadratic_function}
\begin{tikzpicture}
\atoms{linfunc,dot}{0/}
\draw (0)edge[ind=$g$,ar=s]++(180:0.5);
\end{tikzpicture}
=
\exp(q(g))\;.
\end{equation}
\end{myprop}
\begin{proof}
Plugging Eq.~\eqref{eq:group_hopf_algebra} into Eq.~\eqref{eq:hopf_derivatives} yields
\begin{equation}
\begin{tikzpicture}
\atoms{linform}{0/}
\end{tikzpicture}
=
\exp(q^{(0)})\;,\qquad
\begin{tikzpicture}
\atoms{linform}{0/}
\draw (0)edge[ind=$g$,ar=s]++(90:0.5);
\end{tikzpicture}
=
\exp(q^{(1)}(g))\;,\qquad
\begin{tikzpicture}
\atoms{linform}{0/}
\draw (0)edge[ind=$g$,ar=s]++(180:0.5) (0)edge[ind=$h$,ar=s]++(0:0.5);
\end{tikzpicture}
=
\exp(q^{(2)}(g,h))\;,\qquad
\begin{tikzpicture}
\atoms{linform}{0/}
\draw (0)edge[ind=$g$,ar=s]++(180:0.5) (0)edge[ind=$h$,ar=s]++(0:0.5) (0)edge[ind=$i$,ar=s]++(-90:0.5);
\end{tikzpicture}
=
\exp(q^{(3)}(g,h,i))\;.
\end{equation}
For a quadratic function $q$ we have $q^{(3)}=0$ and thus
\begin{equation}
\label{eq:hopf_quadratic_definition}
\begin{tikzpicture}
\atoms{linfunc}{0/}
\draw (0)edge[ar=s,ind=$g$]++(180:0.5) (0)edge[ar=s,ind=$h$]++(0:0.5) (0)edge[ar=s,ind=$i$]++(-90:0.5);
\end{tikzpicture}
=\exp(0)\forall g,h,i
=1\forall g \cdot 1\forall h\cdot 1\forall i
=
\begin{tikzpicture}
\atoms{coalg}{0/, 1/p={0.5,0}, 2/p={1,0}}
\draw (0)edge[ar=s,ind=$g$]++(-90:0.5) (1)edge[ar=s,ind=$h$]++(-90:0.5) (2)edge[ar=s,ind=$i$]++(-90:0.5);
\end{tikzpicture}
\;.
\end{equation}
\end{proof}

\myparagraph{Hopf embedding}
Next, we need to also generalize the embedding $\epsilon$, which is a linear function between abelian groups $E$ and $G$, to the case where $E$ and $G$ are Hopf algebras.

\begin{mydef}
A \emph{Hopf function} from a Hopf algebra $E$ to a Hopf algebra $\textcolor{cyan}{G}$ is 2-index tensor $q$ with one $E$-index and one $\textcolor{cyan}{G}$-index,
\begin{equation}
\begin{tikzpicture}
\atoms{square,small,dcross}{0/}
\draw (0)--++(-90:0.5)  (0)edge[cyan,ar=s]++(90:0.5);
\end{tikzpicture}\;,
\end{equation}
such that
\begin{equation}
\begin{tikzpicture}
\atoms{square,small,dcross}{0/}
\atoms{coalg}{c/p={-90:0.5}}
\draw (0)edge[ar=e](c) (c)--++(-45:0.5) (c)--++(-135:0.5) (0)edge[cyan,ar=s]++(90:0.5);
\end{tikzpicture}
=
\begin{tikzpicture}
\atoms{coalg,astyle=cyan}{c/}
\atoms{square,small,dcross}{0/p={-0.4,-0.6}, 1/p={0.4,-0.6}}
\draw[cyan] (c)edge[out=-135,in=90,ar=e](0) (c)edge[out=-45,in=90,ar=e](1) (c)edge[ar=s]++(90:0.5);
\draw (0)--++(-90:0.4) (1)--++(-90:0.4);
\end{tikzpicture}
\;,\qquad
\begin{tikzpicture}
\atoms{square,small,dcross}{0/}
\atoms{coalg}{c/p={-90:0.5}}
\draw (0)edge[ar=e](c) (0)edge[cyan,ar=s]++(90:0.5);
\end{tikzpicture}
=
\begin{tikzpicture}
\atoms{coalg,astyle=cyan}{c/}
\draw[cyan] (c)edge[ar=s]++(90:0.5);
\end{tikzpicture}
\;.
\end{equation}
We define the zeroth, first and second derivative of $q$ as
\begin{equation}
\begin{tikzpicture}
\atoms{linfunc,all}{0/}
\draw (0)--++(-90:0.5);
\end{tikzpicture}
\coloneqq
\begin{tikzpicture}
\atoms{linfunc,dcross}{0/}
\atoms{alg,astyle=cyan}{1/p={0,0.5}}
\draw (0)edge[ar=s,cyan](1) (0)--++(-90:0.5);
\end{tikzpicture}
\;,\qquad
\begin{tikzpicture}
\atoms{square,small,all}{0/}
\draw (0)--++(-90:0.5)  (0)edge[cyan,ar=s]++(90:0.5);
\end{tikzpicture}
\coloneqq
\begin{tikzpicture}
\atoms{alg}{0/}
\atoms{linfunc,all}{1/p={0.4,0.8}}
\atoms{antipode}{anti/p={0.4,0.4}}
\atoms{linfunc,dcross}{2/p={-0.4,0.8}}
\draw (1)edge[ar=e](anti) (anti)edge[ar=e,out=-90,in=45](0) (2)edge[ar=e,out=-90,in=135](0) (0)--++(-90:0.5) (2)edge[cyan,ar=s]++(90:0.5);
\end{tikzpicture}
\;,\qquad
\begin{tikzpicture}
\atoms{square,small,all}{0/}
\draw (0)--++(-90:0.5) (0)edge[cyan,ar=s]++(0:0.5) (0)edge[cyan,ar=s]++(180:0.5);
\end{tikzpicture}
\coloneqq
\begin{tikzpicture}
\atoms{alg,astyle=cyan}{0/}
\atoms{coalg,astyle=cyan}{1/p={-0.7,0}, 2/p={0.7,0}}
\atoms{alg}{3/p={0,-1.5}}
\atoms{linfunc,all}{a/p={0,-0.5}, b/p={-0.7,-0.5}, c/p={0.7,-0.5}}
\atoms{antipode}{x0/p={-0.7,-0.9}, x1/p={0.7,-0.9}}
\draw[cyan] (0)edge[ar=e](a) (1)edge[ar=e](b) (2)edge[ar=e](c) (0)edge[ar=s](1) (0)edge[ar=s](2) (1)edge[ar=s]++(180:0.5) (2)edge[ar=s]++(0:0.5);
\draw (b)edge[ar=e](x0) (c)edge[ar=e](x1);
\draw[rc,ar=e] (a)--++(-90:0.4)--(3);
\draw[rc,ar=e] (x0)--++(-90:0.4)--(3);
\draw[rc,ar=e] (x1)--++(-90:0.4)--(3);
\draw (3)edge[]++(-90:0.5);
\end{tikzpicture}
\;,
\end{equation}
respectively.
In text, we denote the derivatives as $q^{(0)}$, $q^{(1)}$, and $q^{(2)}$, respectively.
We call Hopf functions \emph{constant} and \emph{linear} if $q^{(1)}$ and $q^{(2)}$ are trivial, respectively in the sense,
\begin{equation}
\begin{tikzpicture}
\atoms{square,small,all}{0/}
\draw (0)--++(-90:0.5)  (0)edge[cyan,ar=s]++(90:0.5);
\end{tikzpicture}
\coloneqq
\begin{tikzpicture}
\atoms{coalg,astyle=cyan}{0/}
\atoms{alg}{1/p={0,-0.5}}
\draw[cyan] (0)edge[ar=s]++(90:0.5);
\draw (1)--++(-90:0.5);
\end{tikzpicture}
\;,\qquad
\begin{tikzpicture}
\atoms{square,small,all}{0/}
\draw (0)--++(-90:0.5) (0)edge[cyan,ar=s]++(0:0.5) (0)edge[cyan,ar=s]++(180:0.5);
\end{tikzpicture}
\coloneqq
\begin{tikzpicture}
\atoms{coalg,astyle=cyan}{0/p={-0.3,0}, x/p={0.3,0}}
\atoms{alg}{1/p={0,-0.5}}
\draw[cyan] (0)edge[ar=s]++(90:0.5) (x)edge[ar=s]++(90:0.5);
\draw (1)--++(-90:0.5);
\end{tikzpicture}
\;.
\end{equation}
We will denote the set of constant and linear Hopf functions from $E$ to $G$ as $F_0[E|G]$ and $F_1[E|G]$, respectively.
We call a Hopf function \emph{normalized} if
\begin{equation}
\begin{tikzpicture}
\atoms{linfunc,all}{0/}
\draw (0)--++(-90:0.5);
\end{tikzpicture}
=
\begin{tikzpicture}
\atoms{alg}{0/}
\draw (0)--++(-90:0.5);
\end{tikzpicture}
\;.
\end{equation}
We denote the set of normalized constant and linear Hopf functions by $\hom_0[E|G]$ and $\hom[E|G]$, respectively.
\end{mydef}
Equivalently to the above definition, we can define a linear Hopf function as one that satisfies
\begin{equation}
\begin{tikzpicture}
\atoms{square,small,dcross}{0/}
\atoms{alg,astyle=cyan}{c/p={90:0.5}}
\draw[cyan] (0)edge[ar=s](c) (c)--++(45:0.5) (c)--++(135:0.5);
\draw (0)edge[]++(-90:0.5);
\end{tikzpicture}
=
\begin{tikzpicture}
\atoms{alg}{c/, {zero/p={90:1.1},astyle=cyan}}
\atoms{square,small,dcross}{0/p={-0.6,0.8}, 1/p={0.6,0.8}, 2/p={90:0.7}}
\atoms{antipode}{a/p={90:0.4}}
\draw (c)edge[out=135,in=-90,ar=s](0) (c)edge[out=45,in=-90,ar=s](1) (c)--++(-90:0.5) (c)edge[ar=s](a) (a)edge[ar=s](2);
\draw[cyan] (0)edge[ar=s]++(90:0.4) (1)edge[ar=s]++(90:0.4) (2)edge[ar=s](zero);
\end{tikzpicture}
\;.
\end{equation}
An equivalent definition for a normalized linear Hopf function is
\begin{equation}
\begin{tikzpicture}
\atoms{square,small,dcross}{0/}
\atoms{alg,astyle=cyan}{c/p={90:0.5}}
\draw[cyan] (0)edge[ar=s](c) (c)edge[ar=s]++(45:0.5) (c)edge[ar=s]++(135:0.5);
\draw (0)edge[]++(-90:0.5);
\end{tikzpicture}
=
\begin{tikzpicture}
\atoms{alg}{c/}
\atoms{square,small,dcross}{0/p={-0.4,0.6}, 1/p={0.4,0.6}}
\draw (c)edge[out=135,in=-90,ar=s](0) (c)edge[out=45,in=-90,ar=s](1) (c)--++(-90:0.5);
\draw[cyan] (0)edge[ar=s]++(90:0.4) (1)edge[ar=s]++(90:0.4);
\end{tikzpicture}
\;.
\end{equation}

Again, linear Hopf functions are a generalization of linear functions.
\begin{myprop}
For any linear function $\epsilon\in F_1[E|G]$ between abelian groups $E$ and $G$, we can define an linear Hopf function between the group Hopf algebras of $E$ and $G$, by
\begin{equation}
\label{eq:group_explicit_1linear}
\begin{tikzpicture}
\atoms{square,small,dcross}{0/}
\draw (0)edge[ind=$g$]++(-90:0.5)  (0)edge[cyan,ind=$e$]++(90:0.5);
\end{tikzpicture}
=
\delta_{\epsilon(e)=g}\;.
\end{equation}
\end{myprop}
Note that this is just the ``linearization'' of $\epsilon$, in the same way as the group Hopf algebra is the linearization of a (abelian) group.

\myparagraph{Double integrals}
Quadratic tensors are obtained by combining the quadratic function $q$ with the linear embedding $\epsilon$.
A first naive attempt to combine the tensors corresponding to $\epsilon$ with $q$ could be the following:
\begin{equation}
\begin{tikzpicture}
\atoms{linfunc,astyle=cyan,dot}{0/}
\atoms{small,square,dcross}{1/p={0.8,0}}
\draw (0)edge[cyan,ar=e,ar=s](1) (1)--++(0:0.5);
\draw[red] (0,-0.4)--(1.3,0.4) (0,0.4)--(1.3,-0.4);
\end{tikzpicture}
\;.
\end{equation}
As indicated, this does not work, since we must not contract an ingoing with another ingoing index.
So in order to be able to combine $\epsilon$ and $q$, we need some structure that allows us to flip between ingoing and outgoing indices.
This structure is given by what we call a \emph{double integral} for $\textcolor{cyan}{E}$ and $G$.
\begin{mydef}
For two (co)commutative Hopf algebras $G$ and $\textcolor{cyan}{E}$, a \emph{double integral} is a 2-index tensor
\begin{equation}
\begin{tikzpicture}
\atoms{square,fflat,small,all}{0/}
\draw (0)edge[ar=s]++(-90:0.5) (0)edge[cyan]++(90:0.5);
\end{tikzpicture}\;,
\end{equation}
such that
\begin{equation}
\label{eq:integral_axiom}
\begin{tikzpicture}
\atoms{square,fflat,small,all}{t/p={-0.4,0.6}}
\atoms{coalg}{a/}
\draw (t)edge[out=-90,in=135,ar=s](a) (a)edge[out=45,in=-90]++(0.4,0.6) (a)edge[ar=s]++(-90:0.5) (t)edge[cyan]++(90:0.5);
\end{tikzpicture}
=
\begin{tikzpicture}
\atoms{square,fflat,small,all}{0/}
\atoms{alg}{1/p={0.5,0}}
\draw (0)edge[ar=s]++(-90:0.5) (0)edge[cyan]++(90:0.5) (1)--++(90:0.5);
\end{tikzpicture}
\;,\qquad
\begin{tikzpicture}
\atoms{square,fflat,small,all}{t/p={-0.4,0.6}}
\atoms{alg,astyle=cyan}{a/}
\draw (t)edge[cyan,out=-90,in=135,ar=e](a) (a)edge[cyan,out=45,in=-90,ar=s]++(0.4,0.6) (a)edge[cyan]++(-90:0.5) (t)edge[ar=s]++(90:0.5);
\end{tikzpicture}
=
\begin{tikzpicture}
\atoms{square,fflat,small,all}{0/}
\atoms{coalg,astyle=cyan}{1/p={0.5,0}}
\draw (0)edge[cyan]++(-90:0.5) (0)edge[ar=s]++(90:0.5) (1)edge[cyan,ar=s]++(90:0.5);
\end{tikzpicture}
\;.
\end{equation}
\end{mydef}
Note that Eq.~\eqref{eq:integral_axiom} is linear in the double integral, and thus the double integrals form a linear subspace.
The double integral can be combined with the comultiplication of $\textcolor{cyan}{E}$ and the multiplication of $G$ in order to effectively invert the direction in which the linear Hopf function $\epsilon$ acts.

\myparagraph{Hopf quadratic tensors}
After introducing quadratic Hopf vectors, linear Hopf functions, and double integrals, we are ready to define Hopf quadratic tensors.
\begin{mydef}
A \emph{Hopf quadratic tensor data} over a Hopf algebra $G$ consists of
\begin{itemize}
\item a Hopf algebra $\textcolor{cyan}{E}$ with and a double integral $\begin{tikzpicture}\atoms{square,rot=90,fflat,small,all}{0/}\draw (0)edge[cyan]++(180:0.4) (0)edge[ar=s]++(0:0.4);\end{tikzpicture}$ for $\textcolor{cyan}{E}$ and $G$,
\item a quadratic Hopf vector $\begin{tikzpicture}\atoms{linfunc,dot,astyle=cyan}{0/}\draw[cyan] (0)--++(0:0.4);\end{tikzpicture}=q\in F_2[\textcolor{cyan}{E}]$,
\item a linear Hopf function $\begin{tikzpicture}\atoms{linfunc,dcross}{0/}\draw (0)edge[ar=s,cyan]++(180:0.4) (0)--++(0:0.4);\end{tikzpicture}=\epsilon\in F_1[\textcolor{cyan}{E}|G]$.
\end{itemize}
The \emph{associated Hopf quadratic tensor} of a Hopf quadratic tensor data is given by
\begin{equation}
\label{eq:hopf_quadratic_tensor_definition}
\begin{tikzpicture}
\atoms{square,rot=90,fflat,small,all}{0/}
\atoms{coalg,astyle=cyan}{c/p={-0.5,-0.5}}
\atoms{alg}{a/p={0.5,-0.5}}
\atoms{linfunc,astyle=cyan,dot}{q/p={-1,-0.5}}
\atoms{small,square,dcross}{e/p={0,-0.5}}
\atoms{rot=90,antipode}{anti/p={1,-0.5}}
\draw[cyan] (q)edge[ar=s](c) (c)edge[ar=e](e) (c)edge[cyan,ar=s,out=90,in=180](0.north);
\draw (e)edge[ar=e](a) (a)edge[ar=s](anti) (anti)edge[ar=s]++(0:0.5) (0.south)edge[out=0,in=90,ar=s](a);
\end{tikzpicture}
\;.
\end{equation}
\end{mydef}

Finally, we need to show that Hopf quadratic tensors are actually a generalization of ordinary quadratic tensors.
\begin{myprop}
Consider a (non-Hopf) quadratic tensor data $(E,\epsilon,q)$ over an abelian group $G$ as defined in Section~\ref{sec:clifford_definition}.
Then the associated quadratic tensor is the same as the associated quadratic tensor in Eq.~\eqref{eq:hopf_quadratic_tensor_definition} for the corresponding group Hopf algebras.
\end{myprop}
\begin{proof}
We start by studying the double integrals when $\textcolor{cyan}{E}$ and $G$ are (abelian) group Hopf algebras.
In this case, Eq.~\eqref{eq:integral_axiom} yields
\begin{equation}
\label{eq:group_double_integral}
\begin{tikzpicture}
\atoms{square,fflat,small,all}{0/}
\draw (0)edge[ind=$g$,ar=s]++(-90:0.4) (0)edge[cyan,ind=$e$]++(90:0.4);
\end{tikzpicture}
=
\begin{tikzpicture}
\atoms{square,fflat,small,all}{0/}
\draw (0)edge[ind=$g$,ar=s]++(-90:0.4) (0)edge[cyan,ind=$e$]++(90:0.4);
\end{tikzpicture}
\delta_{g=0}\;,\qquad
\begin{tikzpicture}
\atoms{square,fflat,small,all}{0/}
\draw (0)edge[ind=$g$,ar=s]++(-90:0.4) (0)edge[cyan,ind=$e-e'$]++(90:0.4);
\end{tikzpicture}
=
\begin{tikzpicture}
\atoms{square,fflat,small,all}{0/}
\draw (0)edge[ind=$g$,ar=s]++(-90:0.4) (0)edge[cyan,ind=$e$]++(90:0.4);
\end{tikzpicture}
\forall e'
\qquad\Rightarrow\quad
\begin{tikzpicture}
\atoms{square,fflat,small,all}{0/}
\draw (0)edge[ind=$g$,ar=s]++(-90:0.4) (0)edge[cyan,ind=$e$]++(90:0.4);
\end{tikzpicture}
\propto
\delta_{g=0}\forall \textcolor{cyan}{e}\in \textcolor{cyan}{E}\;.
\end{equation}
Now plugging Eqs.~\eqref{eq:group_double_integral}, \eqref{eq:group_explicit_1linear}, \eqref{eq:hopf_quadratic_function}, and \eqref{eq:group_hopf_algebra} into Eq.~\eqref{eq:hopf_quadratic_tensor_definition} indeed yields
\begin{equation}
\begin{tikzpicture}
\atoms{square,rot=90,fflat,small,all}{0/}
\atoms{coalg,astyle=cyan}{c/p={-0.5,-0.5}}
\atoms{alg}{a/p={0.5,-0.5}}
\atoms{linfunc,astyle=cyan,dot}{q/p={-1,-0.5}}
\atoms{small,square,dcross}{e/p={0,-0.5}}
\atoms{rot=90,antipode}{anti/p={1,-0.5}}
\draw[cyan] (q)edge[ar=s](c) (c)edge[ar=e](e) (c)edge[cyan,ar=s,out=90,in=180](0.north);
\draw (e)edge[ar=e](a) (a)edge[ar=s](anti) (anti)edge[ar=s,ind=$g$]++(0:0.5) (0.south)edge[out=0,in=90,ar=s](a);
\end{tikzpicture}
=
\sum_{\textcolor{cyan}{e\in E}}
\quad
\begin{tikzpicture}
\atoms{coalg,astyle=cyan}{c/p={-0.5,-0.5}}
\atoms{alg}{a/p={0.5,-0.5}}
\atoms{linfunc,astyle=cyan,dot}{q/p={-1,-0.5}}
\atoms{small,square,dcross}{e/p={0,-0.5}}
\atoms{rot=90,antipode}{anti/p={1,-0.5}}
\draw[cyan] (q)edge[ar=s](c) (c)edge[ar=e](e) (c)edge[cyan,ar=s,ind=$e$]++(90:0.5);
\draw (e)edge[ar=e](a) (a)edge[ar=s](anti) (anti)edge[ar=s,ind=$g$]++(0:0.5) (a)edge[ind=$0$]++(90:0.5);
\end{tikzpicture}
=
\sum_{e\in E} \delta_{\epsilon(e)=g} \exp(q(e))\;.
\end{equation}
\end{proof}

\subsection{Factor decomposition of Hopf vectors and functions}
\label{sec:diagrammatic_decomposition}
In this section, we argue on a purely diagrammatic level that quadratic Hopf vectors on a tensor product of $n$ Hopf algebras can be decomposed into $O(n^2)$ smaller components in analogy with Section~\ref{sec:factor_decomposition}.
The \emph{tensor product} $G\otimes \textcolor{cyan}{H}$ of two Hopf algebras $G$ and $\textcolor{cyan}{H}$ is just the tensor product of the underlying tensors, and defines again a Hopf algebra.
Diagrammatically, $G\otimes \textcolor{cyan}{H}$ can be represented as drawing the tensors of $G$ and $\textcolor{cyan}{H}$ next to another.
In the following, we will draw indices and tensors of $G$ in black and these of $\textcolor{cyan}{H}$ in bright blue:
\begin{equation}
\begin{tikzpicture}
\atoms{alg}{0/lab={t=$\otimes$,p={-150:0.3}}}
\draw (0)edge[ind=$aa'$,ar=s]++(45:0.6) (0)edge[ind=$bb'$,ar=s]++(135:0.6) (0)edge[ind=$cc'$]++(-90:0.6);
\end{tikzpicture}
=
\begin{tikzpicture}
\atoms{alg}{0/, {1/p={0.4,0},astyle=cyan}}
\draw (0)edge[ind=$a$,ar=s]++(45:0.6) (0)edge[ind=$b$,ar=s]++(135:0.6) (0)edge[ind=$c$]++(-90:0.6);
\draw[cyan] (1)edge[ind=$a'$,ar=s]++(45:0.6) (1)edge[ind=$b'$,ar=s]++(135:0.6) (1)edge[ind=$c'$]++(-90:0.6);
\end{tikzpicture}
\;,\qquad
\begin{tikzpicture}
\atoms{antipode}{0/lab={t=$\otimes$,p=180:0.3}}
\draw (0)edge[ind=$aa'$,ar=s]++(90:0.6) (0)edge[ind=$bb'$]++(-90:0.6);
\end{tikzpicture}
=
\begin{tikzpicture}
\atoms{antipode}{0/, {1/p={0.4,0},astyle=cyan}}
\draw (0)edge[ind=$a$,ar=s]++(90:0.6) (0)edge[ind=$b$]++(-90:0.6);
\draw[cyan] (1)edge[ind=$a'$,ar=s]++(90:0.6) (1)edge[ind=$b'$]++(-90:0.6);
\end{tikzpicture}
\;,\qquad
\begin{tikzpicture}
\atoms{alg}{0/lab={t=$\otimes$,p={90:0.3}}}
\draw (0)edge[ind=$aa'$]++(-90:0.6);
\end{tikzpicture}
=
\begin{tikzpicture}
\atoms{alg}{0/, {1/p={0.4,0},astyle=cyan}}
\draw (0)edge[ind=$a$]++(-90:0.6);
\draw[cyan] (1)edge[ind=$a'$]++(-90:0.6);
\end{tikzpicture}\;.
\end{equation}
Note that formally, the tensor product requires index blocking.
Further note that the tensor product involves a non-trivial reordering sign of $(-1)^{|a||b'|}$ as can be seen from the line crossing above.

The components of the decomposition will be bilinear forms, as we define in the following.
\begin{mydef}
A \emph{Hopf bilinear form} for two Hopf algebras $G$ and $\textcolor{cyan}{H}$ is a 2-index tensor
\begin{equation}
\begin{tikzpicture}
\atoms{linform}{0/}
\draw (0)edge[ind=$a$,ar=s]++(180:0.5) (0)edge[cyan,ind=$b$,ar=s]++(0:0.5);
\end{tikzpicture}
\end{equation}
fulfilling
\begin{equation}
\label{eq:hopf_bilinear_form}
\begin{tikzpicture}
\atoms{alg,astyle=cyan}{0/}
\atoms{linform}{1/p={-90:0.6}}
\draw (0)edge[ar=e,cyan](1) (0)edge[ind=$b$,ar=s,cyan]++(45:0.4) (0)edge[ind=$a$,ar=s,cyan]++(135:0.4) (1)edge[ind=$c$,ar=s]++(-90:0.4);
\end{tikzpicture}
=
\begin{tikzpicture}
\atoms{coalg}{0/}
\atoms{linform}{1/p={0.4,0.6}, 2/p={-0.4,0.6}}
\draw[rc,ar=s] (1)--++(-90:0.3)--(0);
\draw[rc,ar=s] (2)--++(-90:0.3)--(0);
\draw (1)edge[ind=$b$,ar=s,cyan]++(90:0.4) (2)edge[ind=$a$,ar=s,cyan]++(90:0.4) (0)edge[ind=$c$,ar=s]++(-90:0.4);
\end{tikzpicture}
\;,\qquad
\begin{tikzpicture}
\atoms{alg}{0/}
\atoms{linform}{1/p={-90:0.6}}
\draw (0)edge[ar=e](1) (0)edge[ind=$b$,ar=s]++(45:0.4) (0)edge[ind=$a$,ar=s]++(135:0.4) (1)edge[ind=$c$,ar=s,cyan]++(-90:0.4);
\end{tikzpicture}
=
\begin{tikzpicture}
\atoms{coalg,astyle=cyan}{0/}
\atoms{linform}{1/p={0.4,0.6}, 2/p={-0.4,0.6}}
\draw[rc,ar=s,cyan] (1)--++(-90:0.3)--(0);
\draw[rc,ar=s,cyan] (2)--++(-90:0.3)--(0);
\draw (1)edge[ind=$b$,ar=s]++(90:0.4) (2)edge[ind=$a$,ar=s]++(90:0.4) (0)edge[ind=$c$,ar=s,cyan]++(-90:0.4);
\end{tikzpicture}
\;,\qquad
\begin{tikzpicture}
\atoms{alg,astyle=cyan}{0/}
\atoms{linform}{1/p={-90:0.6}}
\draw (0)edge[ar=e,cyan](1) (1)edge[ind=$a$,ar=s]++(-90:0.4);
\end{tikzpicture}
=
\begin{tikzpicture}
\atoms{coalg}{0/}
\draw (0)edge[ind=$a$,ar=s]++(-90:0.5);
\end{tikzpicture}
\;,\qquad
\begin{tikzpicture}
\atoms{alg}{0/}
\atoms{linform}{1/p={-90:0.6}}
\draw (0)edge[ar=e](1) (1)edge[ind=$a$,ar=s,cyan]++(-90:0.4);
\end{tikzpicture}
=
\begin{tikzpicture}
\atoms{coalg,astyle=cyan}{0/}
\draw (0)edge[ind=$a$,ar=s,cyan]++(-90:0.5);
\end{tikzpicture}
\;,\qquad
\begin{tikzpicture}
\atoms{antipode,astyle=cyan}{0/}
\atoms{linform}{1/p={-90:0.6}}
\draw (0)edge[ar=e,cyan](1) (1)edge[ind=$b$,ar=s]++(-90:0.4) (0)edge[ind=$a$,ar=s,cyan]++(90:0.4);
\end{tikzpicture}
=
\begin{tikzpicture}
\atoms{antipode}{0/}
\atoms{linform}{1/p={90:0.6}}
\draw (0)edge[ar=e](1) (0)edge[ind=$b$,ar=s]++(-90:0.4) (1)edge[ind=$a$,ar=s,cyan]++(90:0.4);
\end{tikzpicture}
\;.
\end{equation}
We denote the set of Hopf bilinear forms as $\hom^2[G,\textcolor{cyan}{H}]$.
\end{mydef}

Note that the notion of a bilinear can be used to give an equivalent definition of a quadratic Hopf vector $q$:
$q$ is quadratic if $q^{(2)}$ is a Hopf bilinear form in $\hom^2[G,G]$.
The bilinearity axioms in Eq.~\eqref{eq:hopf_bilinear_form} can be derived from Eq.~\eqref{eq:hopf_quadratic_definition} by bringing one of the 2-index tensors to the other side via its inverse, see Eq.~\eqref{eq:hopf_bilinear_group}.
We can also use the inverse property to bring the second derivative Eq.~\eqref{eq:hopf_derivatives} into the form
\begin{equation}
\begin{tikzpicture}
\atoms{alg}{0/}
\atoms{linfunc}{1/p={-90:0.6}}
\draw (0)edge[ar=e](1) (0)edge[ind=$h$,ar=s]++(0:0.5) (0)edge[ind=$g$,ar=s]++(180:0.5);
\end{tikzpicture}
=
\begin{tikzpicture}
\atoms{coalg}{0/, 1/p={1,0}}
\atoms{linfunc}{2/p={0,-0.5}, 3/p={1,-0.5}}
\atoms{linform}{5/p={0.5,0}}
\draw (0)edge[ar=e](5) (1)edge[ar=e](5) (0)edge[ar=e](2) (1)edge[ar=e](3) (1)edge[ind=$h$,ar=s]++(0:0.5) (0)edge[ind=$g$,ar=s]++(180:0.5);
\end{tikzpicture}
\;.
\end{equation}

\begin{myprop}
\label{prop:hopf_quadratic_decomposition}
Let $H$ be a tensor product $H=\otimes_{0\leq i<n} H_i$ of $n$ Hopf algebras.
Then there is a bijection between (1) quadratic Hopf vectors $q\in F_2[H]$ and (2) collections of
\begin{itemize}
\item a number $q^{(0)}\in \cc$,
\item for each $0\leq i<n$, a normalized quadratic Hopf vector $q^{(1)}_i\in \hom_2[H_i]$,
\item for each $0\leq i<j<n$, a Hopf bilinear form $q^{(2)}_{ij}\in \hom^2[H_i,H_j]$.
\end{itemize}
The collection is obtained from $q$ as follows.
$q^{(0)}$ is simply $q^{(0)}$.
$q_i^{(1)}$ is given by contracting all indices of the tensor $q^{(1)}$ with the unit except for the $H_i$ index.
For example, for $n=4$ and $H=H_0\otimes \textcolor{cyan}{H_1}\otimes \textcolor{red}{H_2}\otimes \textcolor{olive}{H_3}$, $q_2^{(1)}$ is given by
\begin{equation}
\begin{tikzpicture}
\atoms{linfunc,astyle=red}{0/}
\draw[red] (0)edge[ar=s]++(90:0.5);
\end{tikzpicture}
\coloneqq
\begin{tikzpicture}
\atoms{square,small,xscale=6}{0/}
\atoms{alg}{0x/p={-0.6,0.6}, {1x/p={-0.2,0.6},astyle=cyan}, {3x/p={0.6,0.6},astyle=olive}}
\draw ([sx=-0.6]0.north)edge[ar=s](0x) ([sx=-0.2]0.north)edge[ar=s,cyan](1x) ([sx=0.2]0.north)edge[ar=s,red]++(90:0.5) ([sx=0.6]0.north)edge[ar=s,olive](3x);
\end{tikzpicture}
\;.
\end{equation}
$q_{ij}^{(2)}$ is given by contracting all components of the first index block of $q^{(2)}$ except the $H_i$ component, and all components of the second index block except the $H_j$ component, with the unit.
For example, $q_{13}^{(2)}$ is given by
\begin{equation}
\begin{tikzpicture}
\atoms{linfunc}{0/}
\draw (0)edge[ar=s,olive]++(90:0.5) (0)edge[ar=s,cyan]++(-90:0.5);
\end{tikzpicture}
\coloneqq
\begin{tikzpicture}
\atoms{square,small,xscale=6}{0/}
\atoms{alg}{0x/p={-0.6,0.6}, {1x/p={-0.2,0.6},astyle=cyan}, {2x/p={0.2,0.6},astyle=red}}
\atoms{alg}{0y/p={-0.6,-0.6}, {2y/p={0.2,-0.6},astyle=red}, {3y/p={0.6,-0.6},astyle=olive}}
\draw ([sx=-0.6]0.north)edge[ar=s](0x) ([sx=-0.2]0.north)edge[ar=s,cyan](1x) ([sx=0.2]0.north)edge[ar=s,red](2x) ([sx=0.6]0.north)edge[ar=s,olive]++(90:0.5);
\draw ([sx=-0.6]0.south)edge[ar=s](0y) ([sx=-0.2]0.south)edge[ar=s,cyan]++(-90:0.5) ([sx=0.2]0.south)edge[ar=s,red](2y) ([sx=0.6]0.south)edge[ar=s,olive](3y);
\end{tikzpicture}
\;.
\end{equation}
Vice versa, $q$ is obtained form the collection by contracting the following tensor-network diagram:
\begin{itemize}
\item For each $0\leq i<n$, there is a $H_i$ coalgebra tensor with with $n$ outgoing indices (see Eq.~\eqref{eq:algeb_coalg_more_indices}), and whose ingoing index is the $i$th index of $q$.
\item For each pair $0\leq i<j<n$ there is one copy of the 2-index tensor $q^{(2)}_{ij}$, whose indices are contracted with an outgoing index of the $H_i$ and the $H_j$ coalgebra tensor, respectively.
\item For each $i$, there is one copy of the 1-index tensor $q^{(1)}_i$, whose index is contracted with an outgoing index of the $H_i$ coalgebra tensor.
\item The whole tensor network is multiplied with the scalar $q^{(0)}$.
\end{itemize}
For example, for $n=4$ we get
\begin{equation}
\label{eq:diagrammatic_decomposition_n4}
\begin{tikzpicture}
\atoms{square,small,xscale=6,dec={xscale=1/6.}{dot}}{0/}
\draw ([sx=-0.6]0.north)edge[ar=s]++(90:0.5) ([sx=-0.2]0.north)edge[ar=s,cyan]++(90:0.5) ([sx=0.2]0.north)edge[ar=s,red]++(90:0.5) ([sx=0.6]0.north)edge[ar=s,olive]++(90:0.5);
\end{tikzpicture}
=
\begin{tikzpicture}
\atoms{coalg}{0/, {1/p={1.5,0},astyle=cyan}, {2/p={0,1.5},astyle=red}, {3/p={1.5,1.5},astyle=olive}}
\atoms{linfunc,rot=45}{x0/p={$(0)+(-45:0.6)$}, {x1/p={$(1)+(-135:0.6)$},astyle=cyan}, {x2/p={$(2)+(45:0.6)$},astyle=red}, {x3/p={$(3)+(135:0.6)$},astyle=olive}}
\atoms{linform}{x01/p={0.75,0}, x02/p={0,0.75}, x13/p={1.5,0.75}, x23/p={0.75,1.5}, {x03/p={0.5,0.5},rot=45}, {x12/p={0.5,1},rot=45}}
\draw (0)edge[ar=e](x0) (1)edge[ar=e,cyan](x1) (2)edge[ar=e,red](x2) (3)edge[ar=e,olive](x3);
\draw (0)edge[ar=e](x01) (0)edge[ar=e](x02) (0)edge[ar=e](x03);
\draw[cyan] (1)edge[ar=e](x01) (1)edge[ar=e](x12) (1)edge[ar=e](x13);
\draw[red] (2)edge[ar=e](x02) (2)edge[ar=e](x12) (2)edge[ar=e](x23);
\draw[olive] (3)edge[ar=e](x03) (3)edge[ar=e](x13) (3)edge[ar=e](x23);
\draw (0)edge[ar=s]++(-135:0.5) (1)edge[ar=s,cyan]++(-45:0.5) (2)edge[ar=s,red]++(135:0.5) (3)edge[ar=s,olive]++(45:0.5);
\atoms{linform}{x/p={2.2,0.75}}
\end{tikzpicture}
\;.
\end{equation}
\end{myprop}
\begin{proof}
We start by showing the proposition for the case $n=2$, for a quadratic Hopf vector $q\in F_2[G\otimes \textcolor{cyan}{H}]$ and its associated Hopf bilinear form in $q^{(2)}$,
\begin{equation}
\label{eq:diagrammatic_decomposition_2factors}
\begin{tikzpicture}
\atoms{square,small,xscale=2.5,dec={xscale=1/2.5}{dot}}{0/}
\draw ([sx=-0.2]0.north)edge[ar=s]++(90:0.5) ([sx=0.2]0.north)edge[ar=s,cyan]++(90:0.5);
\end{tikzpicture}
\;,\qquad
\begin{tikzpicture}
\atoms{square,small,xscale=2.5}{0/}
\draw ([sx=-0.2]0.south)edge[ar=s]++(-90:0.5) ([sx=0.2]0.south)edge[ar=s,cyan]++(-90:0.5) ([sx=-0.2]0.north)edge[ar=s]++(90:0.5) ([sx=0.2]0.north)edge[ar=s,cyan]++(90:0.5);
\end{tikzpicture}
\;.
\end{equation}
The quadratic Hopf vector and its bilinear can indeed be decomposed,
\begin{equation}
\label{eq:2linearfunc_splitting}
\begin{tikzpicture}
\atoms{square,small,xscale=2.5,dec={xscale=1/2.5}{dot}}{0/}
\draw ([sx=-0.2]0.north)edge[ar=s]++(90:0.5) ([sx=0.2]0.north)edge[ar=s,cyan]++(90:0.5);
\end{tikzpicture}
=
\begin{tikzpicture}
\atoms{coalg}{2/, {3/p={1.2,0},astyle=cyan}}
\atoms{linform}{a/p={0,-0.6}, {b/p={1.2,-0.6},astyle=cyan}, c/p={0.6,0}}
\draw (2)edge[ar=e](a) (2)edge[ar=e](c) (2)edge[ar=s]++(90:0.4);
\draw[cyan] (3)edge[ar=e](c) (3)edge[ar=e](b) (3)edge[ar=s]++(90:0.4);
\atoms{linform}{x/p={1.8,0}}
\end{tikzpicture}
\;,\qquad
\begin{tikzpicture}
\atoms{square,small,xscale=2.5}{0/}
\draw ([sx=-0.2]0.south)edge[ar=s]++(-90:0.5) ([sx=0.2]0.south)edge[ar=s,cyan]++(-90:0.5) ([sx=-0.2]0.north)edge[ar=s]++(90:0.5) ([sx=0.2]0.north)edge[ar=s,cyan]++(90:0.5);
\end{tikzpicture}
=
\begin{tikzpicture}
\atoms{coalg}{0/, {1/p={1.2,0},astyle=cyan}, 2/p={0,1.2}, {3/p={1.2,1.2},astyle=cyan}}
\atoms{linform}{a/p={0,0.6}, {b/p={1.2,0.6},astyle=cyan}, c/p={0.4,0.6}, d/p={0.8,0.6}}
\draw (0)edge[ar=e](a) (2)edge[ar=e](a) (0)edge[ar=s]++(-90:0.4) (2)edge[ar=s]++(90:0.4);
\draw[cyan] (1)edge[ar=e](b) (3)edge[ar=e](b) (1)edge[ar=s]++(-90:0.4) (3)edge[ar=s]++(90:0.4);
\draw[ar=s,rc] (c)--++(-90:0.3)--(0);
\draw[rc,ar=s,cyan] (d)--++(-90:0.3)--(1);
\draw[rc,ar=s,cyan] (c)--++(90:0.3)--(3);
\draw[rc,ar=s] (d)--++(90:0.3)--(2);
\end{tikzpicture}
\;,
\end{equation}
where
\begin{equation}
\label{eq:diagrammatic_decomposition_n2}
\begin{tikzpicture}
\atoms{linform}{0/}
\end{tikzpicture}
\coloneqq
\begin{tikzpicture}
\atoms{square,small,xscale=2.5}{0/}
\atoms{alg}{{a/p={0.2,0.5},astyle=cyan},b/p={-0.2,0.5}}
\draw ([sx=-0.2]0.north)edge[ar=s](b) ([sx=0.2]0.north)edge[ar=s,cyan](a);
\end{tikzpicture}
\;,\qquad
\begin{tikzpicture}
\atoms{linform}{0/}
\draw (0)edge[ar=s]++(90:0.5);
\end{tikzpicture}
\coloneqq
\begin{tikzpicture}
\atoms{square,small,xscale=2.5}{0/}
\atoms{alg,astyle=cyan}{a/p={0.2,0.5}}
\draw ([sx=-0.2]0.north)edge[ar=s]++(90:0.5) ([sx=0.2]0.north)edge[ar=s,cyan](a);
\end{tikzpicture}
\;,\qquad
\begin{tikzpicture}
\atoms{linform,astyle=cyan}{0/}
\draw[cyan] (0)edge[ar=s]++(90:0.5);
\end{tikzpicture}
\coloneqq
\begin{tikzpicture}
\atoms{square,small,xscale=2.5}{0/}
\atoms{alg}{a/p={-0.2,0.5}}
\draw ([sx=-0.2]0.north)edge[ar=s](a) ([sx=0.2]0.north)edge[ar=s,cyan]++(90:0.5);
\end{tikzpicture}
\;,\qquad
\begin{tikzpicture}
\atoms{linform}{0/}
\draw (0)edge[cyan,ar=s]++(90:0.5) (0)edge[ar=s]++(-90:0.5);
\end{tikzpicture}
\coloneqq
\begin{tikzpicture}
\atoms{square,small,xscale=2.5}{0/}
\atoms{alg}{a/p={-0.2,0.5}, b/p={0.2,-0.5}}
\draw ([sx=-0.2]0.south)edge[ar=s]++(-90:0.5) ([sx=0.2]0.south)edge[ar=s,cyan](b) ([sx=-0.2]0.north)edge[ar=s](a) ([sx=0.2]0.north)edge[ar=s,cyan]++(90:0.5);
\end{tikzpicture}
\;,\qquad
\begin{tikzpicture}
\atoms{linform}{0/}
\draw (0)edge[ar=s]++(90:0.5) (0)edge[ar=s]++(-90:0.5);
\end{tikzpicture}
\coloneqq
\begin{tikzpicture}
\atoms{square,small,xscale=2.5}{0/}
\atoms{alg,astyle=cyan}{a/p={0.2,0.5}, b/p={0.2,-0.5}}
\draw ([sx=-0.2]0.south)edge[ar=s]++(-90:0.5) ([sx=0.2]0.south)edge[ar=s,cyan](b) ([sx=-0.2]0.north)edge[ar=s]++(90:0.5) ([sx=0.2]0.north)edge[ar=s,cyan](a);
\end{tikzpicture}
\;,\qquad
\begin{tikzpicture}
\atoms{linform,astyle=cyan}{0/}
\draw[cyan] (0)edge[ar=s]++(90:0.5) (0)edge[ar=s]++(-90:0.5);
\end{tikzpicture}
\coloneqq
\begin{tikzpicture}
\atoms{square,small,xscale=2.5}{0/}
\atoms{alg}{a/p={-0.2,0.5}, b/p={-0.2,-0.5}}
\draw ([sx=-0.2]0.south)edge[ar=s](b) ([sx=0.2]0.south)edge[ar=s,cyan]++(-90:0.5) ([sx=-0.2]0.north)edge[ar=s](a) ([sx=0.2]0.north)edge[ar=s,cyan]++(90:0.5);
\end{tikzpicture}
\;.
\end{equation}
This can be derived purely by applying the diagrammatic axioms.
After showing the proposition for $n=2$, we proceed inductively.
For the induction to work, we further need that a Hopf bilinear form in $\hom^2[G,\textcolor{red}{H}\otimes \textcolor{cyan}{I}]$,
\begin{equation}
\begin{tikzpicture}
\atoms{square,small,xscale=2.5}{0/}
\draw (0.south)edge[ar=s]++(-90:0.5) ([sx=-0.2]0.north)edge[ar=s,red]++(90:0.5) ([sx=0.2]0.north)edge[ar=s,cyan]++(90:0.5);
\end{tikzpicture}
\;,
\end{equation}
can be decomposed as
\begin{equation}
\label{eq:hopf_intertwiner_decomposition}
\begin{tikzpicture}
\atoms{square,small,xscale=2.5}{0/}
\draw (0.south)edge[ar=s]++(-90:0.5) ([sx=-0.2]0.north)edge[ar=s,red]++(90:0.5) ([sx=0.2]0.north)edge[ar=s,cyan]++(90:0.5);
\end{tikzpicture}
=
\begin{tikzpicture}
\atoms{coalg}{0/}
\atoms{linform}{a/p={-0.3,0.6}, b/p={0.3,0.6}}
\draw (0)edge[ar=s]++(-90:0.4);
\draw[rc,ar=s] (a)--++(-90:0.3)--(0);
\draw[rc,ar=s] (b)--++(-90:0.3)--(0);
\draw[red,ar=s] (a)--++(90:0.4);
\draw[cyan,ar=s] (b)--++(90:0.4);
\end{tikzpicture}
\;,
\end{equation}
where
\begin{equation}
\begin{tikzpicture}
\atoms{linform}{0/}
\draw (0)edge[ar=s]++(-90:0.5) (0)edge[ar=s,red]++(90:0.5);
\end{tikzpicture}
\coloneqq
\begin{tikzpicture}
\atoms{square,small,xscale=2.5}{0/}
\atoms{alg,astyle=cyan}{a/p={0.2,0.5}}
\draw (0.south)edge[ar=s]++(-90:0.5) ([sx=-0.2]0.north)edge[ar=s,red]++(90:0.5) ([sx=0.2]0.north)edge[ar=s,cyan](a);
\end{tikzpicture}
\;,\qquad
\begin{tikzpicture}
\atoms{linform}{0/}
\draw (0)edge[ar=s]++(-90:0.5) (0)edge[ar=s,cyan]++(90:0.5);
\end{tikzpicture}
\coloneqq
\begin{tikzpicture}
\atoms{square,small,xscale=2.5}{0/}
\atoms{alg,astyle=red}{a/p={-0.2,0.5}}
\draw (0.south)edge[ar=s]++(-90:0.5) ([sx=-0.2]0.north)edge[ar=s,red](a) ([sx=0.2]0.north)edge[ar=s,cyan]++(90:0.5);
\end{tikzpicture}
\end{equation}
are Hopf bilinear forms in $\hom^2[G,\textcolor{red}{H}]$ and $\hom^2[G,\textcolor{cyan}{I}]$, respectively.
By applying Eqs.~\eqref{eq:hopf_intertwiner_decomposition} and \eqref{eq:2linearfunc_splitting} recursively, we arrive at the decomposition such as in Eq.~\eqref{eq:diagrammatic_decomposition_n4}.
\end{proof}

\myparagraph{Decomposition of linear Hopf functions}
The definition of a Hopf quadratic tensor data also includes a linear Hopf function $\epsilon$.
We can also decompose $\epsilon$ into smaller components.
One of these components is a ``Hopf constant'', which we define as follows.
\begin{mydef}
A \emph{Hopf constant} in a Hopf algebra $G$ is a vector
\begin{equation}
\begin{tikzpicture}
\atoms{square,small,all}{0/}
\draw (0)--++(-90:0.5);
\end{tikzpicture}\;,
\end{equation}
which is ``copied'' by the coalgebra,
\begin{equation}
\begin{tikzpicture}
\atoms{square,small,all}{0/}
\atoms{coalg}{c/p={-90:0.5}}
\draw (0)edge[ar=s](c) (c)--++(-45:0.5) (c)--++(-135:0.5);
\end{tikzpicture}
=
\begin{tikzpicture}
\atoms{square,small,all}{0/, 1/p={0.5,0}}
\draw (0)--++(-90:0.5) (1)--++(-90:0.5);
\end{tikzpicture}
\;.
\end{equation}
We denote the set of Hopf constants as $\hom_0[|G]$.
\end{mydef}

With this, a linear Hopf function can be decomposed as follows.
\begin{myprop}
Let $E$ be a product of $m$ Hopf algebras $E_j$, and $G$ be a product of $n$ Hopf algebras $G_i$.
Then there is a bijection between (1) linear Hopf functions $\epsilon\in F_1[E|G]$ and (2) collections of
\begin{itemize}
\item for every $0\leq i<n$, a Hopf constant $\epsilon^{(0)}_i\in \hom_0[|G_i]$,
\item for every $0\leq i<n$ and $0\leq j<m$, a normalized linear Hopf function $\epsilon^{(1)}_{ij}\in \hom[E_j|G_i]$.
\end{itemize}
The collection is obtained from $\epsilon$ as follows.
$\epsilon^{(0)}_i$ is given by contracting all $G$ indices of $\epsilon^{(0)}$ with the respective counit except for the $G_i$ index which remains open.
For example, for $m=2$ and $n=3$ with $E=E_0\otimes \textcolor{cyan}{E_1}$, and $G=G_0\otimes \textcolor{cyan}{G_1}\otimes \textcolor{red}{G_2}$, $\epsilon^{(0)}_1$ is given by
\begin{equation}
\begin{tikzpicture}
\atoms{linfunc,all,astyle=cyan}{0/}
\draw[cyan] (0)--++(90:0.5);
\end{tikzpicture}
\coloneqq
\begin{tikzpicture}
\atoms{square,small,all,xscale=4}{0/}
\atoms{coalg}{0y/p={-0.4,0.5}, {2y/p={0.4,0.5},astyle=red}}
\draw ([sx=-0.4]0-t)edge[ar=e](0y) ([sx=0]0-t)edge[cyan]++(90:0.5) ([sx=0.4]0-t)edge[red,ar=e](2y);
\end{tikzpicture}
\;.
\end{equation}
$\epsilon^{(1)}_{ij}$ is given by contracting all $E$ indices of $\epsilon^{(1)}$ with the respective unit and all $G$ indices with the respective counit, except for the $E_j$ and $G_i$ indices which remain open.
For example, $\epsilon^{(1)}_{20}$ is given by
\begin{equation}
\begin{tikzpicture}
\atoms{linfunc,all}{0/}
\draw (0)edge[red]++(90:0.5) (0)edge[ar=s]++(-90:0.5);
\end{tikzpicture}
\coloneqq
\begin{tikzpicture}
\atoms{square,small,all,xscale=4}{0/}
\atoms{alg}{{1x/p={0.2,-0.5},astyle=cyan}}
\atoms{coalg}{0y/p={-0.4,0.5}, {1y/p={0,0.5},astyle=cyan}}
\draw ([sx=-0.2]0-b)edge[ar=s]++(-90:0.5) ([sx=0.2]0-b)edge[ar=s,cyan](1x);
\draw ([sx=-0.4]0-t)edge[ar=e](0y) ([sx=0]0-t)edge[ar=e,cyan](1y) ([sx=0.4]0-t)edge[red]++(90:0.5);
\end{tikzpicture}
\;.
\end{equation}
Vice versa, $\epsilon$ can be constructed from the components by contracting the following tensor-network diagram:
\begin{itemize}
\item For each $0\leq j<m$, there is one $E_j$ comultiplication tensor with $n$ outgoing indices, whose ingoing index is the $j$th ingoing index of $\epsilon$.
\item For each $0\leq i<n$, there is one $G_i$ multiplication tensor with $m+1$ ingoing indices, whose outgoing index is the $i$th outgoing index of $\epsilon$.
\item For each $0\leq i<n$, there is one copy of $\epsilon^{(0)}_i$, whose index is contracted with an ingoing index of the $G_i$ multiplication tensor.
\item For each $0\leq i<n$ and each $0\leq j<m$, there is one copy of the 2-index tensor $\epsilon^{(1)}_{ij}$, whose $i$ index is contracted with an ingoing index of the $G_i$ multiplication tensor, and whose $j$ index is contracted with an outgoing index of the $E_j$ comultiplication tensor.
\end{itemize}
For example, for $m=2$ and $n=3$ we have
\begin{equation}
\label{eq:linear_hopf_func_decomposition}
\begin{tikzpicture}
\atoms{square,small,dcross,yscale=4}{0/}
\draw ([sy=0.2]0-l)edge[ar=s]++(180:0.5) ([sy=-0.2]0-l)edge[ar=s,cyan]++(180:0.5) ([sy=0.4]0-r)edge[]++(0:0.5) ([sy=0]0-r)edge[cyan]++(0:0.5) ([sy=-0.4]0-r)edge[red]++(0:0.5);
\end{tikzpicture}
=
\begin{tikzpicture}
\atoms{coalg}{0/p={0,0.4}, {1/p={0,-0.4},astyle=cyan}}
\atoms{alg}{2/p={2,0.8}, {3/p={2,0},astyle=cyan}, {4/p={2,-0.8},astyle=red}}
\atoms{linfunc,all}{a/p={1,1}, b/p={1,0.6}, c/p={1,0.2}, d/p={1,-0.2}, e/p={1,-0.6}, f/p={1,-1}}
\atoms{linfunc,all}{x/p={2,1.2}, {y/p={2,0.4},astyle=cyan}, {z/p={2,-0.4},astyle=red}}
\draw[ar=s,rc] (a)--++(180:0.5)--(0);
\draw[ar=s,rc] (c)--++(180:0.5)--(0);
\draw[ar=s,rc] (e)--++(180:0.5)--(0);
\draw (0)edge[ar=s]++(180:0.5);
\draw[ar=s,rc,cyan] (b)--++(180:0.5)--(1);
\draw[ar=s,rc,cyan] (d)--++(180:0.5)--(1);
\draw[ar=s,rc,cyan] (f)--++(180:0.5)--(1);
\draw[cyan] (1)edge[ar=s]++(180:0.5);
\draw[ar=e,rc] (a)--++(0:0.5)--(2);
\draw[ar=e,rc] (b)--++(0:0.5)--(2);
\draw (2)--++(0:0.5);
\draw[ar=e,rc,cyan] (c)--++(0:0.5)--(3);
\draw[ar=e,rc,cyan] (d)--++(0:0.5)--(3);
\draw[cyan] (3)--++(0:0.5);
\draw[ar=e,rc,red] (e)--++(0:0.5)--(4);
\draw[ar=e,rc,red] (f)--++(0:0.5)--(4);
\draw[red] (4)--++(0:0.5);
\draw (x)edge[ar=e](2);
\draw[cyan] (y)edge[ar=e](3);
\draw[red] (z)edge[ar=e](4);
\end{tikzpicture}
\;.
\end{equation}
\end{myprop}

\subsection{Free-fermion Hopf quadratic tensors}
\label{sec:free_fermion_hopf_function}
In this section we investigate what Hopf quadratic tensors involving group Hopf algebras and copies of $\mathcal F$ look like concretely.

\myparagraph{Coefficient groups for quadratic Hopf vectors}
We start by showing how one can express general quadratic Hopf vectors in terms of coefficient groups analogous to Section~\ref{sec:coefficient_groups}, after decomposing them into smaller components as in Section~\ref{sec:diagrammatic_decomposition}.
First of all, we note that the sets of Hopf bilinear forms and quadratic Hopf vectors form abelian groups.
\begin{myprop}
For any two Hopf algebras $G$ and $\textcolor{cyan}{H}$, $\hom^2[G,\textcolor{cyan}{H}]$ forms an abelian group.
The addition, inverse, and identity are given by
\begin{equation}
\label{eq:hopf_bilinear_group}
\begin{tikzpicture}
\atoms{linform}{0/lab={t=$x+y$,p={180:0.6}}}
\draw (0)edge[ar=s,cyan]++(-90:0.5) (0)edge[ar=s]++(90:0.5);
\end{tikzpicture}
\coloneqq
\begin{tikzpicture}
\atoms{linform}{{0/lab={t=$x$,p={180:0.25}}}, {1/p={0.8,0},lab={t=$y$,p={180:0.25}}}}
\atoms{coalg}{{2/p={0.4,-0.6},astyle=cyan}, 3/p={0.4,0.6}}
\draw[rc,ar=s] (0)--++(90:0.3)--(3);
\draw[rc,ar=s] (1)--++(90:0.3)--(3);
\draw[rc,ar=s,cyan] (0)--++(-90:0.3)--(2);
\draw[rc,ar=s,cyan] (1)--++(-90:0.3)--(2);
\draw (2)edge[ar=s,cyan]++(-90:0.5) (3)edge[ar=s]++(90:0.5);
\end{tikzpicture}
\;,\qquad
\begin{tikzpicture}
\atoms{linform}{0/lab={t=$-x$,p={180:0.4}}}
\draw (0)edge[ar=s,cyan]++(-90:0.5) (0)edge[ar=s]++(90:0.5);
\end{tikzpicture}
\coloneqq
\begin{tikzpicture}
\atoms{linform}{0/lab={t=$x$,p={180:0.25}}}
\atoms{antipode}{1/p={90:0.5}}
\draw (0)edge[ar=s](1) (0)edge[ar=s,cyan]++(-90:0.5) (1)edge[ar=s]++(90:0.5);
\end{tikzpicture}
\;,\qquad
\begin{tikzpicture}
\atoms{linform}{0/lab={t=$0$,p={180:0.25}}}
\draw (0)edge[ar=s,cyan]++(-90:0.5) (0)edge[ar=s]++(90:0.5);
\end{tikzpicture}
\coloneqq
\begin{tikzpicture}
\atoms{coalg}{2/astyle=cyan, 3/p={0,0.6}}
\draw (2)edge[ar=s,cyan]++(-90:0.5) (3)edge[ar=s]++(90:0.5);
\end{tikzpicture}
\;.
\end{equation}
We can identify this group with an explicit coefficient group $\homtild^2[G,\textcolor{cyan}{H}]$ via a coefficient isomorphism $\htild^2[G,\textcolor{cyan}{H}]$.
\end{myprop}

\begin{myprop}
For any Hopf algebra $G$, $\hom_2[G]$ forms an abelian group.
The addition, inverse and identity are given by
\begin{equation}
\begin{tikzpicture}
\atoms{linfunc}{0/lab={t=$x+y$,p={90:0.25}}}
\draw (0)edge[ar=s]++(-90:0.5);
\end{tikzpicture}
\coloneqq
\begin{tikzpicture}
\atoms{linfunc}{{0/lab={t=$x$,p={90:0.25}}}, {1/p={0.8,0},lab={t=$y$,p={90:0.25}}}}
\atoms{coalg}{2/p={0.4,-0.6}}
\draw[rc,ar=s] (0)--++(-90:0.3)--(2);
\draw[rc,ar=s] (1)--++(-90:0.3)--(2);
\draw (2)edge[ar=s]++(-90:0.5);
\end{tikzpicture}
\;,\qquad
\begin{tikzpicture}
\atoms{linform}{0/lab={t=$-x$,p={90:0.3}}}
\draw (0)edge[ar=s]++(-90:0.5);
\end{tikzpicture}
\coloneqq
\begin{tikzpicture}
\atoms{linform}{0/lab={t=$x$,p={90:0.3}},lab={t=$-1$,p={0:0.35}}}
\draw (0)edge[ar=s]++(-90:0.5);
\end{tikzpicture}
\;,\qquad
\begin{tikzpicture}
\atoms{linform}{0/lab={t=$0$,p={90:0.3}}}
\draw (0)edge[ar=s]++(-90:0.5);
\end{tikzpicture}
\coloneqq
\begin{tikzpicture}
\atoms{coalg}{0/}
\draw (0)edge[ar=s]++(-90:0.5);
\end{tikzpicture}
\;.
\end{equation}
Note that the existence of an inverse is part of the definition of a Hopf vector, see Eq.~\eqref{eq:hopf_function_inverse_axiom}.
We can identify this group with an explicit coefficient group $\homtild_2[G]$ via a coefficient isomorphism $\htild_2[G]$.
\end{myprop}

For elementary Hopf algebras $G$ and $\textcolor{cyan}{H}$, the coefficient groups $\homtild^2[G,\textcolor{cyan}{H}]$ and $\homtild_2[G]$ are quite simple.
They need to be determined case by case depending on $G$ and $\textcolor{cyan}{H}$.
We can, however, make one general statement about how to find the coefficient group $\homtild_2[G]$ independent of $G$.
Namely, $\homtild_2[G]$ is a twisted product between the group of normalized linear Hopf vectors and the group of Hopf bilinear forms.

\begin{myprop}
\label{prop:hopf_hom2_classification}
Consider $\img(\bullet^{(2)})\subset \hom^2[G,G]$, the subgroup of Hopf bilinear forms $b\in \hom^2[G,G]$ for which there exists $q\in \hom_2[G]$ such that $b=q^{(2)}$.
Note that these bilinear forms must be symmetric in their arguments,
\begin{equation}
\label{eq:hopf_bilinear_symmetry}
\begin{tikzpicture}
\atoms{linform}{0/}
\draw (0)edge[ind=$a$,ar=s]++(180:0.5) (0)edge[ind=$b$,ar=s]++(0:0.5);
\end{tikzpicture}
=
\begin{tikzpicture}
\atoms{linform}{0/}
\draw (0)edge[ind=$b$,ar=s]++(180:0.5) (0)edge[ind=$a$,ar=s]++(0:0.5);
\end{tikzpicture}
\;.
\end{equation}
Then there is an abelian group extension
\begin{equation}
\hom[G]\xrightarrow{\subset} \hom_2[G]\xrightarrow{\bullet^{(2)}} \img(\bullet^{(2)})\;.
\end{equation}
So in order to find all normalized Hopf quadratic functions, it suffices to find all elements of $\hom[G]$ and $\img(\bullet^{(2)})$.
Note that the group extension is not necessarily a direct product, but can be twisted by a group 2-cocycle
\begin{equation}
\Omega[G]\in H^2(B\img(\bullet^{(2)}),\hom[G])\;.
\end{equation}
\end{myprop}

\myparagraph{Coupling abelian groups and $\mathcal F$}
We now explicitly work out the above coefficient groups for the elementary Hopf algebras $G\in \{\zz_k,\rr,\rr/\zz,\zz,\mathcal F\}$.
The first thing we note is that even though a Hopf quadratic tensor can have both abelian-group and free-fermion indices, these two sorts of indices only interact in a trivial way.
\begin{myprop}
Each Hopf quadratic tensor with both abelian-group indices and $\mathcal F$-indices is equal to a tensor product of a quadratic tensor supported on all abelian-group indices, and one supported on all $\mathcal F$-indices.
\end{myprop}
\begin{proof}
We start by determining $\hom^2[\mathcal F,C]$ for an abelian group $C$.
Since $h\in \hom^2[\mathcal F,\textcolor{cyan}{C}]$ is a 2-index fermionic tensor which must be evenly graded, its entries can only be non-zero if the $\mathcal F$-index is in the $0$ configuration,
\begin{equation}
\begin{tikzpicture}
\atoms{linform}{0/}
\draw (0)edge[ind=$f$,ar=s]++(180:0.5) (0)edge[cyan,ind=$g$,ar=s]++(0:0.5);
\end{tikzpicture}
=\delta_{f=0} \textcolor{cyan}{x(g)}\;,
\end{equation}
for some vector $x\in \cc^C$.
Plugging this into the third equation in Eq.~\eqref{eq:hopf_bilinear_form} then yields $x(g)=1\forall g\in C$, corresponding to the trivial Hopf bilinear form.
So we have
\begin{equation}
\hom^2[\mathcal F,C]=0\;,
\end{equation}
which means that any quadratic Hopf vector is a tensor product of a quadratic Hopf vector on the $\mathcal F$ indices, and one on the $C$ indices.
Next, we determine the Hopf linear functions $\hom[\mathcal F|C]$.
We do this by making use of the following equivalence between linear Hopf functions and Hopf bilinear forms:
\begin{equation}
\label{eq:hopf_currying}
\hom[A|B]=\hom^2[A,B^*]\;.
\end{equation}
Using this, we find
\begin{equation}
\hom[\mathcal F|C]=\hom^2[\mathcal F,C^*]=0\;,\qquad
\hom[C|\mathcal F]=\hom^2[C,\mathcal F^*]=0\;.
\end{equation}
So the embedding $\epsilon$ is again the tensor product of an embedding between the $\mathcal F$-factors of $E$ and $G$, and an embedding between the $C$-factors of $E$ and $G$.
Finally, we show that also the double integral does not introduce any coupling between $\mathcal F$ and abelian-group factors.
Consider $\textcolor{cyan}{E=C\otimes \mathcal F^{\otimes n}}$ and $G=D\otimes \mathcal F^{\otimes m}$ for some abelian groups $C$ and $D$.
If $n+m$ is odd, then there are no double integrals (only zero).
If $n+m$ is even, then there is a 1-dimensional space of double integrals, and all of them are proportional to a tensor product:
\begin{equation}
\label{eq:group_f_integral}
\begin{tikzpicture}
\atoms{square,fflat,small,all}{0/}
\draw (0)edge[mark={lab={$(g,\vec x)$},a},ar=s]++(-90:0.5) (0)edge[cyan,mark={lab={$(e,\vec y)$},a}]++(90:0.5);
\end{tikzpicture}
\propto
\delta_{g=0} \prod_{0\leq i<m} \delta_{\vec x_i=1} \prod_{0\leq j<n} \delta_{\vec y_j=1} \forall \textcolor{cyan}{e}\in \textcolor{cyan}{C}\;.
\end{equation}
In other words, the double integral has only one non-zero entry for the $\mathcal F$ components, namely for the $111\ldots$ configuration.
Note that if $m+n$ is odd, then the definition in Eq.~\eqref{eq:group_f_integral} would have odd global parity, and thus would not be a valid fermionic tensor.
So neither $q$ nor $\epsilon$ nor the double integral allow non-trivial mixing between the $C$ and $\mathcal F$ factors.
\end{proof}

\myparagraph{Coefficients on $\mathcal F^n$}
The coefficient groups mixing $\mathcal F$ and abelian groups are trivial, and those purely for abelian groups have already been worked out in Section~\ref{sec:coefficient_groups}.
So all that remains to do is to find coefficient groups for $\mathcal F$ factors.
\begin{myprop}
We have
\begin{equation}
\label{eq:free_fermion_bilinear}
\homtild^2[\mathcal F,\textcolor{cyan}{\mathcal F}]=\cc\simeq \rr\times\rr\;,\qquad
\htild^2(\alpha)
=
\big(
\begin{tikzpicture}
\atoms{linform}{0/}
\draw (0)edge[ind=$a$,ar=s]++(180:0.5) (0)edge[cyan,ind=$b$,ar=s]++(0:0.5);
\end{tikzpicture}
,ab
\big)
\coloneqq
\alpha^a\delta_{ab}
=
\begin{pmatrix}
1&0\\
0&\alpha
\end{pmatrix}
\;.
\end{equation}
\end{myprop}
\begin{proof}
It is easy to see that this choice fulfills all the axioms in Eq.~\eqref{eq:hopf_bilinear_form}.
For example, we find
\begin{equation}
\begin{tikzpicture}
\atoms{alg,astyle=cyan}{0/}
\atoms{linform}{1/p={-90:0.6}}
\draw (0)edge[ar=e,cyan](1) (0)edge[ind=$b$,ar=s,cyan]++(45:0.4) (0)edge[ind=$a$,ar=s,cyan]++(135:0.4) (1)edge[ind=$c$,ar=s]++(-90:0.4);
\end{tikzpicture}
=\delta_{a+b=c} \alpha^{c}
=\delta_{a+b=c} \alpha^a \alpha^b
=
\begin{tikzpicture}
\atoms{coalg}{0/}
\atoms{linform}{1/p={0.4,0.6}, 2/p={-0.4,0.6}}
\draw[rc,ar=s] (1)--++(-90:0.3)--(0);
\draw[rc,ar=s] (2)--++(-90:0.3)--(0);
\draw (1)edge[ind=$b$,ar=s,cyan]++(90:0.4) (2)edge[ind=$a$,ar=s,cyan]++(90:0.4) (0)edge[ind=$c$,ar=s]++(-90:0.4);
\end{tikzpicture}
\;.
\end{equation}
Note that the equation has a reordering sign of $(-1)^{|a||b|}$, but this reordering sign does not show up as the equation is zero if $a=b=1$.
It is also easy to see that the group multiplication in Eq.~\eqref{eq:hopf_bilinear_group} corresponds to addition in $\cc$ and hence $\htild^2$ in Eq.~\eqref{eq:free_fermion_bilinear} is indeed an isomorphism:
\begin{equation}
\begin{tikzpicture}
\atoms{linform}{0/lab={t=$\alpha+\beta$,p={180:0.6}}}
\draw (0)edge[ar=s,cyan]++(-90:0.5) (0)edge[ar=s]++(90:0.5);
\end{tikzpicture}
=
\begin{pmatrix}
1&0\\
0&\alpha+\beta
\end{pmatrix}
=
\begin{pmatrix}
1&0\\
0&\alpha
\end{pmatrix}
+
\begin{pmatrix}
1&0\\
0&\beta
\end{pmatrix}
=
\begin{tikzpicture}
\atoms{linform}{{0/lab={t=$\alpha$,p={180:0.25}}}, {1/p={0.8,0},lab={t=$\beta$,p={180:0.25}}}}
\atoms{coalg}{{2/p={0.4,-0.6},astyle=cyan}, 3/p={0.4,0.6}}
\draw[rc,ar=s] (0)--++(90:0.3)--(3);
\draw[rc,ar=s] (1)--++(90:0.3)--(3);
\draw[rc,ar=s,cyan] (0)--++(-90:0.3)--(2);
\draw[rc,ar=s,cyan] (1)--++(-90:0.3)--(2);
\draw (2)edge[ar=s,cyan]++(-90:0.5) (3)edge[ar=s]++(90:0.5);
\end{tikzpicture}
\;.
\end{equation}
One can also check that the above indeed covers all bilinear forms:
Due to the overall even parity, the most general 2-index tensor is given by
\begin{equation}
\begin{pmatrix}\beta&0\\0&\alpha\end{pmatrix}\;.
\end{equation}
The third or fourth equation in Eq.~\eqref{eq:hopf_bilinear_form} then directly implies that $\beta=1$.
\end{proof}

\begin{myprop}
The only normalized quadratic Hopf vector over $\mathcal F$ is the trivial one:
\begin{equation}
\homtild_2[\mathcal F]=0\;,\qquad
\htild_2(0)=
\begin{tikzpicture}
\atoms{linfunc}{0/lab={t=$0$,p=-90:0.3}}
\draw (0)edge[ind=$a$,ar=s]++(90:0.5);
\end{tikzpicture}
=
\begin{tikzpicture}
\atoms{coalg}{0/}
\draw (0)edge[ind=$a$,ar=s]++(90:0.5);
\end{tikzpicture}
=
\big(\delta_{a=0}, a\big)
\simeq
\begin{pmatrix}
1&0
\end{pmatrix}
\;.
\end{equation}
\end{myprop}
\begin{proof}
This is because any super vector is $\zz_2$-graded and thus any element of $\hom_2[\mathcal F]$ is of the form
\begin{equation}
\mpm{\alpha&0}\;.
\end{equation}
The normalization in Eq.~\eqref{eq:hopf_vector_normalized} implies that $\alpha=1$.
\end{proof}
Note that in accordance with Proposition~\ref{prop:hopf_hom2_classification}, both $\img(\bullet^{(2)})$ and $\hom[\mathcal F]$ are trivial.
$\hom[\mathcal F]$ is trivial for the same reason $\hom_2[\mathcal F]$ is trivial, and $\img(\bullet^{(2)})\subset \hom^2[\mathcal F,\mathcal F]$ is trivial since its elements must be symmetric:
The symmetry condition in Eq.~\eqref{eq:hopf_bilinear_symmetry} for a Hopf bilinear form as in Eq.~\eqref{eq:free_fermion_bilinear} implies
\begin{equation}
\begin{pmatrix}
1&0\\0&\alpha
\end{pmatrix}
=
\begin{pmatrix}
1&0\\0&-\alpha
\end{pmatrix}
\quad\Rightarrow\quad
\alpha=0
\;,
\end{equation}
where the minus sign is due to reordering.

Next we find the coefficient groups for the embedding $\epsilon$.
To this end, we use the identification in Eq.~\eqref{eq:hopf_currying}, and also the following identification between Hopf constants and normalized linear Hopf vectors:
\begin{equation}
\hom_0[|A]=\hom[A^*]\;.
\end{equation}
With this, we find
\begin{equation}
\homtild[\mathcal F|\mathcal F]
=\homtild^2[\mathcal F,\mathcal F^*]
=\cc\;,\qquad
\homtild_0[|\mathcal F]
=\homtild[\mathcal F]
=0\;.
\end{equation}

After finding the coefficient groups $\homtild^2[\mathcal F,\mathcal F]$, $\homtild_2[\mathcal F]$, $\homtild[\mathcal F|\mathcal F]$ and $\homtild_0[|\mathcal F]$, let us summarize which coefficients are needed to specify a Hopf quadratic tensor with $\mathcal F$ indices.
\begin{myprop}
A Hopf quadratic tensor data over $G=\mathcal F^n$ with $E=\mathcal F^m$ is fully specified by
\begin{itemize}
\item a complex $m\times m$ strictly upper triangular matrix $q^{(\tilde2)}$,
\item a complex number $q^{(\tilde0)}\in \cc$, and
\item a complex $n\times m$ matrix $\epsilon^{(\tilde1)}$.
\end{itemize}
For many purposes it is natural to mirror $q^{(\tilde2)}$ into a full anti-symmetric matrix.
\end{myprop}

\myparagraph{Pfaffian representation}
After finding the non-trivial coefficients $q^{(\tilde2)}$, $q^{(\tilde0)}$ and $\epsilon^{(\tilde1)}$, let us show how to explicitly compute the computational-basis entries in terms of these coefficients.
\begin{myprop}
Consider a Hopf quadratic tensor with $G=\mathcal F^n$ and $E=\mathcal F^{n-2l}$, so that $\epsilon^{(\tilde1)}$ is a $n\times (n-2l)$ matrix, and $q^{(\tilde2)}$ is an anti-symmetric $(n-2l)\times (n-2l)$ matrix.
The tensor entry for a length-$n$ bitstring $x$ is given by
\begin{equation}
\label{eq:quadratic_tensor_entries}
T[E,\epsilon,q](x)
= q^{(\tilde0)} \pf \mpm{q^{(\tilde2)} & -(\epsilon^{(\tilde1)}|_{\neg x,\varnothing})^T \\\epsilon^{(\tilde1)}|_{\neg x,\varnothing} & 0}
\;.
\end{equation}
$\neg x$ denotes the bitstring $x$ where we flip every $0$ to a $1$ and vice versa.
$\pf$ denotes the Pfaffian of an anti-symmetric matrix, and $\epsilon^{(\tilde1)}|_{a,\varnothing}$ denotes the submatrix of $\epsilon^{(\tilde1)}$ where we keep all columns but only the rows with $a_i=1$.
\end{myprop}
\begin{proof}
We start by determining the entries $q(x)$ of the quadratic Hopf vector $q$, by evaluating the tensor network in the decomposition of Proposition~\ref{prop:hopf_quadratic_decomposition}.
First we note that all the 1-index tensors in Eq.~\eqref{eq:diagrammatic_decomposition_n4} are trivial, and all the 2-index tensors are of the form $\operatorname{diag}(1,q^{(\tilde2)}_{ij})$ according to Eq.~\eqref{eq:free_fermion_bilinear}.
When evaluating the tensor network, we have to sum over all configurations of the contracted indices.
The two indices of $q^{(2)}_{ij}$, which are contracted with the $E_i$ and $E_j$ comultiplication tensors, must be in the same configuration, which we denote by $y_{ij}\in\{0,1\}$.
Then the evaluation of the tensor network is given by
\begin{equation}
q(x) = q^{(\tilde0)} \sum_{\{y_{ij}\}_{i<j}} \sigma_f(y) \prod_{i<j} (q_{ij}^{(\tilde2)})^{y_{ij}} \prod_i C(\{y_{ij}\}_{j: j\neq i}, x_i)\;.
\end{equation}
Here, $\sigma_f(y)$ is the fermionic reordering sign of the configuration $y$, which is equal to the number of self-intersections of the set of bonds with $y_{ij}=1$.
$(q_{ij}^{(\tilde2)})^{y_{ij}}$ is the tensor entry of the 2-index tensor $q^{(2)}_{ij}=\operatorname{diag}(1,q^{(\tilde2)}_{ij})$.
Finally, $C(a,b,c,\ldots,x)$ denotes the tensor entry of the comultiplication tensor whose input indices are in configurations $a,b,c,\ldots$, and whose output index is in configuration $x$.
$C=1$ if $a=b=c=\ldots=x=0$ or if $x=1$ and exactly one of $a,b,c,\ldots$ is in the $1$ configuration, and $C=0$ otherwise.
If we view the tensor network as a graph, then we have $\prod_i C=1$ if $y$ is a perfect matching of $x$, and $\prod_i C=0$ otherwise.
Note that $y$ is a perfect matching if for every vertex $i$ with $x_i=1$ there is exactly one edge $(i,j)$ with $y_{ij}=1$, and for every vertex $i$ with $x_i=0$, we have $y_{ij}=0$ for all $j$.
So instead of summing over all index configurations $y$ with weight $C$, we can restrict the sum to the perfect matchings of $x$.
Equivalently, we can sum over permutations $\pi\in S_{|x|}$ of the vertices $i$ with $x_i=1$, and take as perfect matching the set of edges $\{(\pi(2f),\pi(2f+1))\}_{0\leq f<|x|/2}$.
For every perfect matching, there are $2^{|x|/2} (|x|/2)!$ such permutations.
With this, the tensor-network contraction reduces to
\begin{equation}
\label{eq:quadratic_hopf_vector_entries}
q(x)
=
\begin{tikzpicture}[cyan]
\atoms{linfunc,dot}{0/}
\draw (0)edge[ind=$x$]++(0:0.5);
\end{tikzpicture}
=q^{(\tilde0)}\frac{1}{2^{|x|/2} (|x|/2)!} \sum_{\pi\in S_{|x|}} \sigma(\pi) \prod_{0\leq f<|x|/2} q^{(\tilde2)}_{\pi(2f)\pi(2f+1)}
=q^{(\tilde0)}\pf(q^{(\tilde2)}|_x)\;.
\end{equation}
$q^{(\tilde2)}|_x$ denotes the matrix $q^{(\tilde2)}$ restricted to the rows and columns $i$ for which $x_i=1$.
$\sigma(\pi)$ denotes the sign of the permutation $\pi$, which is equal to the fermionic reordering sign $\sigma_f(y)$.
To see this, add a transposition between $2f+1$ and $2(f+1)$ to $\pi$.
This corresponds to adding or removing a crossing between two $1$-indices in the tensor-network diagram, which yields a factor of $-1$.
A transposition between $2f$ and $2f+1$ also yields a factor of $-1$ since $q^{(\tilde2)}$ is anti-symmetric, $q^{(\tilde2)}_{\pi(2f)\pi(2f+1)}=-q^{(\tilde2)}_{\pi(2f+1)\pi(2f)}$.

Next, we determine the matrix entries of the linear Hopf function $\epsilon$.
We note that the multiplication and comultiplication tensors of $\mathcal F$ are identical.
Thus, the tensor network in Eq.~\eqref{eq:linear_hopf_func_decomposition} (shown for $n=5$) is a special case of that in Eq.~\eqref{eq:diagrammatic_decomposition_n4} (shown for $n=4$).
The only difference is that the graph in Eq.~\eqref{eq:linear_hopf_func_decomposition} is bipartite and the coupling within each part is zero, whereas the coupling across the bipartition is given by $\epsilon^{(\tilde1)}$.
Thus the matrix entry of $\epsilon$ for a length-$n$ bitstring $x$ and length-$n-2l$ bitstring $z$ is given by
\begin{equation}
\epsilon(x,z)
=
\begin{tikzpicture}
\atoms{small,square,dcross}{0/}
\draw (0)edge[ind=$z$]++(180:0.5) (0)edge[ind=$x$]++(0:0.5);
\end{tikzpicture}
=\pf\mpm{0&\epsilon^{(\tilde1)}\\-(\epsilon^{(\tilde1)})^T&0}\rvert_{x\sqcup z}
=\det(\epsilon^{(\tilde1)}|_{x,z})\;.
\end{equation}
Here, $\det$ denotes the determinant of a matrix and $\epsilon^{(\tilde1)}|_{a,b}$ denotes the matrix $\epsilon^{(\tilde1)}$ restricted to the rows $i$ with $a_i=1$ and the columns $j$ with $b_j=1$.

Finally, we determine the effect of the double-integral.
We have
\begin{equation}
\begin{tikzpicture}
\atoms{coalg}{0/}
\draw (0)edge[ind=$\ket1$,ar=s]++(90:0.5) (0)edge[ind=$x$]++(180:0.5) (0)edge[ind=$y$]++(0:0.5);
\end{tikzpicture}
=
\begin{tikzpicture}
\atoms{alg}{0/}
\draw (0)edge[ind=$\ket1$]++(90:0.5) (0)edge[ind=$x$,ar=s]++(180:0.5) (0)edge[ind=$y$,ar=s]++(0:0.5);
\end{tikzpicture}
=
\delta_{x=\neg y}
\;.
\end{equation}
Using this and Eq.~\eqref{eq:group_f_integral}, we find
\begin{equation}
\begin{tikzpicture}
\atoms{square,rot=90,fflat,small,all}{0/}
\atoms{coalg,astyle=cyan}{c/p={-0.5,-0.5}}
\atoms{alg}{a/p={0.5,-0.5}}
\atoms{small,square,dcross}{e/p={0,-0.5}}
\atoms{rot=90,antipode}{anti/p={1,-0.5}}
\draw[cyan] (c)edge[ind=$z$]++(180:0.5) (c)edge[ar=e](e) (c)edge[cyan,ar=s,out=90,in=180](0.north);
\draw (e)edge[ar=e](a) (a)edge[ar=s](anti) (anti)edge[ar=s,ind=$x$]++(0:0.5) (0.south)edge[out=0,in=90,ar=s](a);
\end{tikzpicture}
=
\begin{tikzpicture}
\atoms{coalg,astyle=cyan}{c/p={-0.5,-0.5}}
\atoms{alg}{a/p={0.5,-0.5}}
\atoms{small,square,dcross}{e/p={0,-0.5}}
\atoms{rot=90,antipode}{anti/p={1,-0.5}}
\draw[cyan] (c)edge[ind=$z$]++(180:0.5) (c)edge[ar=e](e) (c)edge[cyan,ar=s,ind=$\ket1$]++(90:0.5);
\draw (e)edge[ar=e](a) (a)edge[ar=s](anti) (anti)edge[ar=s,ind=$x$]++(0:0.5) (a)edge[ind=$\ket1$]++(90:0.5);
\end{tikzpicture}
=\delta_{x=\neg x'} \det(\epsilon^{(\tilde1)}|_{x',z'}) \delta_{z=\neg z'}
=\det(\epsilon^{(\tilde1)}|_{\neg x,\neg z})\;.
\end{equation}
When we apply the above matrix to the vector $q$ in Eq.~\eqref{eq:quadratic_hopf_vector_entries}, we have to sum over all intermediate configurations $z$:
\begin{equation}
\label{eq:fermion_pfaff_det_formula}
\mathcal T[E,\epsilon,q]= q^{(\tilde0)} \sum_{z\in \{0,1\}^{n-2l}: |z|=|x|-2l} \det(\epsilon^{(\tilde1)}|_{\neg x,\neg z}) \pf(q^{(\tilde2)}|_z)\;.
\end{equation}
Finally, we argue that evaluating the Pfaffian on the right-hand side of Eq.~\eqref{eq:quadratic_tensor_entries} yields the same summation:
The Pfaffian in Eq.~\eqref{eq:quadratic_tensor_entries} is a sum over all pairings of rows/columns.
Since the bottom right block of the matrix is zero, the only pairings that contribute are the ones where the last $n-|x|$ rows/columns are paired with some subset of the first $n-2l$ rows/columns.
The indicator function of this subset is the bitstring $\neg z$.
Summing over all pairings with the fixed subset $\neg z$ yields $\det(\epsilon^{(\tilde1)}|_{\neg x,\neg z})$.
The remaining $2l-|x|$ columns, corresponding to indicator function $z$, are paired among themselves.
Summing over all pairings internal to $z$ yields $\pf(q^{(\tilde2)}|_z)$.
\end{proof}

We now illustrate the proposition above by looking at some special cases and concrete examples.
The first special case is when $l=0$ and the embedding is trivial,
\begin{equation}
E=G=\mathcal F^n\;,\quad \epsilon=\mathbb1\;.
\end{equation}
Then the Hopf quadratic tensor is the same as the quadratic Hopf vector $q$ itself:
\begin{equation}
\begin{tikzpicture}
\atoms{square,rot=90,fflat,small,all}{0/}
\atoms{coalg}{c/p={-0.5,-0.5}}
\atoms{alg}{a/p={0.5,-0.5}}
\atoms{linfunc,dot}{q/p={-1,-0.5}}
\atoms{small,square,dcross}{e/p={0,-0.5}}
\atoms{rot=90,antipode}{anti/p={1,-0.5}}
\draw (q)edge[ar=s](c) (c)edge[ar=e](e) (c)edge[ar=s,out=90,in=180](0.north);
\draw (e)edge[ar=e](a) (a)edge[ar=s](anti) (anti)edge[ar=s]++(0:0.5) (0.south)edge[out=0,in=90,ar=s](a);
\end{tikzpicture}
=
\begin{tikzpicture}
\atoms{linfunc,dot}{q/}
\draw (q)edge[ar=s]++(0:0.5);
\end{tikzpicture}
\;.
\end{equation}
Next, we consider the opposite case of $n=2l$, such that $E$ is the trivial (super) Hopf algebra and $G=\mathcal F^{2l}$.
In this case, the tensor itself is given by the double integral.
That is, up to a global prefactor, the tensor is the all-$1$ state of the $2l$ modes,
\begin{equation}
\begin{tikzpicture}
\atoms{square,rot=90,fflat,small,all}{0/}
\atoms{coalg}{c/p={-0.5,-0.5}}
\atoms{alg}{a/p={0.5,-0.5}}
\atoms{linfunc,dot}{q/p={-1,-0.5}}
\atoms{small,square,dcross}{e/p={0,-0.5}}
\draw (q)edge[ar=s](c) (c)edge[ar=e](e) (c)edge[ar=s,out=90,in=180](0.north);
\draw (e)edge[ar=e](a) (a)edge[ar=s]++(0:0.5) (0.south)edge[out=0,in=90,ar=s](a);
\end{tikzpicture}
=
\begin{tikzpicture}
\atoms{square,rot=90,fflat,small,all}{0/}
\atoms{alg}{a/p={0.5,-0.5}}
\atoms{small,circ}{e/p={0,-0.5}}
\draw (e)edge[ar=e](a) (a)edge[ar=s]++(0:0.5) (0.south)edge[out=0,in=90,ar=s](a);
\end{tikzpicture}
=
\begin{tikzpicture}
\atoms{square,rot=90,fflat,small,all}{0/}
\draw (0)edge[ar=s]++(0:0.5);
\end{tikzpicture}
\propto\ket{\underbrace{11\ldots 1}_{2l}}\;.
\end{equation}
The next example is something in between the trivial embedding and the zero embedding.
Consider $n=3$ and $l=1$, so $E=\mathcal F$ and $G=\mathcal F^3$.
Such a tensor is determined by four numbers $a,b,c,\gamma\in \cc$,
\begin{equation}
q^{(\tilde2)}=\varnothing\;,\quad
q^{(\tilde0)}=\gamma\;,\quad
\epsilon^{(\tilde1)}=\mpm{a\\b\\c}\;.
\end{equation}
The resulting tensor is
\begin{equation}
\mathcal T[E,\epsilon,q]
=\gamma\mpm{\mpm{0&0\\0&a}&\mpm{0&b\\c&0}}\;.
\end{equation}
Note that the tensor entry for the $\ket{000}$ configuration is zero, which is impossible for a tensor with trivial embedding.
We note that in general, $\mathcal T[E,\epsilon,q](x)=0$ for all computational basis configurations $x$ with $|x|<2l$, as then $|x|-2l<0$ and the sum over $z$ in Eq.~\eqref{eq:fermion_pfaff_det_formula} is empty.
As a further example, consider $n=4$ and $l=1$, so $E=\mathcal F^2$ and $G=\mathcal F^4$.
Such a tensor is determined by
\begin{equation}
q^{(\tilde2)}=\mpm{\alpha}\;,\quad
q^{(\tilde0)}=\gamma\;,\quad
\epsilon^{(\tilde1)}=\mpm{a&e\\b&f\\c&g\\d&h}\;.
\end{equation}
The resulting tensor is
\begin{equation}
\mathcal T[E,\epsilon,q]
=\gamma\mpm{\mpm{0&0\\0&af-eb}&\mpm{0&ag-ec\\ah-ed&0}\\\mpm{0&bg-fc\\bh-fd&0}&\mpm{ch-gd&0\\0&\alpha}}\;.
\end{equation}

One interesting final note is that the free-fermion identity matrix on $n$ modes is a quadratic tensor over $\mathcal F^{2n}$ with trivial embedding,
\begin{equation}
E=G=\mathcal F^{2n}\;,\quad
\epsilon=\mathbb1\;,\quad
q^{(\tilde2)}=\mpm{0&1\\-1&0}\;.
\end{equation}
This is in contrast to qudits and continuous variables, where we do need a non-trivial (diagonal) embedding for the identity matrix.

\myparagraph{Free-fermion index contraction}
Ideally, we would like to show that general Hopf quadratic tensors can be efficiently contracted in terms of the underlying data on a purely diagrammatic level.
However, this turns out to be a bit tricky, especially with a non-trivial embedding $\epsilon$ involved.
We thus leave this diagrammatic proof to future work, and comment on how to efficiently perform index contractions specifically for free-fermion quadratic tensors over $\mathcal F^{n+c+c}$, where the contraction is over the last two $c$-tuples of fermionic modes.
Let us for simplicity assume that the embedding $\epsilon$ is trivial, so that the tensor is specified by a complex anti-symmetric $(n+c+c)\times (n+c+c)$ matrix $q^{(\tilde2)}$ which we can write in block form as
\begin{equation}
q^{(\tilde2)}=
\mpm{a&b&c\\-b^T&d&e\\-c^T&-e^T&f}
\;,
\end{equation}
where $a$, $d$, and $f$ are anti-symmetric.
Let us further assume that the matrix
\begin{equation}
\label{eq:fermioncontraction_submatrix}
\mpm{d&e+1\\-e^T-1&f}
\end{equation}
is invertible.
In this case, it was shown in Ref.~\cite{tensor_type} that the contraction is again a quadratic tensor with trivial embedding and
\begin{equation}
q^{(\tilde2)}
=\mpm{a}-\mpm{b&c}\mpm{d&e+1\\-e^T-1&f}^{-1}\mpm{-b^T\\-c^T}\;.
\end{equation}
So we see that the contraction is still based on a Schur complement, just like in the abelian-group case.
Note that if the matrix in Eq.~\eqref{eq:fermioncontraction_submatrix} is not invertible, then the resulting tensor will still be a Hopf quadratic tensor, but with non-trivial embedding.
We will not discuss this case here.

\section{\texorpdfstring{$i$}{i}th order tensors}
\label{sec:ilinear}
In this section, we propose a natural generalization of quadratic tensors to \emph{$i$th order tensors}, where quadratic tensors correspond to 2nd order tensors.
$i$th order tensors with $n$ (elementary) indices can be efficiently represented using $O(n^i)$ many coefficients.
Tensor networks of $i$th order tensors with $i>2$ cannot in general be contracted efficiently, though we believe that contraction can be efficient if the number of $>2$nd order tensors is small, or in special cases where the ``$>2$nd order contributions'' of the different tensors cancel each other.
\subsection{\texorpdfstring{$i$}{i}th order tensors: Definition}
We can easily generalize the definition of quadratic functions $q:G\rightarrow A$ in Section~\ref{sec:clifford_definition} to arbitrary $i$th order functions.
\begin{mydef}
For a function $q:G\rightarrow A$ between abelian groups $G$ and $A$, define the $0$th and $1$st \emph{derivative} $q^{(0)}$ and $q^{(1)}$ as in Eqs.~\eqref{eq:zeroth_derivative} and \eqref{eq:first_derivative}.
For every $i>1$, define the \emph{$i$th derivative} $q^{(i)}:G^{\times i}\rightarrow A$ recursively by
\begin{equation}
q^{(i)}(g_0,\ldots,g_{i-1})
\coloneqq q^{(i-1)}(g_0+g_1,g_2,\ldots,g_{i-1})-q^{(i-1)}(g_0,g_2,\ldots,g_{i-1})-q^{(i-1)}(g_1,g_2,\ldots,g_{i-1})\;.
\end{equation}
We call a function $q:G\rightarrow A$ \emph{$i$th order} if
\begin{equation}
q^{(i+1)}=0\;.
\end{equation}
We denote the group of $i$th order functions under element-wise addition by $F_i[G|A]$.
\end{mydef}
It is easy to see that all $q^{(i)}$ are invariant under permutations of their arguments.
Intuitively, $q^{(i)}$ measures how much $q^{(i-1)}$ fails to be a homomorphism in its first component (or equivalently in any other component, due to the permutation invariance).
Explicitly, we can expand $q^{(i)}$ as
\begin{equation}
\label{eq:derivative_explicit}
q^{(i)}(g_0,\ldots,g_{i-1}) = \sum_{S\subset \{0,\ldots,i-1\}} (-1)^{i-|S|} q(\sum_{x\in S} g_x)\;.
\end{equation}
For example, for $i=3$, we have
\begin{equation}
\begin{multlined}
q^{(3)}(g_0,g_1,g_2) = q^{(2)}(g_0+g_1,g_2)-q^{(2)}(g_0,g_2)-q^{(2)}(g_1,g_2)\\
= q(g_0+g_1+g_2)-q(g_0+g_1)-q(g_0+g_2)-q(g_1+g_2)+q(g_0)+q(g_1)+q(g_2)-q(0)\;.
\end{multlined}
\end{equation}

\begin{mydef}
Let $A$ be an abelian group and $\exp\in \hom[A|\cc^\times]$ as in Definition~\ref{def:quadratic_tensor}.
An \emph{$i$th order tensor data} over an abelian group $G$ is a triple $(E,\epsilon,q)$, where
\begin{itemize}
\item $E$ is an abelian group,
\item $\epsilon\in F_{i-1}[E|G]$ is a $i-1$th order function,
\item and $q\in F_i[E|A]$ is an $i$th order function.
\end{itemize}
The \emph{$i$th order tensor} associated with $(E,\epsilon,q)$ is a vector of the vector space of complex functions over the abelian group $G$,
\begin{equation}
\mathcal T[E,\epsilon,q]:G\rightarrow\cc\;,\qquad \mathcal T[E,\epsilon,q](g)=\sum_{e\in E: \epsilon(e)=g} \exp(q(g))\;.
\end{equation}
\end{mydef}
As for quadratic tensors, the canonical choice for $A$ is $\rr\times\rr/\zz$ such that $\exp$ becomes an isomorphism.

\subsection{Efficient representation of \texorpdfstring{$i$}{i}th order functions}
In this section, we show how to efficiently store $i$th order functions in terms of concrete coefficients.

\myparagraph{Factor decomposition}
We first show how to decompose an $i$th order function over a cartesian product $n$ arbitrary factors into smaller components.
These components are ``$i$th order $k$-ary functions'' as defined in the following.
\begin{mydef}
An \emph{$i$th order normalized $k$-ary function} is a function
\begin{equation}
h: G_0\times G_1\times \ldots\times G_{k-1} \rightarrow A\;,
\end{equation}
such that for all $0\leq x<k$,
\begin{equation}
\label{eq:kary_ithorder_function_definition}
h^{(2,x)}(g_0,\ldots,g_x,g_x',\ldots,g_{k-1})\coloneqq h(g_0,\ldots,g_x+g_x',\ldots, g_{k-1})-h(g_0,\ldots,g_x,\ldots, g_k)-h(g_0,\ldots,g_x',\ldots, g_{k-1})
\end{equation}
is a $k+1$-ary $i-1$th order function.
To terminate this recursive definition, we define a $0$th order normalized $k$-ary function (for $k>0$) as the zero function.
We denote the group of $i$th order normalized $k$-ary functions (under element-wise multiplication) by $\hom_i^k[G_0,\ldots,G_{k-1}|A]$.
With this notation, the condition becomes
\begin{equation}
h^{(2,x)}\in \hom_{i-1}^{k+1}[G_0,\ldots,G_x,G_x,\ldots,G_{k-1}|A]\;.
\end{equation}
\end{mydef}
Note that an $i$th order normalized $k$-ary function is an $i$th order normalized function on each individual argument when keeping all other arguments fixed, but the converse does not hold in general.
A $i$th order normalized $k$-ary function is actually ``normalized'' in the sense that it is zero if any of its arguments are zero.
Also note that the notation $\hom_i^k$ is consistent with earlier notation if we drop a ``$1$'' sub or superscript, $\hom_1^1=\hom$, $\hom_1^2=\hom^2$, $\hom_2^1=\hom_2$.
In general, $\hom_1^i$ are $i$-linear forms.
We also have $\hom_i^0[|A]=A\forall i$ and $\hom_0^i=0 \forall i>0$.

\begin{myprop}
Consider abelian groups $G=\bigtimes_{0\leq i<n} G_i$ and $A$.
Then there is a bijection between (1) $i$th order functions $q\in F_i[G|A]$ and (2) collections
\begin{equation}
\{\{q^{(k)}_{\vec a}\}_{0\leq a_0<a_1<\ldots<a_{k-1}<n}\}_{0\leq k\leq i}\;,
\end{equation}
where each $q^{(k)}_{\vec a}\in\hom_{i-k+1}^k[G_{a_0},\ldots,G_{a_{k-1}}|A]$ is a $i-k+1$th order normalized $k$-ary function.
The components are obtained from $q$ as
\begin{equation}
\label{eq:ilinear_decomposition}
q^{(k)}_{\vec a}(g_0,\ldots, g_{k-1}) = q^{(k)}(g_0|_{a_0},g_1|_{a_1},\ldots, g_{k-1}|_{a_{k-1}})\;,
\end{equation}
using the notation from Eq.~\eqref{eq:argument_position_notation}.
Vice versa, $q$ is obtained from the components via
\begin{equation}
\label{eq:ilinear_composition}
q(g_0,\ldots,g_{n-1}) = \sum_{0\leq k\leq i} \sum_{0\leq a_0<a_1<\ldots<a_{k-1}<n} q^{(k)}_{\vec a}(g_{a_0},\ldots,g_{a_{k-1}})\;.
\end{equation}
\end{myprop}
\begin{proof}
We start by showing that $q^{(k)}_{\vec a}$ obtained from Eq.~\eqref{eq:ilinear_decomposition} is indeed a $i-k+1$th order normalized $k$-ary function.
To see this, we realize that $q^{(k)}$ is by definition a $i-k+1$th order normalized $k$-ary function $G^{\otimes k}\rightarrow A$, as $q^{(x+1)}$ is obtained from $q^{(x)}$ as in Eq.~\eqref{eq:kary_ithorder_function_definition}, and $q^{(i+1)}=0$.
$q^{(k)}_{\vec a}$ is a restriction of $q^{(k)}$ to the subgroup $G_{a_x}$ in the $x$th component.
It is easy to see that the restriction of a $i-k+1$th order normalized $k$-ary function is again a $i-k+1$th order normalized $k$-ary function.
Finally, we need to argue that plugging Eq.~\eqref{eq:ilinear_decomposition} into Eq.~\eqref{eq:ilinear_composition} does indeed yield $q$.
This follows from inverting Eq.~\eqref{eq:derivative_explicit},
\begin{equation}
\sum_{0\leq k\leq i}  \sum_{0\leq a_0<a_1<\ldots<a_{k-1}<n} q^{(k)}(g_{a_0}|_{a_0},g_{a_1}|_{a_1},\ldots,g_{a_{k-1}}|_{a_{k-1}})
= q(g_0|_0+\ldots+g_{k-1}|_{k-1})
=q(g)\;.
\end{equation}
\end{proof}

\myparagraph{Coefficient groups}
Let us next discuss how to represent $i-k+1$th order $k$-ary functions in terms of explicit coefficients, if each $G_i$ is an elementary abelian group.
That is, we want to find explicit coefficient groups $\homtild_i^k[G_0,\ldots,G_{k-1}|A]$ and coefficient isomorphisms
\begin{equation}
\htild_i^k: \homtild_i^k[G_0,\ldots,G_{k-1}|A]\rightarrow \hom_i^k[G_0,\ldots,G_{k-1}|A]\;.
\end{equation}
We can do this by using the following sequence,
\begin{equation}
\label{eq:higher_order_kary_extension}
\hom_1^1[G_0,\hom_i^{k-1}[G_1,\ldots,G_{k-1}|A]] \xrightarrow{\subset} \hom_i^k[G_0,\ldots,G_{k-1}|A] \xrightarrow{\bullet^{(2)}\times \idop^{\times k-1}}\hom_{i-1}^{k+1}[G_0,G_0,G_1,\ldots,G_{k-1}|A]\;.
\end{equation}
Here, $\subset$ is the injective map denoting the interpretation of the left-hand side as functions $G_0\times\ldots\times G_{k-1}\rightarrow A$.
The resulting $k$-ary function is $i$th order, since taking $i+1$ times the derivative in arbitrary arguments yields zero -- in particular, taking the derivative in the first argument yields zero immediately.
If we restrict the right-hand set to $\img(\bullet^{(2)}\times \idop^{\times k-1})$, the sequence becomes exact.
In particular, we have
\begin{equation}
\img(\subset)=\ker(\bullet^{(2)}\times \idop^{\times k-1})\;.
\end{equation}
Namely, any $h\in \hom_i^k[G_0,\ldots,G_{k-1}|A]$ whose derivative on the first argument is zero must be a homomorphism in the first argument.
The exact sequence allows us to identify $\hom_i^k$ as a twisted product,
\begin{equation}
\hom_i^k[G_0,\ldots,G_{k-1}|A] = \img(\bullet^{(2)}\times \idop^{\times k-1}) \times_{\Omega_{i,k}} \hom_1^1[G_0,\hom_i^{k-1}[G_1,\ldots,G_{k-1}|A]]\;.
\end{equation}
twisted by a group 2-cocycle
\begin{equation}
\Omega_{i,k}\in H^2(B\img(\bullet^{(2)}\times \idop^{\times k-1}),\hom_1^1[G_0,\hom_i^{k-1}[G_1,\ldots,G_{k-1}|A]])\;.
\end{equation}
This allows us to systematically find all elements of $\hom_i^k$ given $\hom_i^{k-1}$ and $\hom_{i-1}^{k+1}$.
Proceeding inductively, we can determine all $\hom_i^k$ from the boundary conditions $\hom_0^k=0$ for $k>0$ and $\hom_k^0=A$ for $k\geq 0$.

\myparagraph{Overall representation}
In order to write down an $i$th order function $q$ explicitly, we need to write the list of coefficients $q^{(\tilde k)}_{\vec a}$ for all the $q^{(k)}_{\vec a}$.
Concretely, we write $q^{(\tilde k)}$ as an $n^{\times k}$ array of coefficients, e.g., $q^{(\tilde0)}$ is just a number, $q^{(\tilde1)}$ is a length-$n$ vector, $q^{(\tilde2)}$ is an $n\times n$ matrix, $q^{(\tilde3)}$ is a vector of matrices, $q^{(\tilde4)}$ is a matrix of matrices, etc.
For each array, the indices ordered from the last to the first correspond to (1) the column number of the innermost matrix, (2) the innermost row number, (3) the next-level column number, (4) the next-level row number, etc.
Note however, that we only need fields with $\vec a_0<\vec a_1\ldots<\vec a_{k-1}$, and we will leave all other entries empty.
In particular, we will only write down the upper triangular part of $q^{(\tilde2)}$ with $n(n-1)/2$ entries.
Similarly, $q^{(\tilde3)}$ is a vector with $n-2$ entries, where the $i$th entry is a triangle-shaped section of a matrix with $i(i+1)/2$ entries.
Note that if additionally $A$ consists of $m$ components, then all the arrays have an extra index of length $m$ which we put in front.
This is the case for the $i-1$st order function $\epsilon$, where $E$ and $G$ play the role of $G$ and $A$:
If $E$ consists of $m$ factors and $G$ of $n$ factors, then $\epsilon^{(\tilde i)}$ is an $n\times m^{\times i}$ array.

\subsection{Examples of \texorpdfstring{$i$}{i}th order tensors}
\label{sec:ilinear_examples}
In this section, we give some examples of $i$th order tensors, focusing mostly on 3rd order tensors over qubits, where $E=\zz_2^m$ and $G=\zz_2^n$.
For $q_\phi$ we need to determine $\homtild_i^k[\ldots|\rr/\zz]$ where all groups in $\ldots$ are equal to $\zz_2$.
From applying Eq.~\eqref{eq:higher_order_kary_extension}, we can only get groups of the form $\homtild_i^k[\zz_2,\ldots,\zz_2|\rr/\zz]=\bigtimes_x \zz_{2^{d_x}}$.
Thus, the left-hand side of Eq.~\eqref{eq:higher_order_kary_extension} is always $\zz_2$, $\homtild_1^1[\zz_2,\homtild_i^{k-1}[\zz_2,\ldots,\zz_2|\rr/\zz]]=\zz_2$.
The group extension is always non-trivial, i.e., twisted by the non-trivial 2-cocycle, such that we find
\begin{equation}
\homtild_i^k[\underbrace{\zz_2,}_{\times k}|\rr/\zz]=\zz_{2^i}\;.
\end{equation}
Concretely, the corresponding $n$th order $k$-ary functions are given by
\begin{equation}
\htild_i^k(h)(g_0,\ldots,g_{k-1}) = \frac{\ovl h}{2^i} \prod_{0\leq x<k} \ovl g_x \mod 1\;.
\end{equation}
For $\epsilon$, we need to determine $\homtild_i^k[\zz_2,\ldots,\zz_2|\zz_2]$.
In this case, $\img(\bullet^{(2)}\times \idop^{\times k-1})$ is always trivial, since every function $\zz_2\rightarrow\zz_2$ is linear.
The left-hand side of Eq.~\eqref{eq:higher_order_kary_extension} is always $\zz_2$, so we have
\begin{equation}
\homtild_i^k[\underbrace{\zz_2,}_{\times k}|\zz_2]=\zz_2\;.
\end{equation}
Explicitly, the $n$th order $k$-ary functions are given by
\begin{equation}
\htild_i^k(h)(g_0,\ldots,g_{k-1})=\ovl h\prod_{0\leq x<k} \ovl g_x \mod 2\;.
\end{equation}

After working out all the coefficient groups, we can discuss concrete examples of 3rd order tensors.
The most paradigmatic example is the $T$ state $\ket T=(1,e^{2\pi i\frac18})$.
It is a 3rd order tensor over $G=\zz_2$ with the following data:
\begin{equation}
E=\zz_2\;,\quad \epsilon=1\;,\quad q_\phi^{(\tilde3)}=\varnothing\;,\quad q_\phi^{(\tilde2)}=\varnothing\;,\quad q_\phi^{(\tilde1)}=\mpm{1}\;,\quad q_\phi^{(\tilde0)}=0\;.
\end{equation}
Here, $\varnothing$ symbolizes that $q_\phi^{(\tilde3)}$ and $q_\phi^{(\tilde2)}$ have no entries, respectively, as there are no sequences $\vec a$ with $0\leq \vec a_0<\vec a_1<\vec a_2<1$.
The $T$ gate $T=\operatorname{diag}(1, e^{2\pi i\frac18})$ is a 3rd order tensor over $\zz_2\times \zz_2$ with the same 3rd order function $q$ but a ``diagonal'' embedding $\epsilon$,
\begin{equation}
E=\zz_2\;,\quad \epsilon^{(\tilde2)}=\varnothing\;,\quad\epsilon^{(\tilde1)}=\mpm{1\\1}\;,\quad \epsilon^{(\tilde0)}=\mpm{0\\0}\;,\quad q_\phi^{(\tilde3)}=\varnothing\;,\quad q_\phi^{(\tilde2)}=\varnothing\;,\quad q_\phi^{(\tilde1)}=(1)\;,\quad q_\phi^{(\tilde0)}=0\;.
\end{equation}
The $CS$ state $\ket{CS}=CS\ket{++}$ is a 3rd order tensor over $\zz_2^2$,
\begin{equation}
\begin{gathered}
E=\zz_2^2\;,\quad \epsilon=1\;,\quad
q^{(\tilde3)}_\phi = \varnothing\;,\quad
q^{(\tilde2)}_\phi = \mpm{\mpm{1}}\;,\quad
q^{(\tilde1)}_\phi = \mpm{0&0}\;,\quad
q^{(\tilde0)}_\phi = 0\;.
\end{gathered}
\end{equation}
The $CS$ gate is the same 3rd order function with a ``diagonal'' embedding similar to the $T$ gate.
The $CCZ$ state $\ket{CCZ}=CCZ\ket{+++}$ is a 3rd order tensor over $G=\zz_2^3$ with data
\begin{equation}
\begin{gathered}
E=\zz_2^3\;,\quad
\epsilon=1\;,\\
q_\phi^{(\tilde3)} = \mpm{\mpm{\mpm{1}}}\;,\quad
q_\phi^{(\tilde2)}=\mpm{0&0\\&0}\;,\quad
q_\phi^{(\tilde1)}=\mpm{0&0&0}\;,\quad
q_\phi^{(\tilde0)}=0\;.
\end{gathered}
\end{equation}
The $CCZ$ gate is again the same 3rd order function with diagonal embedding.
Our next example is the $CCX$, or Toffoli gate, a 3rd order tensor over $\zz_2^6$.
It is related to $CCZ$ by conjugating the third qubit with a Hadamard ($H$), but the mechanism to define it as a 3rd order tensor is completely different:
It is the first example, where we need a true 2nd order embedding function $\epsilon$, whereas the 3rd order function $q$ is trivial.
We have $\bra{o_0o_1o_2}CCX\ket{i_0i_1i_2}=\mathcal T[E,\epsilon,q](i_0,i_1,i_2,o_0,o_1,o_2)$ with the following quadratic tensor data.
$E=\zz_2^3$, and the embedding function $\epsilon:\zz_2^3\rightarrow\zz_2^6$ is given by
\begin{equation}
\epsilon(i_0,i_1,i_2)=(i_0,i_1,i_2,i_0,i_1,i_2+i_0i_1)\;.
\end{equation}
That is, only if $i_0=i_1=1$, the third output $o_2=i_2+i_0i_1$ is negated relative to the third input $i_2$.
The corresponding data coefficients are given as follows,
\begin{equation}
\begin{gathered}
E=\zz_2^3\;,\quad
\epsilon^{(\tilde2)}=\mpm{\mpm{0&0\\&0}&\mpm{0&0\\&0}&\mpm{0&0\\&0}&\mpm{0&0\\&0}&\mpm{0&0\\&0}&\mpm{1&0\\&0}}\;,\\
\epsilon^{(\tilde1)}=\mpm{1&0&0\\0&1&0\\0&0&1\\1&0&0\\0&1&0\\0&0&1}\;,\quad
\epsilon^{(\tilde0)}=\mpm{0\\0\\0\\0\\0\\0}\;,\quad
q_\phi=0\;.
\end{gathered}
\end{equation}
As a next example, consider a controlled-Hadamard gate,
\begin{equation}
CH=
\mpm{
\mpm{1&0\\0&1}&\mpm{0&0\\0&0}\\
\mpm{0&0\\0&0}&\mpm{\sqrt\frac12&\sqrt\frac12\\\sqrt\frac12&-\sqrt\frac12}}
\;.
\end{equation}
This is a 3rd order tensor over $\zz_2^4$.
Contrary to all previous examples, there are non-zero entries with different magnitudes, namely either $1$ or $\sqrt{\frac12}$.
So in order for this to be a 3rd order tensor, we need a non-injective embedding $\epsilon$.
The $1$ entries arise as a sum of $\exp(q)$ over two elements $e\in E$, whereas the $\pm\sqrt{\frac12}$ entries correspond to a single $e\in E$.
We have $\bra{o_0o_1}CH\ket{i_0i_1}=\mathcal T[E,\epsilon,q](i_0,i_1,o_0,o_1)$ with the following quadratic tensor data.
The non-injective embedding is given by $E=\zz_2^3$ and
\begin{equation}
\epsilon(i_0,i_1,x)=(i_0,i_1,i_0,i_1+xi_0)\;.
\end{equation}
The index configurations $g$ with a $1$ entry are of the form $g=(0,i_1,0,i_1)$ and have two elements in the pre-image of $\epsilon$, $(0,i_1,0)$ and $(0,i_1,1)$.
The index configurations with a $0$ entry are of the form $(i_0,i_1,i_0+1,o_1)$ or $(0,i_0,0,i_0+1)$, and have no elements in the pre-image.
The index configurations with a $\pm\sqrt{\frac12}$ entry are of the form $(1,i_1,1,o_1)$, and have exactly one element of the pre-image, namely $(1,i_1,i_1+o_1)$.
Now, to get all the phases right requires a bit of fine-tuning, but it can be achieved via the following 3rd order function,
\begin{equation}
q_\phi(i_0,i_1,x)=-\frac18+\frac14 (1-i_0)x+\frac18 i_0 + \frac12 i_0i_1(1-x)\;.
\end{equation}
The representing coefficients are given by
\begin{equation}
\begin{gathered}
E=\zz_2^3\;,\quad
\epsilon^{(\tilde2)}=\mpm{\mpm{0&0\\&0}&\mpm{0&0\\&0}&\mpm{0&0\\&0}&\mpm{0&1\\&0}}\;,\quad
\epsilon^{(\tilde1)}=\mpm{1&0&0\\0&1&0\\1&0&0\\0&1&0}\;,\\
q_\phi^{(\tilde3)} = \mpm{\mpm{\mpm{1}}}\;,\quad
q_\phi^{(\tilde2)}=\mpm{2&3\\&0}\;,\quad
q_\phi^{(\tilde1)}=\mpm{1&0&2}\;,\\
q_\phi^{(\tilde0)}=-\frac18\;,\quad
q_a^{(\tilde0)}=-\frac12\log(2)\;.
\end{gathered}
\end{equation}
Let us next give two simpler examples, without direct physics application, where a non-injective embedding $\epsilon$ can be used to create tensors where not all entries have the same magnitude.
The first example is the vector $(3,1)$.
This is a 3rd order tensor over $\zz_2$ with data
\begin{equation}
E=\zz_2^2\;,\quad
\epsilon^{(\tilde2)}=\mpm{\mpm{1}}\;,\quad
\epsilon^{(\tilde1)}=\mpm{0&0}\;,\quad
\epsilon^{(\tilde0)}=0\;,\quad
q_\phi=0\;.
\end{equation}
The second example is the vector $(4,2)$, a 3rd order tensor over $\zz_2$ with data
\begin{equation}
\begin{gathered}
E=\zz_2^3\;,\quad
\epsilon^{(\tilde2)}=\mpm{\mpm{0&0\\&0}}\;,\quad
\epsilon^{(\tilde1)}=\mpm{1&1&1}\;,\quad
\epsilon^{(\tilde0)}=\mpm{0}\;,\\
q_\phi^{(\tilde3)} = \mpm{\mpm{1}}\;,\quad
q_\phi^{(\tilde2)}=\mpm{0&0\\&0}\;,\quad
q_\phi^{(\tilde1)}=\mpm{0&0&0}\;,\quad
q_\phi^{(\tilde0)}=0\;.
\end{gathered}
\end{equation}
As a last qubit example, consider the quantum Fourier transform on two qubits,
\begin{equation}
\bra{o_0o_1}FT\ket{i_0i_1}=e^{2\pi i(\frac12 i_1o_1+\frac14 i_0o_1+\frac14 i_1o_0)}\;,
\end{equation}
which is a 3rd order tensor over $\zz_2^4$.
We have $\bra{o_0o_1}FT\ket{i_0i_1}=\mathcal T[E,\epsilon,q](i_0,i_1,o_0,o_1)$ with
\begin{equation}
\begin{gathered}
E=\zz_2^4\;,\quad
\epsilon=1\;,\\
q_\phi^{(\tilde3)}=\mpm{\mpm{0}&\mpm{0&0\\&0}}\;,\quad
q_\phi^{(\tilde2)}=\mpm{0&0&1\\&1&2\\&&0}\;,\quad
q_\phi^{(\tilde1)}=\mpm{0&0&0&0}\;,\\
q_\phi^{(\tilde0)}=0\;,\quad
q_a^{(\tilde0)}=-\log(2)\;.
\end{gathered}
\end{equation}

Let us now consider one more example with qudits.
The vector $(1,1,0,0)$ is a 3rd order tensor over $G=\zz_4$.
The 3rd order function $q$ is trivial, and the embedding is given by
\begin{equation}
E=\zz_2\;,\quad \epsilon(g)=\ovl g\mmod 4\;.
\end{equation}
Indeed, $\epsilon$ is a quadratic function $\zz_2\rightarrow \zz_4$.

Our final example is a larger family of operators, namely all diagonal operators in the Clifford hierarchy.
The $i$th level of the Clifford hierarchy is recursively defined as the set of operators that map all Pauli operators to an operator in the $i-1$st level of the Clifford hierarchy under conjugation.
The first level is given by the Pauli operators, and consequently the second level is formed by the Clifford operations.
\begin{myprop}
\label{prop:diagonal_hierarchy}
Consider a diagonal operator $U$ in the $i$th level of the Clifford hierarchy over an arbitrary abelian group $H$.
Then $U$ is an $i$th order tensor over $G=H\times H$, with a ``diagonal'' embedding $\epsilon$:
\begin{equation}
U=\mathcal T[E,\epsilon,q]\;,\qquad
E=H\;,\quad
\epsilon(h)=(h,h)\;,
\end{equation}
where $q$ is an $i$th order function over $H$.
\end{myprop}
\begin{proof}
We will show by induction over $i$ that an operator of the form
\begin{equation}
U_{hh'}=\delta_{h=h'+g} e^{2\pi i q(h')}
\end{equation}
for some $g\in H$ is in the $i$th level of the Clifford hierarchy if and only if $q$ is an $i$th order function, $q\in F_i[H|\rr/\zz]$.
If $q\in F_1[H|\rr/\zz]$, then $q=x+x_0$ for $x\in \hom[H|\rr/\zz]$ and a constant $x_0\in\rr/\zz$, and $U=e^{2\pi i x_0}\rho(g,x)$ is a Pauli operator.
So the statement holds for $i=1$.
Now assume that the statement holds for $i-1$.
The conjugation action of $U$ on a Pauli operator $\rho(h,h')$ is
\begin{equation}
(U^\dagger\rho(h,h')U)_{jk}
= e^{2\pi i q(k)} e^{-2\pi i q(k+h)} e^{2\pi i h'(g+k)} \delta_{j=k+h}
= e^{2\pi i(-q(k+h)+q(k)+h'(k)+h'(g))}\delta_{j=k+h}\;.
\end{equation}
If $q$ is a $i$th order function, then
\begin{equation}
\widetilde q(k)
\coloneqq -q(k+h)+q(k)+h'(k)+h'(g)
= -q^{(2)}(k,h)-q(h)-q(0)+h'(k)+h'(g)
\end{equation}
is a $i-1$th order function, since $q^{(2)}$ is $i-1$st order in the first component and $h'$ is linear (first order).
So $U$ is in the $i$th level, as it maps Pauli operators onto level-$i-1$ operators.
Vice versa, if $U$ is in the $i$th level, then $\widetilde q$ is an $i-1$th order function.
So $q^{(2)}$ is $i-1$th order in the first component, and thus $q$ is $i$th order.
So if the statement holds for $i-1$, then it holds for $i$.
\end{proof}

\section{Conclusion and outlook}
\myparagraph{Conclusion}
We have seen that many mathematical objects related to Clifford circuits, stabilizer codes, or free-fermion and free-boson models can be represented in a unified way as quadratic tensors.
Quadratic tensors are specified by a linear embedding map and a quadratic function over a (co)commutative (super) Hopf algebra, the linearization of an abelian group.
We have shown how to decompose quadratic tensors into smaller components using only the algebraic axioms, both on the level of abelian groups and diagrammatically on the level of Hopf algebras.
We have found that quadratic tensors can be explicitly and systematically manipulated on paper or in computer memory using coefficient groups $\homtild$ and coefficient isomorphisms $\htild$.
Apart from contractions over $\zz$ factors, tensor networks of quadratic tensors can be efficiently contracted.
We have introduced simple algebraic definitions of stabilizer codes and Clifford operations that work for arbitrary abelian groups.
We have shown that the associated stabilizer states, Clifford unitaries, and other objects, are quadratic tensors, and how to explicitly construct the underlying coefficients.
Finally, we have generalized from quadratic (second order) tensors to $i$th order tensors, which are based on an ($i-1$)th order embedding function and an $i$th order entry function.
We have shown that many common non-Clifford gates are 3rd order tensors, and that diagonal operators in the $i$th level of the Clifford hierarchy (for arbitrary abelian groups) are $i$th order tensors.

\myparagraph{Future directions}
Let us discuss some future directions.
A first direction is to implement the contraction of general quadratic tensor networks numerically.
This is especially interesting in the case of mixed degrees of freedom such as composite-dimensional qudits, where software tools are rare.
Our work makes such an implementation relatively straightforward, though it would be beneficial to work out the general composition of a quadratic function $\widetilde q$ with a lift $\rho^\perp$ of a quotient map, which is needed to evaluate Eq.~\eqref{eq:invertiblered}.

Another direction is to further work out the case of free-fermion quadratic tensors.
We have shown that quadratic functions over general (super) Hopf algebras can be decomposed on a purely diagrammatic level.
We have also seen that quadratic tensors over the free-fermion Hopf algebra $\mathcal F$ can be contracted efficiently using the Schur complement, just as in the abelian-group case.
However, it would be nice to show that the contraction is efficient directly on a diagrammatic level, which would unify the contraction for abelian groups and free fermions.
This generalized diagrammatic contraction could also help to unify the two reduction procedures in Propositions~\ref{prop:invert_reduction} and \ref{prop:complex_invert_reduction}.

A further interesting question is whether there is a method to contract tensor networks that also include some higher-order tensors.
Such a contraction cannot be efficient in general since higher-order tensors are universal in the sense that they can approximate arbitrary tensor networks.
It is possible to contract tensor networks with only a constant number of non-quadratic tensors:
We first contract the network without the non-quadratic tensors, and then contract the non-quadratic tensors in the ordinary way without making use of any quadratic structure.
The exponential runtime and memory of the ordinary contraction is not a problem in theory if it is of constant size.
However, there may be a better method if the non-quadratic tensors are $i$th order tensors for some small $i>2$.
For certain tensor networks, the higher-order contributions may also cancel each other in such a way that contraction is still possible with a large number of higher-order tensors.
Examples for this can be found in quantum error correction, where a single logical non-Clifford gate is compiled into a large fault-tolerant circuit.
The logical action of the large circuit is known by construction, despite the fact that it contains a large number of non-Clifford gates.
It would be nice to have a method that implicitly detects when the higher-order contributions cancel each other, and performs contraction efficiently if they do.

Finally, we have shown that Clifford operations correspond to quadratic tensors that are unitary operators.
In Proposition~\ref{prop:diagonal_hierarchy}, we show that all diagonal operators in the $i$th level of the Clifford hierarchy are $i$th order tensors with ``diagonal'' embedding.
It is not true in general that unitary $i$th order tensors correspond to operators in the $i$th level of the Clifford hierarchy:
$\geq 3$rd order tensors are universal, so any operator is a $3$rd order tensor, at least approximately.
However, there could be a correspondence if we only allow a restricted set of embeddings $\epsilon$.
It does not suffice to restrict ourselves to injective embeddings, as we have seen some examples for 3rd level gates that need non-injective embeddings in Section~\ref{sec:ilinear_examples}, such as $CH$.

\myparagraph{Acknowledgements}
We'd like to thank Julio Magdalena de la Fuente for comments on the draft.
This work was supported by the U.S. Army Research Laboratory and the U.S. Army Research Office under contract/grant number W911NF2310255, and by the U.S. Department of Energy, Office of Science, National Quantum Information Science Research Centers, and the Co-design Center for Quantum Advantage (C2QA) under contract number DE-SC0012704.

\bibliography{quadratic_tensors_refs}

\appendix

\section{Verification of \texorpdfstring{Eqs.~\eqref{eq:hom21_precomposition} and \eqref{eq:hom21_precomposition_rr}}{equations}}
\label{sec:precompositon_psi_derivation}
In this appendix, we verify the tables in Eqs.~\eqref{eq:hom21_precomposition} and \eqref{eq:hom21_precomposition_rr}, calculating the value of $\Phi[G,H|A]((h_2,h_1),\gamma)$ as defined in Eq.~\eqref{eq:precomposition_phi}.
In our derivations, we will use $\floor{x}$ for $x\in \rr$ to denote the largest integer smaller or equal to $x$, and we will use the identity
\begin{equation}
\ovl{xk\mmod k} = (x-\floor{x})k\;.
\end{equation}
We also use that in the case $A=\rr/\zz$, the equations are valued $\mmod 1$ and therefore any integer summands can be discarded.
Below we list the non-trivial cases.
\begin{itemize}
\item $A=\rr/\zz$, $H=\zz_m$ even, and $G=\zz_k$ even,
\begin{equation}
\label{eq:precomposition_even_even}
\begin{gathered}
\htild_2(h_2,h_1)(\htild(\gamma)(h))\\
=\frac1k \Big(\frac12 \ovl h_2 \ovl{(\ovl\gamma \ovl h k/\gcd(m,k) \mmod k)}^2 + \ovl h_1 \big(-\ovl{(\ovl\gamma \ovl h k/\gcd(m,k) \mmod k)}^2+\ovl{(\ovl\gamma\ovl hk/\gcd(m,k) \mmod k)}\big)\Big)\\
=\frac1k \Big(\frac12 \ovl h_2 \big(\frac{\ovl\gamma \ovl h}{\gcd(m,k)}-\floor{\frac{\ovl\gamma \ovl h}{\gcd(m,k)}}\big)^2k^2 + \ovl h_1 \big(-\big(\frac{\ovl\gamma \ovl h}{\gcd(m,k)}-\floor{\frac{\ovl\gamma \ovl h}{\gcd(m,k)}}\big)^2k^2+ \big(\frac{\ovl\gamma \ovl h}{\gcd(m,k)}-\floor{\frac{\ovl\gamma \ovl h}{\gcd(m,k)}}\big)k\big)\Big)\\
=\frac1k \Big(\frac12 \ovl h_2 \big(\frac{\ovl\gamma \ovl h}{\gcd(m,k)}\big)^2k^2 + \ovl h_1 \big(-\big(\frac{\ovl\gamma \ovl h}{\gcd(m,k)}\big)^2k^2+ \big(\frac{\ovl\gamma \ovl h}{\gcd(m,k)}\big)k\big)\Big)\\
=\frac1m \Big(\frac12 \big((\ovl h_2-2\ovl h_1)\ovl\gamma^2 \frac{mk}{\gcd(m,k)^2} + 2\ovl h_1\ovl\gamma\frac{m}{\gcd(m,k)}\big)\ovl h^2+\ovl h_1\ovl\gamma \frac{m}{\gcd(m,k)}\big(-\ovl h^2+\ovl h\big)\Big)\\
=\htild_2\Big((\ovl h_2-2\ovl h_1)\ovl\gamma^2 \frac{mk}{\gcd(m,k)^2} + 2\ovl h_1\ovl\gamma\frac{m}{\gcd(m,k)},\quad \ovl h_1\ovl\gamma \frac{m}{\gcd(m,k)}\Big)(h)\;.
\end{gathered}
\end{equation}
\item $A=\rr/\zz$, $H=\zz_m$ odd, and $G=\zz_k$ even,
\begin{equation}
\begin{gathered}
\htild_2(h_2,h_1)(\htild(\gamma)(h))\\
=\frac1k \Big(\frac12 \ovl h_2 \big(\frac{\ovl\gamma \ovl h}{\gcd(m,k)}\big)^2k^2 + \ovl h_1 \big(-\big(\frac{\ovl\gamma \ovl h}{\gcd(m,k)}\big)^2k^2+ \big(\frac{\ovl\gamma \ovl h}{\gcd(m,k)}\big)k\big)\Big)\\
=\frac1m \Big((\frac12 \ovl h_2-\ovl h_1)\ovl\gamma^2 \frac{mk}{\gcd(m,k)^2}\ovl h^2+\ovl h_1\ovl\gamma \frac{m}{\gcd(m,k)}\ovl h\big)\Big)\\
=\htild_2\Big((\ovl h_2-2\ovl h_1)\ovl\gamma^2 \frac{mk}{\gcd(m,k)^2} ,\quad \ovl h_1\ovl\gamma \frac{m}{\gcd(m,k)}\Big)(h)\;.
\end{gathered}
\end{equation}
The first few steps are identical to Eq.~\eqref{eq:precomposition_even_even} and have been omitted.
\item $A=\rr/\zz$, $H=\zz$, and $G=\zz_k$ even,
\begin{equation}
\begin{gathered}
\htild_2(h_2,h_1)(\htild(\gamma)(h))\\
=\frac1k \Big(\frac12 \ovl h_2 \ovl{(\ovl\gamma h \mmod k)}^2 + \ovl h_1 \big(-\ovl{(\ovl\gamma h \mmod k)}^2+\ovl{(\ovl\gamma h \mmod k)}\big)\Big)\\
=\frac1k \Big(\frac12 \ovl h_2 \big(\frac{\ovl\gamma h}{k}-\floor{\frac{\ovl\gamma h}{k}}\big)^2k^2 + \ovl h_1 \big(-\big(\frac{\ovl\gamma h}{k}-\floor{\frac{\ovl\gamma h}{k}}\big)^2k^2+ \big(\frac{\ovl\gamma h}{k}-\floor{\frac{\ovl\gamma h}{k}}\big)k\big)\Big)\\
=\frac1k \Big(\frac12 \ovl h_2 \ovl\gamma^2 h^2 + \ovl h_1 \big(-\ovl\gamma^2 h^2+ \ovl\gamma h\big)\Big)\\
=\frac1k \big((\frac12 \ovl h_2-\ovl h_1) \ovl\gamma^2+\ovl h_1\ovl\gamma\big) h^2 + \frac12\frac2k \ovl h_1 \ovl\gamma (-h^2+ h)\Big)\\
=\htild_2\Big(\frac1k(\frac12 \ovl h_2-\ovl h_1) \ovl\gamma^2+\ovl h_1\ovl\gamma,\quad \frac2k \ovl h_1 \ovl\gamma\Big)(h)\;.
\end{gathered}
\end{equation}
\item $A=\rr/\zz$, $H=\zz_m$ even, and $G=\zz_k$ odd,
\begin{equation}
\label{eq:precomposition_even_odd}
\begin{gathered}
\htild_2(h_2,h_1)(\htild(\gamma)(h))\\
=\frac1k \Big(\ovl{(h_2/2)} \ovl{(\ovl\gamma \ovl h k/\gcd(m,k) \mmod k)}^2 + \ovl h_1 \ovl{(\ovl\gamma\ovl hk/\gcd(m,k) \mmod k)}\Big)\\
=\frac1k \Big(\ovl{(h_2/2)} \big(\frac{\ovl\gamma \ovl h}{\gcd(m,k)}-\floor{\frac{\ovl\gamma \ovl h}{\gcd(m,k)}}\big)^2k^2 + \ovl h_1 \big(\frac{\ovl\gamma \ovl h}{\gcd(m,k)}-\floor{\frac{\ovl\gamma \ovl h}{\gcd(m,k)}}\big)k\Big)\\
=\frac1k \Big(\ovl{(h_2/2)} \big(\frac{\ovl\gamma \ovl h}{\gcd(m,k)}\big)^2k^2 + \ovl h_1 \frac{\ovl\gamma \ovl h}{\gcd(m,k)}k\Big)\\
=\frac1m \Big(\frac12\big(2\ovl{(h_2/2)}\ovl\gamma^2 \frac{mk}{\gcd(m,k)^2} + 2\ovl h_1\ovl\gamma\frac{m}{\gcd(m,k)}\big)\ovl h^2+\ovl h_1\ovl\gamma \frac{m}{\gcd(m,k)}\big(-\ovl h^2+\ovl h\big)\Big)\\
=\htild_2\Big(2\ovl{(h_2/2)}\ovl\gamma^2 \frac{mk}{\gcd(m,k)^2} + 2\ovl h_1\ovl\gamma\frac{m}{\gcd(m,k)},\quad \ovl h_1\ovl\gamma \frac{m}{\gcd(m,k)}\Big)(h)\;.
\end{gathered}
\end{equation}
\item $A=\rr/\zz$, $H=\zz_m$ odd, and $G=\zz_k$ odd,
\begin{equation}
\begin{gathered}
\htild_2(h_2,h_1)(\htild(\gamma)(h))\\
=\frac1k \Big(\ovl{(h_2/2)} \big(\frac{\ovl\gamma \ovl h}{\gcd(m,k)}\big)^2k^2 + \ovl h_1 \frac{\ovl\gamma \ovl h}{\gcd(m,k)}k\Big)\\
=\frac1m \Big(\ovl{(h_2/2)}\ovl\gamma^2 \frac{mk}{\gcd(m,k)^2}\ovl h^2+\ovl h_1\ovl\gamma \frac{m}{\gcd(m,k)}\ovl h\Big)\\
=\htild_2\Big(2\ovl{(h_2/2)}\ovl\gamma^2 \frac{mk}{\gcd(m,k)^2},\quad \ovl h_1\ovl\gamma \frac{m}{\gcd(m,k)}\Big)(h)\;.
\end{gathered}
\end{equation}
Here, we have omitted the first steps which are identical to Eq.~\eqref{eq:precomposition_even_odd}.
\item $A=\rr/\zz$, $H=\zz$, and $G=\zz_k$ odd,
\begin{equation}
\begin{gathered}
\htild_2(h_2,h_1)(\htild(\gamma)(h))\\
=\frac1k \Big(\ovl{(h_2/2)} \ovl{(\ovl\gamma h \mmod k)}^2 + \ovl h_1 \ovl{(\ovl\gamma h \mmod k)}\Big)\\
=\frac1k \Big(\ovl{(h_2/2)} \big(\frac{\ovl\gamma h}{k}-\floor{\frac{\ovl\gamma h}{k}}\big)^2k^2 + \ovl h_1 \big(\frac{\ovl\gamma h}{k}-\floor{\frac{\ovl\gamma h}{k}}\big)k\Big)\\
=\frac1k \Big(\ovl{(h_2/2)} \ovl\gamma^2 h^2 + \ovl h_1 \ovl\gamma h\Big)\\
= \frac1k\big(\ovl{(h_2/2)}\ovl\gamma^2 + \ovl h_1\ovl\gamma\big)h^2+\frac12\frac2k\ovl h_1\ovl\gamma \big(-h^2+h\big)\Big)\\
=\htild_2\Big(\frac1k\big(\ovl{(h_2/2)}\ovl\gamma^2 + \ovl h_1\ovl\gamma\big),\quad \frac2k\ovl h_1\ovl\gamma\Big)(h)\;.
\end{gathered}
\end{equation}
\item $A=\rr/\zz$, $H=\zz$ and $G=\zz$,
\begin{equation}
\begin{gathered}
\htild_2(h_2,h_1)(\htild(\gamma)(h))\\
=\ovl h_2 (\gamma h)^2 + \frac12\ovl h_1 \big(-(\gamma h)^2+\gamma h\big)\\
=\big((\ovl h_2-\frac12 h_1)\gamma^2 +\frac12\ovl h_1\gamma\big) h^2 + \frac12\ovl h_1\gamma (-h^2+ h)\\
=\htild_2\Big((\ovl h_2-\frac12 h_1)\gamma^2 +\frac12\ovl h_1\gamma,\quad \ovl h_1 \gamma\Big)(h)\;.
\end{gathered}
\end{equation}
\item $A=\rr/\zz$, $H=\zz_m$ even, and $G=\rr/\zz$,
\begin{equation}
\begin{gathered}
\htild_2(h_2,h_1)(\htild(\gamma)(h))
= h_1 \ovl{(\frac1m\ovl \gamma \ovl h\mmod 1)}
= \frac1m h_1 \ovl\gamma \ovl h
= \frac12\frac2m h_1 \ovl\gamma \ovl h^2 + \frac1m h_1 \ovl\gamma (-\ovl h^2+\ovl h)\\
=\htild_2\Big(\frac2m h_1 \ovl\gamma,\quad \frac1m h_1 \ovl\gamma\Big)(h)\;.
\end{gathered}
\end{equation}
\item $A=\rr/\zz$, $H=\zz_m$ odd, and $G=\rr/\zz$,
\begin{equation}
\begin{gathered}
\htild_2(h_2,h_1)(\htild(\gamma)(h))
= h_1 \ovl{(\frac1m\ovl \gamma \ovl h\mmod 1)}
= \frac1m h_1 \ovl\gamma \ovl h
=\htild_2\Big(0,\quad \frac1m h_1 \ovl\gamma\Big)(h)\;.
\end{gathered}
\end{equation}
\item $A=\rr/\zz$, $H=\zz$ and $G=\rr/\zz$,
\begin{equation}
\begin{gathered}
\htild_2(h_2,h_1)(\htild(\gamma)(h))
= h_1 \ovl{(\ovl\gamma h\mmod 1)}
= h_1 \ovl\gamma h
= h_1\ovl\gamma h^2+ \frac12 2h_1\ovl\gamma(-h^2+h)
=\htild_2\Big(h_1\ovl\gamma,\quad 2 h_1 \ovl\gamma\Big)(h)\;.
\end{gathered}
\end{equation}
\item $A=\rr/\zz$, $H=\rr/\zz$, and $G=\rr/\zz$,
\begin{equation}
\begin{gathered}
\htild_2(h_2,h_1)(\htild(\gamma)(h))
= h_1 \ovl{(\gamma \ovl h\mmod 1)}
= h_1 \gamma \ovl h
=\htild_2\Big(0,\quad h_1\gamma\Big)(h)\;.
\end{gathered}
\end{equation}
\item $A=\rr/\zz$, $H=\rr$, and $G=\rr/\zz$,
\begin{equation}
\begin{gathered}
\htild_2(h_2,h_1)(\htild(\gamma)(h))
= h_1 \ovl{(\gamma h\mmod 1)}
= h_1 \gamma h
=\htild_2\Big(0,\quad h_1\gamma\Big)(h)\;.
\end{gathered}
\end{equation}
\item $A=\rr/\zz$, $H=\zz$, and $G=\rr$,
\begin{equation}
\begin{gathered}
\htild_2(h_2,h_1)(\htild(\gamma)(h))
= \frac12 h_2 (\gamma h)^2 + h_1 \gamma h\\
= (\frac12 h_2 \gamma^2+h_1 \gamma) h^2 + \frac12 2h_1 \gamma (-h^2+h)
=\htild_2\Big(\frac12 h_2 \gamma^2+h_1 \gamma,\quad 2h_1\gamma\Big)(h)\;.
\end{gathered}
\end{equation}
\item $A=\rr/\zz$, $H=\rr$ and $G=\rr$,
\begin{equation}
\label{eq:precomposition_rr_rr}
\begin{gathered}
\htild_2(h_2,h_1)(\htild(\gamma)(h))
= \frac12 h_2 (\gamma h)^2 + h_1 \gamma h
= \frac12 h_2 \gamma^2 h^2 + h_1 \gamma h
=\htild_2\Big(h_2 \gamma^2,\quad h_1\gamma\Big)(h)\;.
\end{gathered}
\end{equation}
\item $A=\rr$.
There are only three non-trivial cases, $H=\zz$ and $G=\zz$, $H=\zz$ and $G=\rr$, as well as $H=\rr$ and $G=\rr$.
In all of these cases, the computation looks identical to the one in Eq.~\eqref{eq:precomposition_rr_rr}, apart from that some of the variables $h_1$, $h_2$, $h$, and $\gamma$ may be $\zz$ instead of $\rr$ valued.
\end{itemize}

\section{Verification of \texorpdfstring{Eq.~\eqref{eq:precomposition_lambda_table}}{equation}}
\label{sec:lambda_verification}
In this appendix, we confirm the non-trivial entries of the table in Eq.~\eqref{eq:precomposition_lambda_table}, calculating the value of $\Lambda[G_0|A](h)$ defined in Eq.~\eqref{eq:lambda_definition}.
We find
\begin{itemize}
\item $G=\zz_k$ even and $A=\zz_l$ even,
\begin{equation}
\htild^2(h)(g,g)
=\frac{l}{\gcd(k,l)}\ovl h \ovl g \ovl g
=\frac{l}{\gcd(k,\frac{l}{2})}\frac12 \frac{2\gcd(k,\frac{l}{2})}{\gcd(k,l)}\ovl h \ovl g^2
= \htild_2(\frac{2\gcd(k,\frac{l}{2})}{\gcd(k,l)}\ovl h\mmod 2\gcd(k,\frac{l}{2}),0)(g)\;.
\end{equation}
\item $G=\zz_k$ odd and $A=\zz_l$ even,
\begin{equation}
\label{eq:lambda_zzkoddzzleven}
\htild^2(h)(g,g)
=\frac{l}{\gcd(k,l)}\ovl h \ovl g \ovl g
=\frac{l}{\gcd(k,l)}\Big(\ovl{\big((2\ovl h\mmod \gcd(k,l))/2\big)} \ovl g^2+0\ovl g\Big)
= \htild_2(2\ovl h\mmod \gcd(k,l),0)(g)\;.
\end{equation}
\item $G=\zz$ and $A=\zz_l$ even,
\begin{equation}
\htild^2(h)(g,g)
=\ovl h g g
=\frac12 2\ovl h g(g+1)-\ovl h g
=\frac12 \ovl{(2\ovl h \mmod l)}g(g+1)-\ovl h g
= \htild_2(2\ovl h\mmod l,-h)(g)\;.
\end{equation}
\item For $G=\zz_k$ even and $A=\zz_l$ odd, the derivation looks identical to Eq.~\eqref{eq:lambda_zzkoddzzleven}.
\item For $G=\zz_k$ odd and $A=\zz_l$ odd, the derivation looks identical to Eq.~\eqref{eq:lambda_zzkoddzzleven}.
\item $G=\zz$ and $A=\zz_l$ odd,
\begin{equation}
\htild^2(h)(g,g)
=\ovl h g g
=\ovl{\big((2\ovl h\mmod l)/2\big)} g^2+0 g
= \htild_2(2\ovl h\mmod l,0)(g)\;.
\end{equation}
\item $G=\zz$ and $A=\zz$,
\begin{equation}
\htild^2(h)(g,g)
=hgg
=\frac12 (2h)g(g+1)-hg
= \htild_2(2h,-h)(g)\;.
\end{equation}
\item $G=\zz_k$ even and $A=\rr/\zz$,
\begin{equation}
\htild^2(h)(g,g)
= \frac1k \ovl h\ovl g\ovl g
= \frac1k(\frac12 2\ovl h\ovl g^2+0(-\ovl g^2+\ovl g))
= \htild_2(2\ovl h\mmod 2k,0)(g)\;.
\end{equation}
\item $G=\zz_k$ odd and $A=\rr/\zz$,
\begin{equation}
\htild^2(h)(g,g)
= \frac1k \ovl h\ovl g\ovl g
= \frac1k(\ovl{((2\ovl h \mmod k)/2)} \ovl g^2+0\ovl g)
= \htild_2(2\ovl h \mmod k,0)(g)\;.
\end{equation}
\item $G=\zz$ and $A=\rr/\zz$,
\begin{equation}
\htild^2(h)(g,g)
= \ovl h g g
= \ovl h g^2+\frac12 0 (-g^2+g)
= \htild_2(h,0)(g)\;.
\end{equation}
\item $G=\rr$ and $A=\rr/\zz$,
\begin{equation}
\label{eq:quadratic_copy_bilinear_rr_rrzz}
\htild^2(h)(g,g)
= h g g
= \frac12 (2h) g^2+0g
= \htild_2(2h,0)(g)\;.
\end{equation}
\item For $G=\zz$ and $A=\rr$, and $G=\rr$ and $A=\rr$, the derivation looks identical to Eq.~\eqref{eq:quadratic_copy_bilinear_rr_rrzz}.
\end{itemize}

\section{Quadratic functions as twisted product group}
\label{sec:2cocycle}
In this section, we shed some more light on how to determine the coefficient groups $\homtild_2[G|A]$ and coefficient isomorphisms $\htild_2[G|A]$ in Section~\ref{sec:coefficient_groups}.
We start by finding explicit coefficient groups $\homtild^{2s}[G|A]$ and coefficient isomorphisms
\begin{equation}
\htild^{2s}[G|A]: \homtild^{2s}[G|A]\rightarrow \img(\bullet^{(2)})\subset\hom^2[G,G|A]\;.
\end{equation}
The following table lists $\homtild^{2s}[G|A]$ for the cases where $\homtild^{2s}[G|A]= \homtild^2[G,G|A]$ and shows ``$\homtild^{2s}[G|A]\subset \homtild^2[G,G|A]$'' in the one other case ($\zz_k$ even and $\zz_l$ even):
\begin{equation}
\label{eq:hom21_groups}
\begin{tabular}{r|c|c|c|c|c}
\diagbox{$\scriptstyle{G}$}{$\scriptstyle{A}$}
 & $\zz_l$ even & $\zz_l$ odd & $\zz$ & $\rr/\zz$ & $\rr$\\
\hline
$\zz_k$ even &$\zz_{\gcd(k,\frac l2)}\subset \zz_{\gcd(k,l)}$ & $\zz_{\gcd(k,l)}$ & $0$ & $\zz_k$ & $0$\\
\hline
$\zz_k$ odd & $\zz_{\gcd(k,l)}$ & $\zz_{\gcd(k,l)}$ & $0$ & $\zz_k$ & $0$\\
\hline
$\zz$ & $\zz_l$ & $\zz_l$ & $\zz$ & $\rr/\zz$ &  $\rr$\\
\hline
$\rr/\zz$ & $0$ & $0$ & $0$ & $0$ & $0$\\
\hline
$\rr$ & $0$ & $0$ & $0$ & $\rr$ & $\rr$
\end{tabular}\;.
\end{equation}
The corresponding coefficient isomorphism $\htild^{2s}[G|A]$ is given by
\begin{equation}
\label{eq:hom21_morph}
\begin{tabular}{r|c|c|c|c|c}
\diagbox{$\scriptstyle{G}$}{$\scriptstyle{A}$}
 & $\zz_l$ even & $\zz_l$ odd & $\zz$ & $\rr/\zz$ & $\rr$\\
\hline
$\zz_k$ even & $\ovl h \frac{l}{\gcd(k,\frac l2)} \ovl g_0\ovl g_1$ & $\ovl h \frac{l}{\gcd(k,l)} \ovl g_0\ovl g_1$ & $0$ & $\ovl h\frac1k \ovl g_0\ovl g_1$ & $0$\\
\hline
$\zz_k$ odd & $\ovl h \frac{l}{\gcd(k,l)} \ovl g_0\ovl g_1$ & $\ovl h \frac{l}{\gcd(k,l,m)} \ovl g_0\ovl g_1$ & $0$ & $\ovl h\frac1k \ovl g_0\ovl g_1$ & $0$\\
\hline
$\zz$ &  $\ovl h g_0g_1$ &  $\ovl h g_0g_1$ & $hg_0g_1$ & $\ovl h g_0 g_1$ &  $hg_0g_1$\\
\hline
$\rr/\zz$ & $0$ & $0$ & $0$ & $0$ & $0$\\
\hline
$\rr$ & $0$ & $0$ & $0$ & $hg_0g_1$ & $hg_0g_1$
\end{tabular}\;.
\end{equation}
Both tables can be read off from the diagonals of the blocks of Eqs.~\eqref{eq:2hom_rr_table} and \eqref{eq:2hom_iso_rr_table}, except for the top left entry.

Next, we explicitly choose a ``standard quadratic function'' $Q_0(h)\in \hom_2[G|A]$ for each bilinear form $h\in \homtild^{2s}[G|A]$ such that
\begin{equation}
\label{eq:standard_function_definition}
Q_0(h)^{(2)}=\htild^{2s}(h)\;.
\end{equation}
In other words, we choose a lift of $\bullet^{(2)}$ on the level of coefficients.
The following table lists $Q_0(h)(g)$ for different choices of elementary abelian groups $G$ and $A$:
\begin{equation}
\label{eq:standard_function_elementary}
\begin{tabular}{r|c|c|c|c|c}
\diagbox{$\scriptstyle{G}$}{$\scriptstyle{A}$}
 & $\zz_l$ even & $\zz_l$ odd & $\zz$ & $\rr/\zz$ & $\rr$\\
\hline
$\zz_k$ even & $\frac{l}{\gcd(k,\frac{l}{2})}\frac12\ovl h \ovl g^2$ & $\frac{l}{\gcd(k,l)}\ovl{(h/2)} \ovl g^2$ & $0$ & $\frac1k \frac12 \ovl h \ovl g^2$ & $0$\\
\hline
$\zz_k$ odd & $\frac{l}{\gcd(k,l)} \ovl{(h/2)} \ovl g^2$ & $\frac{l}{\gcd(k,l)} \ovl{(h/2)} \ovl g^2$ & $0$ & $\frac{1}{k} \ovl{(h/2)} \ovl g^2$ & $0$\\
\hline
$\zz$ & $\frac12 \ovl h g(g+1)$ & $\ovl{(h/2)} g^2$ & $\frac12 h g(g+1)$ & $\frac12 \ovl h g^2$ & $\frac12 h g^2$\\
\hline
$\rr/\zz$ & $0$ & $0$ & $0$ & $0$ & $0$\\
\hline
$\rr$ & $0$ & $0$ & $0$ & $\frac12 h g^2$ & $\frac12 h g^2$
\end{tabular}
\;.
\end{equation}
We have omitted $\mmod l$ in the first and second, and $\mmod 1$ in the fourth column.
Let us briefly confirm that the above table entries indeed obey Eq.~\eqref{eq:standard_function_definition}.
In the following derivations we will use the abbreviations
\begin{equation}
\label{eq:circle_carry_cycle}
\omega(a,b)\coloneqq (\ovl{a+b}-\ovl a-\ovl b)\in\zz \text{ for } a,b\in \zz/\rr\;,
\end{equation}
and
\begin{equation}
\label{eq:bockstein_cocycle}
\omega_k(a,b)\coloneqq \frac1k (\ovl{a+b}-\ovl a-\ovl b)\in\zz \text{ for } a,b\in \zz_k\;.
\end{equation}
\begin{itemize}
\item Consider the pairs $(G=\zz_k\text{ odd},A=\zz_l\text{ even})$, $(G=\zz_k\text{ even},A=\zz_l\text{ odd})$, $(G=\zz_k\text{ odd},A=\zz_l\text{ odd})$, $(G=\zz,A=\zz_l\text{ odd})$, $(G=\zz_k\text{ odd},A=\rr/\zz)$, $(G=\zz,A=\rr/\zz)$, $(G=\rr,A=\rr/\zz)$, $(G=\zz,A=\rr)$, and $(G=\rr,A=\rr)$.
In all of these cases, $\homtild^{2s}[G|A]$ is equal to $\zz_k$ odd, or $\rr/\zz$, or $\rr$.
In these cases, for any $h\in \homtild^{2s}[G|A]$, there exists $h/2\in \homtild^{2s}[G|A]$ such that
\begin{equation}
h/2+h/2=h\;.
\end{equation}
Then we can set
\begin{equation}
\label{eq:standard_function_half_choice}
Q_0(h)(g)=\htild^{2s}(h/2)(g,g)\;.
\end{equation}
Eq.~\eqref{eq:standard_function_definition} then holds due to
\begin{equation}
\begin{gathered}
(Q_0(h))^{(2)}(g,g')=\htild^{2s}(h/2)(g+g',g+g')-\htild^{2s}(h/2)(g,g)-\htild^{2s}(h/2)(g',g')\\
= \htild^{2s}(h/2)(g,g')+\htild^{2s}(h/2)(g',g) = \htild^{2s}(h/2+h/2)(g,g')=\htild^{2s}(h)(g,g')\;.
\end{gathered}
\end{equation}
The standard functions in the table in Eq.~\eqref{eq:standard_function_elementary} are indeed of this form.
Note that in the case $(G=\zz,A=\rr/\zz)$ with $\homtild^{2s}[G|A]=\rr/\zz$, there are two possible choices of $h/2$, namely $\frac12 \ovl h$ and $\frac12\ovl h+\frac12$, and we have chosen the former.
\item For $G=\zz_k$ even, $A=\rr/\zz$, we find
\begin{equation}
(Q_0(h))^{(2)}(g,g')
=\frac{\overline{h}}{2k}\big(\overline{g+g'}^2-\overline g^2-\overline g'^2\big)
\overset{\eqref{eq:bockstein_cocycle}}{=}\frac{\overline{h}}{2k}\big(2\overline g\overline g' + (k\omega_k(g,g'))^2\big)
\overset{*}{=}\frac{\overline h}{k} \overline g\overline g'
=\htild^{2s}(h)(g,g')\;,
\end{equation}
where for $*$ we have used the fact that $\frac{k^2}{2k}\in \zz$ for $k$ even.
\item For $G=\zz_k$ even, $A=\zz_l$ even, we find
\begin{equation}
\begin{gathered}
(Q_0(h))^{(2)}(g,g')
=\frac{l}{\gcd(2k,l)} \ovl h (\ovl{(g+g')}^2-\ovl{g}^2-\ovl{g'}^2)
= \frac{l}{\gcd(2k,l)} \ovl h \big((\ovl g + \ovl g' + k \omega_k(g,g'))^2-\ovl{g}^2-\ovl{g'}^2\big) \\
= \frac{l}{\gcd(2k,l)} \ovl h \big(2\ovl g\ovl g' +2k (\ovl g + \ovl g')\omega_k(g,g') + k^2\omega_k(g,g')^2\big) \\
\overset{*}{=} \frac{2l}{\gcd(2k,l)} \ovl h \ovl g\ovl g'
= \frac{l}{\gcd(k,\frac l2)}\ovl h \ovl g\ovl g'
=\htild^{2s}(h)(g,g')\;.
\end{gathered}
\end{equation}
In the equation labeled $*$, we have used that $\frac{2k}{\gcd(2k,l)}$ and $\frac{k^2}{\gcd(2k,l)}$ are integral for even $k$.
\item For $G=\zz$, $A=\zz$, we find
\begin{equation}
(Q_0(h))^{(2)}(g,g')
=\frac12 h (g+g')(g+g'+1) - \frac12 h g(g+1) - \frac12 h g'(g'+1)
=\frac12 h (gg'+g'g)
= h gg'
=\htild^{2s}(h)(g,g')\;.
\end{equation}
\item $G=\zz$, $A=\zz_l$ is obtained from taking $\mmod l$ of the equation above.
\end{itemize}
We can see how the standard quadratic functions appear in the coefficient isomorphism $\htild_2$:
Setting $h_1=0$ and $h_2=h$ in Eq.~\eqref{eq:coefficient_iso_2} yields Eq.~\eqref{eq:standard_function_elementary}.
On the other hand, removing the quadratic part in Eq.~\eqref{eq:coefficient_iso_2} yields the coefficient isomorphism $\htild[G|A]$ in Eq.~\eqref{eq:elementary_homomorphisms} applied to $h=h_1$.
In some cases, the quadratic part in Eq.~\eqref{eq:coefficient_iso_2} depends on both $h_2$ and $h_1$, which is due to the fact that the subset of standard functions $\{Q_0(h)\}_{h\in \homtild^{2s}[G|A]}\subset \hom_2[G|A]$ is not closed under element-wise addition, but generates a larger subgroup of $\hom_2[G|A]$.
$h_2$ takes values in this larger subgroup.

We can make the above a bit more systematic by viewing $\hom_2[G|A]$ as an abelian group extension, corresponding to a short exact sequence
\begin{equation}
\hom[G|A]\xrightarrow{\subset}\hom_2[G|A]\xrightarrow{\bullet^{(2)}} \img(\bullet^{(2)})\subset\hom^2[G,G|A]\;.
\end{equation}
Here, the ``$\subset$'' over the left arrow denotes the natural inclusion of linear functions into quadratic functions, and $\bullet^{(2)}$ is the second derivative mapping a quadratic function to a (symmetric) bilinear form.
This implies that we can write $\hom_2[G|A]$ as a twisted product of the groups on the left and right,
\begin{equation}
\hom_2[G|A] = \img(\bullet^{(2)}) \times_{\Omega} \hom[G|A]\;,
\end{equation}
where $\Omega$ is a function
\begin{equation}
\Omega: \img(\bullet^{(2)})\times \img(\bullet^{(2)})\rightarrow \hom[G|A]\;,
\end{equation}
namely a group 2-cocycle representing a cohomology class in $H^2(B\img(\bullet^{(2)}),\hom[G|A])$.
As a set, $\hom_2[G|A]$ is equal to the cartesian product, but the group multiplication (written additively) is given by
\begin{equation}
(h_2,h_1)+(h_2',h_1') = (h_2+h_2', h_1+h_1'+\Omega(h_2,h_2'))\;.
\end{equation}
Applying the coefficient isomorphism to the above, we obtain
\begin{equation}
\hom_2[G|A] \simeq \homtild^{2s}[G|A] \times_{\omegatild} \homtild[G|A]\;,
\end{equation}
where $\omegatild$ is given by
\begin{equation}
\label{eq:omega_definition}
\omegatild(h_2,h_2') = \htild^{-1} \big(Q_0(\htild^{2s}(h_2+h_2'))-Q_0(\htild^{2s}(h_2))-Q_0(\htild^{2s}(h_2'))\big)\;.
\end{equation}
The following table lists $\omegatild(h,h')$ for different elementary abelian groups $G$ and $A$:
\begin{equation}
\label{eq:addition_cocycle_table}
\begin{tabular}{r|c|c|c|c|c}
\diagbox{$\scriptstyle{G}$}{$\scriptstyle{A}$}
 & $\zz_l$ even & $\zz_l$ odd & $\zz$ & $\rr/\zz$ & $\rr$\\
\hline
$\zz_k$ even & $\frac{\gcd(k,l)}{2}\omega_{\gcd(k,\frac{l}{2})}(h,h')$ & $0$ & $0$ & $\frac{k}{2} \omega_k(h,h')$ & $0$\\
\hline
$\zz_k$ odd & $0$ & $0$ & $0$ & $0$ & $0$\\
\hline
$\zz$ & $0$ & $0$ & $0$ & $\frac12\omega(h,h')$ & $0$\\
\hline
$\rr/\zz$ & $0$ & $0$ & $0$ & $0$ & $0$\\
\hline
$\rr$ & $0$ & $0$ & $0$ & $0$ & $0$
\end{tabular}
\;.
\end{equation}
As usual, $\mmod l$ has been omitted in the first and second column and $\mmod 1$ in the fourth column.
We verify this table below:
\begin{itemize}
\item Consider the pairs $(G=\zz_k\text{ odd},A=\zz_l\text{ even})$, $(G=\zz_k\text{ even},A=\zz_l\text{ odd})$, $(G=\zz_k\text{ odd},A=\zz_l\text{ odd})$, $(G=\zz,A=\zz_l\text{ odd})$, $(G=\zz_k\text{ odd},A=\rr/\zz)$, $(G=\rr,A=\rr/\zz)$, $(G=\zz,A=\rr)$, and $(G=\rr,A=\rr)$.
In these cases we have defined $Q_0$ via Eq.~\eqref{eq:standard_function_half_choice}, and additionally, $\bullet/2$ is a homomorphism, $(h+h')/2=h/2+h'/2$, since there is a unique choice of $h/2$ for each $h$.
Thus, we indeed find
\begin{equation}
\htild(\omegatild(h,h'))=\htild^{2s}((h+h')/2)-\htild^{2s}(h/2)-\htild^{2s}(h'/2)=0\;.
\end{equation}
\item For $G=\zz_k$ even and $A=\rr/\zz$, we find
\begin{equation}
\htild(\omegatild(h,h'))(g) = \frac{1}{2k} (\ovl{h+h'}-\ovl h-\ovl h') \ovl g^2
\overset{\eqref{eq:bockstein_cocycle}}{=} \frac12 \omega_k(h,h') \ovl g^2
\overset{\eqref{eq:square_mod_two_trivial}}{=} \frac12 \omega_k(h,h') \ovl g
\overset{\eqref{eq:elementary_homomorphisms}}{=} \htild(\frac{k}{2} \omega_k(h,h'))(g)\;,
\end{equation}
where we have used that the equation is valued $\mmod 1$.
For $*$ we have used that
\begin{equation}
\label{eq:square_mod_two_trivial}
x\in \zz\quad\Rightarrow\quad \frac12 x^2\mmod 1=\frac12 x\mmod 1.
\end{equation}
\item For $G=\zz$ and $A=\rr/\zz$, we find
\begin{equation}
\htild(\omegatild(h,h'))(g) = \frac{1}{2} (\ovl{h+h'}-\ovl h-\ovl h') g^2
\overset{\eqref{eq:circle_carry_cycle}}{=} \frac12 \omega(h,h') g^2
\overset{\eqref{eq:square_mod_two_trivial}}{=} \frac12 \omega(h,h') g
\overset{\eqref{eq:elementary_homomorphisms}}{=} \htild(\frac12 \omega(h,h'))(g)\;.
\end{equation}
\item For $G=\zz$ and $A=\zz$, we find
\begin{equation}
\htild(\omegatild(h,h'))(g) = \frac12 (h+h'-h-h') g(g+1)=0.
\end{equation}
\item For $G=\zz_k$ even and $A=\zz_l$ even, we find
\begin{equation}
\begin{gathered}
\htild(\omegatild(h,h'))(g)
= \frac{l}{\gcd(2k,l)} (\ovl{h+h'}-\ovl h-\ovl h') \ovl g^2
\overset{\eqref{eq:bockstein_cocycle}}{=} \frac{l}{\gcd(2k,l)} \gcd(k,\frac{l}{2})\omega_{\gcd(k,\frac{l}{2})}(h,h') \ovl g^2\\
=\frac{l}{2}\omega_{\gcd(k,\frac{l}{2})}(h,h') \ovl g^2
\overset{\eqref{eq:square_mod_two_trivial}}{=}\frac{l}{2}\omega_{\gcd(k,\frac{l}{2})}(h,h') \ovl g
=\frac{l}{\gcd(k,l)}\frac{\gcd(k,l)}{2}\omega_{\gcd(k,\frac{l}{2})}(h,h') \ovl g\\
\overset{\eqref{eq:elementary_homomorphisms}}{=}\htild(\frac{\gcd(k,l)}{2}\omega_{\gcd(k,\frac{l}{2})}(h,h'))(g)\;.
\end{gathered}
\end{equation}
\end{itemize}
The cohomology class of $\omegatild$ determines whether the group extension is a direct product or not.
If $\omegatild=0$, then $\homtild_2[G|A]$ in Eq.~\eqref{eq:hom2_groups} is equal to the direct product $\homtild^{2s}[G|A]\times\homtild[G|A]$.
If $\omegatild\neq 0$, then the cocycle twist mixes up the group multiplications of $\homtild^{2s}[G|A]$ and $\homtild[G|A]$, generally by moving a $\zz_2$ subgroup from $\homtild[G|A]$ to $\homtild^{2s}[G|A]$.

\end{document}